\documentclass[a4paper,11pt]{article}
\usepackage{wrapfig}
\usepackage{jcappub}
\usepackage{hyperref} 
\usepackage[T1]{fontenc}
\usepackage{ulem}
\usepackage{adjustbox} 
\usepackage{scalerel} 
\usepackage{slashed}
\usepackage{mathtools}
\usepackage{amsmath}
\usepackage{environ}
\usepackage{comment}
\usepackage{feynmp-auto}

\usepackage{xspace}
\newcommand*{\GW}{\ensuremath{\mathrm{GW}}\xspace}

\newcommand*{\DBPL}{\ensuremath{\mathrm{DBPL}}\xspace}

\newcommand{\U}[1]{\mathrm{U}(1)_{\mathrm{#1}}}			
\newcommand{\SU}[2]{\mathrm{SU}(#1)_{\mathrm{#2}}}		
\newcommand{\g}[1]{g_{\mathrm{#1}}}

\usepackage{physics}
\usepackage{tensor}
\usepackage[caption=false]{subfig}
\usepackage{graphicx}
\usepackage{caption}
\usepackage{bm}

\usepackage{graphicx}
\usepackage{wrapfig}
\usepackage{lipsum}

\usepackage{array}
\usepackage{hhline}
\usepackage{booktabs}   
\usepackage{makecell} 

\usepackage{cancel}

\usepackage{multirow}

\usepackage[dvipsnames]{xcolor}

\definecolor{ForestGreen}{rgb}{0.13, 0.55, 0.13}
\definecolor{Mulberry}{rgb}{0.77, 0.29, 0.55}
\definecolor{bostonuniversityred}{rgb}{0.8, 0.0, 0.0}
\definecolor{amber}{rgb}{1.0, 0.49, 0.0}

\newcommand{\SUD}[0]{\ensuremath{\mathrm{SU(2)_D}}}
\newcommand{\Lag}[0]{\ensuremath{\mathcal{L}}}

\newcommand{\RPrep}[2]

\usepackage[capitalise]{cleveref}

\crefname{section}{Sec.}{Secs.}
\crefname{table}{Tab.}{Tabs.}
\crefname{figure}{Fig.}{Figs.}
\crefname{equation}{Eq.}{Eqs.}
\crefname{appendix}{Appendix\ }{Appendix\ }

\newcommand{\Z}{{\mathbb Z}}

\newcommand{\com}[1]{\iffalse #1 \fi}

\title{\boldmath Gravitational Waves from Dark Gauge Sectors}

\author[a]{Alexander Belyaev,}
\author[b]{M\r{a}rten Bertenstam,}
\author[c,d,1]{Jo\~{a}o Gon\c{c}alves,\note{Corresponding author.}}
\author[d,e]{Ant\'{o}nio P.~Morais,}
\author[b]{Roman Pasechnik}
\author[a,f]{and Nakorn Thongyoi}

\affiliation[a]{School of Physics and Astronomy, University of Southampton, Highfield, Southampton SO17 1BJ, UK}
\affiliation[b]{Department of Physics, Lund University, SE-223 62 Lund, Sweden}
\affiliation[c]{Departamento de F\'{i}sica da Universidade de Aveiro and Centre  for  Research  and  Development  in  Mathematics  and  Applications  (CIDMA), Campus de Santiago, 3810-183 Aveiro, Portugal.}
\affiliation[d]{Laborat\'{o}rio de Instrumenta\c{c}\~{a}o e F\'{i}sica Experimental de Part\'{i}culas (LIP), Universidade do Minho, 4710-057 Braga, Portugal.}
\affiliation[e]{Departamento de F\'{i}sica, Escola de Ci\^{e}ncias, Universidade do Minho, 4710-057 Braga, Portugal}
\affiliation[f]{Khon Kaen Particle Physics and Cosmology Theory Group (KKPaCT),\\
Department of Physics, Faculty of Science, Khon Kaen University,\\
123 Mitraphap Rd, Khon Kaen 40002, Thailand}

\emailAdd{a.belyaev@soton.ac.uk}
\emailAdd{marten.bertenstam@fysik.lu.se}
\emailAdd{jpedropino@ua.pt}
\emailAdd{amorais@fisica.uminho.pt}
\emailAdd{roman.pasechnik@fysik.lu.se}
\emailAdd{nakorn.thongyoi@gmail.com}

\abstract{
We explore gravitational-wave (GW) signatures from a strong first-order phase transition in a non-Abelian dark sector, which naturally gives rise to vector dark matter (DM). We consider a general class of models featuring a new dark gauge sector communicating with the Standard Model (SM) through a Higgs portal and a vector-like fermionic portal. We also study the scenario where the dark sector interacts with the SM only via gravity. In all cases, we scan the full parameter space and analyse GW production and highlight the regions with visible GW signatures. Notably, the fermionic portal yields distinctive GW signals at LISA with peak frequencies of 1--10 mHz, reaching up to 1 Hz for future interferometers like BBO and DECIGO, while the Higgs portal scenario remains limited to around 1 mHz. Both frameworks account for the observed DM abundance and predict detectable LISA signals for dark vector bosons near 1--4 TeV, with a $\sim$10 GeV dark Higgs. Finally, we identify a unique six-top final state from pair-produced vector-like fermions, offering a striking collider signature within HL-LHC reach. Its detection would provide a smoking-gun signal for the fermionic portal, establishing complementarity between collider, GW, and DM signals.
}

\begin{document} 
\maketitle
\flushbottom
\newpage

%
\section{Introduction}
\label{sec:intro}

The observed asymmetry between matter and antimatter in the Universe, along with the presence of an unknown component commonly referred to as dark matter (DM), which constitutes about 85\% of the total matter in the Universe, are among the most significant unresolved problems in modern cosmology. Among the possible mechanisms to explain the former, electroweak (EW) baryogenesis stands out as one of the most promising solutions \cite{Canetti_2012,Elor:2022hpa,Farrar:1993hn,Shaposhnikov:1986jp}. A key element of this mechanism, \textit{i.e.} a departure from thermal equilibrium \cite{Sakharov:1967dj}, can be achieved through a strong first-order phase transition (FOPT) that may have occurred in the early post-inflationary universe.

It is well known that the Standard Model (SM) of particle physics lacks the necessary components to explain the baryon asymmetry of the Universe (BAU). Specifically, both the EW and Quantum Chromodynamics (QCD) phase transitions are smooth crossovers \cite{Kajantie:1996mn,Karsch:1996yh,DOnofrio:2015gop,Fodor:2004nz}, making them unsuitable for accounting for the observed BAU. This inadequacy is due to the measured mass of the SM Higgs boson being too high for an EW FOPT to occur. Many beyond the SM (BSM) frameworks feature strong FOPTs that can successfully explain the observed BAU. Furthermore, FOPTs can potentially lead to the generation of a stochastic gravitational wave background (SGWB). Recently, pulsar timing arrays (PTAs) have reported evidence for the existence of such a signal \cite{NANOGrav:2023gor,NANOGrav:2023hvm,EPTA:2023fyk,Reardon:2023gzh,Xu:2023wog} in the nHz frequencies, although current data is still insufficient to fully determine its origin. If FOPTs were responsible for the gravitational wave (GW) signal, the transition scale would likely occur at or near the QCD scale, $T \sim 0.17~\mathrm{GeV}$ \cite{Fujikura:2023lkn, Bringmann:2023opz, Borah:2021ftr}. Alternatively, if the phase transition is driven by heavier fields, placing the scale around $100~\mathrm{GeV}$ to a few TeV, then these transitions could produce an SGWB at much higher frequencies than those probed by NANOGrav. Consequently, other experiments, such as LISA \cite{LISA:2017pwj}, DECIGO \cite{Kawamura:2006up}, and BBO \cite{Harry:2006fi}, may potentially detect them. Among these experiments, LISA has been approved by the European Space Agency and is scheduled for launch in 2035.

Recently, a novel framework for Vector DM has been proposed, where communication between the dark and SM sectors is achieved via a portal of vector-like (VL) fermions~\cite{Belyaev:2022shr,Belyaev:2022zjx}. This fermionic portal to vector DM (FPVDM) has significant theoretical and phenomenological implications for the DM sector. The dark sector of the FPVDM model introduces a new non-Abelian gauge group, $\rm{SU}(2)_{\rm{D}} \times \rm{U}(1)_{Y_\mathrm{D}}^{\text{Global}}$, and a complex dark doublet scalar, $\Phi_{\rm{D}}$. When $\Phi_{\rm{D}}$ acquires a vacuum expectation value (vev), the symmetry group $\rm{SU}(2)_{\rm{D}} \times \rm{U}(1)_{Y_\mathrm{D}}^{\text{Global}}$ is spontaneously broken to $\mathrm{U(1)}_{Q_\mathrm{D}}^{\text{Global}}$, leading to the existence of a conserved dark charge, $Q_\mathrm{D} = Y_\mathrm{D} + I_\mathrm{3D}$, where $Y_\mathrm{D}$ is the dark hypercharge and $I_\mathrm{3D}$ is the dark isospin. In this framework, the new vector bosons, $V_{\rm{D}}$, serve as DM candidates stabilised by $Q_\mathrm{D}$.

The connection between the SM and dark sectors is established through a Yukawa interaction between SM right-handed weak-singlet fermions ($f_{\rm{R}}^{\rm SM}$), the dark doublet of VL fermions ($\Psi$) and the dark scalar doublet ($\Phi_{\rm{D}}$). Therefore, the FPVDM model does not rely on the conventional Higgs portal to explain DM. In a previous study \cite{Belyaev:2022shr,Belyaev:2022zjx}, the authors examined a scenario where the VL fermion doublet couples to the right-handed top quark, and they comprehensively explored both the cosmological and collider phenomenology of this setup.

In this work, we analyse the impact of FOPTs in the early universe originating from the dark sector of the FPVDM model and demonstrate how they can potentially lead to visible signatures in the form of an observable SGWB. To consistently assess the influence of fermions on the phase transitions, we explore three distinct scenarios. Scenario I consists of a pure dark sector with only gauge and Higgs fields. Scenario II includes the SM, dark gauge, and Higgs sectors, as well as the Higgs portal (and only the Higgs portal) between them. Scenario III builds on Scenario II by incorporating the VL fermionic portal. This approach allows us to understand how each component of the model affects the SGWB spectrum.

To construct an effective thermal potential and reduce substantial theoretical uncertainties arising from the renormalisation scale and gauge dependence \cite{Gould:2021oba,Croon:2020cgk,Carena:2019une,Kainulainen:2019kyp,Guo:2021qcq},  we employ the dimensional reduction (DR) method, which results in a 3D effective field theory (EFT). Using this approach the uncertainties mentioned above are greatly reduced. One should also note that recent works \cite{Gould:2023ovu,Ekstedt:2024etx} have shown that 2- and 3-loop matched 3D EFTs provide a good fit to non-perturbative lattice calculations.

The paper is organised as follows. In \cref{sec:fopt_gws}, we provide an overview of the current state of the art in the physics of phase transitions and GWs. In \cref{sec:model}, we discuss the theoretical structure of the FPVDM with a focus on the scalar potential and the dimensionally reduced thermal theory framework. In \cref{sec:scan} we set up the scan strategy, phase transition framework and discuss DM Constraints.
Then in \cref{sec:results}, we present the numerical results for all scenarios and discuss them. Finally, in \cref{sec:conclusion} we summarise our work and discuss its future implications.

\section{Gravitational waves from cosmological phase transitions}\label{sec:fopt_gws}

\subsection{Dynamics of first-order phase transitions}\label{subsec:FOPT}

As the early Universe expands and cools down, the initially symmetric potential undergoes thermal evolution leading to the generation of different vacua that might trigger spontaneous symmetry breaking.  Consequently, phase transitions may occur between local (false vacuum) and global (true vacuum) minima. These transitions can be classified either as second-order transitions, smooth crossovers or FOPTs. The latter is particularly relevant in the context of GWs \cite{Kosowsky:1992rz} and in addressing the problem of BAU \cite{Bochkarev:1990fx,Cohen:1990py,Cohen:1990it} if a FOPT occurs at the EW scale.

The dynamics of FOPTs is well understood. At low-temperatures, they are primarily driven by quantum tunnelling between different vacua, while for high temperatures they are dominated by thermal fluctuations. The starting point is the 3-dimensional (3D) Euclidean action \cite{Linde:1981zj,Coleman:1977py}
\begin{equation}\label{eqn:euclidean_action}
    S_3(T) = 4\pi \int_0^\infty dr~r^2 \qty[\frac{1}{2} \qty(\frac{d\hat{\phi}}{dr})^2 + V_\mathrm{eff}(\hat{\phi},T)]\,,
\end{equation}
where $\hat{\phi}$ is the bounce solution that minimises the action, and $V_\mathrm{eff}$ corresponds to the thermal effective potential of the model, whose details are described in \cref{subsec:VEff} and \cref{sec:DR_eff}.

Given the total $\mathrm{O(3)}$ symmetry of the action, the equation of motion for $\hat{\phi}$ reads
\begin{equation}\label{eqn:bounce_eq}
    \frac{d^2\hat{\phi}}{dr^2} + \frac{2}{r}\frac{d\hat{\phi}}{dr} - \frac{\partial V_\mathrm{eff}(\hat{\phi},T)}{\partial \hat{\phi}} = 0\,.
\end{equation}
In general, analytical solutions are not available, so one must rely on numerical methods. For this work, we used the \texttt{Python} package \texttt{CosmoTransitions}\footnote{Specifically, we utilised a modified version of \texttt{CosmoTransitions} in the soon-to-be-released \texttt{Dratopi} \cite{Dratopi} package. This is a tool to interface the \texttt{Mathematica} package \texttt{DRAlgo} \cite{Ekstedt:2022bff} with \texttt{Python} and a modified version of \texttt{CosmoTransitions}, for phase transition analysis in the DR formalism. Some further details are given in \cref{subsec:VEff} below.} to numerically find bounce solutions \cite{Wainwright:2011kj}.

The total decay rate between the true and false vacuum at a given finite temperature $T$ is calculated as follows \cite{Linde:1981zj}:
\begin{equation}\label{eqn:decay_rate}
\Gamma(T) = T \qty[\frac{\det^\prime[-\nabla^2 + \partial^2_{\hat{\phi}} V(\hat{\phi},T)]}{\det[-\nabla^2 + \partial^2_{\hat{\phi}} V(\phi_F,T)]}]^{-1/2} \qty(\frac{S_3(T)}{2\pi T})^{3/2} \exp{-S_3(T)/T} \, ,
\end{equation}
where $\det^\prime$ indicates that the zero-modes are not included in the calculation of the determinant and $\phi_F$ is the false vacuum. The determinant pre-factor can be simplified by noting that it scales with $T^3$, allowing us to write \cite{Linde:1981zj}:
\begin{equation}\label{eqn:decay_rate_1}
\Gamma(T) \approx T^4\qty(\frac{S_3(T)}{2\pi T})^{3/2} \exp{-S_3(T)/T} \, ,
\end{equation}
which is implemented in our numerical analysis\footnote{We note, however, that there is a publicly available code \cite{Ekstedt:2023sqc} tailored to numerically calculate this pre-factor.}.

For a complete description of phase transitions, various temperatures are considered corresponding to different intermediate stages of the cosmological evolution. As noted above, at high temperatures, the Universe is fully symmetric, with a single vacuum. As the temperature decreases and the symmetry gets broken, various degenerate minima begin to emerge in the potential, with a potential barrier forming between them. We define the temperature for which this happens as the critical temperature of a given transition, $T_c$. As the temperature continues to drop, some of the minima are no longer degenerate. At this stage, we introduce the nucleation temperature $T_n$ for which thermal fluctuations have become large enough that at least one bubble of the true vacuum nucleates per cosmological horizon. We can interpret this as the temperature where the phase transition effectively begins. With this, $T_n$ is defined as the solution of
\begin{equation}\label{eqn:nucletion_temperature}
\int_{T_n}^{T_c}~\frac{dT}{T} \frac{\Gamma(T)}{H(T)^4} = 1 \, ,
\end{equation}
where $H(T)$ is the Hubble rate at a temperature $T$. Taking both the radiation and vacuum energy densities into account, the Hubble rate can be determined as 
\begin{equation}\label{eqn:hubble_rate}
    H(T) = \frac{1}{\sqrt{3}\; \overline{M}_{\mathrm{PL}}}\sqrt{ \frac{T^4}{\xi_g^2} + \Delta V_\mathrm{eff}(T)}\,,
\end{equation}
where $\overline{M}_{\mathrm{PL}} \approx 2.4 \times 10^{18}~\mathrm{GeV}$ is the reduced Planck mass, and $\Delta V_\mathrm{eff}(T)$ is the potential energy difference between the true and false vacua at a given temperature $T$. Here, $\xi_g = \sqrt{30/[\pi^2 g^*(T)]}$, with $g^*(T)$ representing the number of relativistic degrees of freedom at temperature $T$. Following previous literature, we treat the number of relativistic degrees of freedom as a constant with $g^* \approx 100$ at $T \sim 100$~GeV. In the absence of supercooled transitions, the contribution from $\Delta V_\mathrm{eff}(T)$ can be safely neglected.

In the literature, various approximations exist that help simplify the calculation of the nucleation temperature \cite{Huber:2007vva,JoseRedondo}, with the most commonly used condition $S_3(T_n)/T_n \sim 140$ suitable for EW transitions (for a discussion on its range of applicability, see e.g.~Ref.~\cite{Athron:2022mmm}), which corresponds to the default setting in \texttt{CosmoTransitions} \cite{Wainwright:2011kj}. For the majority of the parameter space analysed here, this approximation holds since no supercooling is expected for most cases. In the supercooled case, however, one needs to use \cref{eqn:nucletion_temperature} to consistently obtain the nucleation temperature.

The percolation temperature, $T_p$, is defined when the true vacuum bubbles become causally connected, preventing the Universe from reverting to the false vacuum phase. Some authors use the nucleation temperature as the reference scale for calculating GW observables instead of the percolation temperature. However, as noted in \cite{Athron:2023rfq, Athron:2022mmm}, the choice of temperature scale in these calculations affects the predicted GW spectrum's amplitude and frequency. For weakly supercooled transitions, this variation is typically only a few percent, but for strongly supercooled ones, it can reach up to an order of magnitude. Although the scenarios considered in this article are closer to the former, we use the percolation temperature in our numerical analysis.

Quantitatively, the percolation temperature, $T_p$, is defined as the point at which $34\%$ of the false vacuum has transitioned to the true phase \cite{Stauffer_Aharony_2014}. At this stage, the probability of finding a region still in the false vacuum can be described by an exponential probability distribution, $\mathcal{P}(T) = e^{-I(T)}$, where $I(T)$ represents the volume of true vacuum per unit of comoving volume and is given by \cite{Guth:1981uk}
\begin{equation}\label{eqn:It_percolation}
    I(T) = \frac{4\pi v^3_w}{3} \int_T^{T_c} dT^\prime \frac{\Gamma(T^\prime)}{T^{\prime 4} H(T^\prime)} \qty(\int_T^{T^\prime} \frac{d\tilde{T}}{H(\tilde{T})})^3\,,
\end{equation}
where $v_w$ is the bubble wall velocity. This expression assumes that the Universe expands adiabatically in a cosmology following the Friedmann–Lemaître–Robertson–Walker (FLRW) background metric \cite{Guth:1981uk}. Percolation is then defined when $I(T) = 0.34$, or equivalently when $\mathcal{P}(T) = 0.71$.

The characteristics and dynamics of FOPTs, and consequently the GW spectrum, are uniquely defined by four distinct thermodynamic parameters:
\begin{equation}\label{eqn:parameters}
    T_p\,, \quad v_w\,, \quad \alpha\,, \quad \beta/H(T_p)\,.
\end{equation}
The first two parameters, $T_p$ (the percolation temperature) and $v_w$ (the bubble wall velocity), have been introduced above. The remaining parameters are $\alpha$, a dimensionless measure of the strength of the phase transition, and $\beta/H(T_p)$, which represents the inverse timescale of the transition normalised to the Hubble parameter $H(T_p)$ at the time of percolation. Each of these parameters can introduce uncertainties in the predicted GW spectrum. While most of these parameters are well-defined theoretically, they all depend (either directly or indirectly) on the thermal effective potential. As shown in previous studies \cite{Gould:2021oba,Croon:2020cgk,Athron:2022jyi,Martin:2001vx,McKeon:2015zxa}, the primary source of uncertainty arises from the choice of renormalisation scale. Although the existence of a FOPT is a robust prediction in the sense that changes in the renormalisation scale do not eliminate the transition, different choices of this scale can significantly affect GW predictions, potentially by orders of magnitude. This effect is particularly pronounced for the parameter $\alpha$, which depends on the difference in potential energy between the true and false vacua and is inversely proportional to the fourth power of the temperature. Moreover, for $\alpha < 1$, the SGWB amplitude scales approximately with the square of $\alpha$, further amplifying these theoretical uncertainties. A robust and theoretically consistent analysis can be achieved through DR, which we adopt in this work.

The percolation temperature, $T_p$, is well-defined from a theoretical standpoint and is expected to be relatively unaffected by uncertainties, although small variations can significantly impact the SGWB. Its determination is intrinsically linked to the bubble wall velocity, $v_w$. A recent study \cite{Ai:2023see} provided model-independent analytical expressions for $v_w$. Depending on the type of fluid motion—deflagration, detonation, or hybrids—different expressions and approximations apply. For more details, we refer the reader to \cite{Ai:2023see}.

For the parameter space of interest in this article, \textit{i.e.}, within the sensitivity reach of GW interferometers, we approach the supercooling region where, alongside $\alpha \gtrsim 1$, the wall velocity tends towards unity. We verified that the code snippet provided in \cite{Ai:2023see} correctly reproduces this behaviour. However, for weaker transitions, the code did not converge and consistently outputs a warning that the phase transition strength, $\alpha$, was too small. Nonetheless, such a parameter space region is beyond our interest here as it falls outside the planned experimental sensitivities. Furthermore, a recent work by some of the authors \cite{Addazi:2023ftv} noted that for a Majoron model, variations in the bubble wall velocity, ranging from $v_w \sim 0.45$ to $v_w \sim 0.95$, did not result in significant changes for SGWB predictions. Additionally, we refer the reader to a recent study \cite{Krajewski:2024gma} that performed real-time hydrodynamical lattice simulations. Their analysis revealed that most solutions tended towards a runaway scenario (\textit{i.e.}, with $v_w$ approaching the speed of light, $v_w \simeq 1$). Scenarios leading to a steady-state solution, where the analytical results from \cite{Ai:2023see} align with \cite{Krajewski:2024gma}, were found to be rare and required fine-tuning of the nucleation temperature. Considering this, we will treat the bubble wall velocity as a free parameter and subsequently analyse its impact on the predictions for the GW spectrum.

The strength of the phase transition, $\alpha$, is determined by the vacuum free energy difference, which is proportional to the potential energy difference, and the entropy change, which is proportional to the temperature variation of the free energy. This relationship is expressed as \cite{Hindmarsh:2017gnf,Hindmarsh:2015qta}:
\begin{equation}\label{eqn:alpha_param}
\alpha = \frac{\Delta V_\mathrm{eff}}{\rho_R}\Biggr|_{\substack{T = T_p}} - \frac{T}{\rho_R} \frac{\partial \Delta V_\mathrm{eff}}{\partial T}\Biggr|_{\substack{T = T_p}}\,,
\end{equation}
where $\rho_R$ represents the radiation energy density, defined as $\rho_R =T_p^4/\xi_g^2$.

The inverse duration of the phase transition, expressed in units of the Hubble rate at percolation, $\beta/H(T_p)$, is given by:
\begin{equation}
\frac{\beta}{H(T_p)} = T\frac{d}{dT} \qty(\frac{\hat{S}_3}{T}) \Biggr|_{\substack{T = T_p}}\,.
\end{equation}
The main source of uncertainty in this calculation stems from determining the Euclidean action, which may be influenced by numerical artifacts. To reduce the impact of these uncertainties, we employed a smoothing process using the spline interpolation of the action as implemented in \texttt{CosmoTransitions}.

\subsection{Stochastic gravitational wave spectrum}\label{subsec:pGWs}

Phase transitions in the early Universe are dynamic events that occurred during the post-inflationary era. If these transitions are strong enough, they can leave detectable imprints in the form of a SGWB. Due to their cosmological origins, the resulting signals are anticipated to be isotropic, stationary and unpolarised \cite{Maggiore:1999vm}. In this section, we provide an overview of the key concepts related to the derivation of the GW power spectrum as presented in the literature. 

Typically, the SGWB is derived from the transverse and traceless (TT) components of the metric perturbations. Using the FLRW metric \cite{Figueroa:2012kw,Caprini:2007xq}, we have
\begin{equation}\label{eqn:FLRW}
    ds^2 = a^2(\tau)\qty[ \eta_{\mu\nu} + h_{\mu\nu}] dx^\mu dx^\nu\,,
\end{equation}
where $a$ is the scale factor, $h_{\mu\nu}$ is the metric perturbation, and $\tau$ is the conformal time. Expanding the Einstein equations to first order in the TT part of $h_{\mu\nu}$, one obtains \cite{Figueroa:2012kw}
\begin{equation}
    \ddot{\bar{h}}^{\mathrm{TT}}_{ij} - \qty( \nabla^2 + \frac{\ddot{a}(\tau)}{a(\tau)} )\bar{h}^{\mathrm{TT}}_{ij} = \frac{16 \pi a(\tau)}{M^2_{\mathrm{PL}}} \Pi^{\mathrm{TT}}_{ij}\,,
\end{equation}
where $\bar{h}^{\mathrm{TT}}_{ij} = a h^{\mathrm{TT}}_{ij}$, and $\Pi^{\mathrm{TT}}_{ij}$ is the TT component of the spatial part of the energy-momentum tensor $T_{ij}$. During FOPTs, GWs are sourced by scalar fields $\Phi$, which are assumed to behave like a perfect fluid \cite{Athron:2023xlk}
\begin{equation}
    T_{\mu\nu} = (\partial_\mu\Phi)(\partial_\nu \Phi) - \frac{1}{2} \eta_{\mu\nu} (\partial_\alpha \Phi)^2 + (\rho + p)v_\mu v_\nu - \eta_{\mu\nu} p\,,
\end{equation}
where $\rho$ is the energy density, $p$ is the pressure, and $v_\mu$ is the four-velocity in the bubble center's reference frame. The GW power spectrum can then be calculated as \cite{Figueroa:2012kw,Caprini:2007xq}
\begin{equation}
    h^2 \Omega_{\mathrm{GW}} \equiv \frac{h^2}{\rho_c} \frac{d \rho_{\mathrm{GW}}}{d ~\mathrm{ln} f} = \frac{h^2}{\rho_c} \frac{M^2_{\mathrm{PL}} k^3 \abs{\dot{h}_k(\tau)}^2}{64\pi^3 a^2(\tau)}\,,
\end{equation}
where $\abs{\dot{h}_k (\tau)}$ is the power spectrum of $\dot{h}^{\mathrm{TT}}_{ij}$, and $\rho_c = 3H_0^2/(8\pi G)$ is the critical density today. Here, $H_0 = 100h~\mathrm{km/s/Mpc}$ is the Hubble parameter at present, and $G$ is the gravitational constant. The GW spectra can then be estimated by determining $\abs{\dot{h}_k(\tau)}$, which intricately depends on the dynamics of the scalar fields as described by $T_{\mu\nu}$. While analytical solutions can be derived in specific cases, such as with topological defects \cite{Figueroa:2012kw}, in the case of FOPTs, one typically relies on fits made to numerical simulations.

In general, three contributions affect the GW spectrum: sound waves \cite{Hindmarsh:2013xza,Hindmarsh:2015qta,Hindmarsh:2017gnf}, bubble wall collisions \cite{Kosowsky:1991ua,Kosowsky:1992vn,Cutting:2018tjt,Kamionkowski:1993fg,Lewicki:2020azd}, and magnetohydrodynamic turbulence in the plasma \cite{Kamionkowski:1993fg,Caprini:2009yp,RoperPol:2019wvy,Kahniashvili:2020jgm,RoperPol:2021xnd,Auclair:2022jod}. 
First, we exclude the effects of turbulence from our analysis, as it remains the greatest source of uncertainty in predicting primordial GW spectra. This uncertainty stems in part from the limited understanding of the efficiency factor, $\kappa_{\mathrm{turb}}$, which describes the fraction of vacuum energy converted into turbulence \cite{Athron:2023xlk}. Some studies in the literature assume $\kappa_{\mathrm{turb}} = \varepsilon \kappa_{\mathrm{SW}}$, where $\kappa_{\mathrm{SW}}$ is the efficiency factor for sound wave contributions, and $\varepsilon$ ranges between $1\%$ and $10\%$ \cite{Caprini:2015zlo,Azatov:2019png,Alves:2018jsw}. Based on these values, the turbulence contribution is expected to be consistently subdominant compared to the sound wave contribution. Additionally, we do not consider the bubble wall collision component, as it is typically dominant only in scenarios involving strong supercooling \cite{Kierkla:2022odc}, which we do not anticipate in our case. Nonetheless, as a sanity check for benchmarks with $\alpha > 1$, we estimated the peak frequency and amplitude using the formulas from \cite{Lewicki:2020azd} and confirmed that this contribution is indeed subdominant compared to that of sound waves. Therefore, in the remainder of this article, we focus exclusively on the sound wave contribution to the primordial GW spectra. 

The sound wave contribution is described by the following double-broken power law \cite{Caprini:2024hue}:
\begin{align}
\Omega_{\GW}^{\DBPL}(f, \Omega_2, f_1, f_2) &= \Omega_\text{int} \times S(f) \,, \label{eq:DBPL}
\\
S(f) &= N \left( \frac{f}{f_1} \right)^{n_1}
\left[
1 + \left( \frac{f}{f_1} \right)^{a_1}
\right]^{\frac{-n_1 + n_2}{a_1}}
\left[
1 + \left( \frac{f}{f_2} \right)^{a_2}
\right]^{\frac{-n_2 + n_3}{a_2}} \,. \nonumber
\end{align}
The fitting parameters are $n_1 = 3$, $n_2 = 1$, $n_3 = -3$, $a_1 = 2$, and $a_2 = 4$. The normalisation factor $N$ is determined by ensuring that $\int_{-\infty}^{+\infty} S(f) d(\ln f) = 1$. The geometric frequencies $f_1$ and $f_2$ are calculated as follows
\begin{align}
f_1 &\simeq 0.2 \, H_{*,0} \, (H(T_p) R_*)^{-1} \,, \label{eq:geom_soundwave}
\\
f_2 &\simeq 0.5 \, H_{*,0} \, \Delta_w^{-1} (H(T_p) R_*)^{-1} \,, \label{eq:sw_shape}
\end{align}
where $\Delta_w = v_\mathrm{shell}/\mathrm{max}(v_w, c_s)$, with the speed of sound $c_s=1/\sqrt{3}$, and $v_\mathrm{shell} = |v_w - c_s|$ represents the dimensionless thickness of the sound shell. The bubble radius $R_*$ is related to $\beta/H(T_p)$ through the relation
\begin{equation}
\frac{\beta}{H(T_p)} = (8\pi)^{1/3} \frac{\mathrm{max}(v_w, c_s)}{H(T_p) R_*} \,.
\end{equation}
The integrated amplitude $\Omega_{\rm int}$ is defined by \cite{Jinno:2022mie}
\begin{align}
h^2 \Omega_\mathrm{int} &= 0.11 h^2 F_{\GW,0} \, K^2 \left( H(T_p) \tau_{\rm SW} \right) \left( H(T_p) R_* \right) \,, \label{eq:sw_amplitude}
\end{align}
where $H(T_p) \tau_{\mathrm{SW}} = \mathrm{min} \left( \frac{2H(T_p) R_*}{\sqrt{3K}}, 1 \right)$ denotes the lifetime of sound waves in units of Hubble time. The parameter $K = 0.6 \kappa_{\mathrm{SW}} \alpha / (1 + \alpha)$ represents the fraction of kinetic energy converted into sound waves. 
The parameters $F_{\GW,0}$ and $H_{*,0}$ accounting for redshift read as follows:
\begin{equation}\label{eq:redshit_H_FGW0}
    \begin{aligned}
        &H_{*,0} \simeq 1.65\times 10^{-5}~\mathrm{Hz}\,\left(\frac{g_*(T_p)}{100}\right)^{1/6} \left(\frac{T_p}{\mathrm{GeV}}\right)  \,, \\
        &h^2 \, F_{\GW,0}  \simeq 1.65\times 10^{-5} \left(\frac{100}{g_{*}(T_p)}\right)^{1/3}\,.
    \end{aligned}
\end{equation}
The efficiency factor $\kappa_{\mathrm{SW}}$ is estimated based on the formalism of \cite{Espinosa:2010hh}, which is summarised in \cref{app:efficiency_factor}.

\section{Scenarios of vector dark matter models}\label{sec:model}

In this section, we discuss various scenarios for a dark gauge sector with a central focus on the \textit{Fermionic Portal to Vector Dark Matter (FPVDM)} framework. FPVDM  introduces a class of models where the DM candidates are massive gauge bosons associated with a non-Abelian symmetry group, $\mathrm{SU(2)_D}$. These gauge bosons acquire mass through a spontaneous symmetry breaking mechanism in the dark sector, mediated by a scalar doublet $\Phi_D$. The interaction between the dark sector and the SM occurs via new fermions that transform non-trivially under $\mathrm{SU(2)_D} \times \mathrm{U(1)_Y}$. Unlike models requiring a substantial quartic coupling between the Higgs and the dark doublets at tree level, FPVDM leverages a global $\mathrm{U(1)_D}$ symmetry, ensuring DM stability---a consequence of the pseudo-real nature of the fundamental representation of $\mathrm{SU(2)}$. In the absence of these fermions, the Higgs portal would become the sole interaction channel between the dark and SM sectors, with DM stability instead ensured by custodial symmetry within the scalar sector~\cite{Hambye:2008bq}. 
The symmetry breaking pattern is $\mathrm{SU(2)_D}\times \mathrm{U(1)_D} \to \mathrm{U(1)_D^d}$. With the $\mathrm{U(1)_D}$ phase assignments $Y_\mathrm{D}=\frac{1}{2}$ for dark scalar and fermion doublets, while $Y_\mathrm{D}=0$ for vector triplet, there is still an invariance under the subgroup $\Z_2 \equiv (-1)^{Q_\mathrm{D}}$, where $Q_\mathrm{D}=T^3_\mathrm{D}+Y_\mathrm{D}$.
The summary of the quantum numbers for the particles is given in \cref{tab:particlesQN}.
\begin{table}[htbp]
\centering
\begin{tabular}{c|cc|c||c|r}
\hline
&&&&\\[-8pt]
 & $\mathrm{SU(2)_L}$ & $\mathrm{U(1)_Y}$ & $\SUD$ & $\Z_2$ & $Q_\mathrm{D}$\\
&&&&&\\[-9pt]
\hline
&&&&\\[-8pt]
$\Phi_\mathrm{D}=\left(\begin{array}{c} \varphi^0_{\mathrm{D}+ \frac{1}{2} } \\ \varphi^0_{\mathrm{D}-\frac{1}{2} } \end{array}\right)$ & $\mathbf{1}$ & $0$ & $\mathbf{2}$ 
& $\begin{array}{c} - \\ + \end{array}$ 
& $\begin{array}{r} +1 \\ 0 \end{array}$ 
\\[10pt]
\hline
&&&&\\[-8pt]
\multirow{2}{*}{$\Psi=\left(\begin{array}{c} \psi_{\mathrm{D}} \\ \psi \end{array}\right)$} & \multirow{2}{*}{$\mathbf{1}$} & \multirow{2}{*}{$Q$} & \multirow{2}{*}{$\mathbf{2}$} & $-$ & $+1$\\
& & & & $+$ & $0$\\[2pt]
\hline
&&&&\\[-8pt]
$V^\mathrm{D}_{\mu}=\left(\begin{array}{c} V^0_{\mathrm{D}+\mu} \\ V^0_{\mathrm{D}0\mu} \\ V^0_{\mathrm{D}-\mu} \end{array}\right)$ & $\mathbf{1}$ & $0$ & $\mathbf{3}$ 
& $\begin{array}{c} - \\ + \\ - \end{array}$ 
& $\begin{array}{r} +1 \\ 0 \\ -1 \end{array}$ 
\\[12pt]
\hline
\end{tabular}
\caption{\label{tab:particlesQN}The quantum numbers  of the new particles under the EW and $\SUD$ gauge groups.}
\end{table}

In this paper, we explore the potential of an FPVDM 
fermionic doublet that mixes with the SM top-quark,
which was suggested as an exemplary model in~\cite{Belyaev:2022shr,Belyaev:2022zjx}.
This dark fermionic doublet can naturally be identified as a VL top-quark doublet, $\Psi = (t_\mathrm{D}, T)$, which
for the sake of generality we will be referring to as 
$(f_\mathrm{D}, F)$ hereafter.
The mixing between the SM and fermions arises from the Yukawa interaction term $y^\prime$ we detail below.

The most general Lagrangian for this scenario takes the following form:
\begin{eqnarray}
\Lag &\supset& - \frac{1}{4} (V_{\mu\nu}^i)^2|_{B,W^i,V^i_\mathrm{D}} + \bar{f}^{\rm SM} i \slashed{D} f^{\rm SM} + \bar{\Psi} i \slashed{D} \Psi + | D_\mu \Phi_\mathrm{H} |^2 + | D_\mu \Phi_\mathrm{D} |^2 - V(\Phi_\mathrm{H}, \Phi_\mathrm{D}) \nonumber \\
&-& (y \bar{f}^{\rm SM}_\mathrm{L} \Phi_\mathrm{H} f^{\rm SM}_\mathrm{R} + y^\prime \bar{\Psi}_\mathrm{L} \Phi_\mathrm{D} f^{\rm SM}_\mathrm{R} + h.c.) - m_{f_\mathrm{D}} \bar{\Psi} \Psi \;,
\label{eq: L of D sector}
\end{eqnarray}
where the scalar potential $V(\Phi_H, \Phi_D)$  is given by
\begin{eqnarray}
 V(\Phi_\mathrm{H},\Phi_\mathrm{D}) &=& - \mu_\mathrm{H}^2 \Phi_\mathrm{H}^\dagger \Phi_\mathrm{H} - \mu_\mathrm{D}^2 \Phi_\mathrm{D}^\dagger \Phi_\mathrm{D} + \lambda_\mathrm{H} (\Phi_\mathrm{H}^\dagger \Phi_\mathrm{H})^2 + \lambda_\mathrm{D} (\Phi_\mathrm{D}^\dagger \Phi_\mathrm{D})^2 + \lambda_{\mathrm{HD}} (\Phi_\mathrm{H}^\dagger \Phi_\mathrm{H})(\Phi_\mathrm{D}^\dagger \Phi_\mathrm{D})\;.
 \label{eq:scalarpotential}
\end{eqnarray}
In this study we consider both the  fermionic and  the Higgs portals, which  could play an important role  to produce strong FOPTs and related GW signals. 

\subsection{The Fermions}

Expanding the fermion kinetic terms of both the visible and dark sectors around the vacuum of the theory yields the following mass terms:
\begin{equation}
  \Lag_m^f = (\bar f_{\rm D_L} F_{\rm L}) \mathcal M_F^d \left(\begin{array}{c} f_{\rm D_R} \\ F_R \end{array}\right) = (\bar f_{\rm D_L} F_{\rm L}) U_{\rm L}^\dagger \mathcal M_F U_{\rm R} \left(\begin{array}{c} f_{\rm D_R} \\ F_R \end{array}\right)\,, 
\end{equation}
where $\mathcal{M}_F^d$ denotes the diagonal mass matrix expressed in the physical basis and 
\begin{equation}
\mathcal{M}_F = \left(\begin{array}{cc} y \dfrac{v}{\sqrt 2} & 0 \\[1.2em] y^\prime \dfrac{v_{\rm D}}{\sqrt 2} & m_{f_{\rm D}} \end{array}\right)\,.
\end{equation}
with $U_{\rm L,R}$ representing orthogonal matrices given by 
\begin{align}
    U_{\rm L,R}=\left(\begin{array}{cc} \cos\theta_{\rm L,R} & \sin\theta_{\rm L,R} \\ -\sin\theta_{\rm L,R} & \cos\theta_{\rm L,R} \end{array} \right)\,,
\end{align}
that diagonalize $\mathcal{M}_F$ upon a bi-unitary transformation. Note that $U_\mathrm{R}$ is physical in the presence of a dark charged current. The Yukawa term with $y^\prime$  mixes the SM fermion and the even component $T$ of the $\Psi$ doublet. Their masses are given by 
\begin{equation}
 m_{f^{\text{SM}},F}^2=\frac{1}{4} 
 \left[y^2 v^2 + y^{\prime2} v_\mathrm{D}^2 + 2 m_{f_\mathrm{D}}^2 \mp \sqrt{(y^2 v^2 + y^{\prime2} v_\mathrm{D}^2 + 2 m_{f_\mathrm{D}}^2)^2-8y^2v^2m_{f_\mathrm{D}}^2}\right]\; ,
 \label{eq: fermionic masses}
\end{equation}
where $v$ and $v_\mathrm{D}$ are the SM and Dark Higgs vevs,  respectively.
The fermions  obey the following hierarchy
\begin{align}
m_{f^{\text{SM}}} &< m_{f_{\rm D}} \leq m_{F}\;.
\label{eqn:fermionic mass hierarchy}
\end{align}
The mixing between SM and dark(D)-fermions is given by right and left mixing angles:
\begin{equation}
\sin\theta_{\mathrm{R}} = \sqrt{\frac{m_{F}^2 - m_{f_\mathrm{D}}^2}{m_{F}^2 - m_f^2}}, \quad \sin\theta_{L} = \frac{m_f^{\text{SM}}}{m_{f_\mathrm{D}}}\sin\theta_{\mathrm{R}}\;.
\end{equation}
The masses and the mixing angles are related to  Yukawa couplings $y$ and $y^\prime$ as follows:
\begin{equation}
\label{eqn:yukawas}
y = \sqrt{2} \frac{m_{f^{\text{SM}}} m_{F}}{m_{f_{\rm D}} v},\quad y^\prime = \sqrt2 \frac{\sqrt{(m_{F}^2 - m_{f_{\rm D}}^2)(m_{f_{\rm D}}^2 - m_{f^{\text{SM}}}^2)}}{m_{f_{\rm D}} v_{\rm D}}\;.
\end{equation}
The new fermion sector is exactly decoupled in the limit $m_{F} = m_{f_{\rm D}}$, for which $y = y^f_{\rm SM} = \sqrt{2} \frac{m_{f^\text{SM}}}{v}$, 
$y^\prime = 0$, and $\sin\theta_{\rm L} = \sin\theta_{\rm R} = 0$, thus restoring the pure SM scenario. 
\subsection{The Scalars}

When the complex scalar $\Phi_{\rm{D}}$ acquires a vev, $\SU{2}{D}$ undergoes spontaneous symmetry breaking, which gives masses to the dark gauge bosons, $V^0_{D\pm}$ and $\mathrm{V^0_D}$. Simultaneously, the SM Higgs state $\Phi_{\rm{H}}$ also acquires a vev, giving masses to the SM particles through the breaking of the EW gauge group. Minimising the potential, we obtain
\begin{equation}
v\qty(- \mu_{\rm H}^2 +\lambda_{\rm H} v^2 + {\frac{1}{2}} \lambda_{\rm HD} v_{\rm D}^2) = 0 \quad\text{and}\quad
v_{\rm D}\qty(- \mu_{\rm D}^2 + \lambda_{\rm D} v_{\rm D}^2 + {\frac{1}{2}}\lambda_{\rm HD} v^2) = 0 \,,
\end{equation}
yielding two non-trivial stationary points
\begin{equation}
\label{eqn:vev1}
 v=\pm\sqrt{\frac{4\lambda_{\rm D} \mu_{\rm H}^2-2\lambda_{\rm HD}\mu_{\rm D}^2} {4\lambda_{\rm H}\lambda_{\rm D}-\lambda_{\rm HD}^2}}
 \quad\text{and}\quad 
 v_{\rm D}=\pm\sqrt{\frac{4\lambda_{\rm H} \mu_{\rm D}^2 - 2\lambda_{\rm HD}\mu_{\rm H}^2}{4\lambda_{\rm H}\lambda_{\rm D}-\lambda_{\rm HD}^2}}\,,
\end{equation}
for the vacuum of the theory, provided that the Hessian matrix,
\begin{equation}
\label{eqn:Hess matrix}
\mathcal{M_S}=\left(
\begin{array}{cc}
  \lambda_{\rm H} v^2 & \frac{\lambda_{\rm HD}}{2} v v_{\rm D} \\ \frac{\lambda_{\rm HD}}{2} v v_{\rm D} & \lambda_{\rm D} v_{\rm D}^2
\end{array}
\right)\,,
\end{equation}
is positive definite. The physical masses are obtained by diagonalizing $\mathcal{M_S}$ via an orthogonal transformation parametrised by the rotation matrix
\begin{equation}
    V_S=\left(\begin{array}{cc} \cos\theta_S & \sin\theta_S \\ -\sin\theta_S & \cos\theta_S \end{array} \right)\,.
\end{equation}
The mass eigenvalues for the scalar sector read
\begin{equation}
 M_{H,H_\mathrm{D}}^2=\lambda_H v^2+\lambda_\mathrm{D} v_\mathrm{D}^2\mp\sqrt{(\lambda_\mathrm{H} v^2-\lambda_\mathrm{D} v_\mathrm{D}^2)^2+\lambda_{\mathrm{HD}}^2v^2 v_\mathrm{D}^2} 
\end{equation}
with the mixing angle 
\begin{equation}
\sin\theta_S = \sqrt{2 \frac{m_{H_\mathrm{D}}^2 v^2 \lambda_\mathrm{H} - M_H^2 v_\mathrm{D}^2 \lambda_\mathrm{D}}{M_{H_\mathrm{D}}^4 - M_H^4}}\;.
\end{equation}
In the above expression, the $H$-$H_{\mathrm{D}}$ mixing depends implicitly on the portal coupling which vanishes in the limit of $\lambda_\mathrm{\mathrm{HD}} = 0$. However, even in the absence of mixing induced by the quartic term at tree-level, the SM and Dark Higgs doublets mix at one-loop via their interactions with the dark fermions.

\subsection{Dark bosons}
At tree level, the masses of the dark gauge bosons read
\begin{equation}
M_{V^0} = M_{V^\pm} \equiv M_{V_\mathrm{D}} = g_\mathrm{D}\frac{v_\mathrm{D}}{2} \label{eq:VPmass} \;.
\end{equation}
At loop level, the mass degeneracy is broken by the kinetic mixing of $\gamma$-$Z$-$\mathrm{V_D}$ states and the mass correction of $V^{0}_{\mathrm{D}\pm}$ and $V^0_\mathrm{D}$ themselves, controlled by the mass difference of the dark fermions. To distinguish between the two, we introduce the notation
\begin{equation}
    \begin{cases}
        &\hspace*{-0.8em} V^{0}_{\mathrm{D}\pm} \equiv \mathrm{V_D}\,,\quad \mathrm{with~mass~}M_{\mathrm{V_D}}\,, \\
        &\hspace*{-0.8em} V^0_\mathrm{D} \equiv V^\prime\,,\quad \mathrm{with~mass~}M_{V^\prime}\,.
    \end{cases}
\end{equation}

The mass difference between $V_\mathrm{D}$ and $V^\prime$ due to the one-loop mass correction is given by
\begin{align}
M_{V_\mathrm{D}}-M_{V^\prime}=\frac{g_\mathrm{D}^2m_{F}^2}{32 \pi^2 M_{V_\mathrm{D}}}\left(\frac{m_{F}^2-m_{f_\mathrm{D}}^2}{m_{F}^2}\right)^2.
	\label{eq:simple_mass_splitting_2}
\end{align}
This radiative mass splitting between the $\mathrm{V_D}$ and $V^{\prime}$ bosons plays a very important role in the determination of DM relic density and DM direct and indirect detection rates.

\subsection{Model scenarios for phase transition analysis}\label{subsec:scenarios}

The emergence of FOPTs critically depends on the presence of extended scalar sectors. These can be either coupled to the SM, where interactions with the visible sector are mediated by non-negligible portal couplings, or decoupled, if such couplings are tiny, or if there is a large hierarchy between the Higgs and the BSM sectors \cite{Borah:2021ocu,Borah:2021ftr,Kierkla:2022odc,Lewicki:2021xku,Marzo:2018nov}. In this article, we study three versions of the FPVDM model in order to comprehensively cover the possibilities outlined above.
\begin{itemize}
    \item Scenario I: First, by switching off both the Higgs and fermion portal couplings, we consider a pure $\mathrm{SU(2)_D}$ dark sector, assuming that it does not reach thermal equilibrium with the SM. This allows us to neglect the SM sector for this scenario. The corresponding Lagrangian reads:
    \begin{align}\label{eq:scenario_I_lag}
    \mathcal{L}_{\mathrm{I}} = - {\frac{1}{4}} (V^{i}_{\mu\nu})^2 + | D_\mu \Phi_{\rm{D}} |^2 - \mu^2_{\rm D} \Phi_{\rm{D}}^\dagger \Phi_{\rm{D}} - \lambda_{\rm D} (\Phi_{\rm{D}}^\dagger \Phi_{\rm{D}})^2\,.
    \end{align}

    We emphasise that our primary goal is to understand the dynamics of FOPTs in this minimal model and to compare them with two alternative scenarios discussed below. As it stands, however, this minimal setup is cosmologically excluded. After the phase transition, the dark-sector particles become non-relativistic and the dark sector starts to dominate the total energy density, eventually overclosing the Universe. To avoid such an outcome, one must consider extensions of the model in which the energy stored in the dark sector is efficiently transferred into some form of dark radiation.
    A well-motivated extension is to gauge an additional abelian symmetry, $\mathrm{U(1)_D}$, in the dark sector. This leads to the symmetry-breaking pattern
    $ 
    \mathrm{SU(2)_D} \times \mathrm{U(1)_D} \to \mathrm{U(1)'},
    $ 
    where the vacuum expectation value $v_{\rm D}$ leaves a residual unbroken $\mathrm{U(1)'}$ symmetry. The presence of this unbroken $\mathrm{U(1)'}$ naturally ensures the conservation of a dark charge, which would otherwise have to be imposed by hand. The residual gauge symmetry gives rise to a massless dark photon, $\gamma'$, which behaves as dark radiation. In this framework, the energy density of the massive dark-sector states, $V_{\rm D}$ and $H_{\rm D}$, is efficiently transferred into that of $\gamma'$, which, being massless, redshifts more rapidly and therefore does not lead to overclosure of the Universe.
    \\
    For the model to be cosmologically viable, it is further required that the dark photon does not thermalise with the SM. In this Scenario I, this condition is naturally satisfied due to the absence of any portal coupling between the dark and visible sectors. In particular, no kinetic mixing with $\mathrm{U(1)_Y}$ is present, and the resulting dark-photon-to-SM temperature ratio remains below the cosmological bound of $0.6$~\cite{Breitbach:2018ddu}.
    \\
    A detailed investigation of this extended framework lies beyond the scope of the present work. Since our interest is focused on the main qualitative features of the phase transition itself, these are fully captured by the simplified Lagrangian given in Eq.~\eqref{eq:scenario_I_lag}. We therefore assume that the minimal scenario under consideration remains cosmologically viable due to the existence of a consistent and well-motivated extension, as outlined above. Moreover, this simplified setup allows for a more direct and transparent comparison with the two scenarios discussed below.

    \item Scenario II: At finite temperatures, phase transitions involving the Higgs direction can occur when both sectors are in thermal equilibrium and the Higgs portal interaction is significant. Furthermore, the interplay among different sectors drives intriguing DM phenomenology. We will therefore consider the $\mathrm{G_{SM}}\times \mathrm{SU(2)_D}$ model, where $\mathrm{G_{SM}}$ is the SM gauge group, incorporating the SM particle content alongside the bosonic part of the dark sector, such that the second version of the model is described by the Lagrangian
     \begin{align}
        \mathcal{L}_{\mathrm{II}} =& \mathcal{L}_{\mathrm{I}}+\mathcal{L}_{\mathrm{SM}} 
         - \lambda_{\rm HD}(\Phi_{\rm{H}}^\dagger \Phi_{\rm{H}})(\Phi_{\rm{D}}^\dagger \Phi_{\rm{D}}) \,.
    \end{align}
    \item Scenario III: Finally, we consider the complete FPVDM $\mathrm{G_{SM}}\times \mathrm{SU(2)_D}$ model, which incorporates both the Higgs and Yukawa portal couplings, assuming a top-partner. The Lagrangian is as follows
    \begin{align}
        \mathcal{L}_{\mathrm{III}} =& \mathcal{L}_{\mathrm{II}} - (y^\prime \bar \Psi_{\rm L} \Phi_{\rm{D}} f^{\rm SM}_{\rm R} + \mathrm{h.c.}) +
       \bar{\Psi}\left( i\slashed{D} - m_{f_{\rm D}}\right)  \Psi \,.
    \end{align}
\end{itemize}

\subsection{Thermal effective potential}\label{subsec:VEff}

Having presented the model, we now discuss the calculation of the thermal effective potential, as needed for the phase transition analysis. As mentioned in \cref{sec:fopt_gws}, theoretical uncertainties arising from renormalisation scale and gauge dependence can be significantly reduced through DR (see, for example, Table 3 in \cite{Croon:2020cgk} for a comparison between 3D and 4D effective potentials). Despite these advantages, the exploration of 3D effective potentials in the context of SGWB studies remains somewhat limited in the literature. This is partly due to the complex calculations required, particularly their numerical implementation for phase tracing and bounce solvers, which contrasts with the standard 4D methods that are relatively straightforward. In this work we use the DR formalism. A detailed derivation is provided in \cref{sec:DR_eff}, but here we outline the most important steps and provide some further details on the software implementation. For reference, see \cite{Ekstedt:2022bff} but also the other references given in \cref{sec:DR_eff}. 

For a system in thermal equilibrium, we may study the thermal effects in the imaginary time formalism, with $it \in [-\beta,\beta]$ and $\beta=1/T$. Due to the compactification of the (imaginary) time dimension, an infinite tower of modes with squared masses of the form $\omega_n^2 + m^2$ emerges from the perspective of 3D Euclidean space, as discussed in Appendix~\ref{sec:DR_eff}.
Here, $m$ is the ordinary mass of the field in question and $\omega_n, n \in \mathbb{Z}$, is the Matsubara frequency, with $\omega_n = 2n\pi T$ for bosons and $\omega_n = (2n+1)\pi T$ for fermions. When $T$ is large, in the sense that $T \gg m$, all the fermionic modes plus the bosonic modes with $n \neq 0$ are heavy, and we are justified in integrating them out. By doing so, we are matching the 4D theory, said to live at the hard scale, to a 3D EFT, said to be living at the soft scale, where all the thermal effects have been absorbed into the effective parameters of the 3D EFT; this process is referred to as \textit{dimensional reduction}. 

In the soft-scale 3D EFT, the temporal modes of the vector fields exist as scalar fields, decoupled from the spatial part and with associated Debye masses. These Debye masses are often large compared to the scale of interest for the phase transition, an assumption made throughout this work. Thus, we are justified in further integrating out also these temporal modes from the theory. By doing so, we are matching the soft-scale 3D EFT to yet another 3D EFT, said to be living at the ultrasoft scale. 

In the ultrasoft 3D EFT, we can  calculate the effective potential, $V_\mathrm{eff}^{3D}$. It is a function of the 3D fields $\varphi$, with mass units $[M^{1/2}]$, and the parameters of the 3D ultrasoft EFT, which in turn depend on the temperature and are collectively denoted $\textbf{p}^\mathcal{US}(T)$. Hence, we indicate the arguments of $V_\mathrm{eff}^{3D}$ as follows: $V_\mathrm{eff}^{3D}(\varphi;\textbf{p}^\mathcal{US}(T))$. Note that in the appendix, we will also use the abbreviated notation $V_\mathrm{eff}^{3D}(T)$. Finally, we define the 4D thermal effective potential $V_\mathrm{eff}=V^{4D}$, as used in \cref{eqn:euclidean_action}, by 
\begin{equation}
V^{4D}(\Phi,T) = TV_\mathrm{eff}^{3D}(\Phi/\sqrt{T};\textbf{p}^\mathcal{US}(T)). \label{eq:V4D_vs_V3D}    
\end{equation}
Here, $\Phi$ denotes the ordinary 4D fields, with mass dimension $[M^{1}]$.

For the implementation, we use the soon-to-be released package \texttt{Dratopi} \cite{Dratopi}, which interfaces \texttt{DRalgo} \cite{Ekstedt:2022bff} to \texttt{Python} and a modified version of \texttt{CosmoTransitions} \cite{Wainwright:2011kj}. \texttt{Dratopi} provides a script to export from \texttt{DRalgo} into \texttt{Python}, among other things, the beta functions for the 4D theory, the results of the hard-to-soft and the soft-to-ultrasoft matchings, as well as the effective potential in the ultrasoft 3D EFT. Moreover, \texttt{Dratopi} provides the necessary routines to calculate the 4D thermal effective potential, which can then be used for the phase transition analysis in a slightly modified version of \texttt{CosmoTransitions}. Further details on \texttt{Dratopi} will be provided in the manual accompanying its upcoming release. Below, we summarise the crucial steps for calculating the 4D thermal effective potential at field values $\Phi$ and temperature $T$. Note that the first two steps are only done once, during setup, while the other steps are done for each new value of $\Phi$ and $T$:
\begin{enumerate}
    \item Define the model by specifying its 4D parameters, collectively denoted $\textbf{p}^{4D}$, at some given reference energy scale $\mu_\mathrm{ref}$.
    \item Using the beta functions, solve the renormalisation group (RG) equations, to obtain an interpolated solution of $\textbf{p}^{4D}$ as a function of the RG scale/energy scale (over some specified range).
    \item Set the hard matching scale to $\mu_{4D} = \pi\kappa T$, where $\kappa$ is a prefactor that defaults to 1.
    \item Construct the soft 3D EFT, by matching the 4D theory to the 3D EFT, at the scale $\mu_{4D}$.
    \item Set the soft matching scale $\mu_{3D}^\mathcal{S}$ equal to the smallest Debye mass, at temperature $T$.
    \item Construct the ultrasoft 3D EFT, by integrating out the temporal modes, thus obtaining the 3D parameters $\textbf{p}^\mathcal{US}(T)$ in the ultrasoft 3D EFT.
    \item Calculate the 4D thermal effective potential through $V^{4D}(\Phi,T) = TV_\mathrm{eff}^{3D}(\Phi_i/\sqrt{T};\textbf{p}^\mathcal{US}(T))$.
\end{enumerate}

\section{Scan Strategy, Phase Transition Framework, and Dark Matter Constraints}\label{sec:scan}

\subsection{Parameter Setup and Scanning Strategy}
As detailed in \cref{subsec:scenarios}, we explore the dynamics of phase transitions across three distinct scenarios. Scenario I is the simplest among them and involves only three free parameters: the physical masses of the dark scalar and dark vector bosons, $M_\mathrm{H_D}$ and $M_\mathrm{V_D}$, respectively, as well as the dark gauge coupling, $\g{D}$. The specific ranges for these parameters are provided in the first three columns of \cref{tab:sample1}.
\begin{table}[htb!]
	\centering
    \captionsetup{justification=raggedright}
    \begin{tabular}{c|c|c|c}
		\toprule
		$M_{\rm V_{D}}$ (GeV) & $M_{\rm H_{D}}$ (GeV) & $\g{D}$ & - \\
		\midrule
		$\left[ 10 , 10~000 \right]$ & $\left[10^{-3}, 10~000 \right]$ &  $\left[10^{-3}, 4.0 \right]$ & \makecell{$\lambda_{\rm D}=\tfrac{m^2_{\rm H_{D}}}{8M^2_{\mathrm{V_D}}} \g{D}^2$ \\[0.8em] $v_{\rm D} = \tfrac{2 M_{\rm V_{D}}}{\g{D}}$ \\[0.8em] $\mu^2_{\rm D} = \frac{1}{2} m_{\rm H_{D}}^2$ }
		\\
		\bottomrule
	\end{tabular}
	\caption{\footnotesize Ranges of parameters used in the numerical scan for \textbf{Scenario~I}. In the last column, we list the relationships used to calculate the model parameters that are not free.}
	\label{tab:sample1}
\end{table}
In the fourth column, $\lambda_\mathrm{D}$, $v_\mathrm{D}$ and $\mu^2_\mathrm{D}$ are derived from the free parameters according to the displayed expressions.

\subsection{Coupling Structure and Parameter Relations in the Scalar Sector}
In both Scenarios II and III, the SM is incorporated, and scalar mixing with the dark sector, induced by the portal coupling $\lambda_\mathrm{HD}$, is considered. Alongside the free parameters of Scenario I, these scenarios also include the scalar mixing angle $\theta_S$, whose range complies with the current LHC data \cite{ATLAS:2021vrm, Papaefstathiou:2022oyi}.
The Higgs boson mass is fixed at $m_\mathrm{H} = 125.1~\mathrm{GeV}$, while the EW symmetry breaking vev is set according to $v = \tfrac{2 M_\mathrm{W}}{\g{W}}\simeq 246$ GeV, where $M_\mathrm{W} = 80.37~\mathrm{GeV}$ \cite{ParticleDataGroup:2024cfk} is the mass of the $W$ boson, and $\g{W}= 0.65$ \cite{ParticleDataGroup:2024cfk} is the weak gauge coupling. With these definitions, we can express the scalar quadratic and quartic couplings in the gauge eigenbasis in terms of the physical parameters as follows
\begin{equation}\label{eqn:coups_inverted}
    \begin{aligned}
    & \lambda_{\rm H} = \frac{\g{W}^2}{8M^2_{\rm W}}\qty(M^2_{\rm H} \cos^2\theta_S + M^2_{\rm H_{D}} \sin^2\theta_S)\,, \\
    & \lambda_{\rm D} = \frac{\g{D}^2}{8M^2_{\mathrm{V_D}}}\qty(M^2_{\rm H} \sin^2\theta_S + M^2_{\rm H_{D}} \cos^2\theta_S)\,, \\
    & \lambda_{\rm HD} = \frac{\g{D} \g{W}}{8M_{\rm V_{D}} M_{\rm W}}\qty(M^2_{\rm H_{D}} - M^2_{\rm H})\sin 2\theta_S\,, \\
    & \mu^2_{\rm H} = \frac{1}{2}\qty(M_{\rm H}^2\cos^2\theta_S + M_{\rm H_{D}}^2\sin^2\theta_S + \frac{1}{2}\frac{M_{\rm V_{D}} \g{W}}{M_{\rm W} \g{D}}(M^2_{\rm H_{D}} - M^2_{\rm H})\sin 2\theta_S)\,, \\
    & \mu^2_{\rm D} = \frac{1}{2}\qty(m_{\rm H}^2\sin^2\theta_S + M_{\rm H_{D}}^2\cos^2\theta_S + \frac{1}{2}\frac{M_{\rm W} \g{D}}{M_{\rm V_{D}} \g{W}}(M^2_{\rm H_{D}} - M^2_{\rm H})\sin 2\theta_S)\,.
    \end{aligned}
\end{equation}
The primary distinction between Scenario II and Scenario III is that the latter includes the complete FPVDM model. This encompasses the additional consideration of dark and visible VL fermion masses, $m_{f_{\rm D}}$ and $m_F$, as free input parameters. The objective is to investigate the impact of the extended fermion sector, particularly through the portal described by the Yukawa coupling $y^\prime$, on the phase transition dynamics. We also analyse the conditions under which visible SGWB predictions align with viable DM phenomenology for both scenarios. The parameter ranges for the scan are presented in \cref{tab:sample2}. Note that the fourth and fifth columns, as well as the last line in the seventh column, apply only to Scenario III.
\begin{table}[htb!]
	\centering
    \captionsetup{justification=raggedright}
	\resizebox{\textwidth}{!}{%
    \begin{tabular}{c|c|c|c|c|c|c}
		\toprule
		$M_{\rm V_{D}}$ (GeV) & $M_{\rm H_{D}}$ (GeV) & $\g{D}$ & $m_{f_{\rm D}}$ (GeV) & $m_{F}$ (GeV) & $\sin\theta_S$ & - \\
		\midrule
		$\left[ 10 , 50~000 \right]$ & $\left[10^{-3} , 10~000 \right]$ &  $\left[10^{-3}, 4.0 \right]$ & $\left[500, 65~000\right]$ & $\left[500, 65~000\right]$ & $\left[ -0.2 , 0.2 \right]$ & \makecell{$\lambda_{\rm H},~\lambda_{\rm D},~\lambda_{\rm HD} =$ \cref{eqn:coups_inverted} \\[0.5em] $v_{\rm D} = 2 M_{\rm V_{D}}/\g{D}$ \\[0.5em] $\mu^2_{\rm H},~\mu^2_{\rm D} =$ \cref{eqn:coups_inverted} \\[0.5em] $y_t,~y^\prime =$ \cref{eqn:yukawas}}
		\\
		\bottomrule
	\end{tabular}}
	\caption{\footnotesize Ranges of parameters used in the numerical scan for \textbf{Scenario~II} and \textbf{Scenario~III}. The fourth and fifth columns pertain only to Scenario III. The last column references the relations used to calculate the dependent parameters, with the last line being applicable solely to Scenario III.}
	\label{tab:sample2}
\end{table}
%

\subsection{Renormalisation, Matching, and Bounce Action Setup}
In the complete FPVDM model (Scenario III), we only accept numerical solutions where the hierarchy $M_{\rm V_{D}} < m_{f_{\rm D}}, m_{F}$ is satisfied. Otherwise, $f_\mathrm{D}$ would emerge as the DM candidate instead of $V_\mathrm{D}$, which would be inconsistent as $f_\mathrm{D}$ is electrically charged.

In our numerical analysis of Scenario I, the 4D reference scale is set to the vector boson's mass, \textit{i.e.}, $\mu_\mathrm{ref} = M_{\rm V_{D}}$. For Scenarios II and III, the BSM parameters rely on the values of SM physical parameters defined at the EW scale, as explicit in \cref{eqn:coups_inverted}. Consequently, we use the Z-boson mass scale as the reference taking $\mu_\mathrm{ref} = 91~\mathrm{GeV}$. The hard-to-soft matching scale is defined as $\mu_{\mathrm{4D}} = \kappa \pi T$, where $\kappa$ is a dimensionless parameter fixed to $\kappa = 1$. 
The evolution of theory couplings across these scales is governed by the RG flow, defined by their $\beta$ functions in \cref{app:rges_4d}. The matching scale between the soft and ultrasoft theories is determined by the smallest Debye mass at a given temperature $T$, \textit{i.e.}, $\mu_{\mathrm{3D}}(T) = \mathrm{min}[\mu^2_{\rm Debye}(T)]$, with $\mu^2_{\rm Debye}(T) \in \{ \mu^2_{\SU{3}{C}}, \mu^2_{\SU{2}{D}}, \mu^2_{\SU{2}{L}}, \mu^2_{\U{Y}} \}$ being calculated at two-loop order according to the expressions in \cref{app:debye_masses}. The bounce action is computed numerically using our modified version of \texttt{CosmoTransitions} \cite{Wainwright:2011kj} as discussed above.

\subsection{Uncertainty Estimation and Transition Criteria}
To minimise uncertainties associated with the numerical tracing of the action, we perform a spline fit using numerical arrays containing 200, 220, 240, and 260 points, analogous to the polynomial fit strategy adopted in \cite{Freitas:2021yng}. To ensure that the spline function accurately captures the divergent behaviour around the critical temperature, $T_c$, we use an adaptive temperature range with a higher density of points near $T_c$. The final values are obtained by averaging the results from the four different fits. Since the primary source of numerical errors stems from estimating the $\beta/H(T_p)$ parameter, we define the uncertainty as
\begin{equation}\label{eqn:uncertainty}
    \Delta\left[\beta/H(T_p)\right] = \frac{\left\{\mathrm{max}\left[\beta/H(T_p)\right] - \mathrm{min}\left[\beta/H(T_p)\right]\right\}}{\mathrm{mean}\left[\beta/H(T_p)\right]} \,,
\end{equation}
where the values of $\beta/H(T_p)$ are derived from each fit. For all numerical results presented in this paper, we enforce that $\Delta\left[\beta/H(T_p)\right] < 0.25$. Additionally, we only consider transitions where $\beta/H(T_p) < 10^5$, as larger values correspond to transitions occurring almost instantaneously, rendering them physically equivalent to second-order phase transitions \cite{Kajantie:1996mn}.

We would like to highlight two aspects regarding the calculation of temperatures. First, we do not adopt the nucleation temperature provided by default in \texttt{CosmoTransitions}, which relies on the approximation $S_3/T = 140$, where $S_3$ is the Euclidean action, valid only for transitions at and around the EW scale. Instead, we use the generic definition given in \cref{eqn:nucletion_temperature}. Second, for very strongly supercooled transitions, the percolation condition may be insufficient, necessitating explicit verification that the transition completes. This condition is expressed as
\begin{equation}
    H(T)\left(3 + T\frac{dI}{dT}\right)\Biggr|_{\substack{T = T_p}} < 0\,.
\end{equation}
However, this constraint becomes significant only for parameter points with large values of $\alpha$ (orders of magnitude above 1). Indeed, the vast majority of our points have $\alpha < 100$, and numerical tests showed that this condition was always satisfied.

\subsection{Constraints from Cosmology and Dark Matter Observables}
\label{sec:cosmo-dm}

Scenarios II and III include portal interactions between the dark and visible sectors, whereas Scenario I does not contain such a portal.

As a result, only Scenario I avoids thermal contact. Therefore, if sufficiently light (below $\sim$ MeV), any additional dark-sector degrees of freedom would otherwise contribute to the effective number of neutrino species, $\Delta N_\mathrm{eff}$, constrained to $\Delta N_{\mathrm{eff}}(T_{\mathrm{BBN}}) < 0.55$ at 95\% confidence level (CL)~\cite{Planck:2018vyg} where $T_\mathrm{BBN} \approx 1~\mathrm{MeV}$ is the temperature at the big-bang nucleosynthesis (BBN) epoch. Since $\Delta N_\mathrm{eff} \propto (T_\mathrm{D}/T_\mathrm{SM})^4$ (where $T_\mathrm{D}$ and $T_\mathrm{SM}$ are the dark and SM temperatures after the phase transition), we have:
\begin{equation}\label{eq:deltaNeff}
    \Delta N_\mathrm{eff} = \frac{4}{7} 
    \left(\frac{11}{4}\right)^{4/3} g_*^\mathrm{D}(T_\mathrm{D}) 
    \left(\frac{T_\mathrm{D}}{T_\mathrm{SM}}\right)^4\,,
\end{equation}
where $g_*^\mathrm{D}(T_\mathrm{D})$ is the number of dark-sector degrees of freedom at $T_\mathrm{D}$. Eq.~\eqref{eq:deltaNeff} applies only at temperatures relevant for BBN and below, $T_{\rm SM} \sim \mathcal{O}(1~{\rm MeV})$, and only if the corresponding dark-sector degrees of freedom are relativistic at that epoch. In particular, if a dark sector field ($H_{\rm D}$ or $V_{\rm D}$) satisfies $M_{\rm H_D},M_{\rm V_D} > T_D$ at $T_{\rm SM} \sim 1~{\rm MeV}$, it does not contribute to $\Delta N_{\rm eff}$ and Eq.~\eqref{eq:deltaNeff} is not applicable. In this case, such parameter points are not excluded by $\Delta N_{\rm eff}$ constraints.

In the following, we make the conservative assumption that prior to the phase transition the SM temperature $T_{\rm SM}$ and the dark-sector temperature $T$ are the same, $T_{\rm SM} \simeq T$. This assumption is motivated by the possibility that at some sufficiently high energy scale the SM and dark sectors were in thermal equilibrium, leading to comparable temperatures at the time of the phase transition. 

With this in mind, in the aftermath of the phase transition energy is injected into the plasma, heating the dark sector \cite{Marfatia:2021twj}. Due to energy conservation the total energy density must remain the same before and after the phase transition. Assuming that the Universe remains in a radiation dominated epoch after the FOPT, we write that
\begin{equation}\label{eqn:rho_conserved}
    \rho_R(T_D) = \rho_R(T_p) + \epsilon(T_p)\,, 
    \qquad 
    \epsilon(T_p) = \left(\Delta V - T\frac{\partial}{\partial T}\Delta V\right)\Bigg|_{T = T_p}\,,
\end{equation}
where $\epsilon(T_p)$ denotes the latent heat released during the phase transition, which is converted into dark radiation. Using Eq.~\eqref{eqn:alpha_param} we can write this expression as
\begin{equation}
    \rho_R(T_D) = \rho_R(T_p) + \rho_R^\mathrm{tot}\,\alpha\,, 
    \qquad 
    \rho_R^\mathrm{tot} = \frac{\pi^2 T_\mathrm{SM}^4}{30} 
    \left[ g_*^\mathrm{SM}(T_{\rm SM}) + g_*^D(T_p)\left(\frac{T_p}{T_\mathrm{SM}}\right)^4 \right]\,,
\end{equation}
where $\rho_R^\mathrm{tot}$ is the total radiation energy density, including contributions from both the SM and dark sectors and $g_*^\mathrm{SM}(T_p)( g_*^\mathrm{D}(T_p) )$ is the SM (dark) sector's degrees of freedom at the percolation temperature, $T_p$. Solving this with respect to $T_D$ yields
\begin{equation}\label{eq:darksector}
    T_\mathrm{D}^4 = T_p^4 \frac{g^D_*(T_p)}{g^D_*(T_D)}(\alpha + 1) + \alpha T_\mathrm{SM}^4 \frac{g_*^\mathrm{SM}(T_{\rm SM})}{g_*^\mathrm{D}(T_D)}\,,
\end{equation}
where all degrees of freedom are evaluated at the broken phase.

In the absence of portal interactions in Scenario~I, the latent heat released during the phase transition reheats only the dark sector. After reheating from the FOPT and under the assumption $T_{\rm SM} \simeq T_p$, one generically finds $T_D > T_{\rm SM}$. In this conservative setup, if the dark sector remains relativistic at $T_{\rm SM} \sim 1~{\rm MeV}$, Eq.~\eqref{eq:deltaNeff} applies and such parameter points are excluded by $\Delta N_{\rm eff}$ constraints. We stress, however, that $T_{\rm SM}$ and $T_p$ need not coincide in general. In particular, scenarios with $T_{\rm SM} > T_p$ prior to the phase transition can be realised, in which case the reheating of the dark sector does not necessarily imply $T_D > T_{\rm SM}$. In such model-dependent scenarios, a larger region of parameter space may survive the $\Delta N_{\rm eff}$ constraint. In practice, one may assume a given relation between $T_{\rm SM}$ and $T_p$, use Eq.~\eqref{eq:darksector} to determine $T_D$ iteratively, and then assess whether Eq.~\eqref{eq:deltaNeff} excludes or allows the corresponding parameter point. To remain on the conservative side, throughout this work we adopt the assumption $T_{\rm SM} \simeq T_p$, which generically leads to $T_D > T_{\rm SM}$ after the phase transition in Scenario~I. Under this assumption, parameter points for which the dark sector masses are relativistic at $T_{\rm SM} \sim 1~{\rm MeV}$ are excluded by Eq.~\eqref{eq:deltaNeff}. Relaxing this assumption in a model-dependent way could reopen additional regions of parameter space. We implement this condition by imposing the constraint $M_{\rm V_{D}}, M_{\rm H_{D}} > 1 \ \mathrm{MeV}$, as seen in \cref{tab:sample1}.

In scenarios II and III, the dark sector interacts with the visible sector through both the Higgs and fermion portals. Consequently, we must consider constraints from the DM relic density and direct detection, following the methodology outlined in \cite{Belyaev:2022shr, Belyaev:2022zjx}. In particular, measurements from the \textit{Planck} satellite determine the DM relic density to be $h^2 \Omega_{\mathrm{DM}} = 0.12 \pm 0.0012$ \cite{Planck:2018vyg}. Parameter points predicting a relic density exceeding this value are excluded, as they would overclose the Universe. However, we retain points yielding a lower relic density, under the assumption that the total DM abundance may arise from multiple sources beyond the vector DM candidate considered in this work. Therefore, FPVDM may either account for the entire DM content or contribute as a subcomponent within a multi-component dark sector. In addition, to avoid overclosure of the Universe, we impose that $T_n > T_f$, where $T_f \approx M_{V_D}/30$ is the freeze-out temperature.

To ensure solutions in which the bubble walls accelerate to relativistic velocities, we impose the B\"{o}deker--Moore criterion \cite{Bodeker:2009qy}. This condition relates the leading-order friction acting on the bubble wall, $\mathcal{P}_\mathrm{LO}$, to the potential energy released during the phase transition, $\Delta V_\mathrm{eff}(T=0)$. In particular, if $\Delta V_\mathrm{eff}(T=0) > \mathcal{P}_\mathrm{LO}$, the bubbles enter a runaway regime and reach relativistic velocities. Conversely, if $\Delta V_\mathrm{eff}(T=0) < \mathcal{P}_\mathrm{LO}$, the bubbles do not run away and instead approach a terminal, typically non-relativistic, velocity. In the latter case, the resulting GW signal is expected to be strongly suppressed, as the GW amplitude scales with the bubble wall velocity $v_w$. Moreover, in such scenarios, DM production via bubble filtering \cite{Baker:2019ndr} can dominate over the standard freeze-out mechanism. Since our primary goal is to correlate GW signatures with DM production from freeze-out, we restrict our analysis to parameter points that satisfy the B\"{o}deker--Moore criterion.

Keeping only the leading contributions, the zero-temperature potential difference and the friction pressure for scenarios II and III are given by
\begin{equation}\label{eq:delta_pressure}
\begin{aligned}
    &\Delta V_\mathrm{eff}(T=0) = \frac{\lambda_{\rm H}}{4}v^4 
    + \frac{\lambda_{\rm HD}}{4}v^2 v_{\rm D}^2 
    + \frac{\lambda_{\rm D}}{4}v_{\rm D}^4\,,\\
    &\mathcal{P}_\mathrm{LO} = \sum_{i} k_i \frac{\Delta m_i^2}{24} T_p^2 
    = \frac{T_p^2}{8}\left(2\Delta m^2_t + 2\Delta M_{\rm W}^2 + \Delta M_Z^2 + 3 \Delta M_{V_{\rm D}}^2\right)\,,
\end{aligned}
\end{equation}
where $k_i$ denote the internal degrees of freedom and $\Delta m$ is the mass difference between the true and false vacuum.

\subsection{Mapping the shape of the effective potential with physical parameters}\label{sec:FOPT}

The most generic scalar potential of the ultrasoft effective field theory studied in this article  is expressed as follows
\begin{equation}\label{eqn:US_pot}
    \begin{aligned}
        &~V^{3D}_{\mathrm{LO}}(T) = \frac{1}{2}[\mu^{\mathcal{US}}_{\rm D}]^2 \varphi_{\rm D}^2 + \frac{1}{2}[\mu^{\mathcal{US}}_{\rm H}]^2 \varphi_{\rm H}^2  + \frac{1}{4}\lambda^{\mathcal{US}}_{\rm D} \varphi_{\rm D}^4 + \frac{1}{4}\lambda^{\mathcal{US}}_{\rm H} \varphi_{\rm H}^4 + \frac{1}{4}\lambda^{\mathcal{US}}_{\rm HD} \varphi_{\rm H}^2 \varphi_{\rm D}^2 \,, \\
        &V^{3D}_{\mathrm{NLO}}(T) = -\frac{1}{12 \pi} \sum_{i \subset \mathrm{scl.}} M_i^{3}(\varphi_{\rm H}, \varphi_{\rm D}, T) -\frac{2}{12 \pi} \sum_{i \subset \mathrm{vec.}} M_i^{3}(\varphi_{\rm H}, \varphi_{\rm D}, T) \,, \\
    \end{aligned}
\end{equation}
where $i$ sums over the scalar fields (first term of $V^{3D}_{\mathrm{NLO}}$) and the vector fields (second term of $V^{3D}_{\mathrm{NLO}}$). Here
\begin{equation}\label{eqn:US_pot_tot}
    \begin{aligned}
        &V^{4D}_{\mathrm{eff.}} = T\left[(V^{3D}_{\mathrm{LO}}(T) + V^{3D}_{\mathrm{NLO}}(T)\right]\,.
    \end{aligned}
\end{equation}
The masses entering $V^{3D}_{\mathrm{NLO}}(T)$ are detailed in \cref{app:mass_matrices}. In this section, we focus on the following vector boson contributions
\begin{equation}\label{eqn:massV}
    \begin{aligned}
        \mathcal{M}^2_{\mathcal{V}_{1,2,3}} = \frac{1}{4} [g^{\mathcal{US}}_{\rm D}]^2\varphi_{\rm D}^2 \,, \qquad \mathcal{M}^2_{\mathcal{V}_{4,5}} = \frac{1}{4} [g^{\mathcal{US}}_{\rm W}]^2\varphi_{\rm H}^2 \,,\qquad \mathcal{M}^2_{\mathcal{V}_6} = \frac{1}{4} \left([g^{\mathcal{US}}_{\rm W}]^2 + [g^{\mathcal{US}}_{\rm Y}]^2 \right)\varphi_{\rm H}^2\,
    \end{aligned}
\end{equation}
which play a crucial role in the development of a potential barrier, thus driving a FOPT. The temperature-dependent ultrasoft parameters in \cref{eqn:US_pot} and \cref{eqn:massV} are comprehensively detailed in \cref{app:Ultrasoft_matching_couplings} and \cref{app:soft_matching_scalar}, with precision up to next-to-leading order (NLO) accuracy for couplings, and next-to-next leading-order (NNLO) for scalar mass parameters.

For a clear analysis, we reformulate the temperature-dependent vacuum representation using polar coordinates
\begin{equation}
\begin{aligned}
    \varphi_{\rm H} (T) = \phi(T) \cos \delta(T)\,, \qquad \varphi_{\rm D} (T) = \phi(T) \sin \delta(T)\,.
\end{aligned}
\end{equation}
We further redefine the background 3D field $\phi$ in terms of a dimensionless parameter as $\phi \to \varphi \sqrt{T}$, such that in the high-$T$ approximation, the potential can be expressed as
\begin{equation}\label{eqn:Vdim-1}
    \tilde{V}_\varphi(\delta,T) = \tilde{c}_0(\delta,T) + \frac{1}{2!}\tilde{d}(\delta,T) \varphi^2 + \frac{1}{3!}\tilde{e}(\delta,T) \varphi^3 + \frac{1}{4!}\tilde{\lambda}(\delta,T) \varphi^4 + \mathcal{O}(\varphi^6)\,.
\end{equation}
The aim of this analysis is to examine the behaviour of the potential concerning input parameters such as the gauge coupling $\g{D}$ and the dark vector mass $M_{\rm V_{D}}$. However, in the expansion of \cref{eqn:Vdim-1}, the temperature dependence of the coefficients, which scales with $T^4$, obscures these effects. Therefore, for the purpose of this analysis, we define a dimensionless potential by normalising it as
\begin{equation}
    V_\varphi(\delta,T) = \frac{\tilde{V}_\varphi(\delta,T)}{T^4}
    \label{eqn:Vdim}
\end{equation}
where the dimensionless coefficients, denoted without a tilde, are given as follows 
\begin{itemize}
    \item constant term:
        \begin{equation}
            c_0(\delta,T) = \frac{[\mu_{\rm D}^\mathcal{US}]^2 \sqrt{[-\mu_{\rm D}^\mathcal{US}]^2} + [\mu_{\rm H}^\mathcal{US}]^2 \sqrt{-[\mu_{\rm D}^\mathcal{US}]^2}}{3 \pi T^3}\,.
        \end{equation}
    \item quadratic term:
    \begin{equation}\label{eqn:quad}
        \begin{aligned}
            d(\delta,T) &= \frac{1}{2T^2} \Big( [\mu_{\rm D}^\mathcal{US}]^2 \sin^2\delta  + [\mu_{\rm H}^\mathcal{US}]^2 \cos^2\delta   \Big) + \cdots\,,    
        \end{aligned}
    \end{equation}
    with the ellipses denoting the sub-dominant contributions from $T V^{3D}_{\mathrm{NLO}}(T)$.
    \item cubic term:\\
    \\
    The LO potential does not contribute to the cubic term. The dominant effect appears at NLO and is given by
        \begin{equation}\label{eqn:cub}
            e(\delta,T) = -\frac{3 [g^{\mathcal{US}}_{\rm D}]^3 \sin ^3\delta 
            +\,\left[2 [g^{\mathcal{US}}_{\rm W}]^3+\left([g^{\mathcal{US}}_{\rm W}]^2
   +[g^{\mathcal{US}}_{\rm Y}]^2\right)^{3/2} \right] \cos^3\delta}{48 \pi T^{3/2}}\,,
    \end{equation}
    originating purely from the gauge sector.
    \item quartic term:\\
    \\
    Similarly to $d(\delta,T)$, the dominant contribution to the coefficient of the quartic term is calculated at LO and is given by
    \begin{equation}\label{eqn:quart}
        \begin{aligned}
            \lambda(\delta,T) &= \frac{1}{4 T} \left[\lambda_{\rm D}^{\mathcal{US}} \sin^4\delta +\lambda_{\rm H}^{\mathcal{US}} \cos^4 \delta + \lambda_{\rm HD}^{\mathcal{US}} \cos^2 \delta\,\sin^2 \delta \right] + \cdots\,.
        \end{aligned}
    \end{equation}    
\end{itemize}
An FOPT requires the formation of a potential barrier that separates the true vacuum from the false vacuum. This scenario occurs only if the condition $$ e(\delta,T) < 0 $$ is satisfied, which, according to \cref{eqn:cub}, takes place at NLO and is driven by the gauge sector. In the following analysis, we will derive four analytical expressions to assess the position and depth of the true vacuum, as well as the position and height of the potential barrier. This analysis will aid in understanding the numerical results and the general behaviour of the potential across all sampled points.

At large field values, neglecting the quadratic term contribution, the minimisation condition yields
\begin{equation}\label{eqn:min_pos}
    \begin{aligned}
       \varphi_\mathrm{min} &\approx -\frac{3 e(\delta,T)}{4 \lambda(\delta,T)}\\
       &=\frac{3 [g^{\mathcal{US}}_{\rm D}]^3 \sin ^3(\delta ) + \left[2 [g^{\mathcal{US}}_{\rm W}]^3+\left([g^{\mathcal{US}}_{\rm W}]^2
   +[g^{\mathcal{US}}_{\rm Y}]^2\right)^{3/2} \right] \cos ^3(\delta )}{16 \sqrt{T} \pi \left[ \lambda_{\rm D}^{\mathcal{US}} \sin^4\delta +\lambda_{\rm H}^{\mathcal{US}} \cos^4 \delta + \lambda_{\rm HD}^{\mathcal{US}} \cos^2 \delta\,\sin^2 \delta \right] }\,,
    \end{aligned}
\end{equation}
which represents a dimensionless quantity that provides a measure for the position of the true vacuum. This can be further simplified to the case of scenario I by taking $\delta = \pi/2$ which yields
\begin{equation}\label{eqn:min_pos_I}
       \varphi_\mathrm{min}
       \approx \frac{3 [g^{\mathcal{US}}_{\rm D}]^3}{16 \sqrt{T} \pi  \lambda_{\rm D}^{\mathcal{US}}}\,.
\end{equation}
At large field values, by neglecting the quadratic term contribution and substituting the first line of \cref{eqn:min_pos} into the dimensionless potential of \cref{eqn:Vdim}, we derive a measure to quantify the depth of the true vacuum as follows
\begin{equation}\label{eqn:min_depth}
    \begin{aligned}
       V_{\varphi_\mathrm{min}}(\delta,T) &\approx -\frac{27 e(\delta,T)^4}{256 \lambda(\delta,T)^3}\\
       &=-\frac{\left(3 [g^{\mathcal{US}}_{\rm D}]^3 \sin ^3(\delta ) + \left[2 [g^{\mathcal{US}}_{\rm W}]^3+\left([g^{\mathcal{US}}_{\rm W}]^2
   +[g^{\mathcal{US}}_{\rm Y}]^2\right)^{3/2} \right] \cos ^3(\delta )\right)^4}{786432 \; T^3 \pi^4 \left[ \lambda_{\rm D}^{\mathcal{US}} \sin^4\delta +\lambda_{\rm H}^{\mathcal{US}} \cos^4 \delta + \lambda_{\rm HD}^{\mathcal{US}} \cos^2 \delta\,\sin^2 \delta \right]^3 }\,.
    \end{aligned}
\end{equation}
For the case of scenario I, the latter takes the form
\begin{equation}\label{eqn:min_depth_I}
       V_{\varphi_\mathrm{min}}(\tfrac{\pi}{2},T)
       \approx -\frac{27 [g^{\mathcal{US}}_{\rm D}]^{12}}{262144 \; T^3 \pi^4  [\lambda_{\rm D}^{\mathcal{US}}]^3 }\,.
\end{equation}
The development of a potential barrier arises from the interplay between the quadratic and cubic terms, with the latter eventually dominating as $\varphi$ grows. By neglecting the quartic term contribution, which is a good approximation for field values around the potential barrier, the maximisation condition yields
\begin{equation}\label{eqn:max}
 \begin{aligned}
      \varphi_\mathrm{max} &\approx -\frac{2 d(\delta,T)}{3 e(\delta,T)}\\
       &=\frac{16 \pi \left( [\mu_{\rm D}^\mathcal{US}]^2 \sin^2\delta  + [\mu_{\rm H}^\mathcal{US}]^2 \cos^2\delta \right)}{\sqrt{T} \left( 3 [g^{\mathcal{US}}_{\rm D}]^3 \sin ^3(\delta ) + \left[2 [g^{\mathcal{US}}_{\rm W}]^3+\left([g^{\mathcal{US}}_{\rm W}]^2
   +[g^{\mathcal{US}}_{\rm Y}]^2\right)^{3/2} \right] \cos ^3(\delta ) \right) }\,.
    \end{aligned}
\end{equation}
For scenario I the latter simplifies to
\begin{equation}\label{eqn:max_I}
       \varphi_\mathrm{max} \approx \frac{16 \pi [\mu_{\rm D}^\mathcal{US}]^2}{3 \sqrt{T}  [g^{\mathcal{US}}_{\rm D}]^3}\,.
\end{equation}
Last but not least, the height of the potential barrier can be quantified as
\begin{equation}\label{eqn:height}
    \begin{aligned}
       V_{\varphi_\mathrm{max}}(\delta,T) &\approx \frac{4 d(\delta,T)^3}{27 e(\delta,T)^2}\\
       &=\frac{128 \pi^2 \left( [\mu_{\rm D}^\mathcal{US}]^2 \sin^2\delta  + [\mu_{\rm H}^\mathcal{US}]^2 \cos^2\delta \right)^3}{3\,T^3 \left( 3 [g^{\mathcal{US}}_{\rm D}]^3 \sin ^3(\delta ) + \left[2 [g^{\mathcal{US}}_{\rm W}]^3+\left([g^{\mathcal{US}}_{\rm W}]^2
   +[g^{\mathcal{US}}_{\rm Y}]^2\right)^{3/2} \right] \cos ^3(\delta ) \right)^2}\,,
    \end{aligned}
\end{equation}
whereas for scenario I
\begin{equation}\label{eqn:height_I}
       V_{\varphi_\mathrm{max}}(\tfrac{\pi}{2},T)
       \approx\frac{128 \pi^2 [\mu_{\rm D}^\mathcal{US}]^6}{27\, T^3 [g^{\mathcal{US}}_{\rm D}]^6}\,.
\end{equation}
The potential barrier and the true vacuum are highly sensitive to variations in $d(\delta,T)$, $e(\delta,T)$, and $\lambda(\delta,T)$, as evidenced by the exponents in \cref{eqn:min_depth,eqn:height}. Concerning the ultrasoft parameters, the true vacuum scales with the twelfth power of $g^{\mathcal{US}}_{\rm D}$ and the inverse third power of $\lambda^{\mathcal{US}}_{\rm D}$. In contrast, the height of the potential barrier scales with the sixth power of $\mu^{\mathcal{US}}_{\rm D}$ and the inverse sixth power of $g^{\mathcal{US}}_{\rm D}$. Consequently, small variations in these parameters will be significantly amplified, suggesting that the parameter space region featuring a FOPT is expected to be relatively narrow, especially concerning the values of $\g{D}$, as we will see below.

\section{Results}\label{sec:results}

\subsection{Scenario I} \label{subsec:Scenario_I}

\begin{figure}[htb!]
    \includegraphics[width=0.50\textwidth]{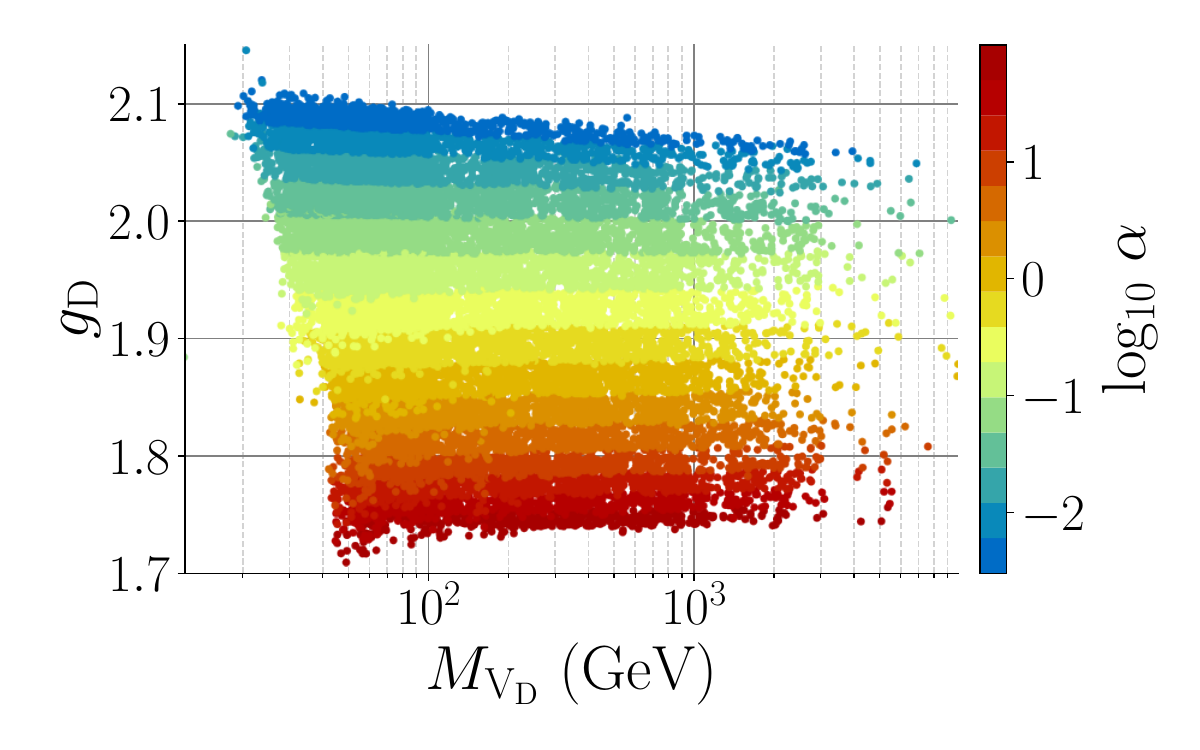}
    \includegraphics[width=0.50\textwidth]{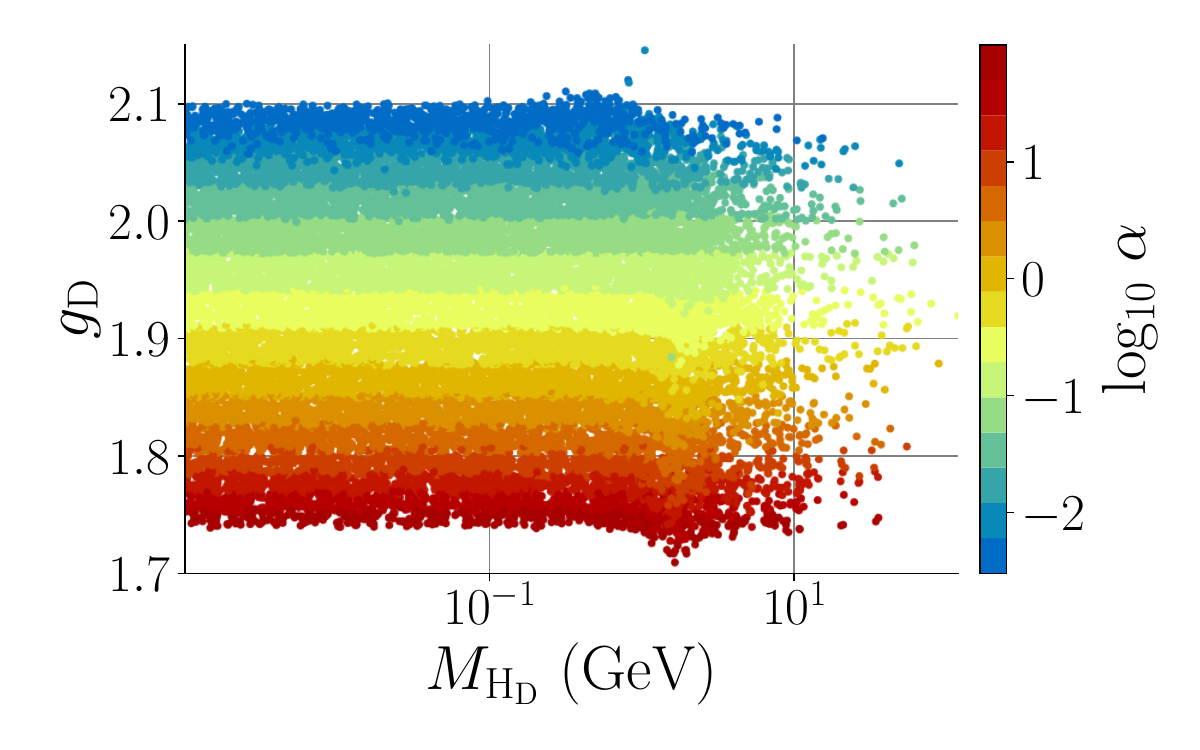}
    \caption{\footnotesize The colour map of the value of the phase transition strength, $\log_{10}\alpha$, for the 2D projections of the three-dimensional parameter scan for Scenario I: the $\g{D}$ versus $M_{\rm V_{D}}$ (left) and the $\g{D}$ versus $M_{\rm H_{D}}$ (right).}
	\label{fig:GW_plots_couplings}
\end{figure}

\begin{figure*}[htb!]
    \centering
    {\includegraphics[width=0.50\textwidth]{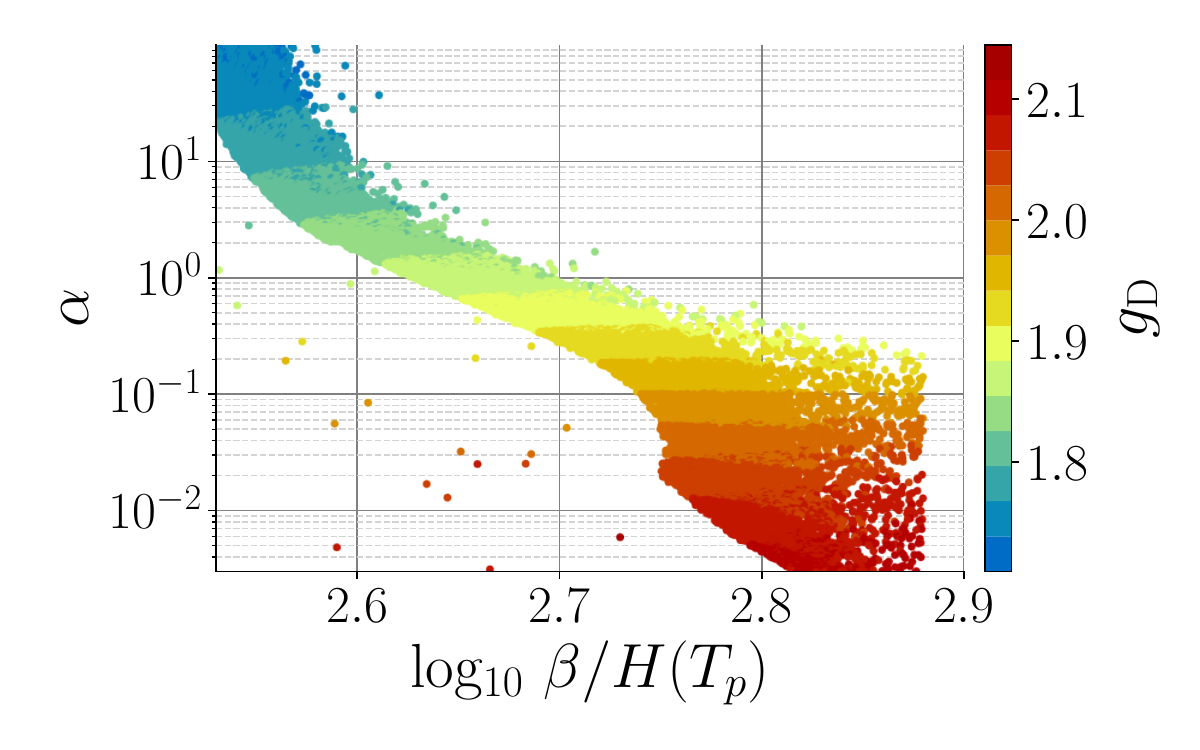}}{\includegraphics[width=0.50\textwidth]{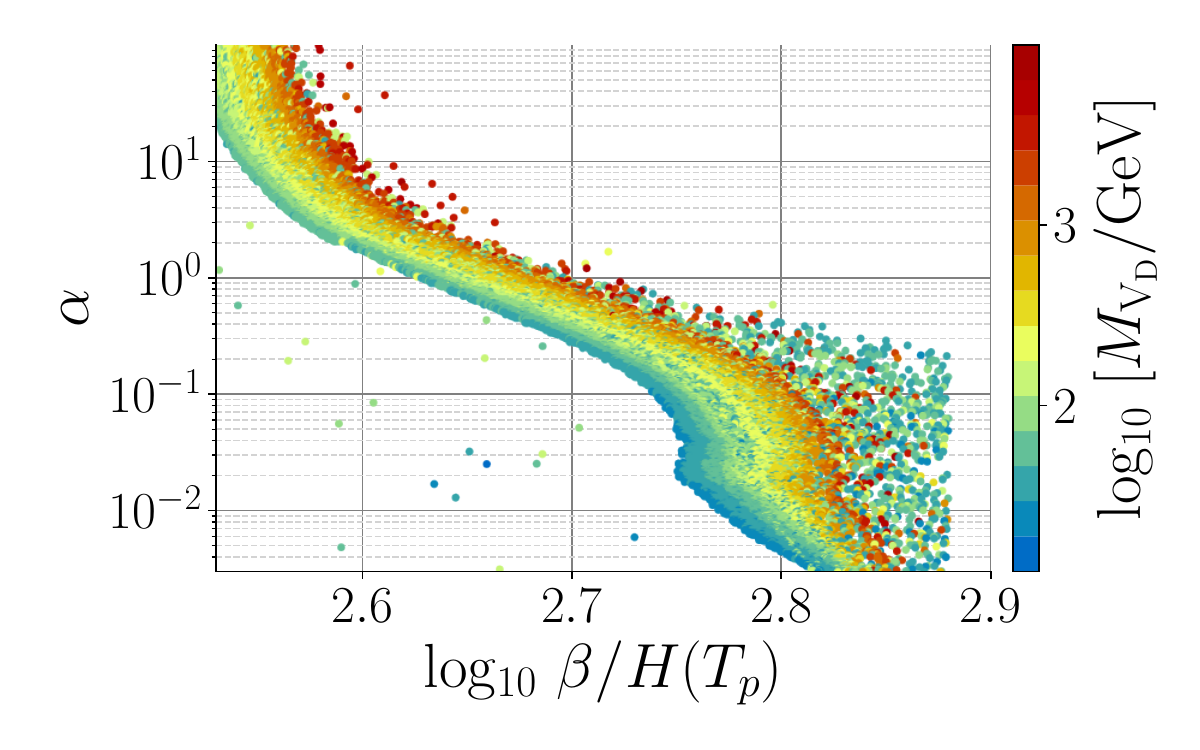}}
    \caption{\footnotesize The colour map of the 2D projections of the three-dimensional parameter scan for Scenario I onto the $(\alpha, \beta/H(T_p))$ plane: $g_\mathrm{D}$ (left) and $M_{\rm V_{D}}$ (right).}
	\label{fig:beta_alpha}
\end{figure*}

The results of our parameter space scan are illustrated in \cref{fig:GW_plots_couplings}, with the $\g{D}$ vs.~$M_\mathrm{V_D}$ projection shown in the left panel and the $\g{D}$ vs.~$M_\mathrm{H_D}$ plane in the right panel. The colour scale indicates the strength of the phase transition, $\alpha$, highlighting that larger values of $\alpha$ correspond to smaller values of the gauge coupling $\g{D}$. This underscores that the phase transition dynamics is primarily driven by gauge interactions.
In general, for the three input parameters in \cref{tab:sample1}, we have found that FOPTs can occur within a relatively narrow range of the dark gauge coupling, \textit{i.e.}, $1.7 \lesssim \g{D} \lesssim 2.1$. The range for the dark boson masses is much wider for the scalar, spanning over nine orders of magnitude, $10^{-8} < M_\mathrm{H_D}/\mathrm{GeV} \lesssim 10$, compared to the vector, $20 \lesssim M_\mathrm{V_D} /\mathrm{GeV}\lesssim 10^4$. Note that the lower limit on the dark Higgs mass corresponds to the smallest value considered in the sampling. In \cref{fig:beta_alpha}, we show the regions featuring FOPTs in the plane of the thermodynamic parameters $\beta/H(T_p)$ and $\alpha$. 
On the left, the colour scale represents the $\SU{2}{D}$ gauge coupling, while on the right, it shows the dark vector mass $M_\mathrm{V_D}$. It can be observed that, for a fixed value of $\beta/H(T_p)$, the phase transition strength is controlled by $\g{D}$, whereas for a fixed value of $\alpha$, the phase transition timescale is determined by $M_\mathrm{V_D}$.

\begin{figure*}[htb!]
	\centering
	\subfloat{\includegraphics[width=1.0\textwidth]{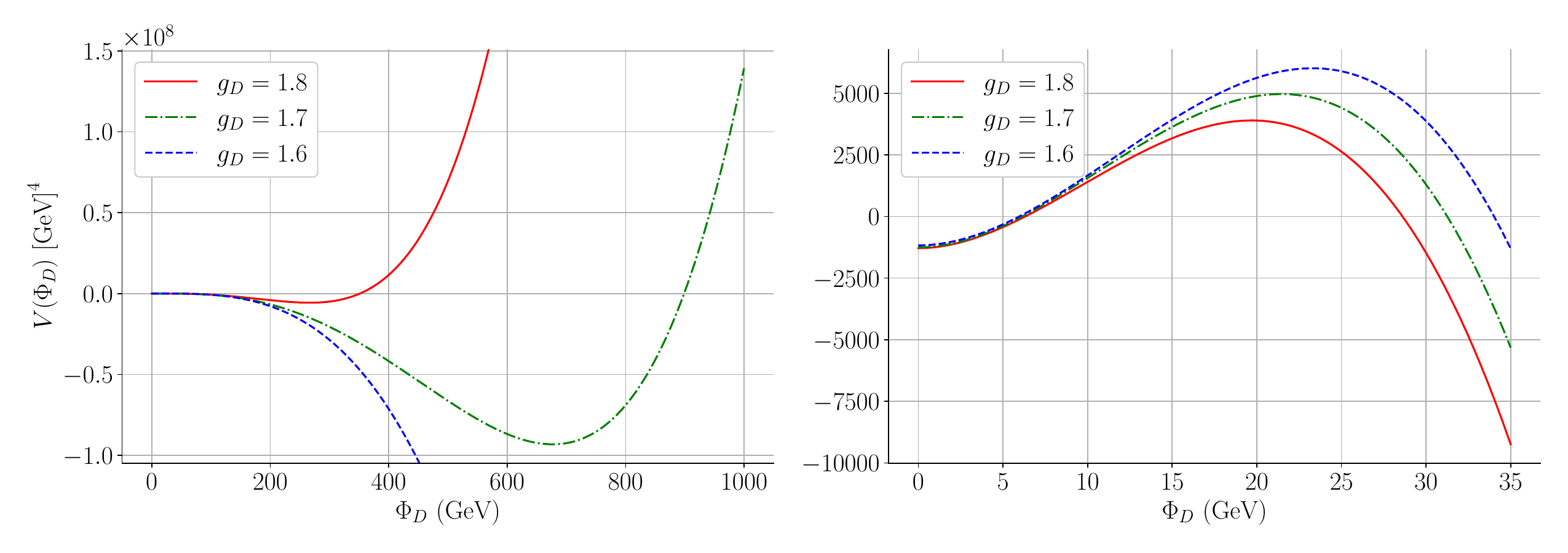}} 
    \caption{\footnotesize Snapshots of the 4D effective potential ($V^{\mathrm{4D}} = TV^{\mathrm{3D}}$) for different values of the gauge coupling in Scenario I at $T=15.6~\mathrm{GeV}$. The left panel shows the potential near the false vacuum and close to the barrier, while the right panel depicts it near the true vacuum. Here, the free parameters are fixed to $M_{\mathrm{V_D}} = 98.75~\mathrm{GeV}$ and $M_\mathrm{H_D} = 0.17~\mathrm{GeV}$.}
	\label{fig:GW_potential_gD}
\end{figure*}

To better understand the impact of the gauge coupling on the phase transition, we examine three snapshots of the potential for distinct values of the gauge coupling at a temperature $T=15.6~\mathrm{GeV}$ in \cref{fig:GW_potential_gD}. The right panel zooms in on small field values to visualise the potential barrier, while the left panel shows the entire range of field values to capture different minima, if they exist at finite field values. The latter is not the case for $\g{D} = 1.6$ where the potential becomes unbounded. 

From this analysis, we observe that as the gauge coupling decreases, the potential around the true vacuum deepens, resulting in an increased potential difference between the true and false vacua, $\Delta V_\mathrm{eff}$, which implies a larger value of the $\alpha$ parameter.  Additionally, examining the behaviour of the barrier reveals that lower coupling values lead to a higher barrier, consequently lowering the $\beta/H(T_p)$ parameter. This trend is consistent with other models where the gauge coupling plays a dominant role in phase transitions (see, $e.g.$, \cite{Ellis:2020nnr,Kierkla:2022odc,Fujikura:2023lkn,Marzo:2018nov}) and aligns with the discussion in the previous section. 

\begin{figure*}[htb!]
	\centering
	\subfloat{\includegraphics[width=0.6\textwidth]{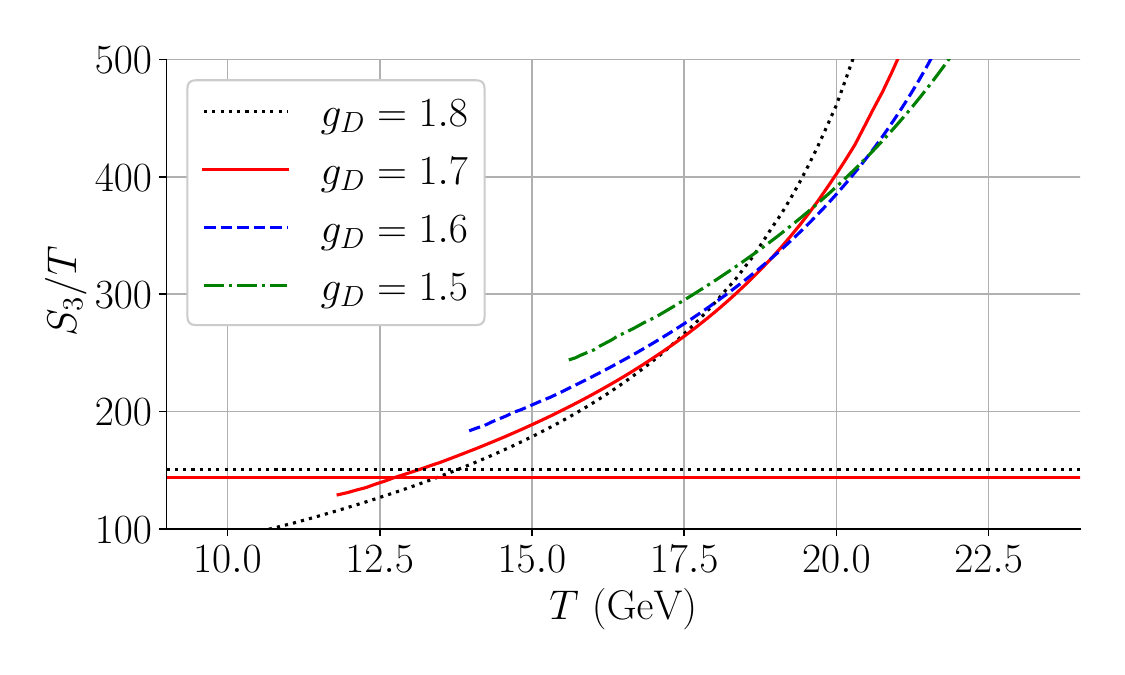}} 
    \caption{\footnotesize Snapshots of the bounce action normalised to the temperature, $S_3/T$, as a function of $T$ for three distinct values of $\g{D}$. The initial value of each curve represents the temperature below which the potential becomes unbounded from below. Here, the free parameters are fixed to $M_{\mathrm{V_D}} = 78.48~\mathrm{GeV}$ and $M_\mathrm{H_D} = 7.96\times 10^{-7}~\mathrm{GeV}$.}
	\label{fig:Action_snapshots}
\end{figure*}

Even for smaller values of $\g{D}$, a minimum in the scalar potential can develop at higher temperatures. However, the bounce action $S_3/T$ becomes increasingly larger, preventing the nucleation and percolation conditions from being met at any temperature. This is illustrated in \cref{fig:Action_snapshots}, where we plot the action as a function of temperature for different values of the gauge coupling $\g{D}$. 
The initial values of each curve correspond to the temperature below which the potential becomes unbounded from below. The horizontal lines represent the values of $S_3/T$ at which nucleation occurs, with their colour and line style matching those of the corresponding $S_3/T$ curve. Specifically, for $\g{D} = 1.5$ (green dot-dashed curve), the ultra-soft $\lambda^\mathcal{US}_D$ coupling becomes negative at $T \approx 15.5~\mathrm{GeV}$, while for $\g{D} = 1.6$ (blue dashed curve), the inflection point occurs at $T \approx 13~\mathrm{GeV}$. In accordance with our numerical results, we find that only for $\g{D} \gtrsim 1.7$ is there a solution for nucleation at $12.6~\mathrm{GeV}$, as indicated by the horizontal line. Conversely, perturbativity limits $\g{D}$ from above, \textit{i.e.}, at the scale $\mu = \pi T_p$, where $\g{D}(\pi T_p) > 4\pi$, rendering our calculation no longer reliable. 

\begin{figure*}[htb!]
    {\includegraphics[width=0.50\textwidth]{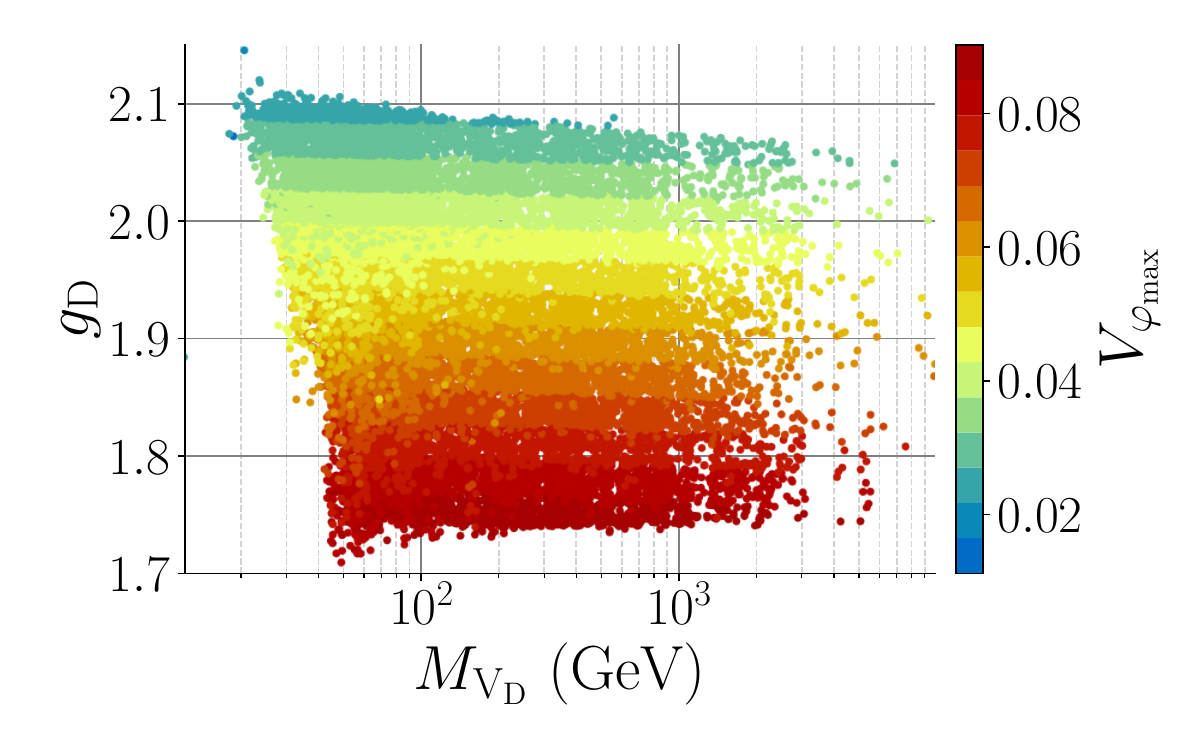}}
    {\includegraphics[width=0.50\textwidth]{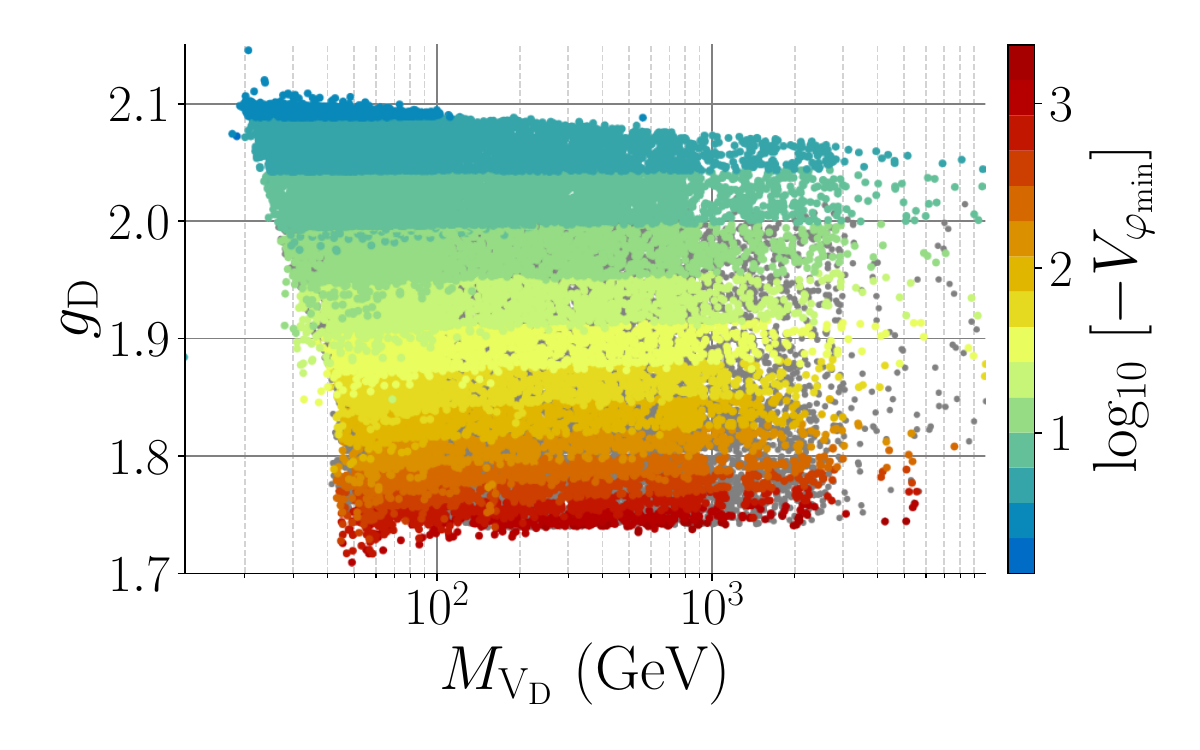}} 
    \caption{\footnotesize The colour map for the height of the potential barrier $V_{\varphi_\mathrm{max}}$ (left) and the depth of the true vacuum $V_{\varphi_\mathrm{min}}$ (right) projected in the $g_\mathrm{D}$ versus $M_{V_\mathrm{D}}$ plane.}
	\label{fig:projections_VmaxVmin}
\end{figure*}

To demonstrate that the discussion above represents the general behaviour at any given scale, we present in \cref{fig:projections_VmaxVmin} the height of the potential barrier, $V_{\varphi_\mathrm{max}}$ from \cref{eqn:height_I} (left panel), and the depth of the broken phase minimum, $V_{\varphi_\mathrm{min}}$ from \cref{eqn:min_depth_I} (right panel), projected onto the $M_\mathrm{V_D}$ vs.~$\g{D}$ plane.
As anticipated from the snapshots in \cref{fig:GW_potential_gD,fig:Action_snapshots}, the potential value at the barrier increases for smaller values of the $\SU{2}{D}$ gauge coupling, which in turn implies a lower $\beta/H(T_p)$, while the true vacuum becomes deeper, resulting in a larger $\alpha$. Recall that both $V_{\varphi_\mathrm{max}}$ and $V_{\varphi_\mathrm{min}}$ are defined as dimensionless quantities, with their values normalised for the different temperature scales across the entire $M_\mathrm{V_D}$ range. Indeed, the FOPT temperature is determined by the mass of the dark vector. Otherwise, these images would primarily reflect the temperature dependence of the scalar potential with a horizontal colour gradient, obscuring the relationship between the height of the potential barrier, the depth of the true minimum and the gauge coupling. One can also observe that small changes in $\g{D}$ lead to significant variations in the depth of the true minimum and, consequently, in the potential energy difference.

\begin{figure*}[htb!]
    \centering
	\subfloat[]{\includegraphics[width=0.50\textwidth]{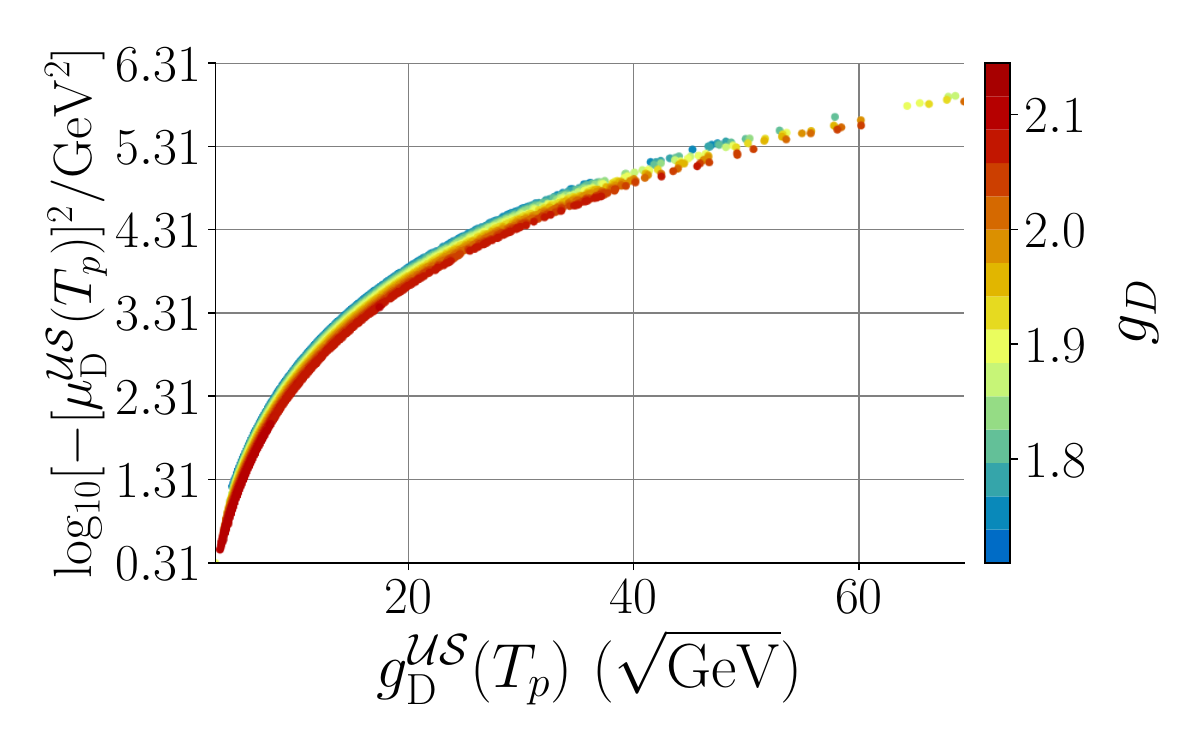} }
    \subfloat[]{\includegraphics[width=0.50\textwidth]{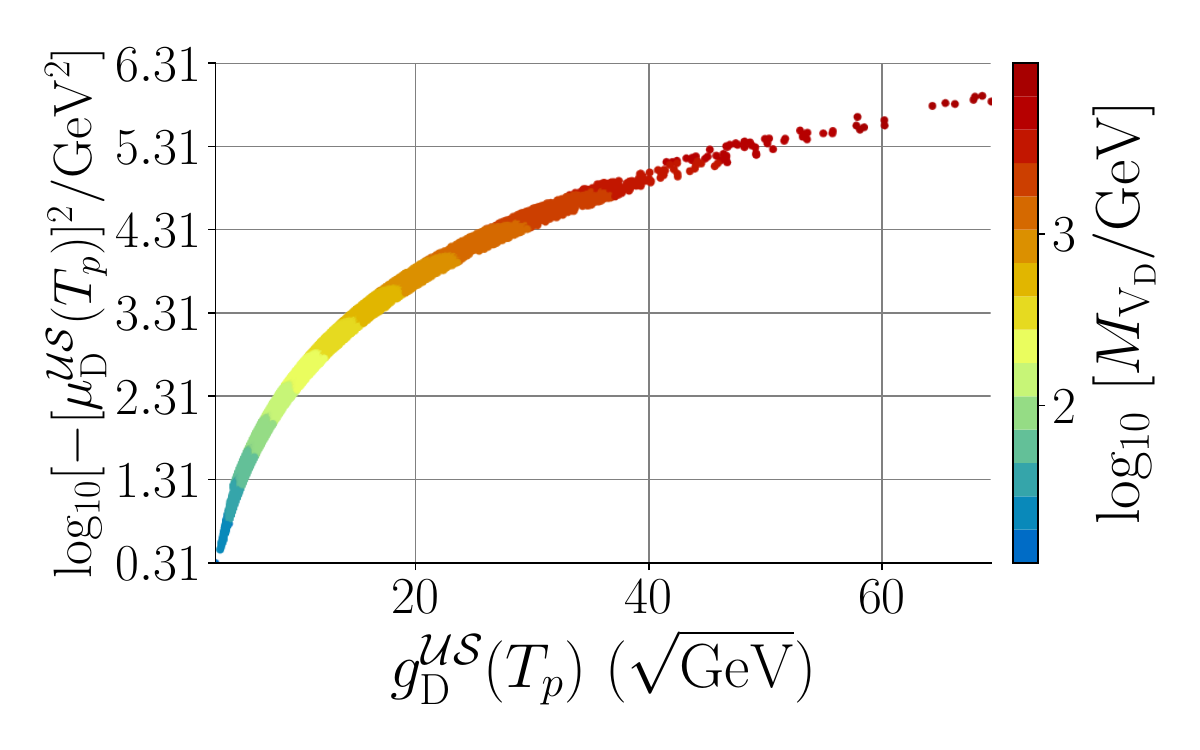}} \\
    \subfloat[]{\includegraphics[width=0.50\textwidth]{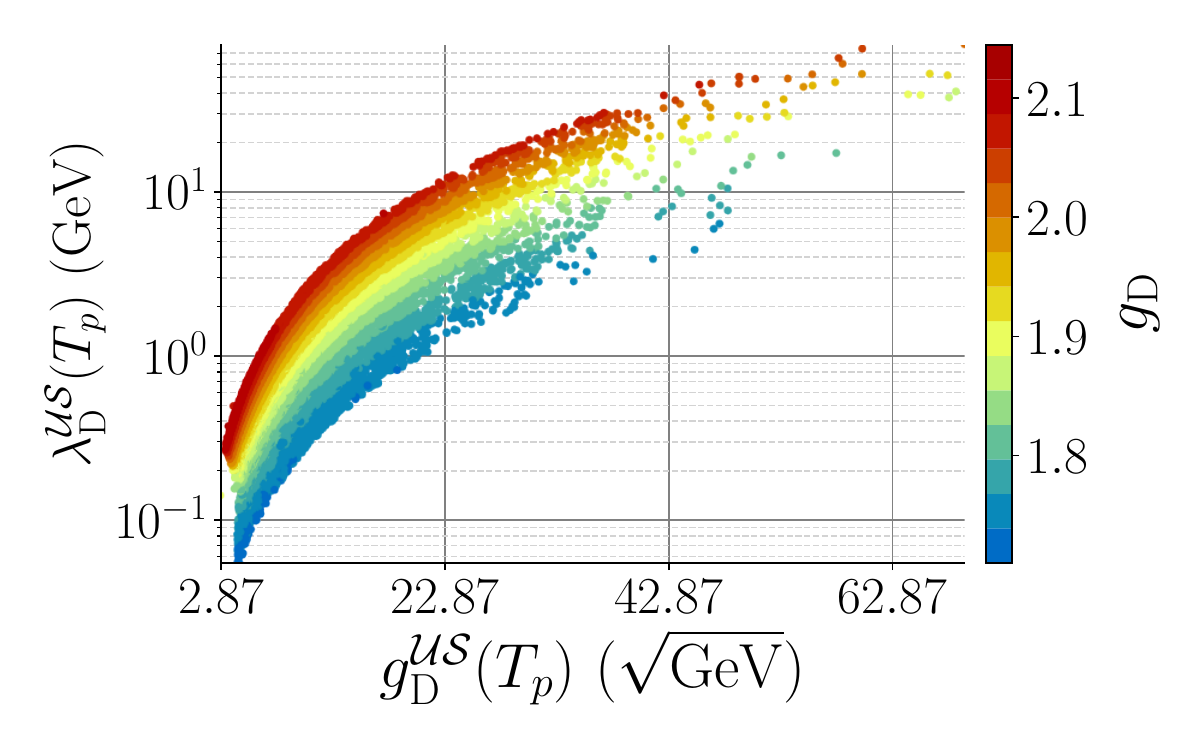}}
    \subfloat[]{\includegraphics[width=0.50\textwidth]{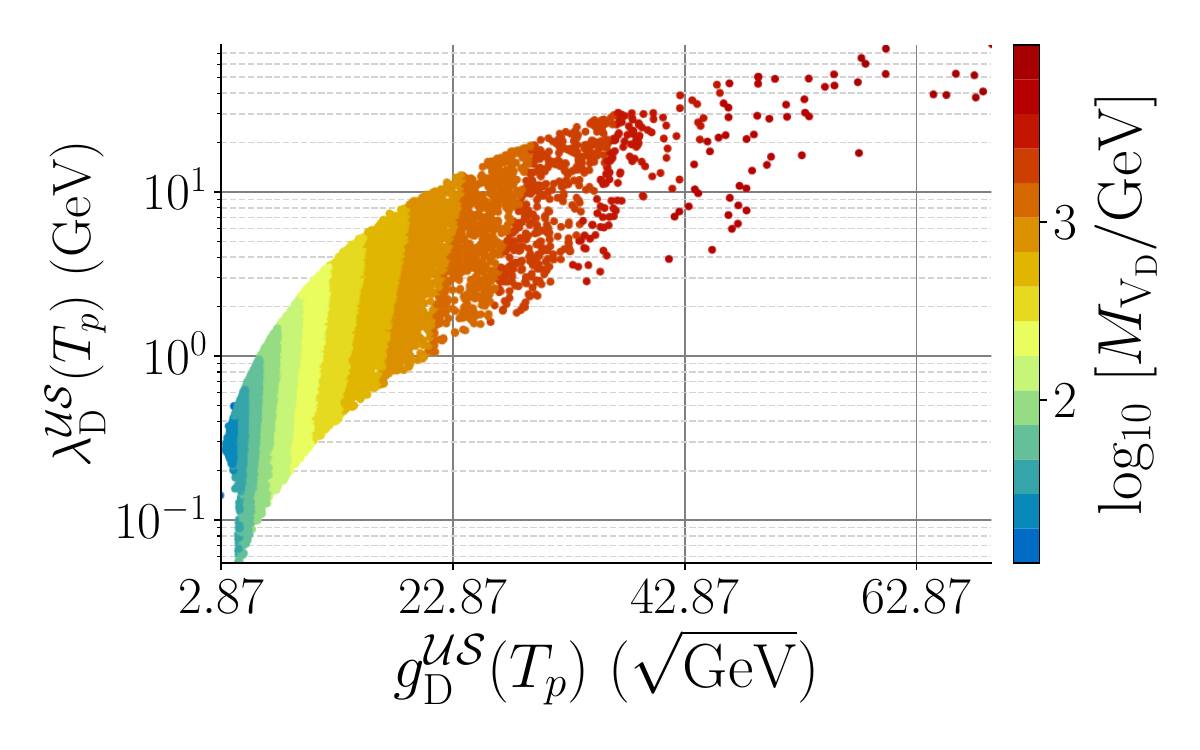}} 
    \caption{\footnotesize 
    The colour map for the value of the gauge coupling $g_\mathrm{D}$ (panels a and c) and the value of the $M_{V_\mathrm{D}}$ (panels b and d) projected onto different temperature-dependent ultrasoft parameters of Scenario I. All ultrasoft parameters are evaluated at the percolation temperature.}
	\label{fig:GW_plots_US}
\end{figure*}

A question that still remains open at this stage is how the gauge coupling $\g{D}$ and the dark vector mass $M_\mathrm{V_D}$ relate to the temperature dependent 3D ultrasoft EFT parameters, which directly drive the phase transition. A detailed understanding of the observed behaviour can be further scrutinised in the context of Scenario-I, whose relatively simple structure is particularly convenient for this purpose. The scalar potential used in our numerical analysis is expressed in terms of the ultrasoft parameters discussed in \cref{sec:FOPT}. In the first row of \cref{fig:GW_plots_US}, we illustrate how $\g{D}^{\mathcal{US}}$ and $[\mu^\mathcal{US}_{\rm D}]^2$ (evaluated at $T_p$) depend on the dark gauge coupling $\g{D}$ (left panel) and the mass of the dark gauge boson $M_\mathrm{V_D}$ (right panel). The colour gradient in panel (b) indicates that the mass of the vector field determines the magnitude of both the ultrasoft dark gauge coupling and $[\mu^\mathcal{US}_\mathrm{D}]^2$. This is because $M_\mathrm{V_D}$ sets the FOPT temperature. As the ultrasoft parameters are temperature-dependent, they are expected to increase with $M_\mathrm{V_D}$. Additionally, this dependence scales as $[\mu^\mathcal{US}_{\rm D}]^2 \propto T^2$ and $\g{D}^{\mathcal{US}} \propto \sqrt{T}$, which explains why the former grows faster with temperature.
In panel (a), for a given fixed $\g{D}^{\mathcal{US}}$, which is equivalent to fixing the temperature as in \cref{fig:GW_potential_gD}, increasing the fundamental 4D dark gauge coupling $\g{D}$ results in a smaller value of $[\mu^\mathcal{US}_{\rm D}]^2$, as indicated by the colour gradient. Consequently, $V_{\varphi_\mathrm{max}}$ in \cref{eqn:height_I}, with a fixed denominator, decreases with increasing $\g{D}$, indicating that the potential barrier becomes shallower. As a result, the FOPT will proceed more quickly, leading to a larger $\beta/H(T_p)$.

In the bottom row of \cref{fig:GW_plots_US}, we show the dependence of $\g{D}^{\mathcal{US}}$ and $[\lambda^\mathcal{US}_{\rm D}]$ on the physical parameters $\g{D}$ (left panel) and the mass of the dark gauge boson $M_{\rm V_D}$ (right panel). In panel (d), we again observe that $\g{D}^{\mathcal{US}}$ increases with increasing $M_{\rm V_{D}}$. The main difference compared to the top row is that, for a fixed value of $\g{D}^{\mathcal{US}}$, decreasing the gauge coupling results in an increase in $\lambda^{\mathcal{US}}_{\rm D}$. Thus, according to \cref{eqn:min_depth_I} with fixed numerator, a smaller value of $\g{D}$ yields a deeper true vacuum, $V_{\varphi_\mathrm{min}}$, which increases the potential energy difference between the true and false vacua, $\Delta V_\mathrm{eff}$, thereby enhancing the strength of the phase transition $\alpha$, as noted in \cref{fig:GW_plots_couplings,fig:beta_alpha}.

\begin{figure*}[htb!]
    \centering
 	\subfloat[]{\includegraphics[width=0.50\textwidth]{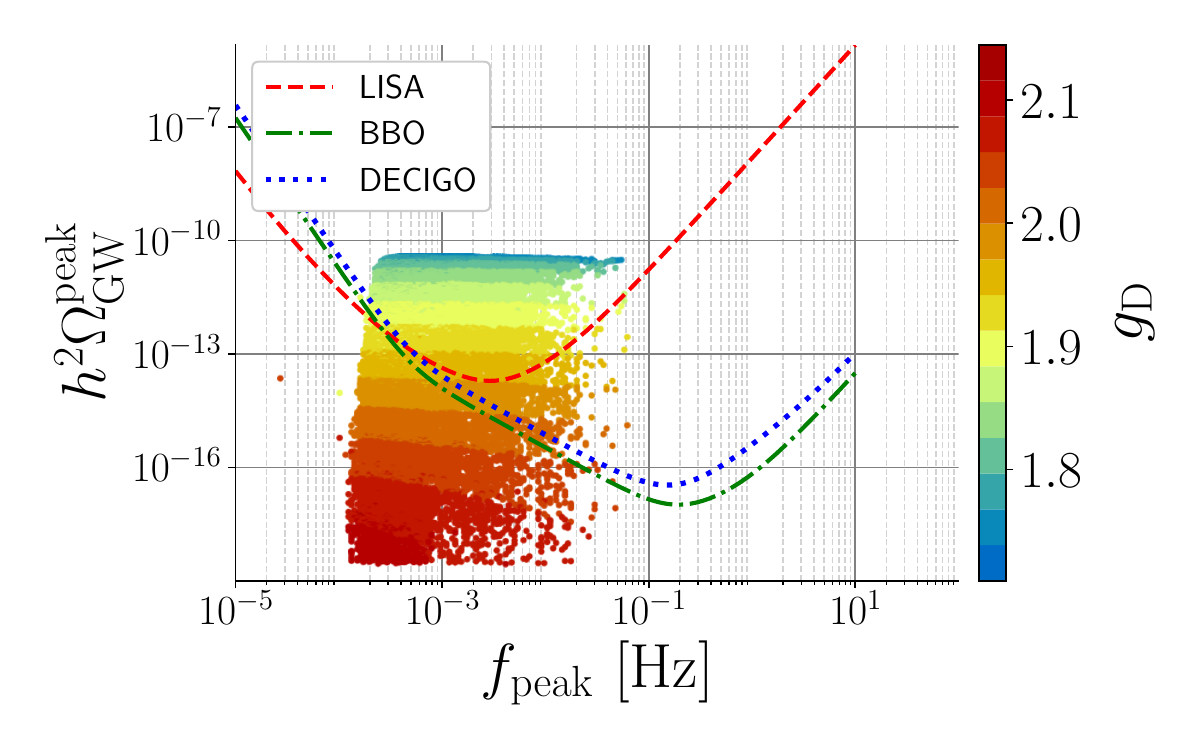}}
    \subfloat[]{\includegraphics[width=0.50\textwidth]{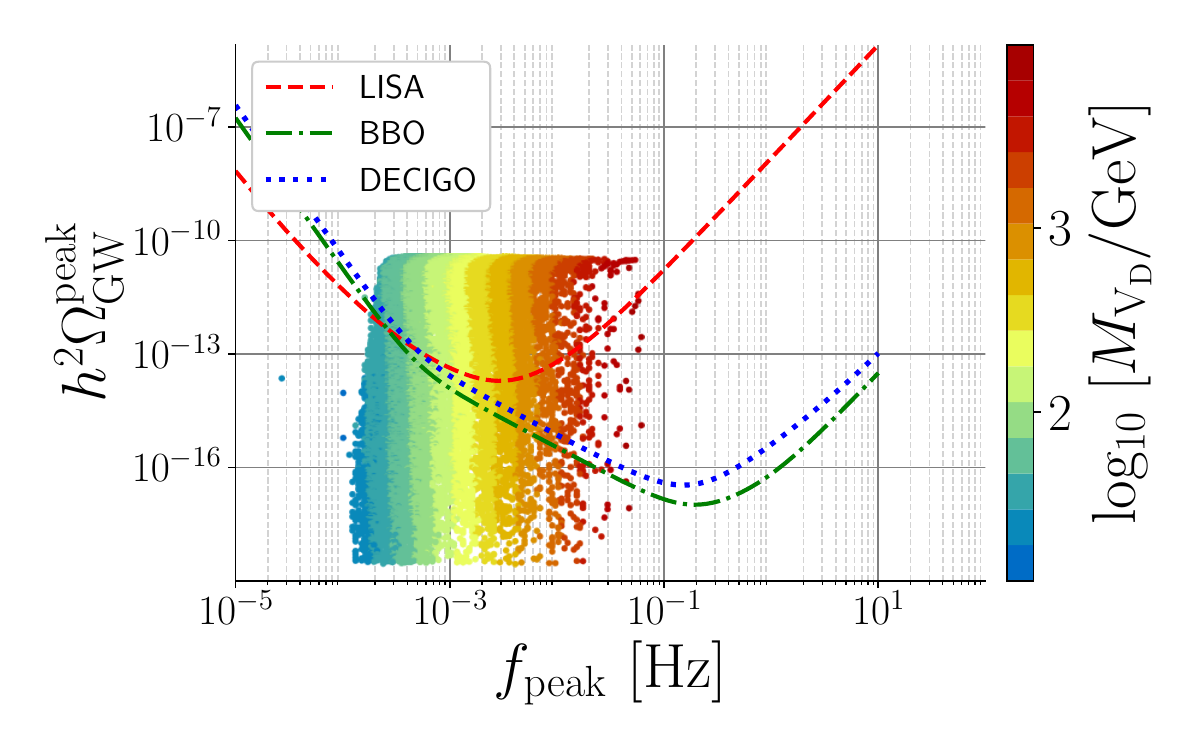}}\\
    \subfloat[]{\includegraphics[width=0.50\textwidth]{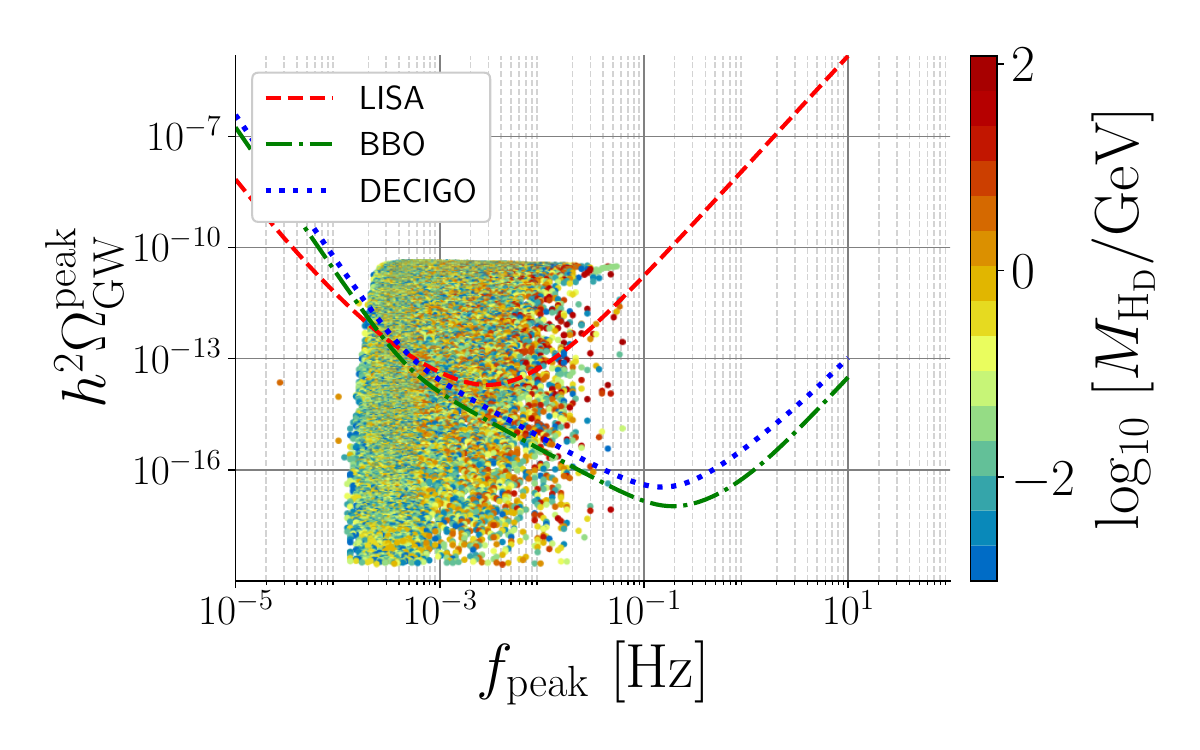}}
    \subfloat[]{\includegraphics[width=0.50\textwidth]{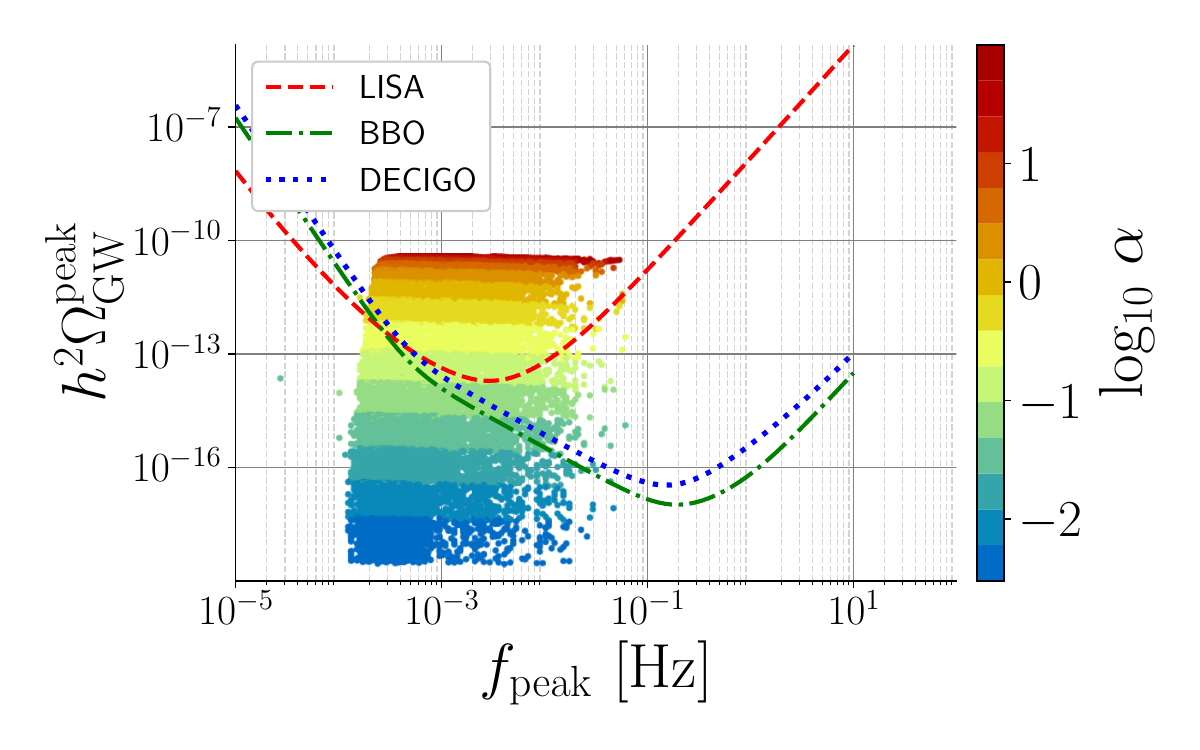}} \\
    \caption{\footnotesize Predictions for the SGWB geometric parameters $h^2 \Omega^\mathrm{peak}_\mathrm{GW}$ and $f_{\mathrm{peak}}$ for scenario I in terms of the $\SU{2}{D}$ gauge coupling (a), the dark vector boson mass (b), the dark Higgs mass (c) and the logarithm of the phase transition strength $\alpha$ (d). The red dashed, green dot-dashed, and blue dotted curves represent the $s$-channel peak-integrated sensitivity curves (PISCs) for LISA, BBO, and DECIGO, respectively \cite{Schmitz:2020syl}.}
	\label{fig:GW_plots_spectra_modelI}
\end{figure*}

We have so far understood how the fundamental theory parameters $\g{D}$ and $M_\mathrm{V_D}$ affect the vacuum structure and the phase transition thermodynamics. The ultimate goal of this analysis is to determine how these parameters translate into SGWB predictions. In \cref{fig:GW_plots_spectra_modelI}, we present scatter plots of the SGWB peak frequency $f_\mathrm{peak}$ and the peak energy density amplitude $h^2 \Omega^\mathrm{peak}_\mathrm{GW}$ in terms of the fundamental parameters $\g{D}$ (a) and the logarithm of $M_\mathrm{V_D}$ (b), the logarithm of $M_{\rm H_D}$ (c) and the logarithm of $\alpha$ (d). 
In essence, we conclude that the dark gauge coupling primarily controls the peak amplitude of the SGWB, while the dark vector boson mass dictates its peak frequency. Furthermore, we observe that for $1.7 \lesssim \g{D} \lesssim 2.0$ and $100~\mathrm{GeV} \lesssim M_\mathrm{V_D} \lesssim 10~\mathrm{TeV}$, the resulting SGWB falls within the sensitivity range of LISA and future planned interferometers such as BBO and DECIGO. We find that the dark Higgs mass is largely uncorrelated with the GW spectrum, although a very mild dependency with the frequency is observed, as seen in panel (c).

\subsection{Scenario II} \label{subsec:Scenario_II}

In this section, we focus on Scenario II, where the main difference is that the SM is fully incorporated and communicates with the dark sector through the Higgs portal quartic coupling. The inclusion of the SM in our calculations requires a consistent matching of the 3D EFT to ensure compliance with the SM phenomenology at the EW scale. As shown in \cref{tab:sample2}, the considered size of the dark boson masses in our scan is comparable to the EW scale. Consequently, the phase transition analysis must take into account both the dark and visible Higgs directions.

From here onwards for the sake of brevity we will use the same notations for the finite temperature vevs as for zero temperature ones:
$$v_T\to v \  , \  v_{\mathrm{D}T}\to v_\mathrm{D} \ .$$
In particular, the possible vacuum configurations are: 
\begin{itemize}
    \item Fully symmetric -- $(0,0)$,
    \item Fully broken -- $(v,v_\mathrm{D})$,
    \item EW broken -- $(v,0)$,
    \item EW symmetric -- $(0,v_\mathrm{D})$.
\end{itemize}

\begin{figure*}[htb!]
	\centering
	\subfloat[]{\includegraphics[width=0.50\textwidth]{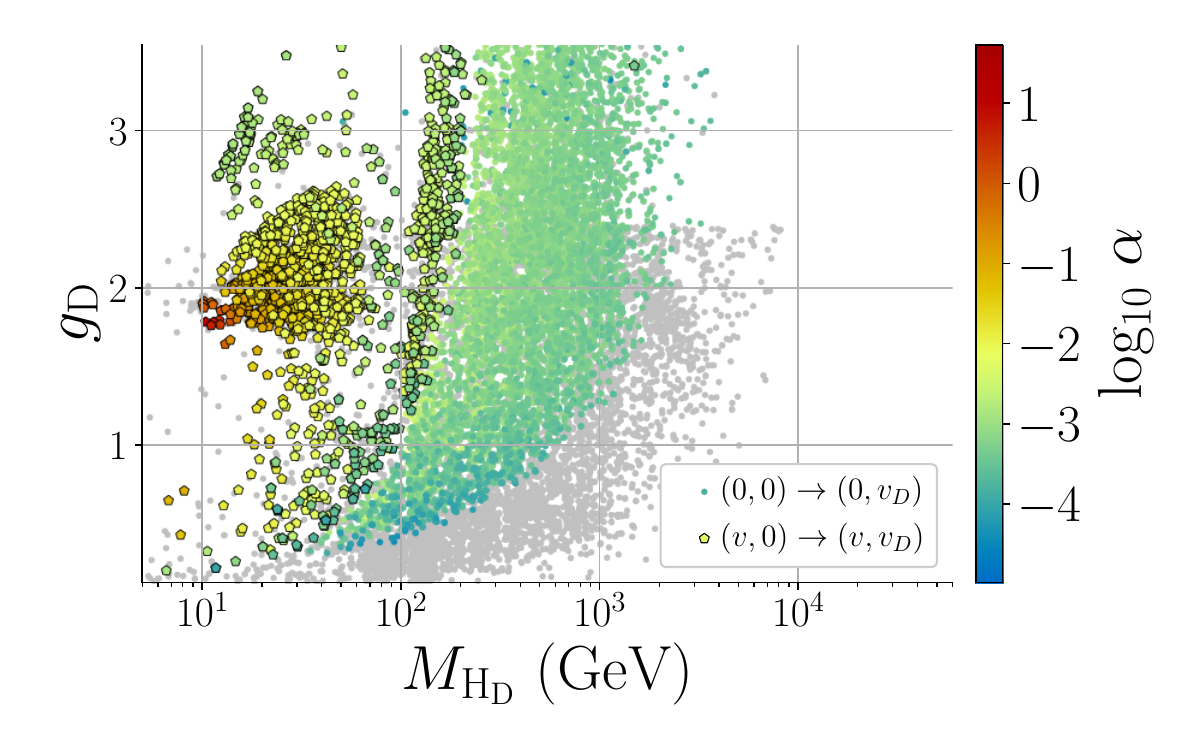}}
    \subfloat[]{\includegraphics[width=0.50\textwidth]{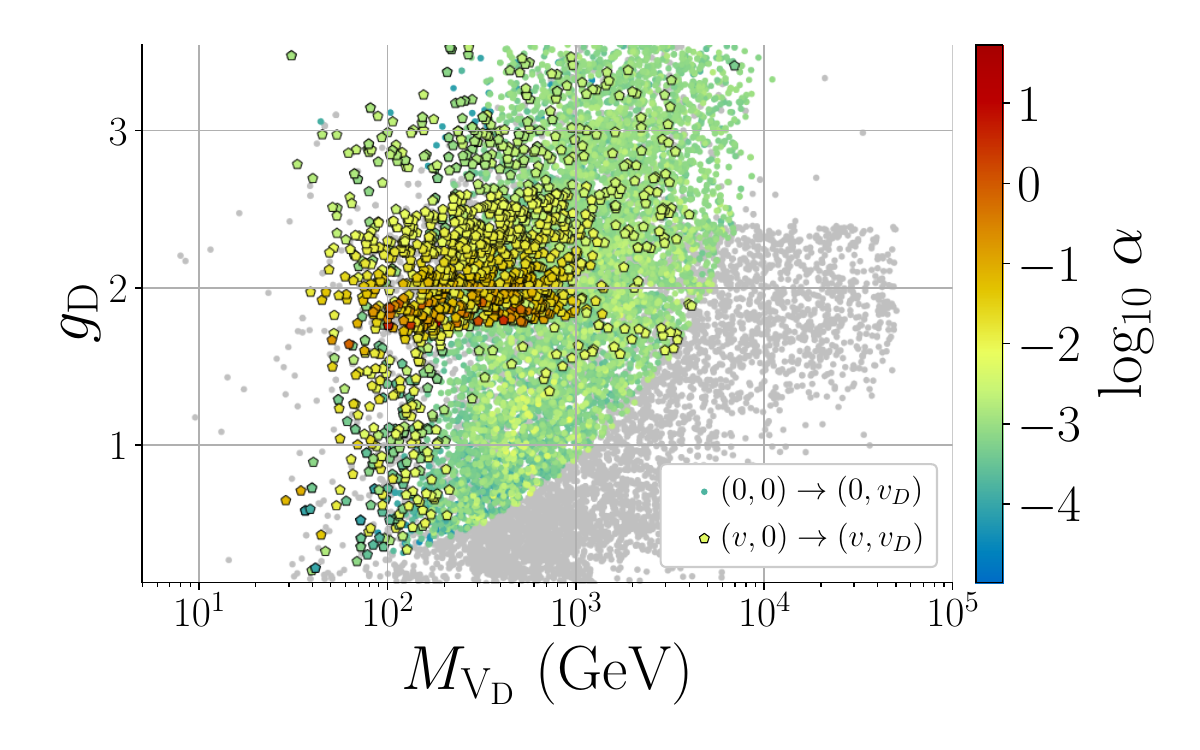}}
    \caption{\footnotesize 
    The colour map for the value of the phase transition strength, $\mathrm{log_{10}}~\alpha$, for the 2D projections of the four-dimensional parameter scan for Scenario II: the $g_\mathrm{D}$ versus $M_{H_\mathrm{D}}$ (left) and $g_\mathrm{D}$ versus $M_{V_\mathrm{D}}$. Points marked in grey are excluded due to DM constraints, while circled and uncircled points represent different phase transition patterns as indicated in the legend.}
	\label{fig:GW_scenarioII}
\end{figure*}

In \cref{fig:GW_scenarioII}, we present the results from our numerical scan, considering two distinct projections of the fundamental 4D theory parameter space.
In both panels, we display the $\SU{2}{D}$ gauge coupling $\g{D}$ against the physical dark-Higgs mass $M_\mathrm{H_D}$ on the left (panel a) and against the vector mass $M_\mathrm{V_D}$ on the right (panel b). The colour gradient represents the phase transition strength $\alpha$. The grey points are predominantly excluded due to excessive DM relic abundance, while direct-detection constraints are also taken into account using our model's \texttt{micrOMEGAs} implementation. In addition, we retain only parameter points satisfying $T_n > T_f$, where $T_f \approx M_{\rm V_D}/30$, since in our scenario the phase transition must occur before DM freeze-out. If $T_n < T_f$, the phase transition occurs too late for the cosmological history assumed in our relic-density analysis, and such points generally lead to  excessive DM abundance. We have explicitly checked that the condition $T_n > T_f$ is satisfied for all parameter points that simultaneously yield potentially observable GW signals and fulfil the DM constraints.

Notably, all grey points in panel (b) that lie to the right of the coloured area exhibit $h^2\Omega_\mathrm{DM} > 0.12$, establishing an upper limit on the dark vector mass of $M_\mathrm{V_D} \lesssim 10~\mathrm{TeV}$.

Two sets of points are identified in both panels, reflecting distinct patterns of two-step phase transitions:
\begin{enumerate}
    \item $(0,0) \xrightarrow[]{\textbf{FOPT}} (0,v_\mathrm{D}) \xrightarrow[]{\text{SOPT}} (v,v_\mathrm{D})$ \,,
    \item $(0,0) \xrightarrow[]{\text{SOPT}} (v,0) \xrightarrow[]{\textbf{FOPT}} (v,v_\mathrm{D})$ \,,
\end{enumerate}
where SOPT denotes a second order phase transition. The first pattern occurs when the dark Higgs is heavy, typically above the EW scale, unless the gauge coupling is small, below $\g{D} < 1.0$. Here, the FOPT involves only the dark direction before the SOPT occurs for the SM-like Higgs field. The second pattern, identified in the figures with a black pentagon around the data-point, occurs in the presence of light dark Higgs bosons with masses below the EW scale, where both dark and visible directions in the field space participate in the FOPT. This pattern reproduces features similar to Scenario I, where stronger FOPTs occur at $\g{D} \approx 1.7$ becoming weaker as $\g{D}$ increases. This is visible in panel (b), which contains the region obtained in panel (a) of \cref{fig:GW_plots_couplings}.

Focusing again on \cref{fig:GW_scenarioII}, we observe in panel (a) that the strength of the phase transition reaches its maximum at $\alpha \sim \mathcal{O}(10)$ (red points) when the dark Higgs boson mass and the gauge coupling converge to a small region where $M_\mathrm{H_D} \approx 10 \ \mathrm{GeV}$ and $1.7 \lesssim \g{D} \lesssim 2.0$. Conversely, weaker FOPTs are found for $M_\mathrm{H_D} \sim 100 \ \mathrm{GeV}$ and $\g{D} \approx 0.5$ with $\alpha \approx 10^{-4}$ (blue points).
\begin{figure*}[htb!]
	\centering
    \subfloat[]{\includegraphics[width=0.50\textwidth]{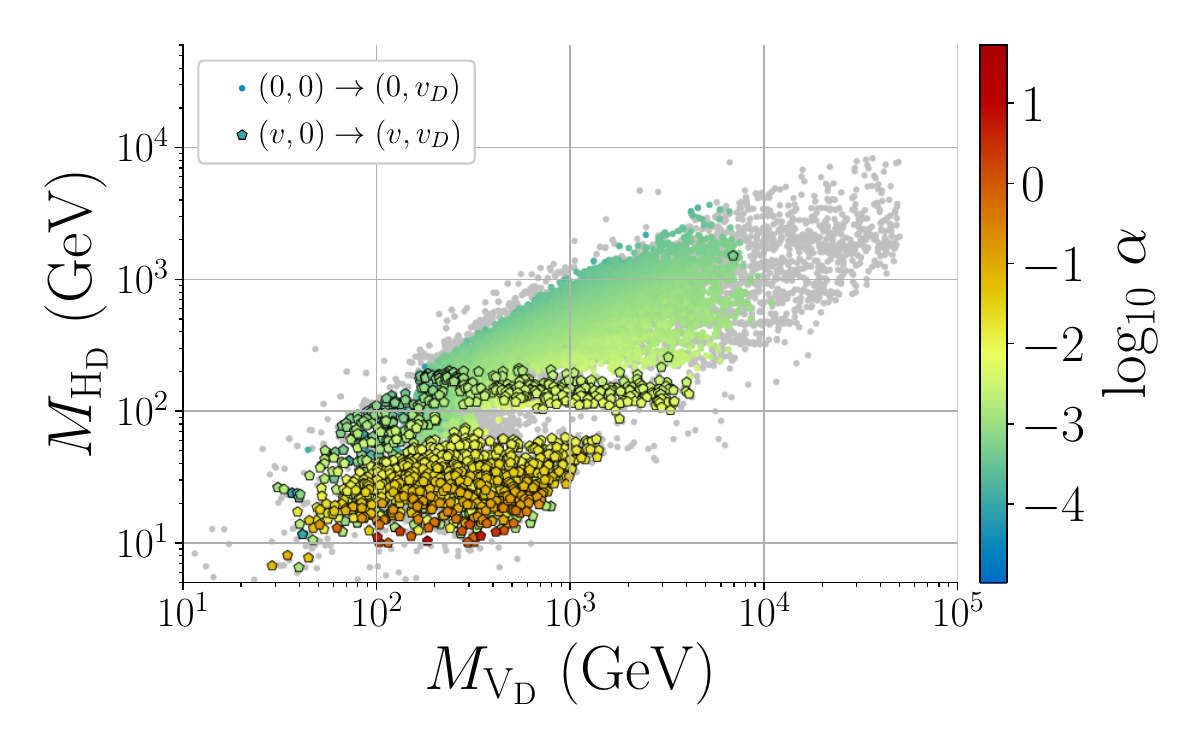}}
    \subfloat[]{\includegraphics[width=0.50\textwidth]{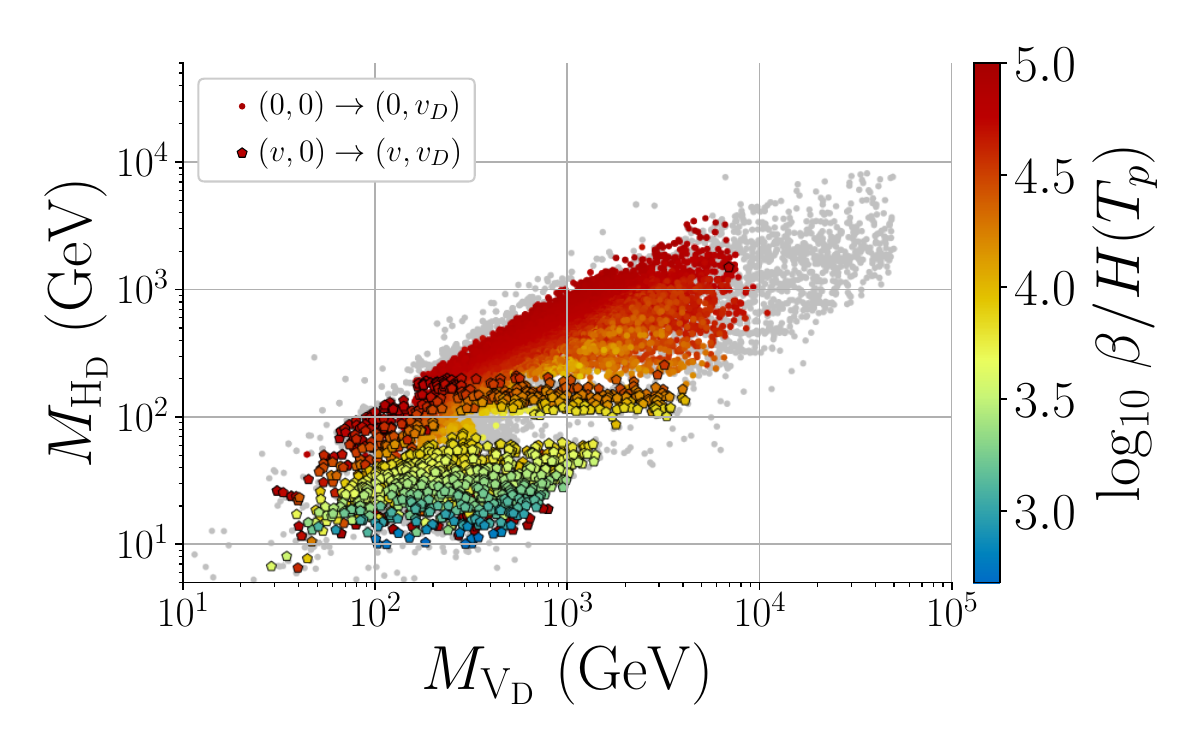}} \\
	\subfloat[]{\includegraphics[width=0.50\textwidth]{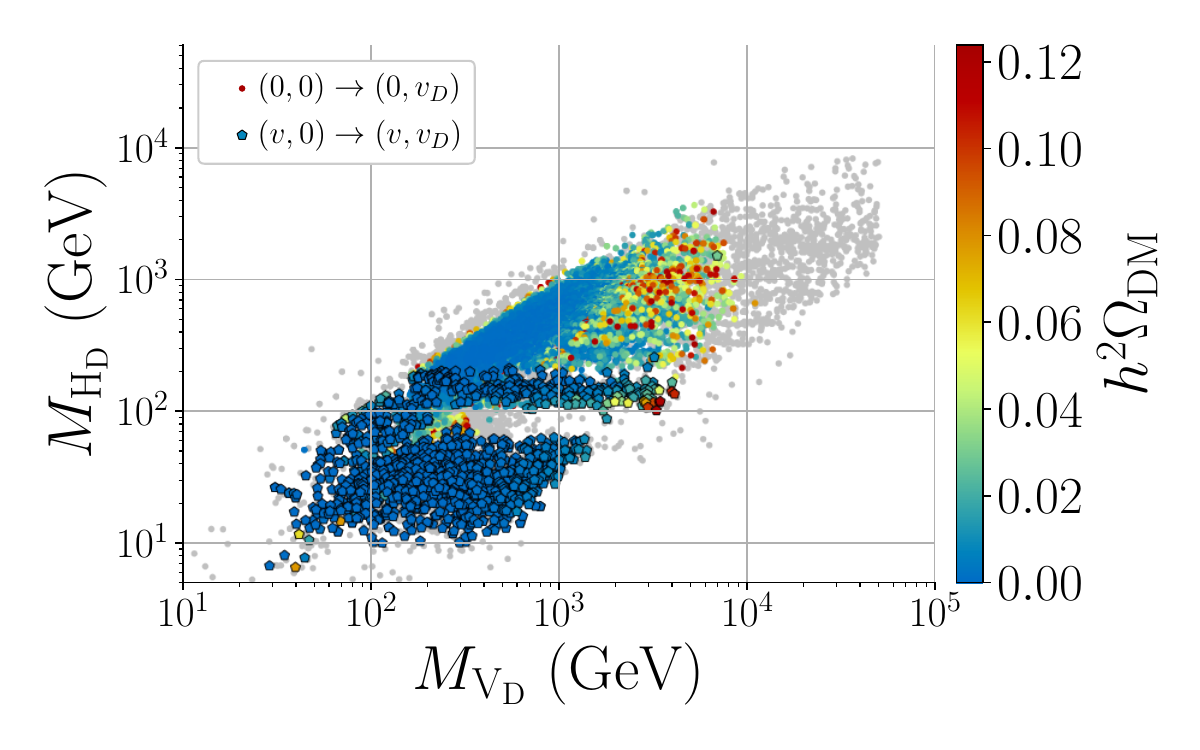}}
    \subfloat[]{\includegraphics[width=0.50\textwidth]{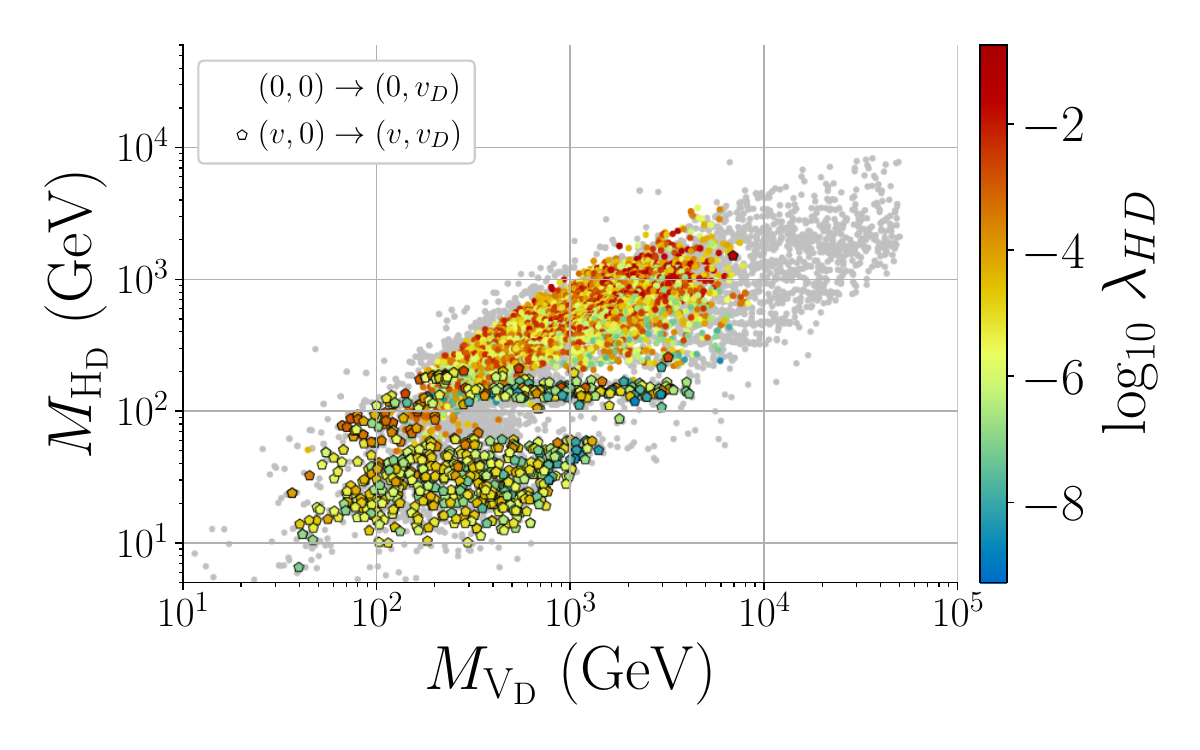}} \\
    \subfloat[]{\includegraphics[width=0.50\textwidth]{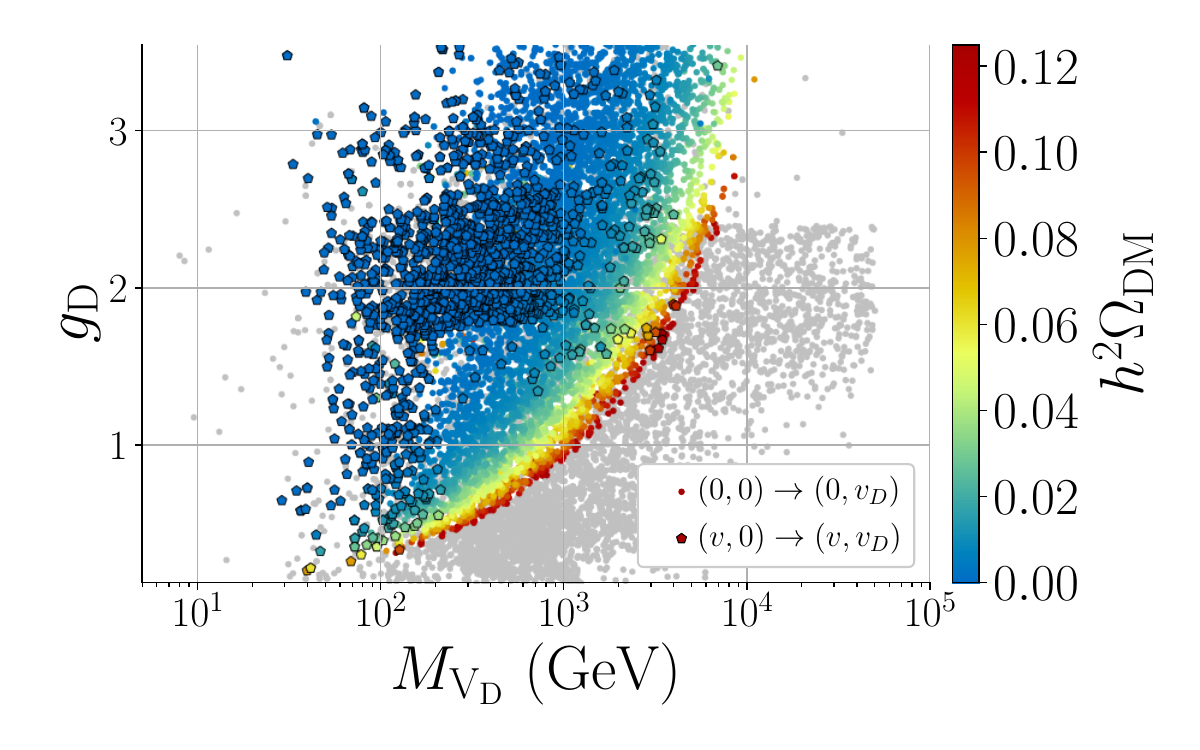}} 
    \subfloat[]{\includegraphics[width=0.50\textwidth]{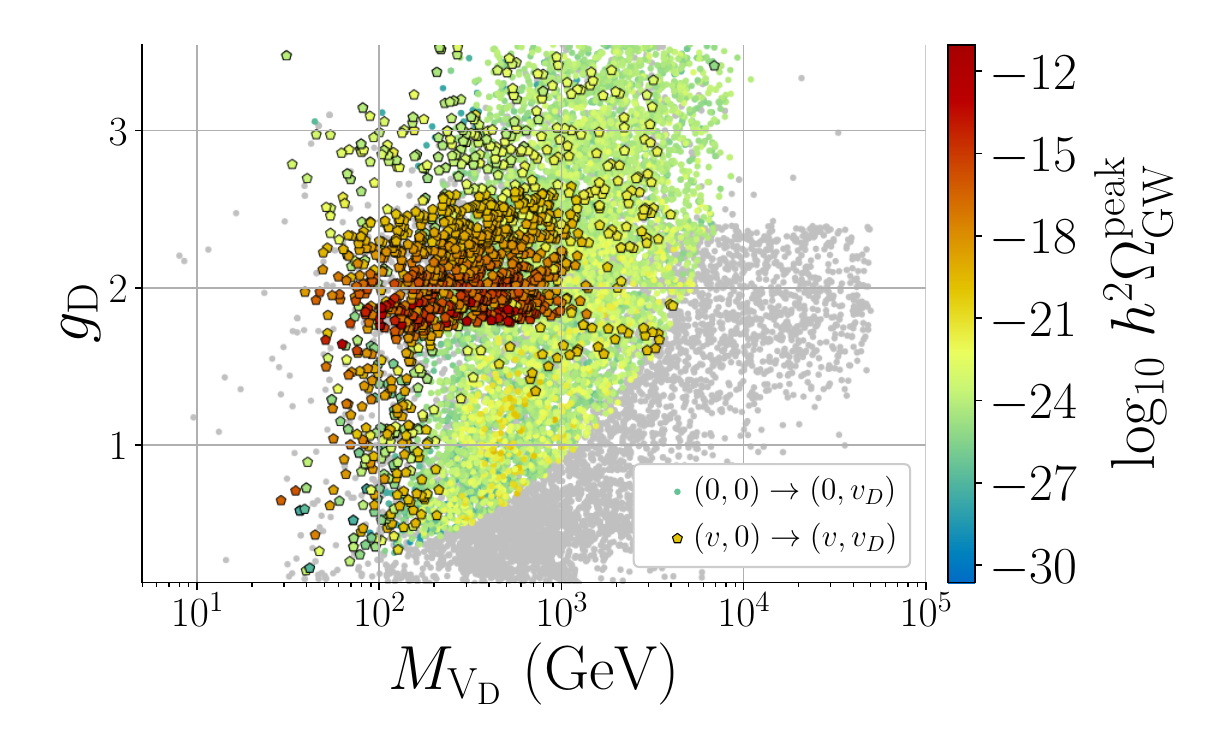}}
    \caption{\footnotesize 
    The colour map for the value of the phase transition strength, $\mathrm{log_{10}}~\alpha$ (panel a), the value of the inverse time duration $\mathrm{log_{10}}~\beta/H(T_p)$ (panel b), the DM relic abundance $h^2 \Omega_\mathrm{DM}$ (panels c and e), the magnitude of the Higgs portal coupling $\lambda_\mathrm{HD}$ (panel d) and the SGWB peak amplitude $h^2 \Omega_\mathrm{GW}^\mathrm{peak}$ (panel f) for the 2D projections of the four-dimensional parameter scan for Scenario II:  $M_{H_\mathrm{D}}$ versus $M_{V_\mathrm{D}}$ (panels a, b, c and d) and $g_\mathrm{D}$ versus $M_{V_\mathrm{D}}$ (panels e and f). Gray, circled, and uncircled points have the same meanings as in \cref{fig:GW_scenarioII}.}
	\label{fig:GW_and_DM_scenarioII_A}
\end{figure*}

In \cref{fig:GW_and_DM_scenarioII_A}, we present the $M_\mathrm{H_D}$ vs.~$M_\mathrm{V_D}$ (top and middle rows) and the $\g{D}$ vs.~$M_\mathrm{V_D}$ (bottom row) projections of the parameter space. The colour gradient in the top-left (a) and top-right (b) panels depicts the thermodynamic parameters $\alpha$ and $\beta/H(T_p)$, respectively. In the middle-left (c) and middle-right (d) panels, it represents the DM relic density $h^2 \Omega_\mathrm{DM}$ and the magnitude of the Higgs portal coupling $\lambda_\mathrm{HD}$, respectively. In the bottom row, the colour scale describes the relic abundance in panel (e) and the SGWB energy density peak amplitude $h^2 \Omega_\mathrm{GW}^\mathrm{peak}$ in panel (f). We also note that, in the region of parameter space where strong first-order phase transitions occur in Scenario II, the portal and gauge couplings are sufficiently large to ensure thermal equilibration between the dark and visible sectors. In particular, as can be seen from \cref{fig:GW_and_DM_scenarioII_A}(d,e), this region corresponds to $\lambda_{\rm HD} \gtrsim 10^{-8}$ and $g_{\rm D} \gtrsim 0.1$. We have verified that in this region the relevant interaction rates satisfy $\Gamma > H$, thereby ensuring thermal equilibration between the two sectors. This justifies the use of a common temperature for the two sectors in the relic-density analysis in the region of interest.
\begin{figure*}[htb!]
    \centering
 	\subfloat[]{\includegraphics[width=0.33\textwidth]{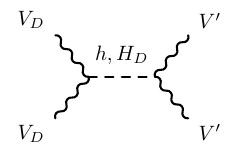}}
    \subfloat[]{\includegraphics[width=0.33\textwidth]{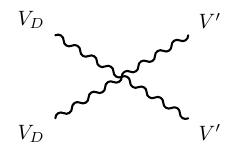}}
    \subfloat[]{\includegraphics[width=0.33\textwidth]{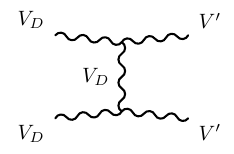}}
    \caption{\footnotesize The relevant Feynman diagrams for DM annihilation in Scenario II into a pair of $V^\prime,V^\prime$: (a) the $h$ or $\mathrm{H_D}$ resonant process, (b) the quartic gauge interaction and (c) the t-channel propagation of $V^\prime$.}
    \label{fig:DM diagrams}
\end{figure*}

Once again, points with a black pentagon identify the $(v,0) \to (v,v_\mathrm{D})$ transitions. As seen in the panels of the top row, these correspond to scenarios with larger $\alpha$, \textit{i.e.}, where the released latent heat is maximised, favoured by the lightest dark Higgs bosons as evident in (a), and where the inverse duration of the phase transition $\beta/H(T_p)$ is minimised, as indicated by the red region in (b). In panel (c), the dark-orange and red points, where the observed relic density is reproduced, favour the smallest $\lambda_\mathrm{HD}$ values, as evident by comparison with panel (d). This is due to the resonant contribution to DM annihilation \cite{Belyaev:2022shr,Belyaev:2022zjx}, which becomes less efficient for smaller portal couplings, thereby increasing $h^2 \Omega_\mathrm{DM}$. For the $(0,0)\to (0,v_\mathrm{D})$ structure, DM annihilates with each other effectively through the {\it resonant process} of Higgs ($H$) and new scalar ($H_\mathrm{D}$), as indicated by diagram (a) in \cref{fig:DM diagrams}, which appears as a diagonal band formed by red points in panel (c) and corresponds to a region where $M_{V_\mathrm{D}}\to M_{H_\mathrm{D}}/2$. On the other hand, the DM process for the $(v,0)\to (v,v_\mathrm{D})$ structure is dominated by the so-called {\it generic DM annihilation} where a pair of DM annihilate into a pair of $V^\prime V^\prime$ through gauge interactions (diagram (b) in \cref{fig:DM diagrams}), and the t-channel propagation of $V_\mathrm{D}$ (diagram (c) in \cref{fig:DM diagrams}). These processes require small $\lambda_{\rm HD}$ and large $g_\mathrm{D}$ as in panel (d), (e) and (f), respectively. The correlation between $g_\mathrm{D}$ and $M_{V_\mathrm{D}}$ can be clearly seen from panel (e). For a fixed value of the relic density, $g_\mathrm{D}$ appears as a parabolic function of $M_{V_\mathrm{D}}$ in the log-linear scale. This can be understood from a naive dimensional analysis where $h^2  \Omega_{\rm DM}=8\pi G \rho_{\rm DM}/3 H_0^2$ and $\rho_{\rm DM} \sim m_{\rm DM} n_{\rm DM} \sim m_{\rm DM} T_F^3/g_\mathrm{D}^2$ which leads to $g_\mathrm{D} \propto \sqrt{M_{V_\mathrm{D}}}$.

The SGWB energy density amplitude is maximised for large values of $\alpha$ and small $\beta/H(T_p)$. From panels (a) and (b) in \cref{fig:GW_and_DM_scenarioII_A}, we observe that this criterion can be satisfied for $M_\mathrm{H_D} \approx 10~\mathrm{GeV}$, regardless of the vector DM mass. However, if full compliance with the DM relic abundance is also required, panel (c) constrains $3 \lesssim M_\mathrm{V_D}/\mathrm{TeV} \lesssim 10$, with a few points in that range showing $\alpha \sim 10$ and $\beta/H(T_p) \sim 400$. 

\begin{figure*}[htb!]
    \subfloat[]{\includegraphics[width=0.50\textwidth]{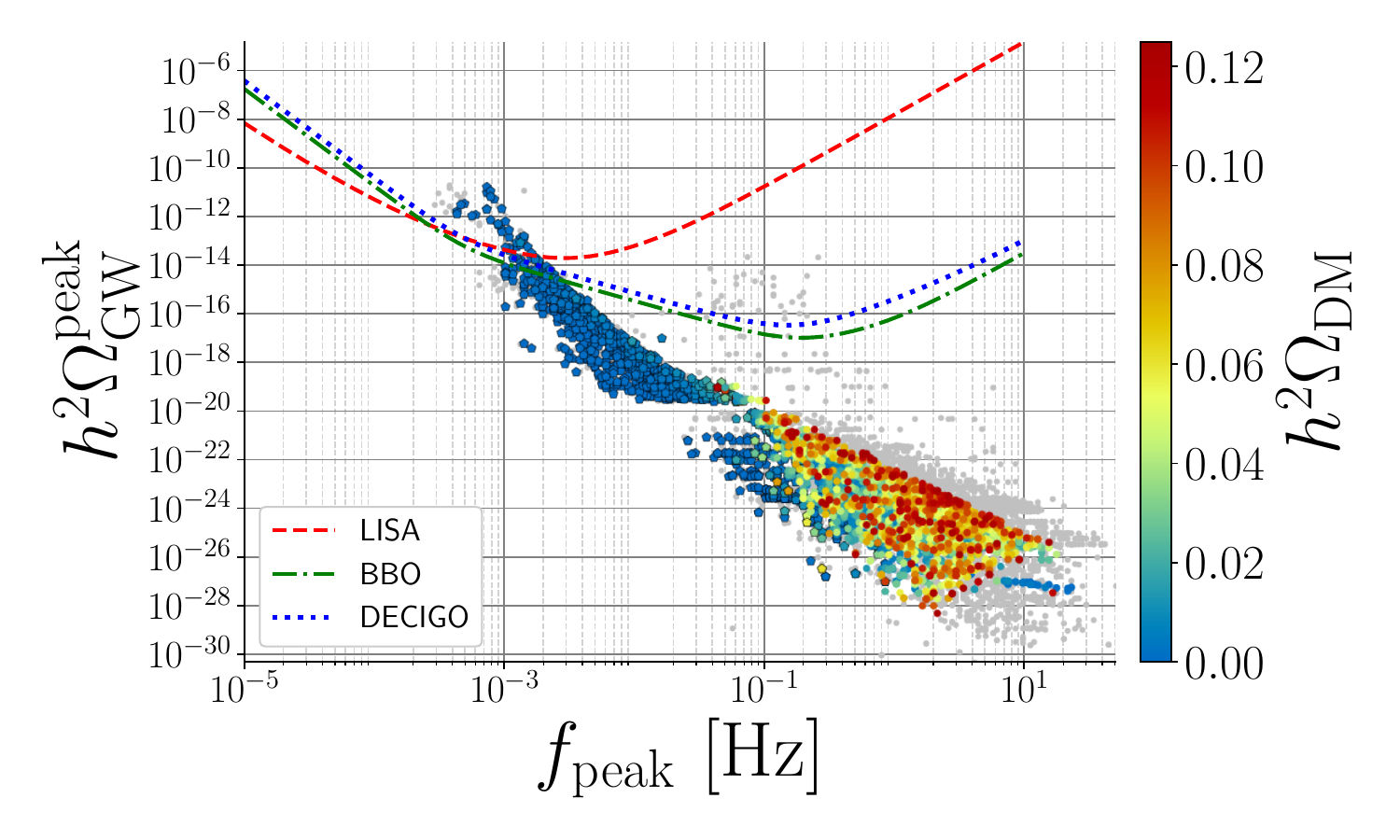}}
 	\subfloat[]{\includegraphics[width=0.50\textwidth]{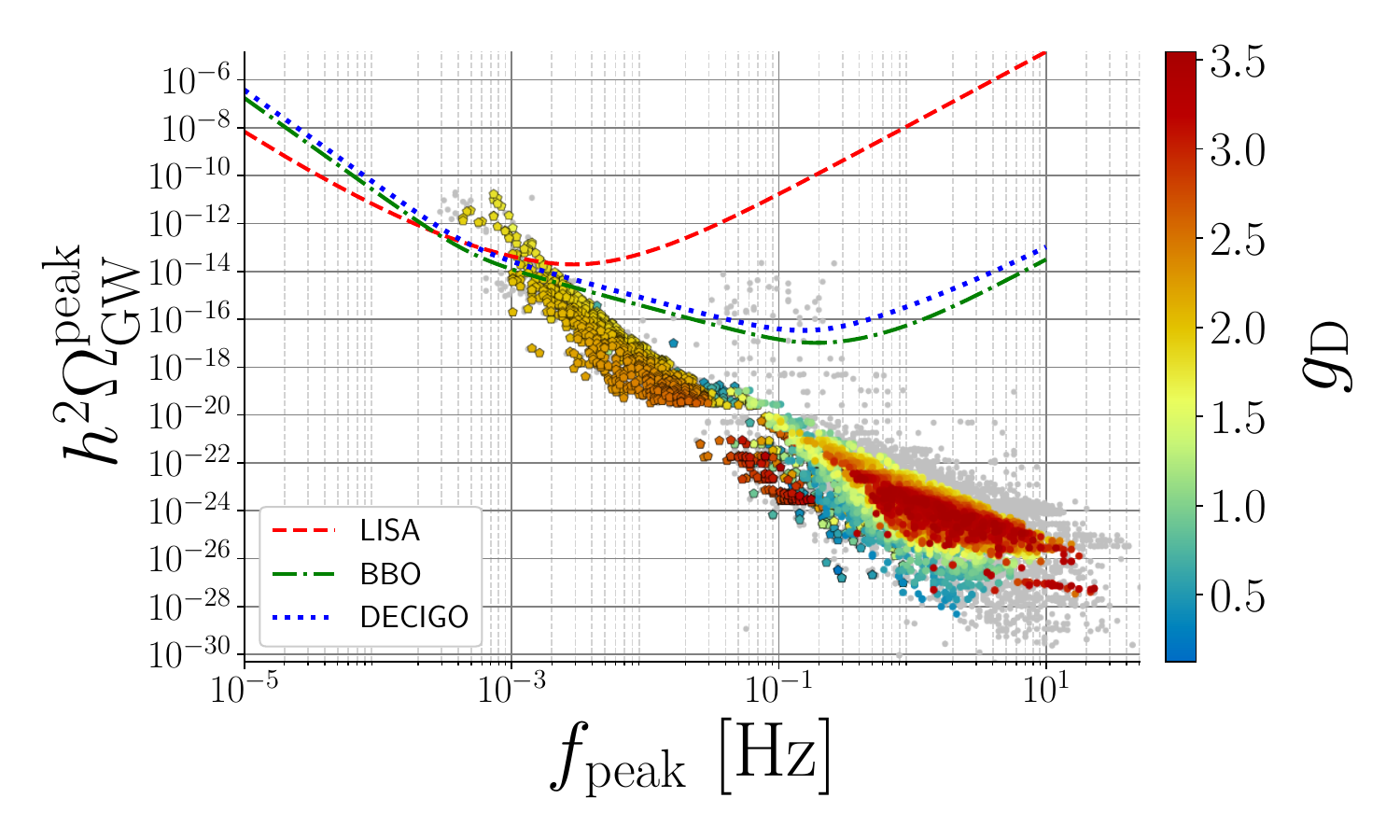}}  \vspace*{-0.4cm}\\
    \subfloat[]{\includegraphics[width=0.50\textwidth]{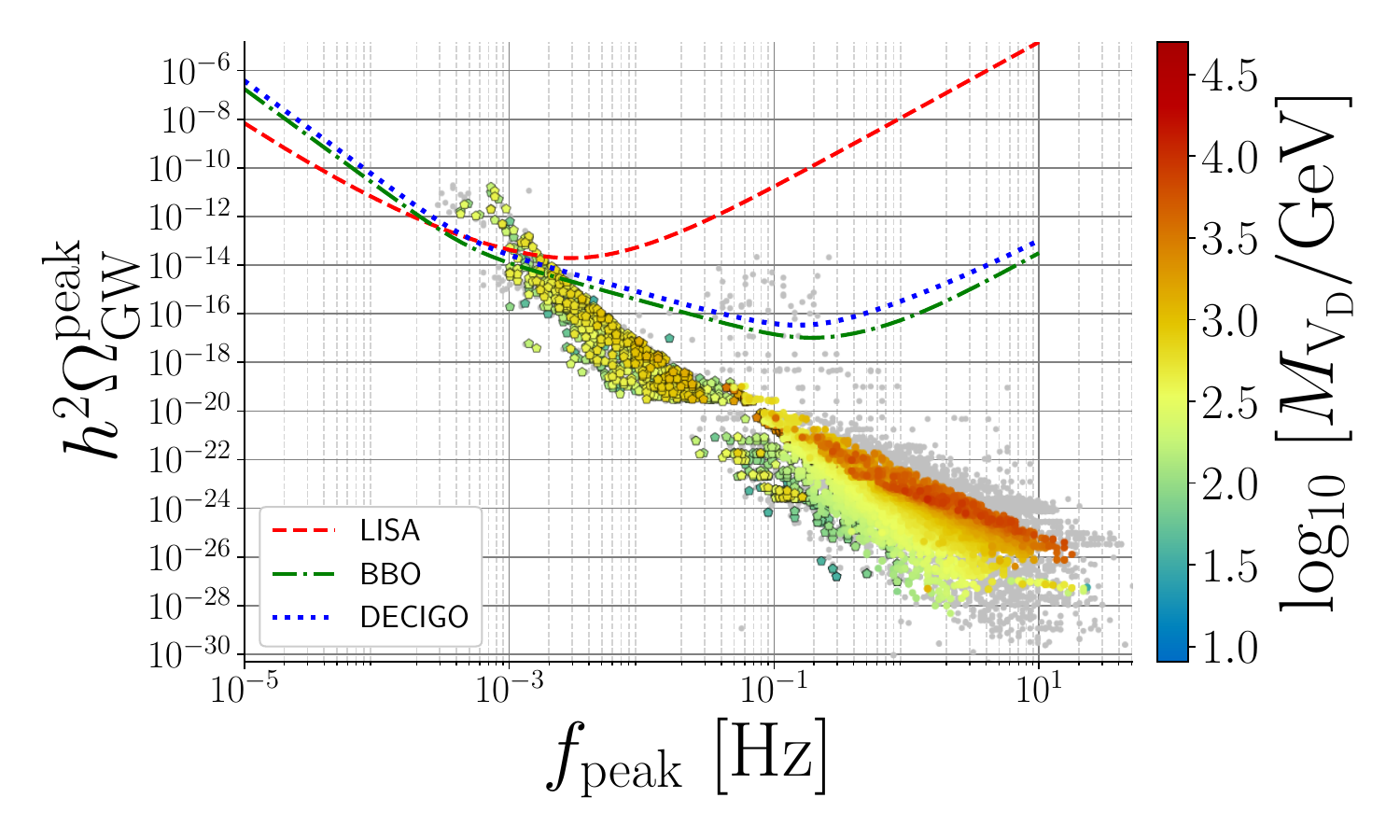}}
    \subfloat[]{\includegraphics[width=0.50\textwidth]{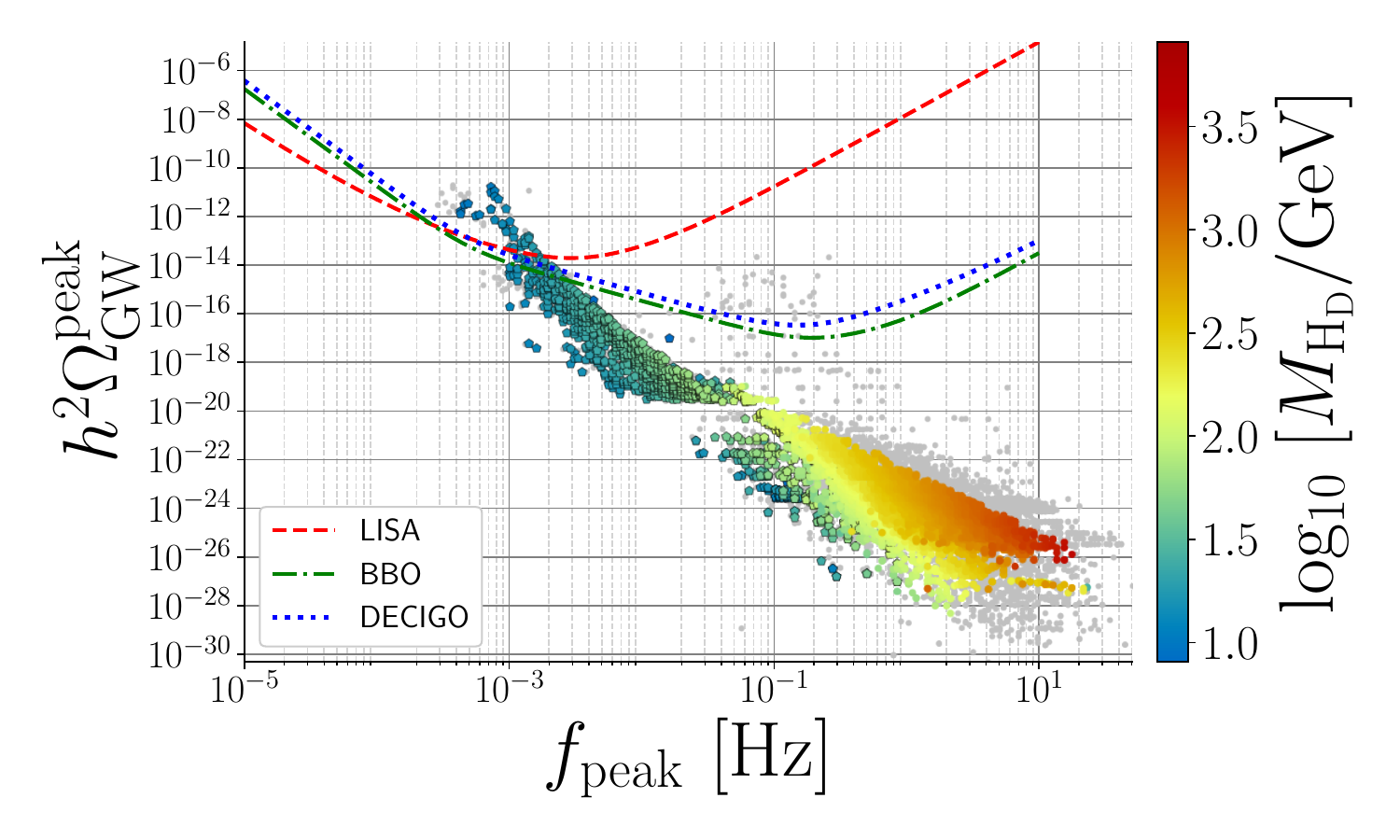}}
    \caption{\footnotesize 
    Predictions for the SGWB geometric parameters $h^2 \Omega^\mathrm{peak}_\mathrm{GW}$ and $f_{\mathrm{peak}}$ for Scenario II in terms of the DM relic abundance $h^2 \Omega_\mathrm{DM}$ (a), the $\SU{2}{D}$ gauge coupling (b), the dark gauge boson mass (c), and the dark Higgs mass (d). The points' markers follow the same legend as in \cref{fig:GW_scenarioII}. The sensitivity curves are the same as in \cref{fig:GW_plots_spectra_modelI}.}
	\label{fig:GW_plots_spectra}
\end{figure*}

To complete the analysis, we present our results for the SGWB peak frequency and peak amplitude in \cref{fig:GW_plots_spectra}.
In panel (a), the colour scale indicates the DM relic abundance, where no direct correlation with the SGWB geometric parameters is observed. For the points that can explain DM, the associated SGWB signal is far below the region probed by current and future-planned experiments. More concretely, we find that in the LISA region for signals with SNR around 100\footnote{Recall that the SNR for a given point can be determined from its vertical distance to the PISCs for a given interferometer \cite{Schmitz:2020syl}.}, the maximum DM relic density that we have obtained is $h^2\Omega_\mathrm{DM} = 0.008$. Generally, we conclude that Scenario II possesses strong predictive power, as observable SGWB within LISA's sensitivity range necessitates $\g{D} \approx 1.7$ (panel b) and a dark Higgs mass of approximately $10 \ \mathrm{GeV}$, with heavier ones falling outside the reach of planned GW experiments and in the high-frequency region (panel d). Conversely, the dark vector mass can vary between $100~\mathrm{GeV}$ and $4~\mathrm{TeV}$ for points within LISA's reach, as illustrated in panel (c) and in \cref{fig:GW_and_DM_scenarioII_A}.

\subsection{Scenario III} \label{subsec:ScenarioIII_IV}

To finalise our analysis, we examine the complete FPVDM model, where communication between the dark and visible sectors is established through both the Higgs and the fermionic portal. We first performed an inclusive scan of the parameter space, as outlined in \cref{tab:sample2}. Subsequently, we selected two benchmark points with specific physical properties, such as mass spectra and mixing angles, requiring proximity to LISA's sensitivity range and consistent DM phenomenology. The aim is to gain a deeper understanding of how the parameter space behaves near these phenomenologically interesting regions.

\subsubsection{Inclusive scan analysis}

\begin{figure*}[htb!]
    {\includegraphics[width=0.50\textwidth]{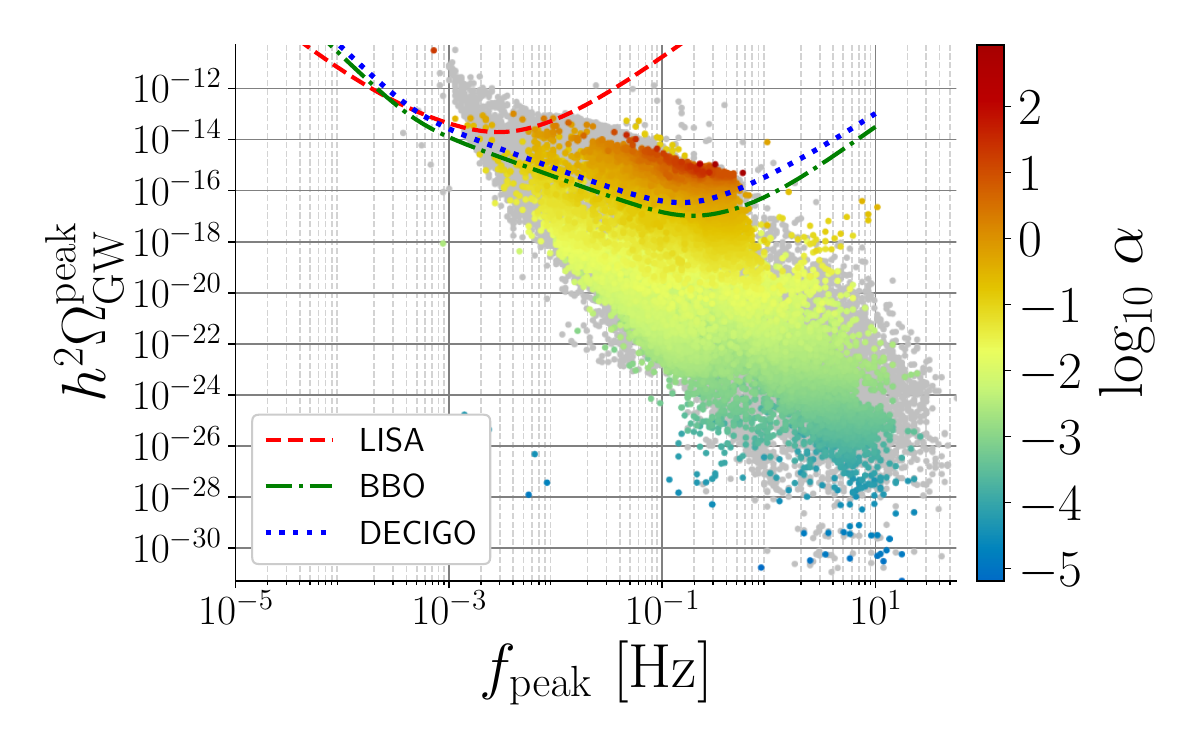}}
    {\includegraphics[width=0.50\textwidth]{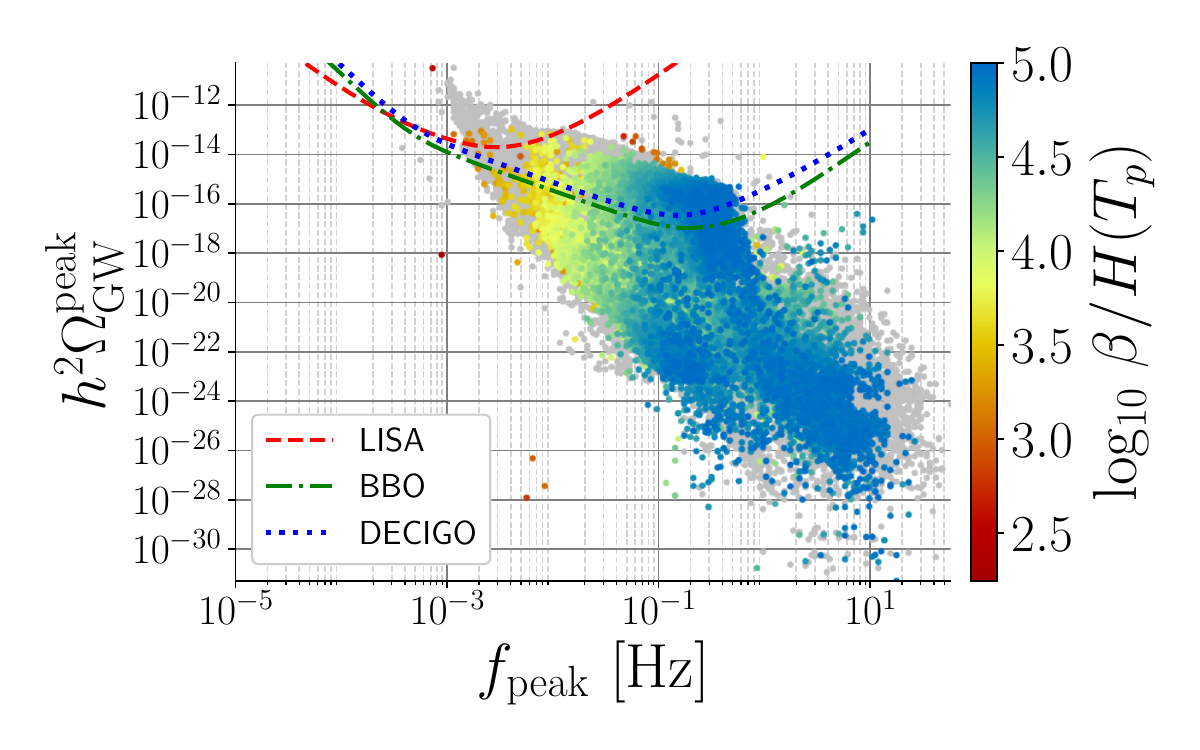}} 
      \vspace*{-0.4cm}
    \caption{\footnotesize Predictions for the SGWB geometric parameters $h^2 \Omega^\mathrm{peak}_\mathrm{GW}$ and $f_{\mathrm{peak}}$ for Scenario III in terms of the phase transition strength $\mathrm{log_{10}}~\alpha$ (left) and the inverse time duration $\mathrm{log_{10}}~\beta/H(T_p)$ (right). The points' markers follow the same legend as in \cref{fig:GW_scenarioII}. The sensitivity curves are the same as in \cref{fig:GW_plots_spectra_modelI}.
    }
	\label{fig:GW_plots_spectra_scenarioIII}
\end{figure*}

\begin{figure*}[htb!]
    \centering
 	\includegraphics[width=0.8\textwidth]{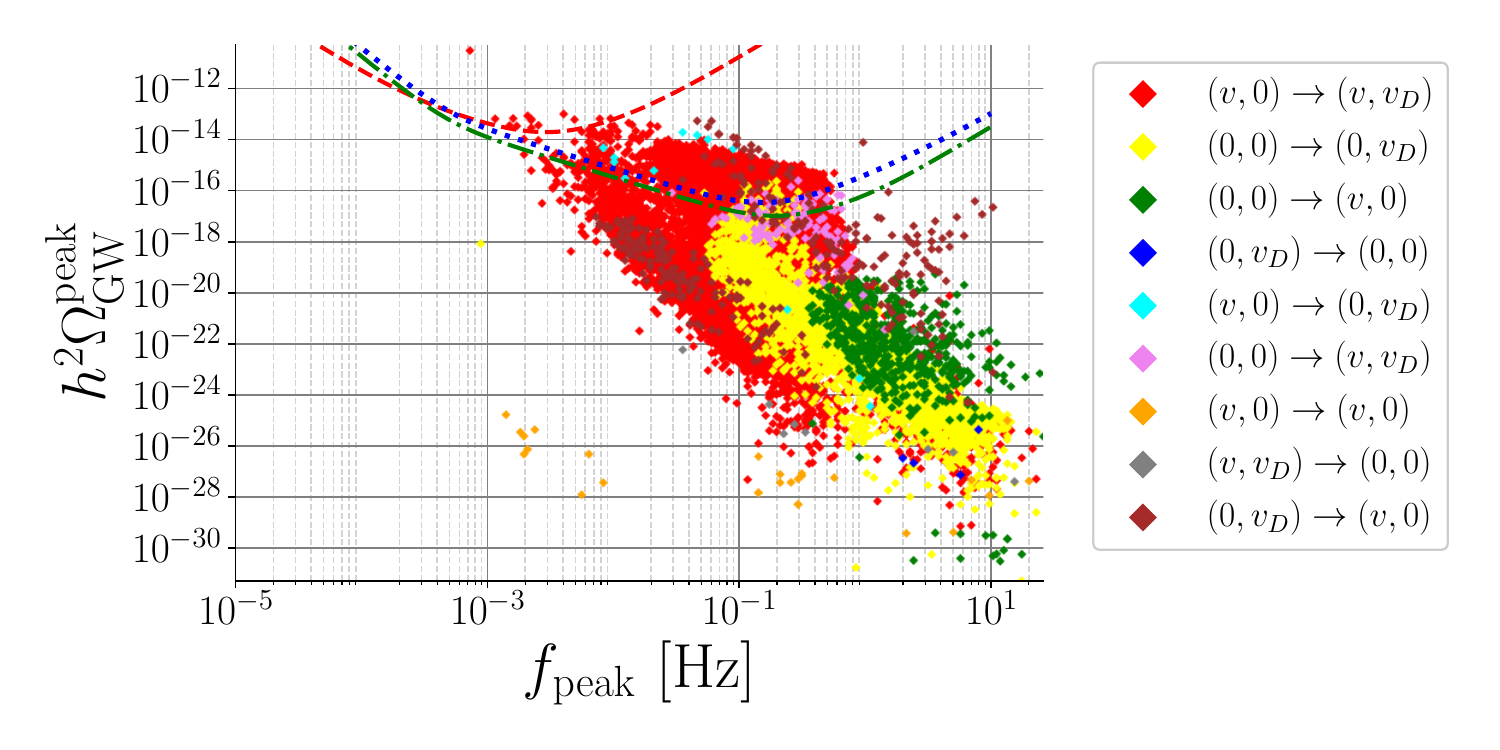}
    \caption{\footnotesize
    Predictions for the SGWB geometric parameters $h^2 \Omega^\mathrm{peak}_\mathrm{GW}$ and $f_{\mathrm{peak}}$ for Scenario III with the colour coding representing different phase transition patterns. The sensitivity curves are the same as in \cref{fig:GW_plots_spectra}.}
	\label{fig:GW_plots_phasepattern}
\end{figure*}

The SGWB is a physical observable that is strongly correlated with thermodynamic parameters of a given BSM theory. In \cref{fig:GW_plots_spectra_scenarioIII}, we show the distribution of the FOPT strength $\alpha$ and its inverse duration $\beta/H(T_p)$ in terms of the SGWB geometric parameters. 
Here we see that for a given frequency, the value of $\alpha$ shifts the SGWB vertically; that is, larger $\alpha$ implies a larger $h^2 \Omega_\mathrm{GW}^\mathrm{peak}$, and vice versa. Conversely, the inverse duration of the phase transition shifts the SGWB diagonally, such that smaller values of $\beta/H(T_p)$ lead to a larger $h^2 \Omega_\mathrm{GW}^\mathrm{peak}$ and a higher $f_\mathrm{peak}$. We define the early observability region as that covered by the sensitivity curve of LISA, whereas the late observability region is characterised by the reach of future planned interferometers such as BBO and DECIGO. In these regions, the full FPVDM model must generate FOPTs with $\mathcal{O}(1) \lesssim \alpha \lesssim \mathcal{O}(100)$, as shown by the orange and red points in panel (a). Meanwhile, the values of $\beta/H(T_p)$ in panel (b) can be as large as $\mathcal{O}(10^5)$ around the dHz regime and as small as $\mathcal{O}(10^3)$ as we approach the mHz range. However, within the late observability region, we also find a number of points that do not follow this trend, where $\alpha \sim \mathcal{O}(0.1)$ and $\beta/H(T_p) \sim \mathcal{O}(100)$. These correspond to the orange points in panel (a) and red points in panel (b) found within frequencies $0.03 < f_\mathrm{peak}/\mathrm{Hz} < 0.1$ and for amplitudes varying within $10^{-14} < h^2 \Omega_\mathrm{GW}^\mathrm{peak} < 10^{-13}$. To understand this behaviour, we must examine the phase transition patterns allowed in the full model, as shown in \cref{fig:GW_plots_phasepattern}.
Among the nine possibilities highlighted in the legend, one pattern clearly stands out: $(v,0) \to (v,v_\mathrm{D})$, represented by red squares, already identified as prevalent in Scenario II. This pattern achieves the largest SGWB peak amplitudes, enters both the early and late observability regions, and corresponds to the general trend observed for $\alpha$ and $\beta/H(T_p)$ in \cref{fig:GW_plots_spectra_scenarioIII}. Another interesting pattern that approaches the early observability region is marked by brown squares, corresponding to $(0,v_\mathrm{D}) \to (v,0)$. Here, the FOPT breaks the EW symmetry and restores that of the dark sector, with the latter being subsequently broken via either a SOPT or a crossover. This phase transition pattern explains the points that deviate from the dominant $(\alpha, \beta/H(T_p))$ trend, where weaker FOPTs with $\alpha \approx 0.3$ are compensated by their long-lasting nature, with $\beta/H(T_p) \sim \mathcal{O}(100)$. A third transition pattern, defined by $(v,0) \to (0,v_\mathrm{D})$ and represented in cyan, can also be identified. Although this pattern is rare in our scan, it has the potential to approach the LISA sensitivity region.
\begin{figure*}[htb]
	\subfloat[]{\includegraphics[width=0.49\textwidth]{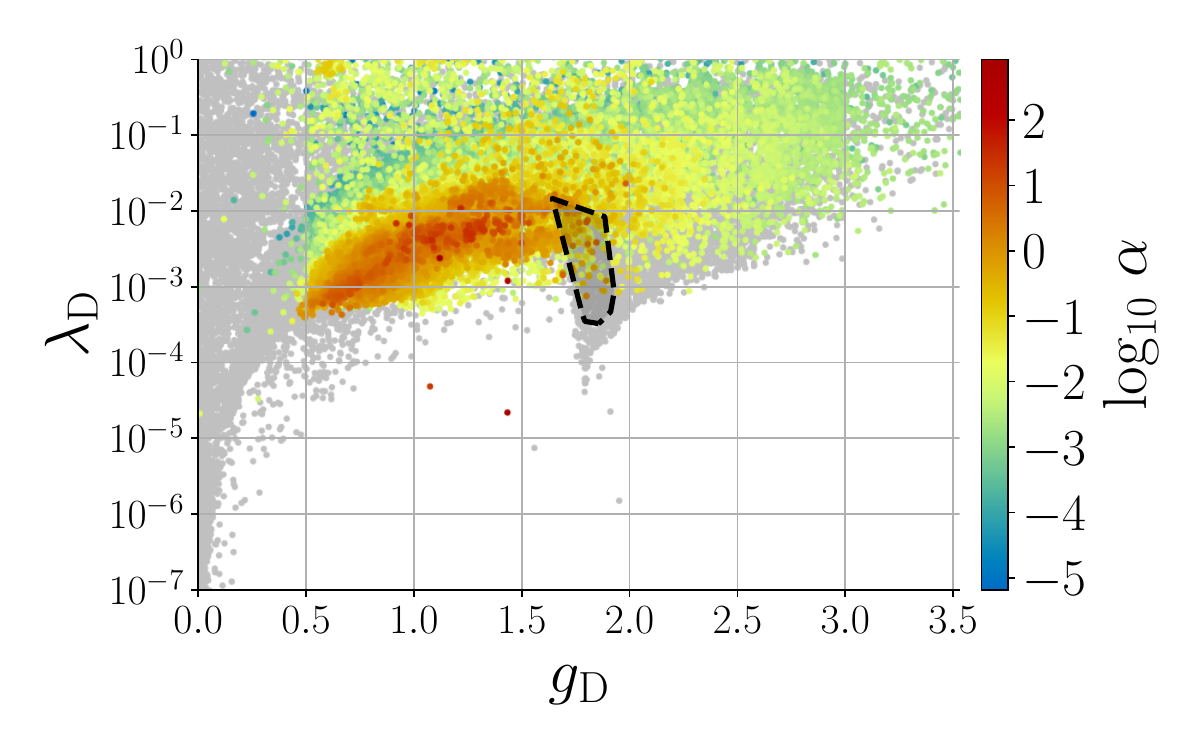}}
    \subfloat[]{\includegraphics[width=0.49\textwidth]{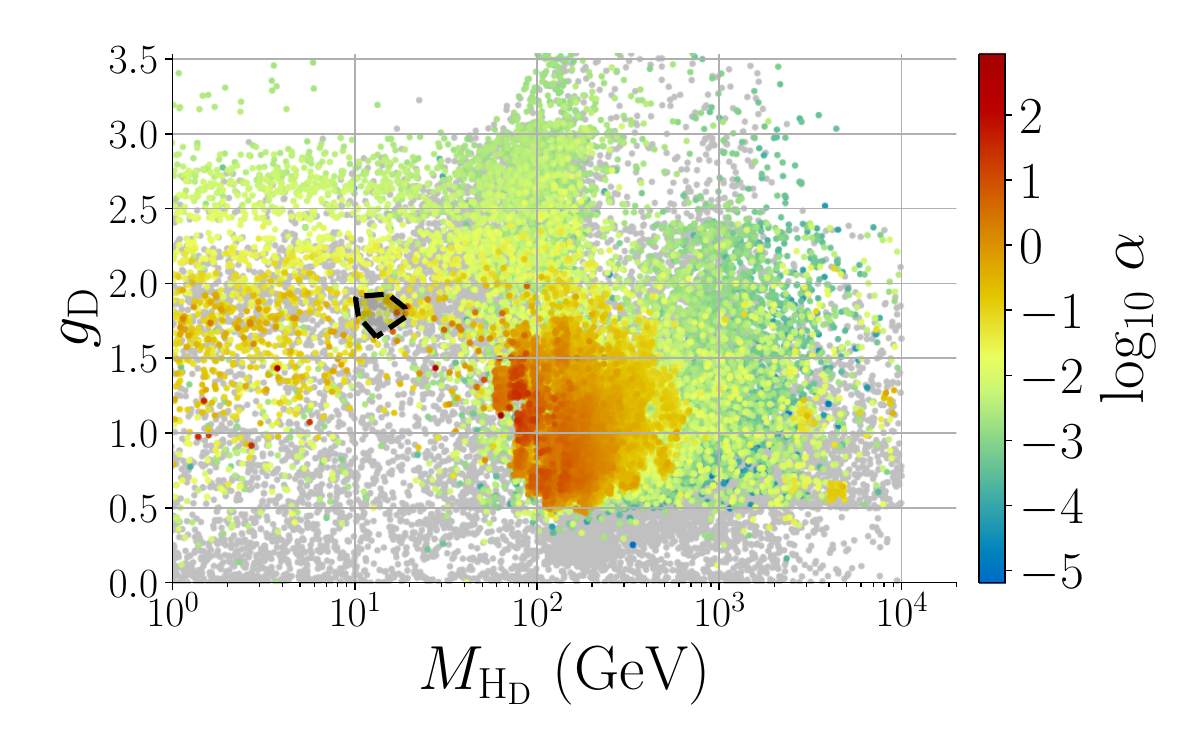}} 
    \vspace*{-0.4cm}\\
    \subfloat[]{\includegraphics[width=0.49\textwidth]{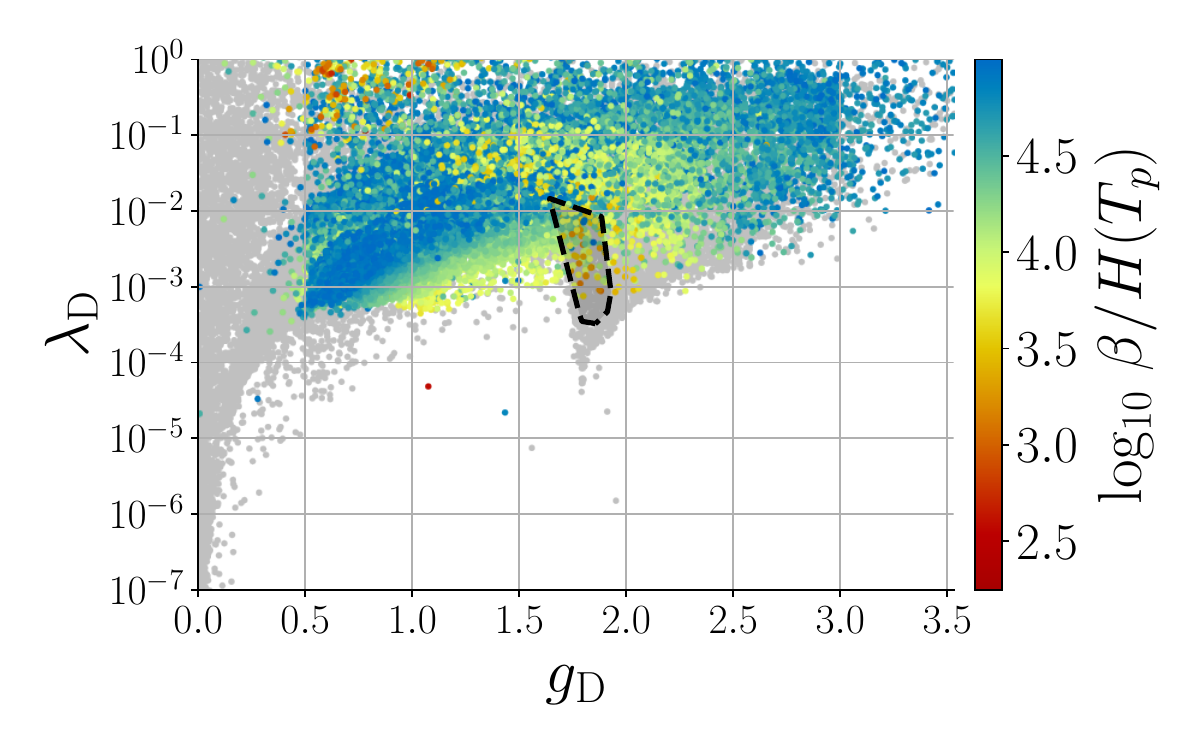}} 
    \subfloat[]{\includegraphics[width=0.49\textwidth]{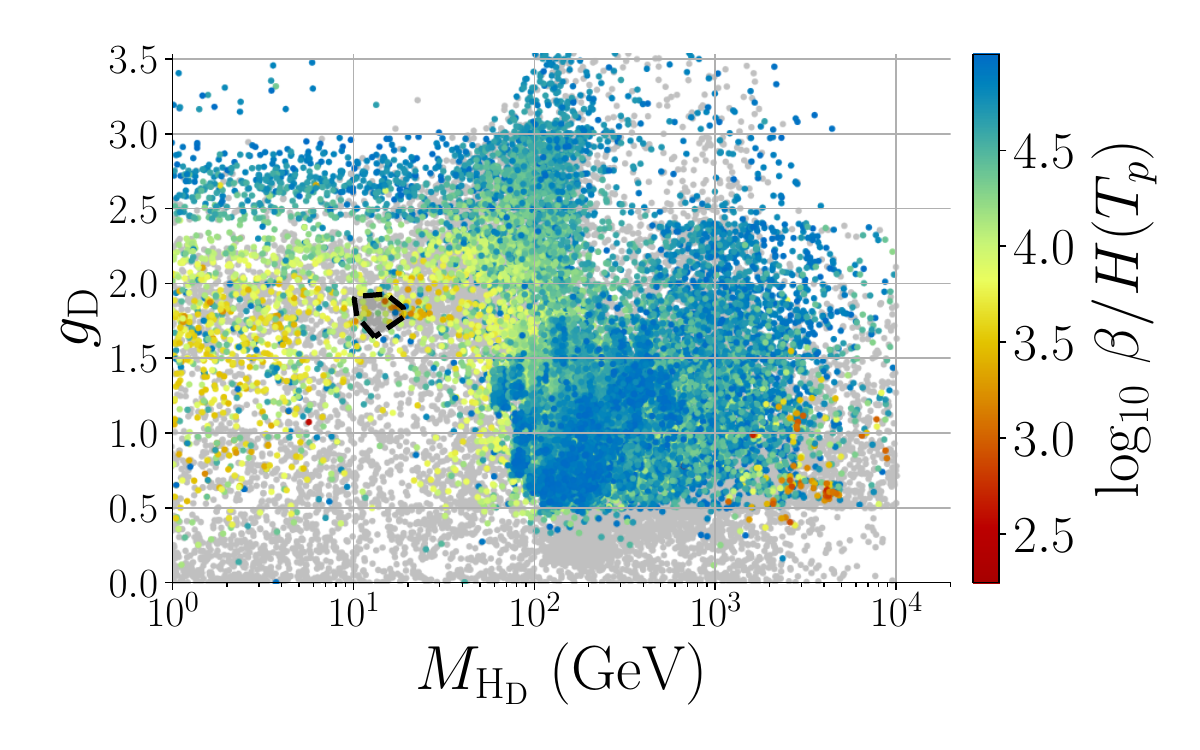}}   \vspace*{-0.4cm}\\
    \subfloat[]{\includegraphics[width=0.49\textwidth]{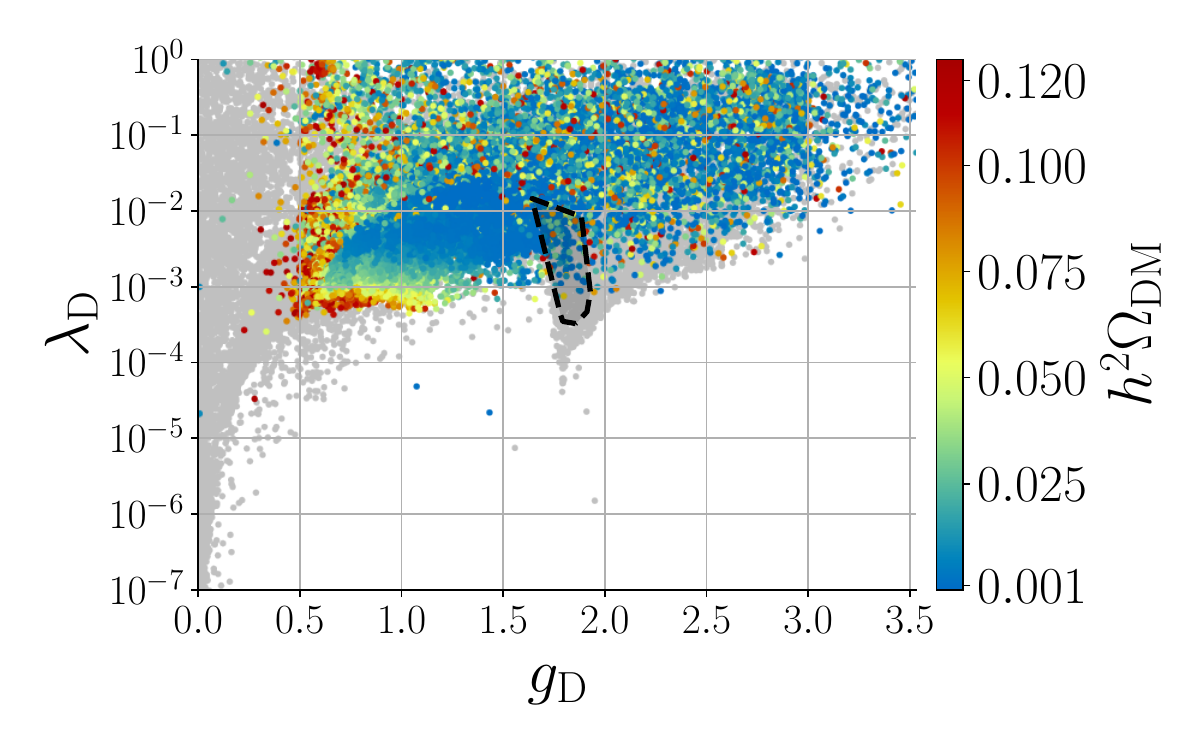}}
    \subfloat[]{\includegraphics[width=0.49\textwidth]{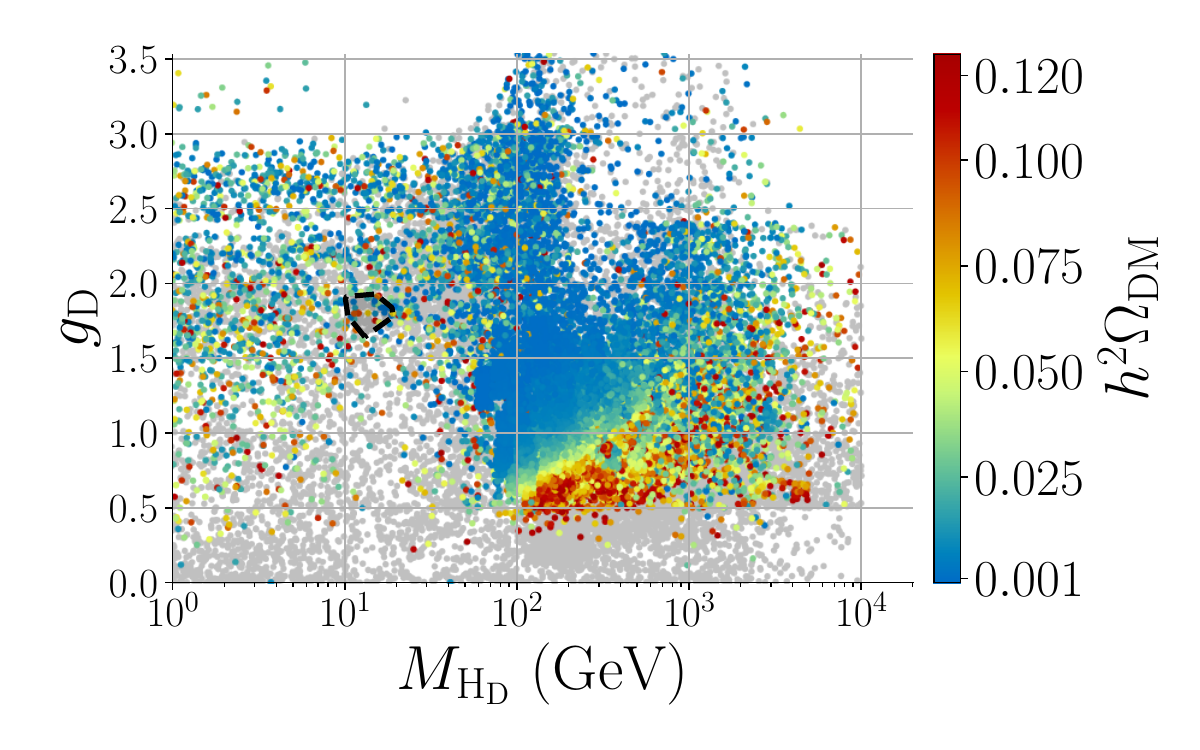}}
    \caption{\footnotesize 
    The colour map for the value of the phase transition strength, $\mathrm{log_{10}}~\alpha$ (panels a and b), the value of the inverse time duration $\mathrm{log_{10}}~\beta/H(T_p)$ (panels c and d) and the DM relic abundance $h^2 \Omega_\mathrm{DM}$ (panels e and f) for the 2D projections of the six-dimensional parameter scan for Scenario III:  $\lambda_\mathrm{D}$ versus $g_\mathrm{D}$ (panels a, c and e) and $g_\mathrm{D}$ versus $M_{H_\mathrm{D}}$ (panels b, d and f). Gray, circled, and uncircled points have the same meanings as in \cref{fig:GW_scenarioII}. The area inside the black dashed contour highlights the region with a potentially observable SGWB at LISA or future interferometers compatible with Scenario II.}
	\label{fig:GW_scenarioIII_1}
\end{figure*}

Thus far, we have identified the preferred values of $\alpha$ and $\beta/H(T_p)$ necessary to enter the observability regions. The goal now is to map this information into the FPVDM model's parameters and DM predictions. We first present in \cref{fig:GW_scenarioIII_1} the results obtained for the $\lambda_\mathrm{D}$ vs. $\g{D}$ (left column) and $\g{D}$ vs. $M_\mathrm{H_D}$ (right column) parameter space projections.

In the panels of the top and middle rows the colour scale indicates the strength of the phase transition and its inverse duration time, respectively, while in the bottom row, it represents the DM relic abundance. Points inconsistent with DM phenomenology, either due to overabundant DM or exclusion by direct detection cross-section limits, are marked in gray. For comparison purposes, we add a black dashed contour in each parameter space projection highlighting the regions where strong FOPTs were identified in Scenario II. As expected, the larger dimensionality of Scenario III allows for a significantly broader area of the parameters space with large $\alpha > 0.1$ (red and orange points), although most of the $\beta/H(T_p)$ values suggest that the majority of it fall within the late observability region. The bottom panels confirm that this region includes points that saturate the DM relic abundance.

Comparing the top and bottom row panels, we observe that for large $\alpha$, represented by the red and orange points in panels (a) and (b), there is a partial overlap with the red points in panels (e) and (f), where the DM abundance saturates experimental bounds. The main difference compared to Scenario II is the wider range of $\g{D}$, approximately, between 0.5 and 2.0, and a heavier dark Higgs boson mass ranging from $M_\mathrm{H_D} \sim \mathcal{O}(100~\mathrm{GeV})$ to a few TeV. Furthermore, for $\lambda_\mathrm{D} < 10^{-5}$, we find a point with $\log_{10} \beta/H(T_p) < 3.5$ entering the LISA sensitivity region. 

This description primarily corresponds to the dominant FOPT pattern. However, another version of the model is found for $\lambda_\mathrm{D} \sim 1$ and $\g{D}$ slightly above 0.5. In this case, we observe a few orange points in panel (a) with $\alpha \sim \mathcal{O}(0.1)$, which correspond to the red points in panel (c) where $\beta/H(T_p) \sim \mathcal{O}(100)$. In the right column plots, we also find that $M_\mathrm{H_D} \approx 4~\mathrm{TeV}$. Additionally, panels (e) and (f) show that DM can also be accounted for in this setup. Referring back to \cref{fig:GW_plots_spectra_scenarioIII,fig:GW_plots_phasepattern}, these scenarios are associated with the $(0,v_\mathrm{D}) \to (v,0)$ FOPT pattern that falls within the late observability region.

\begin{figure*}[htb!]
	\centering
    \subfloat[]{\includegraphics[width=0.50\textwidth]{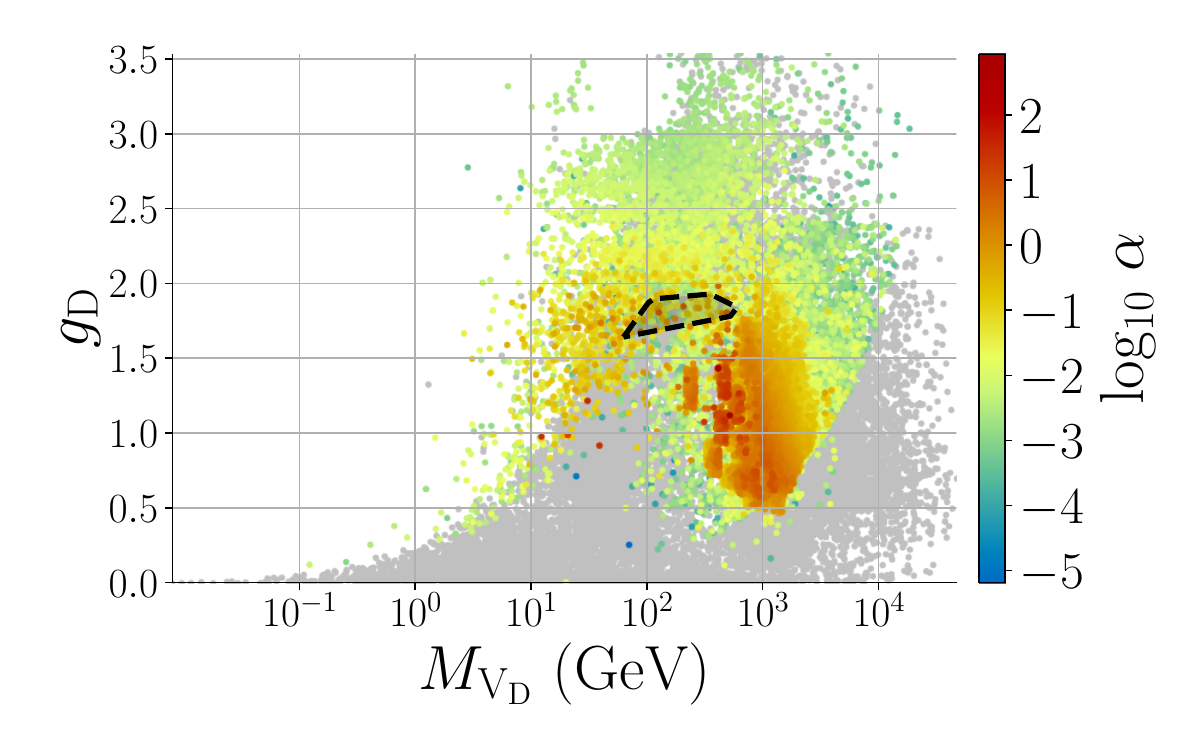}}
    \subfloat[]{\includegraphics[width=0.50\textwidth]{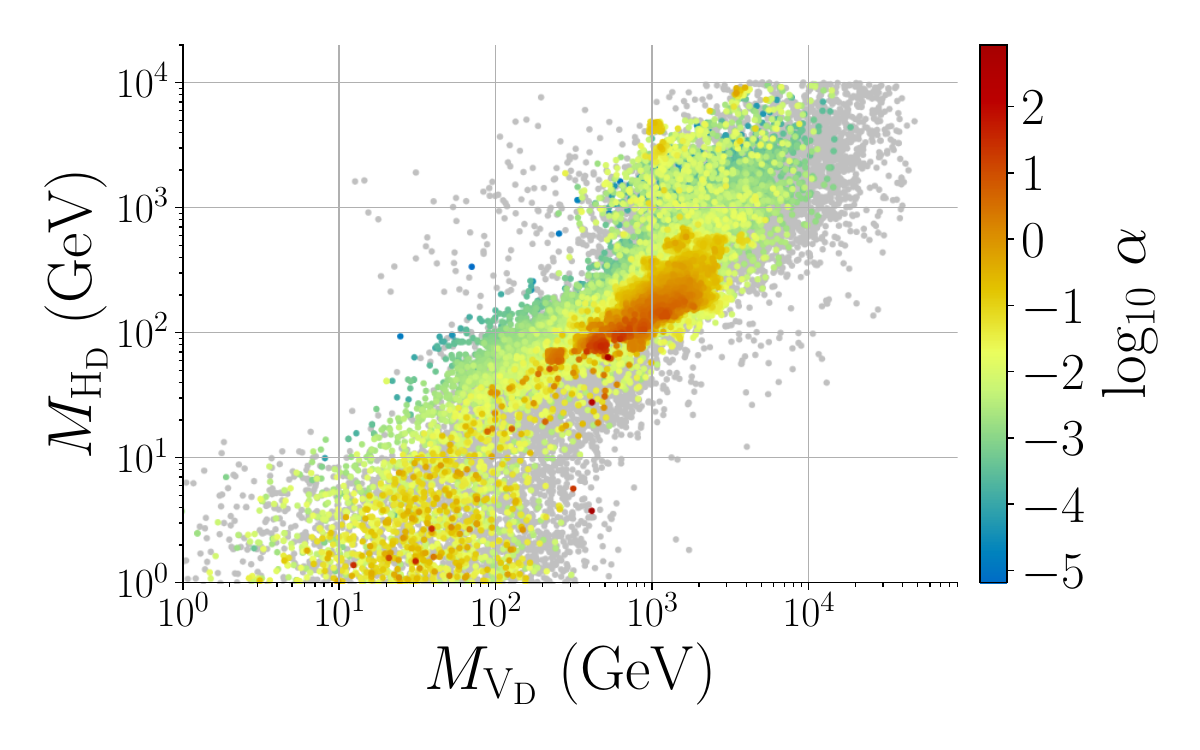}} \\
    \subfloat[]{\includegraphics[width=0.50\textwidth]{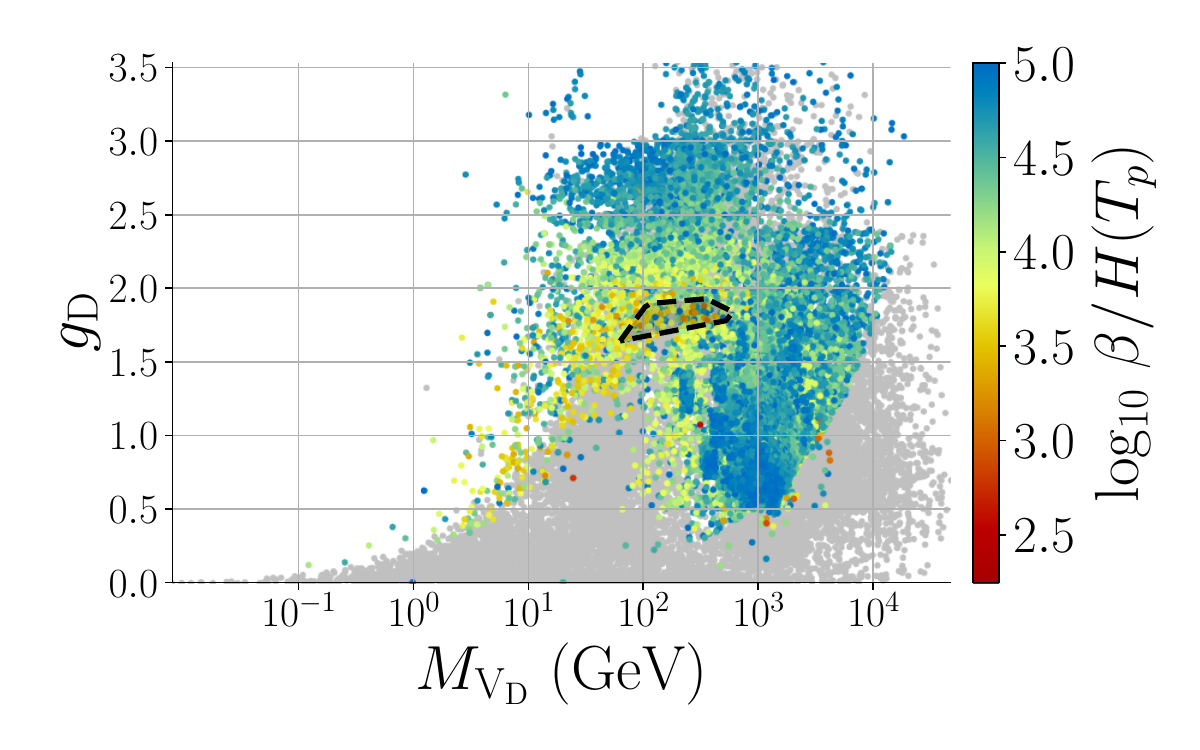}}
    \subfloat[]{\includegraphics[width=0.50\textwidth]{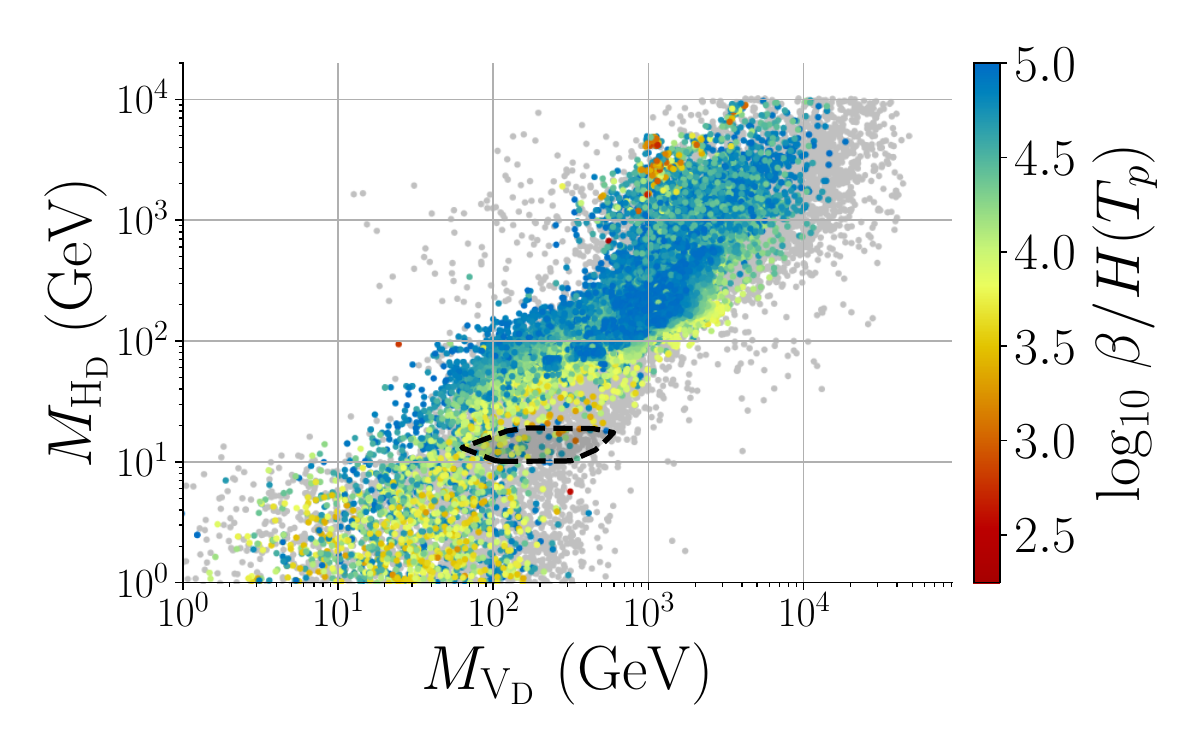}} \\
    \subfloat[]{\includegraphics[width=0.50\textwidth]{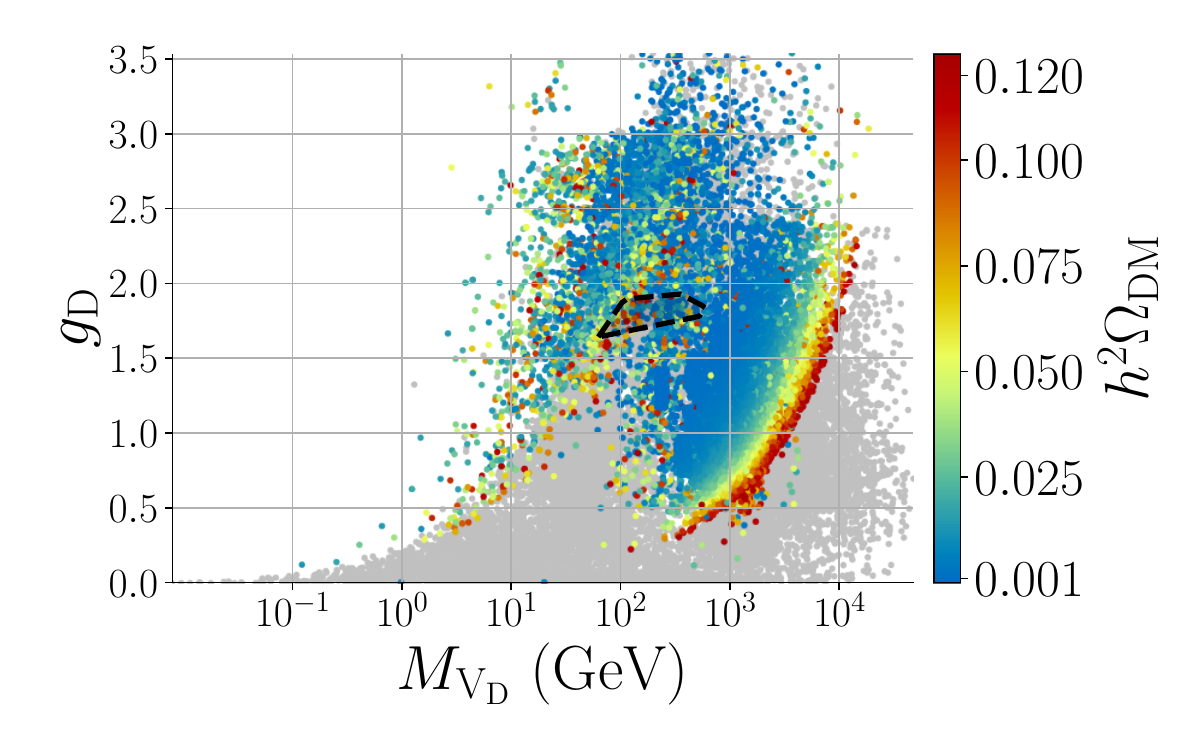}}
    \subfloat[]{\includegraphics[width=0.50\textwidth]{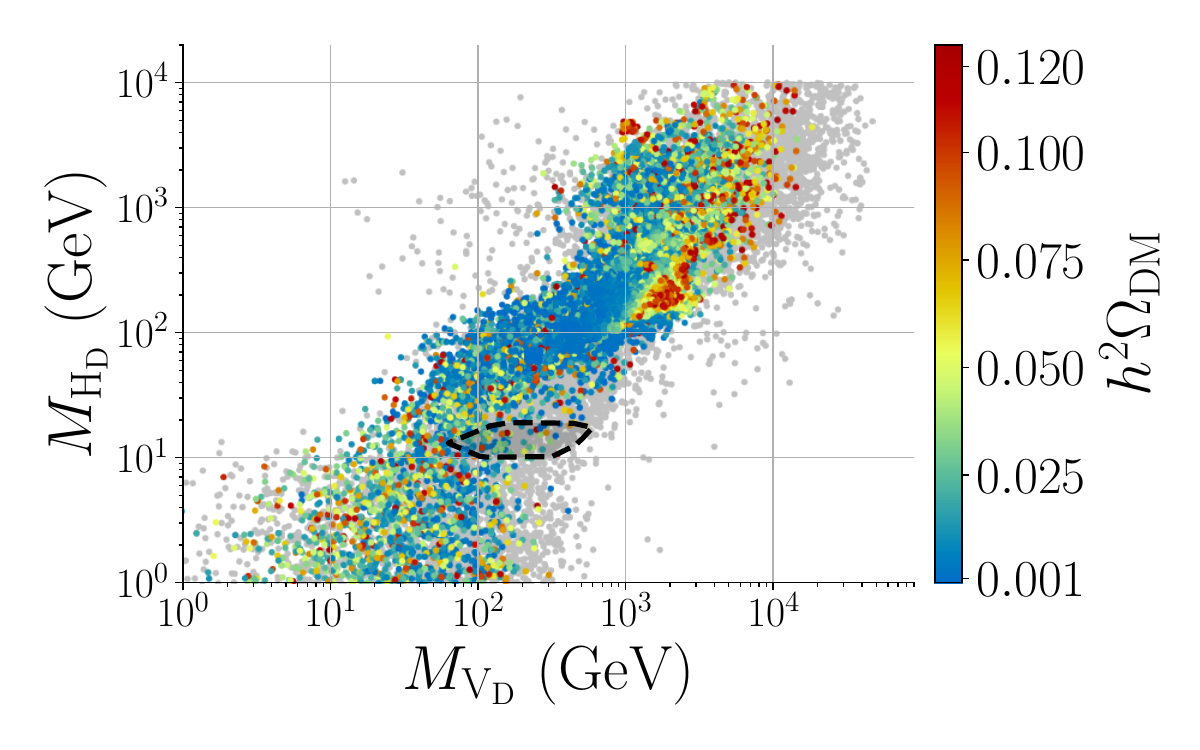}}
    \caption{\footnotesize 
    The colour map for the value of the phase transition strength, $\mathrm{log_{10}}~\alpha$ (panels a and b), the value of the inverse time duration $\mathrm{log_{10}}~\beta/H(T_p)$ (panels c and d) and the DM relic abundance $h^2 \Omega_\mathrm{DM}$ (panels e and f) for the 2D projections of the six-dimensional parameter scan for Scenario III:  $g_\mathrm{D}$ versus $M_{V_\mathrm{D}}$ (panels a, c and e) and $M_{H_\mathrm{D}}$ versus $M_{V_\mathrm{D}}$ (panels b, d and f). Gray, circled, and uncircled points have the same meanings as in \cref{fig:GW_scenarioII}. The area inside the black dashed contour highlights the region with a potentially observable SGWB at LISA or future interferometers compatible with Scenario II.}
	\label{fig:GW_scenarioIII_2}
\end{figure*}

\begin{figure*}[htb!]
    \centering
 	\subfloat[]{\includegraphics[width=0.33\textwidth]{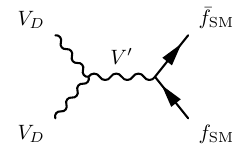}}
    \subfloat[]{\includegraphics[width=0.33\textwidth]{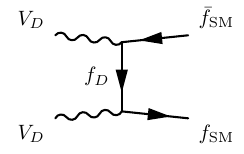}}
    \subfloat[]{\includegraphics[width=0.33\textwidth]{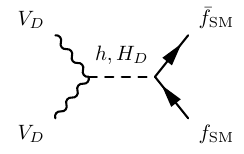}}
    \caption{\footnotesize The relevant Feynman diagrams for DM annihilation in Scenario III into a pair of $f_{\rm SM}\bar{f}_{\rm SM}$.}
    \label{fig:DM diagrams 2}
\end{figure*}

In \cref{fig:GW_scenarioIII_2}, we present our results for the $\g{D}$ vs.~$M_\mathrm{V_D}$ parameter space projection in panels (a), (c) and (e), while panels (b), (d) and (f) display the $M_\mathrm{H_D}$ vs.~$M_\mathrm{V_D}$ plane. 
Note that the difference between Scenarios II and III is evident by the lack of correspondence between the areas encompassed by the dashed contours across the parameter space projections considered in \cref{fig:GW_scenarioIII_1,fig:GW_scenarioIII_2}. We also observe that it does not overlap with the red band in panel (f), where DM is fully accounted for in the complete FPVDM model. This arises from the inclusion of the fermion portal expanding available DM annihilation products. With additional annihilation channels as depicted in \cref{fig:DM diagrams 2}, the relic density is saturated with smaller $g_\mathrm{D}$ for a value of $M_{V_\mathrm{D}}$ while the functional form of $g_\mathrm{D}(M_{V_\mathrm{D}})$ is not significantly different from that of scenario II. This can be clearly seen by comparing the red points in \cref{fig:GW_and_DM_scenarioII_A} (e) with \cref{fig:GW_scenarioIII_2} (e). Regardless of the nature of DM, there is a much wider parameter space region encompassing strong FOPTs with $\alpha \gtrsim 1$, as shown in panels (a) and (b). While these are characterised by the standard $(v,0) \to (v,v_\mathrm{D})$ FOPT pattern, panels (b), (d), and (f) reveal that the points associated with the $(0,v_\mathrm{D}) \to (v,0)$ transition also feature $M_\mathrm{V_D} \approx 1~\mathrm{TeV}$.

\begin{figure*}[htb!]
 	\subfloat[]{\includegraphics[width=0.50\textwidth]{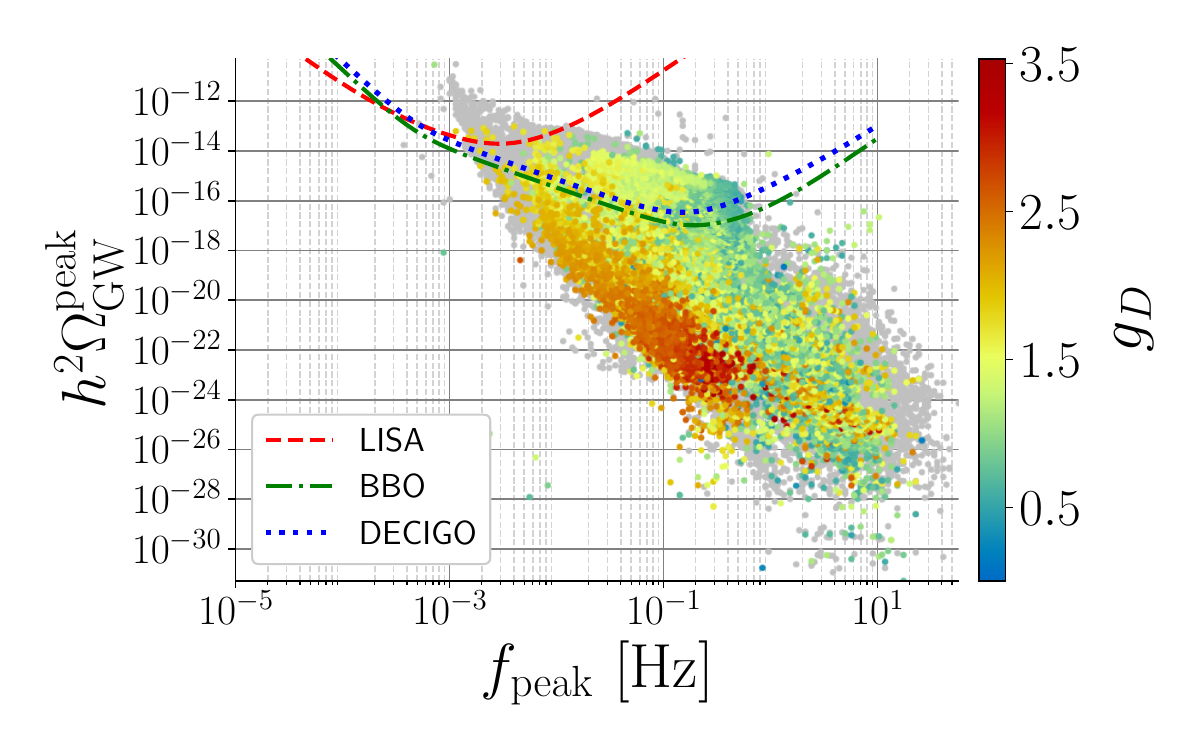}}
    \subfloat[]{\includegraphics[width=0.50\textwidth]{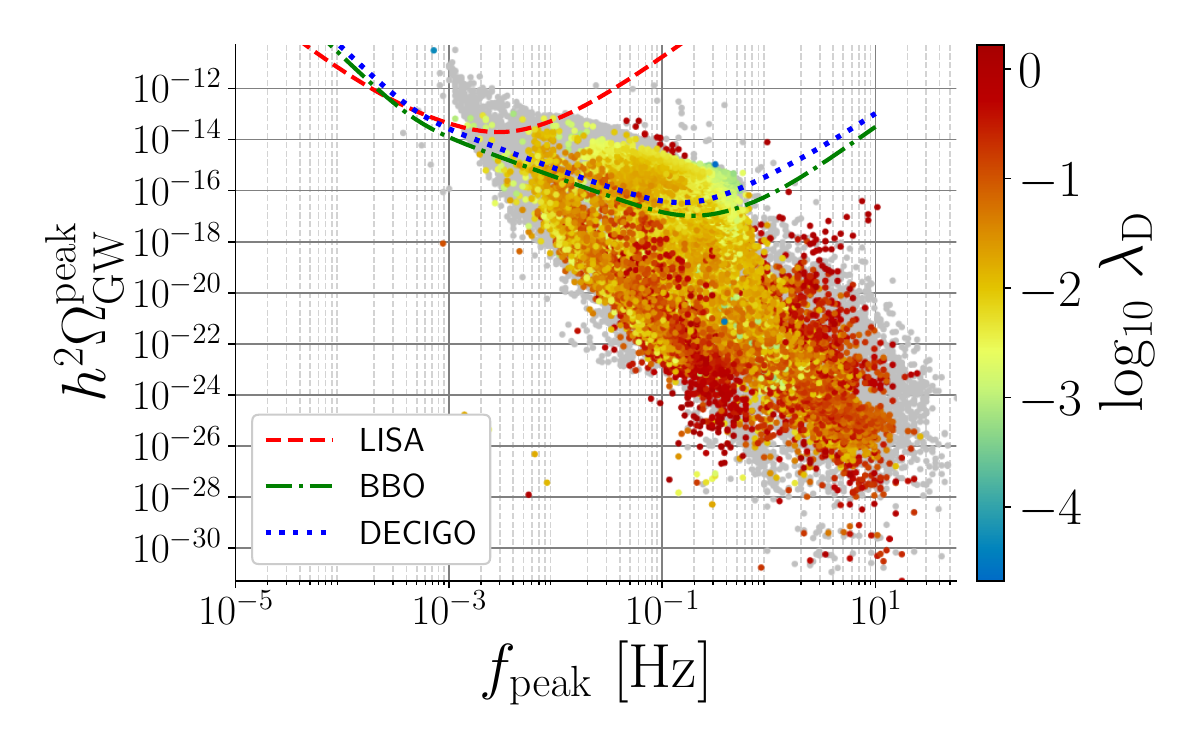}} \\ 
    \subfloat[]{\includegraphics[width=0.50\textwidth]{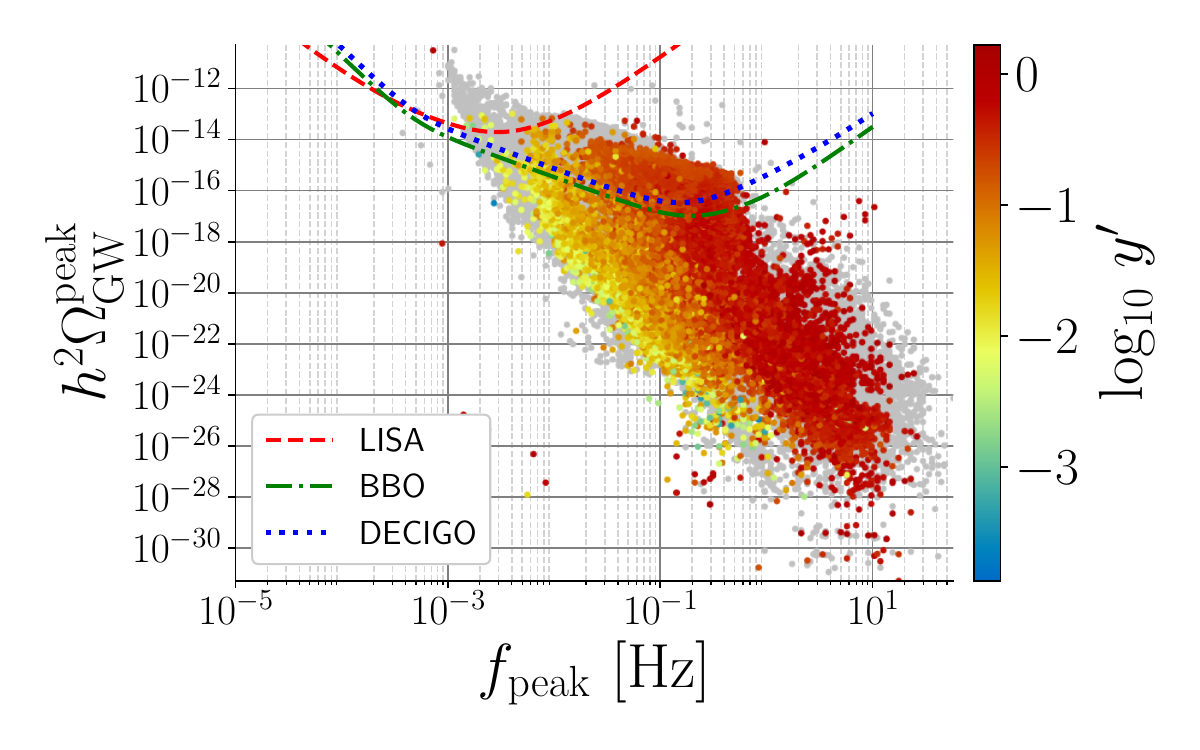}}
    \subfloat[]{\includegraphics[width=0.50\textwidth]{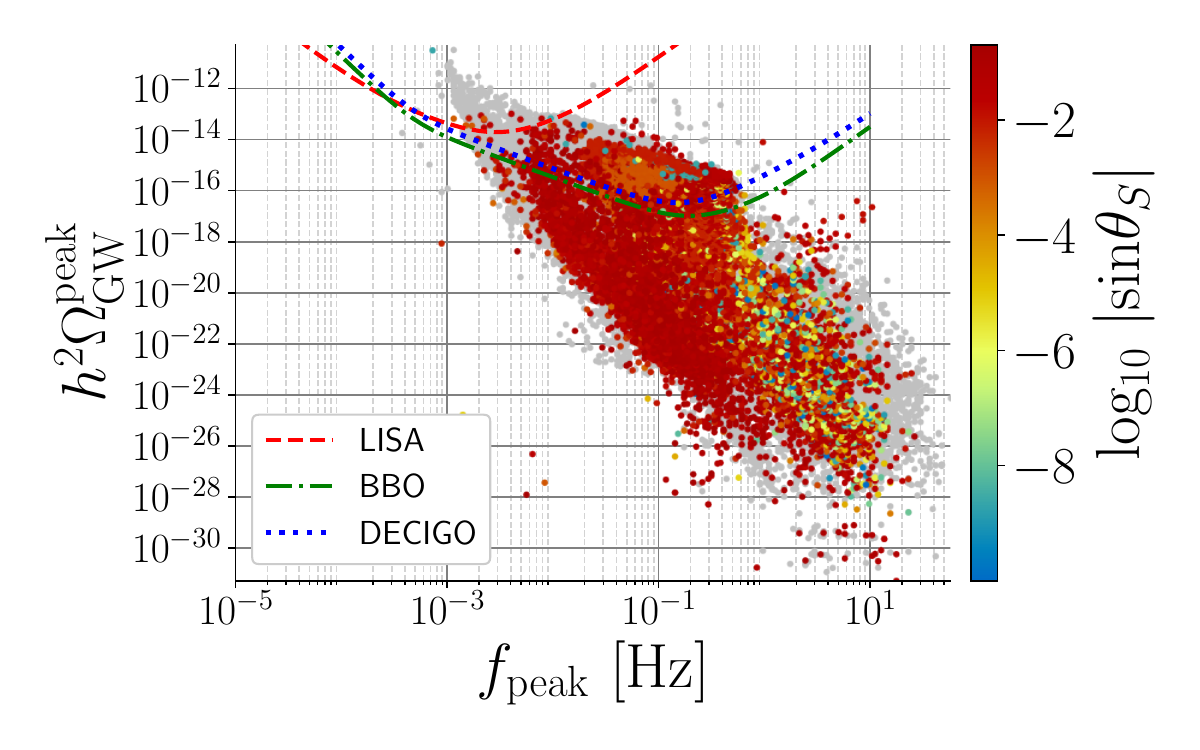}}\\
    \subfloat[]{\includegraphics[width=0.50\textwidth]{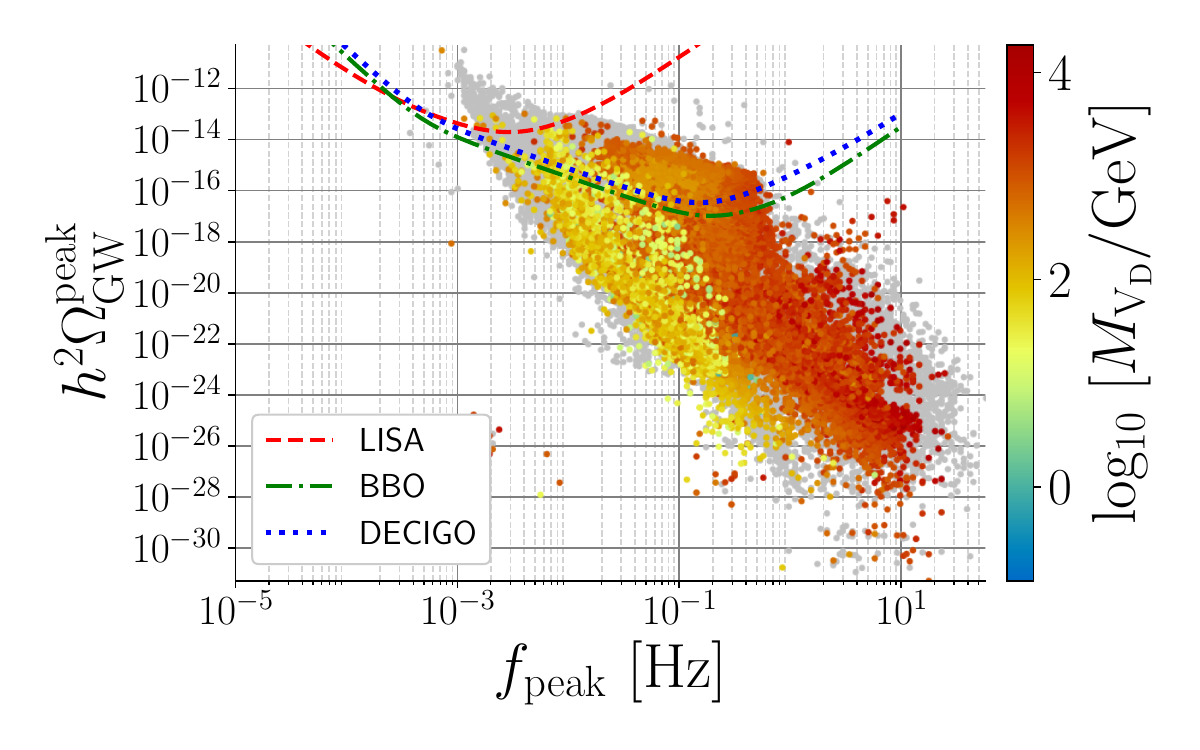}}
    \subfloat[]{\includegraphics[width=0.50\textwidth]{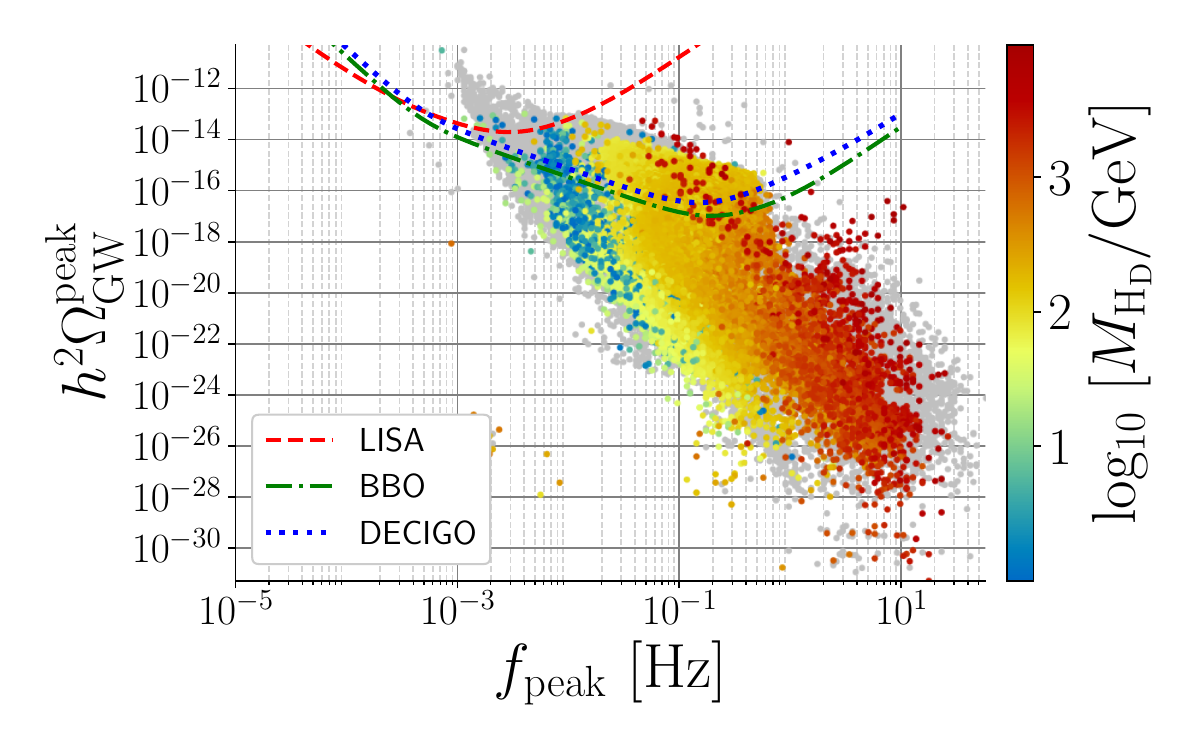}}
    \caption{\footnotesize
    Predictions for the SGWB geometric parameters $h^2 \Omega^\mathrm{peak}_\mathrm{GW}$ and $f_{\mathrm{peak}}$ for Scenario III in terms of the $\SU{2}{D}$ gauge coupling in panel (a), the dark scalar doublet quartic coupling, the Yukawa fermionic portal, the fermion portal Yukawa coupling in panel (c), the absolute value of the scalar mixing angle in panel (d), the vector DM mass in panel (e), and the dark Higgs boson mass in panel (f). The sensitivity curves are the same as in \cref{fig:GW_plots_spectra_modelI}.}
	\label{fig:GW_plots_spectra_scenarioIII_params}
\end{figure*}

In \cref{fig:GW_plots_spectra_scenarioIII_params}, we present the relevant model parameters projected onto the $(h^2\Omega^{\mathrm{peak}}_\mathrm{GW}, f_\mathrm{peak})$ plane. Note that in the observability region, the colour distribution is sufficiently uniform to extract the preferred sizes of the parameters that feature strong FOPTs as shown in \cref{tab:summary}. We also note that, in the region of the generic Scenario III scan where strong first-order phase transitions occur, the dark gauge and fermion-portal Yukawa couplings are sufficiently large to ensure thermal equilibration between the dark and visible sectors. In particular, as can be seen from \cref{fig:GW_plots_spectra_scenarioIII_params}, this region corresponds to $g_{\rm D} \gtrsim 0.2$ and $y' \gtrsim 10^{-3}$. We have verified that in this region the relevant interaction rates satisfy $\Gamma > H$, thereby ensuring thermal equilibration between the two sectors. This justifies the use of a common temperature for the two sectors in the relic-density analysis in the region of interest.

\begin{table*}[ht!]
\centering
\begin{tabular}{@{}rccccr@{}}
& \multicolumn{3}{c}{} \\
\hline
& FOPT pattern & $(v,0) \to (v,v_\mathrm{D})$ & $(0,v_\mathrm{D}) \to (v,0)$ &\\ \hline
&  & &  & &\\
& $\alpha$ & $1$ to $10^2$ & $1$ &\\
&  & &  & &\\
& $\beta/H(T_p)$ & $10^{3}$ to $10^{5}$ & $10^2$ to $10^3$ &\\
&  & &  & &\\
& $\g{D}$ & $0.5$ to $2.0$ & $0.5$ &\\
&  & &  & &\\
& $\lambda_\mathrm{D}$ & $10^{-4}$ to $10^{-2}$ & $1$ &\\
&  & &  & &\\
& $y^\prime$ & $10^{-2}$ to $1$ & $1$ &\\
&  & &  & &\\
& $\sin \theta_S$ & $10^{-4}$ to $10^{-2}$ & $10^{-2}$ to $10^{-1}$ &\\
&  & &  & &\\
& $M_\mathrm{V_D}/\mathrm{GeV}$ & $10^2$ to $10^{4}$ & $10^{3}$ &\\
&  & &  & &\\
& $M_\mathrm{H_D}/\mathrm{GeV}$ & $10$ to $100$ & $10^{3}$ &\\
&  & &  & &\\
\hline
\end{tabular}
\caption{ \footnotesize Approximate magnitude of the theory and thermodynamic parameters in the complete FPVDM model focusing on the observability region and two possible phase transition patterns. Masses are given in GeV. }
\label{tab:summary}
\end{table*}

%
In general, the coloured areas in each panel exhibit a consistent pattern across all six panels, highlighting the connection between the FPVDM model parameters and the predicted SGWB peak amplitude and frequency. Within the observability region, we identify two distinct areas that correspond to the two columns of \cref{tab:summary}. The scalar potential is also sensitive to the fermion sector, as suggested by panel (c). This sensitivity arises from the relative magnitudes of $\g{D}$, $\lambda_\mathrm{D}$, and $y^\prime$, and their effects on the ultrasoft parameters in \cref{eqn:US_pot}. In particular, the same qualitative features discussed in relation to \cref{fig:GW_plots_US} apply. However, instead of a single parameter dependence of $[\mu^{\mathcal{US}}_{\rm D}]^2$ and $[\lambda^{\mathcal{US}}_{\rm D}]^2$ on the gauge coupling\footnote{Recall that in Scenario I we have $\lambda_\mathrm{D} \ll \g{D}$.}, the influence of the portal Yukawa coupling, and in some cases, the dark sector self quartic coupling $\lambda_\mathrm{D}$, becomes relevant and may compete with $\g{D}$. 

For most points within the observability region, the shape of the 3D ultrasoft potential is primarily influenced by the dark gauge coupling, since $\g{D} > y^\prime \gg \lambda_\mathrm{D}$. At leading order, the dominant contributions can be approximated as\footnote{Refer to \cref{sec:3d_eff} for the exact expressions.}
\begin{equation}
    \begin{aligned}
        [\mu^{\mathcal{US}}_{\rm D}]^2_{_1} &\approx \mu_\mathrm{D}^2 +\frac{3}{16} \g{D}^2 T^2 \\
        [\lambda^{\mathcal{US}}_{\rm D}]_{_1} &\approx  \frac{T}{256 \pi^2} \g{D}^4 (6-9L_b)\,.
    \end{aligned}
\end{equation}
$L_b$ and $L_f$ are numerical factors that appear in the matching procedure. They are defined in Eq.~\eqref{eq:lb_lf_factors}. This corresponds to the phase transition pattern $(v,0) \to (v,v_\mathrm{D})$ in the first column of \cref{tab:summary}, which also favours the largest vector masses of up to $\mathcal{O}(10\ \mathrm{TeV})$ and the lightest dark Higgs mass of approximately $\mathcal{O}(10\ \mathrm{GeV})$. In this case, the scalar mixing angle is typically very small, with most points featuring $\abs{\sin \theta_S} \sim \mathcal{O}(10^{-4}) - \mathcal{O} (10^{-2})$, remaining unconstrained by direct searches at the LHC. This also indicates that the FOPT proceeds primarily along the dark direction, with negligible influence from the visible Higgs sector.

Within this set, we also have transitions in which the fermion portal $y^\prime$ can compete with the gauge coupling $\g{D}$, \textit{i.e.} $g_D \sim y^\prime$, implying that the vacuum structure is governed by both the gauge and Yukawa sectors. At leading order, the dominant contributions to the scalar potential can be approximated as
\begin{equation}
    \begin{aligned}
        [\mu^{\mathcal{US}}_{\rm D}]^2_{_2} &\approx [\mu^{\mathcal{US}}_{\rm D}]^2_{_1} + \frac{1}{4} {y^\prime}^2 T^2 \\
        [\lambda^{\mathcal{US}}_{\rm D}]_{_2} &\approx  [\lambda^{\mathcal{US}}_{\rm D}]_{_1}  + \frac{3}{16 \pi^2} L_f {y^\prime}^4 T\,.
    \end{aligned}
\end{equation}

We have also encountered a few points with $M_\mathrm{V_D} \sim M_\mathrm{H_D} \sim \mathcal{O}(1~\mathrm{TeV})$ where $\g{D} \lesssim y^\prime \sim \lambda_\mathrm{D} \sim \mathcal{O}(1)$. However, the most significant difference lies in a larger scalar mixing, making the portal coupling $\lambda_\mathrm{HD}$ relevant for the vacuum structure. The leading contributions to the ultrasoft 3D scalar potential can now be expressed as
\begingroup
\allowdisplaybreaks
\begin{align}
    [\mu^{\mathcal{US}}_{\rm D}]^2_{_3} &\approx [\mu^{\mathcal{US}}_{\rm D}]^2_{_2} + T^2\left(\frac12 \lambda_\mathrm{D} + \frac16 \lambda_\mathrm{HD} \right) \\
    [\lambda^{\mathcal{US}}_{\rm D}]_{_3} &\approx  [\lambda^{\mathcal{US}}_{\rm D}]_{_2}  - \frac{T}{4 \pi^2} \left( 3 L_b \lambda_\mathrm{D}^2 +\frac32 \lambda_\mathrm{D} {y^\prime}^2 + \frac14 L_b \lambda_\mathrm{HD}^2 \right) \\
    [\mu^{\mathcal{US}}_{\rm H}]^2_{_3} &\approx [\mu^{\mathcal{US}}_{\rm H}]^2_{_\mathrm{SM}} + \frac16 \lambda_\mathrm{HD} T^2 \\
    [\lambda^{\mathcal{US}}_{\rm H}]_{_3} &\approx [\lambda^{\mathcal{US}}_{\rm H}]_{_\mathrm{SM}} - \frac{T}{16 \pi^2} L_b \lambda_\mathrm{HD}^2 \\
    [\lambda^{\mathcal{US}}_{\rm HD}]_{_3} &\approx \frac{T}{8 \pi^2} \left( 3 L_f {y^\prime}^2 y_t^2 - \frac32 L_f \lambda_\mathrm{HD} {y^\prime}^2 -L_b \lambda_\mathrm{HD}^2 - 3 L_b \lambda_\mathrm{HD} \lambda_\mathrm{D} + \frac98 L_b \lambda_\mathrm{HD} \g{D}^2 \right)\,,
\end{align}
\endgroup
where the subscript SM in the third and fourth lines indicates contributions solely from the visible sector. The immediate consequence is a FOPT that involves both the visible and dark directions, which explains the transition pattern in the second column of \cref{tab:summary}, \textit{i.e.}, $(0,v_\mathrm{D}) \to (v,0)$. With the scalar mixing typically on the order of $0.01$, the trans-TeV scale for the dark Higgs mass makes it likely unconstrained by direct searches for new scalars at the LHC \cite{ATLAS:2020zms,ATLAS:2020jgy,ATLAS:2020tlo,ATLAS:2021uiz}. A dedicated analysis of this is left for future work.

\begin{figure*}[t!]
    \centering
 	\includegraphics[width=0.74\textwidth]{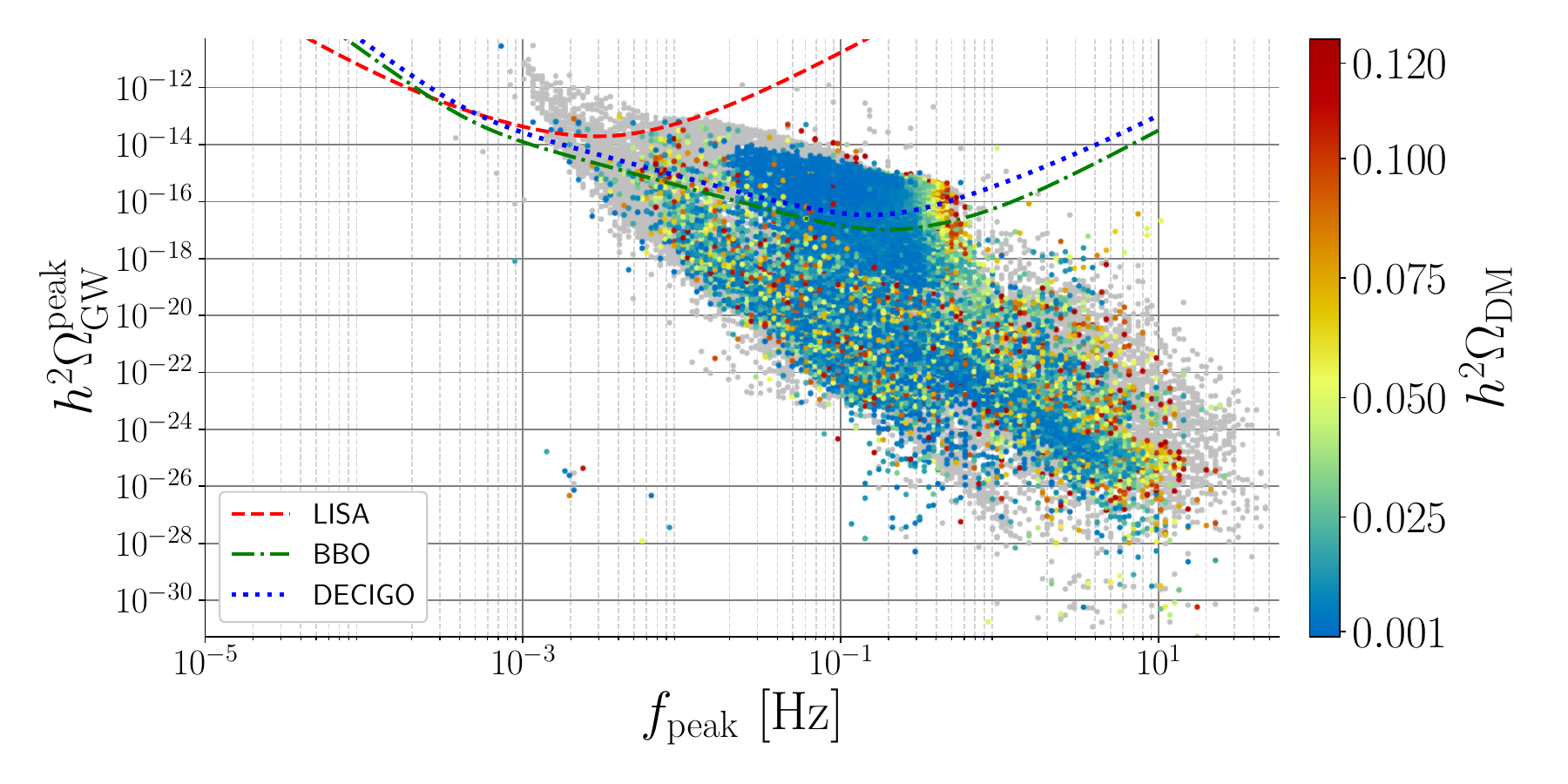}
    \caption{\footnotesize
    Predictions for the SGWB geometric parameters $h^2 \Omega^\mathrm{peak}_\mathrm{GW}$ and $f_{\mathrm{peak}}$ for Scenario III in terms of the DM relic abundance. The sensitivity curves are the same as in \cref{fig:GW_plots_spectra_modelI}.}
	\label{fig:GW_plots_spectraDM_scenarioIII}
\end{figure*}

Having established the connection between the 4D theory parameters, the phase transition patterns, and the predicted SGWB peak amplitude and frequency, the remaining question is whether the measured DM relic abundance can be accommodated within the observability region, as suggested in \cref{fig:GW_scenarioIII_1,fig:GW_scenarioIII_2}. This is indeed confirmed in \cref{fig:GW_plots_spectraDM_scenarioIII} where  there is a noticeable clustering of points with $0.05 \lesssim h^2 \Omega_\mathrm{DM} \lesssim 0.12$, which accounts for $40\%$ or more of the total DM abundance at the sensitivity reach of LISA, BBO, and DECIGO.
In particular, we identify several points that saturate the DM relic density within the $\mathrm{dHz}$ frequency range, featuring an SNR\footnote{The SNR is calculated as the vertical distance from the point to the PISC curve \cite{Schmitz:2020syl}.} at BBO/DECIGO ranging from approximately $10$ to $100$. Notably, a hypothetical observation of a SGWB at BBO/DECIGO would likely favour scenarios that account for a significant fraction or even the entirety of the DM relic abundance, if interpreted in the scope of the FPVDM model.

\subsubsection{Two-dimensional scan analysis}

The multi-dimensional nature of inclusive scans can easily obscure the continuous connection between different regions of the parameter space when viewed in two-dimensional projections. For a clearer picture, we select a two benchmark points based on the regions identified in \cref{tab:summary} and perform a scan varying $\g{D}$ and the DM mass $M_{\rm V_{D}}$, while keeping all other parameters fixed. The two selected benchmark points are shown in \cref{tab:benchs}.
\begin{table*}[h!]
    \centering
    \begin{tabular}{@{}rccccr@{}}
        & \multicolumn{3}{c}{} \\
        \hline
        & Parameter & BP1 - \cref{fig:2DParamScan_scenario3} & BP2 - \cref{fig:2DParamScan_scenario3_BigGW} & \\ \hline
        & &  & &\\
        & $M_{\mathrm{H_D}}/\mathrm{GeV}$ & $152.99$ & $2569.54$ & \\
        &  & &  & &\\
        & $M_{\mathrm{V_D}}/\mathrm{GeV}$ & $2323.50$ & $1050.19$ & \\
        &  & &  & &\\
        & $m_{f_{\rm D}}/\mathrm{GeV}$ & $2750.53$ & $1305.17$ & \\
        &  & &  & &\\
        & $m_{F}/\mathrm{GeV}$ & $2784.39$ & $1616.83$ & \\
        &  & & &\\
        & $y^\prime$ & $0.109$ & $0.345$ & \\
        &  & &  & &\\
        & $\sin\theta_S$ & $0.092$ & $0.042$ & \\
        &  & & &\\
        & $g_{\rm D}$ & $0.83$ & $0.54$ & \\
        &  & &  & &\\
        & $h^2 \Omega^\mathrm{GW}_{\mathrm{peak}}$ & $9.53\times 10{-17}$ & $3.62\times 10^{-15}$ & \\
        &  & & &\\
        & $f_{\mathrm{peak}}/\mathrm{Hz}$ & $0.108$ & $0.068$ &\\
        &  & &  & &\\
        & $\alpha$ & $ 0.22$ & $0.084$ & \\
        &  & &  & &\\
        & $\beta/H(T_p)$ & $12061.98$ & $574.36$ &\\
        &  & &  & &\\
        & $h^2 \Omega_\mathrm{DM}$ & $0.120$ & $0.120$ &\\
        &  & &  & &\\
        \hline
    \end{tabular}
    \caption{\footnotesize Model and thermodynamic parameters for the benchmark points used to produce \cref{fig:2DParamScan_scenario3,fig:2DParamScan_scenario3_BigGW}. The masses of the fields are given in GeV and the frequency $f_\mathrm{peak}$ is given in Hz.}
    \label{tab:benchs}
\end{table*}

\begin{figure}[htbp]
    \centering
 	\subfloat{\includegraphics[width=0.48\textwidth]{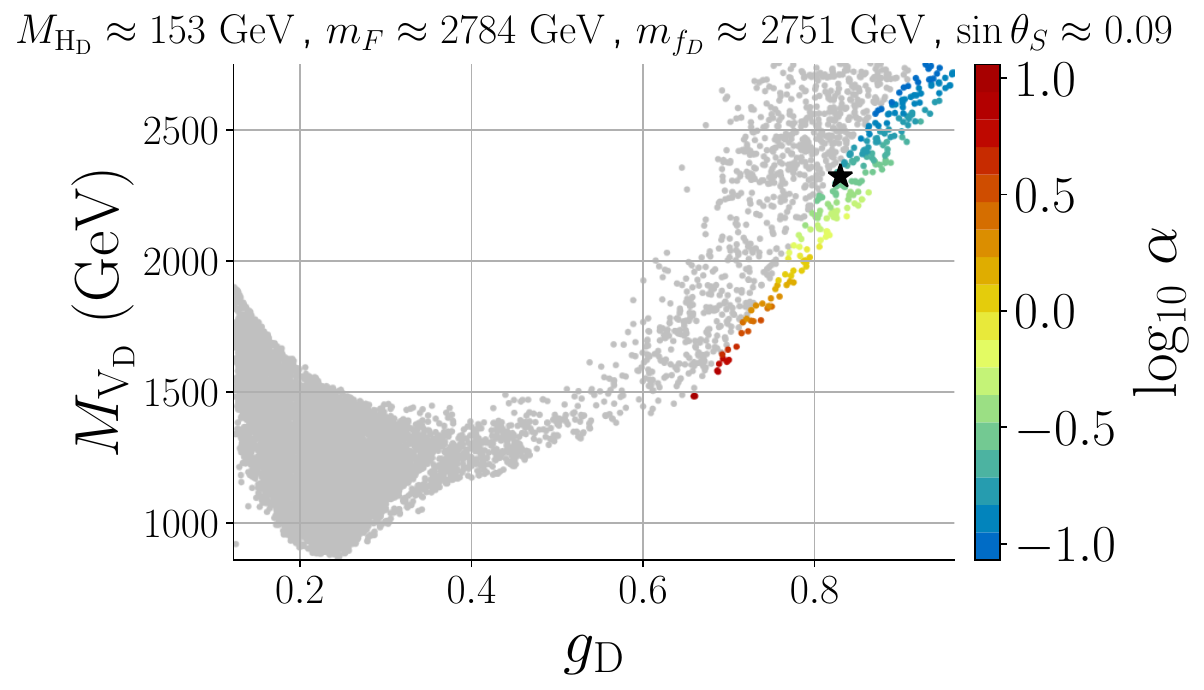}}
    \subfloat{\includegraphics[width=0.48\textwidth]{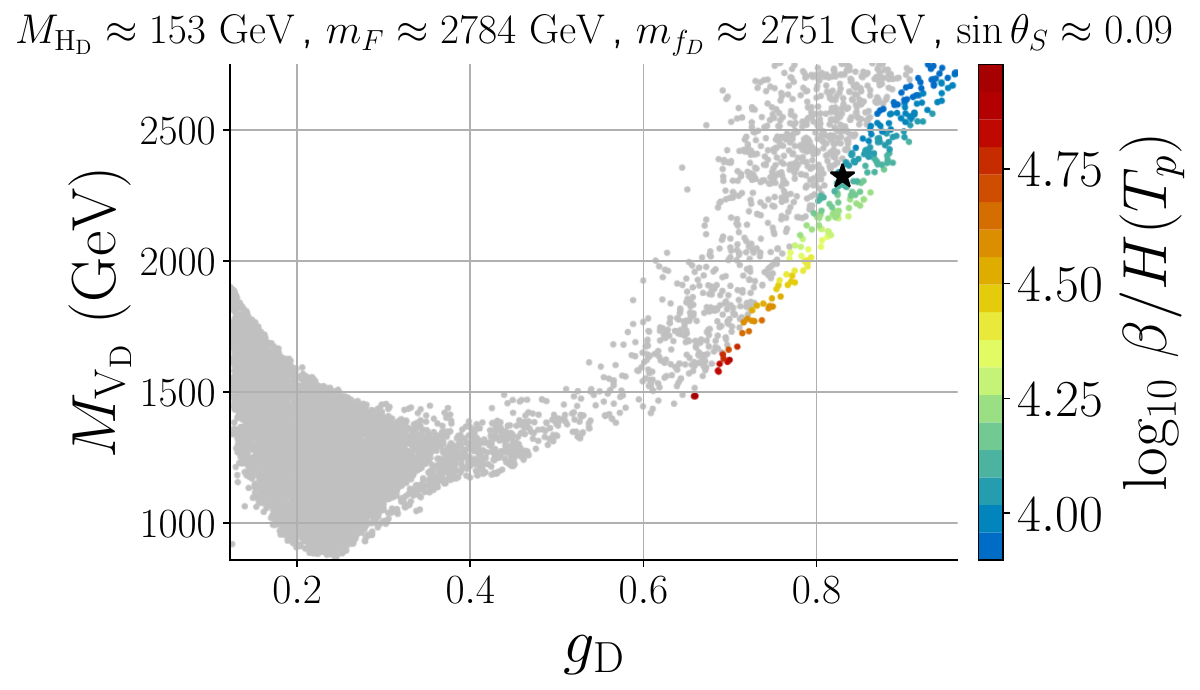}} \\
    \subfloat{\includegraphics[width=0.48\textwidth]{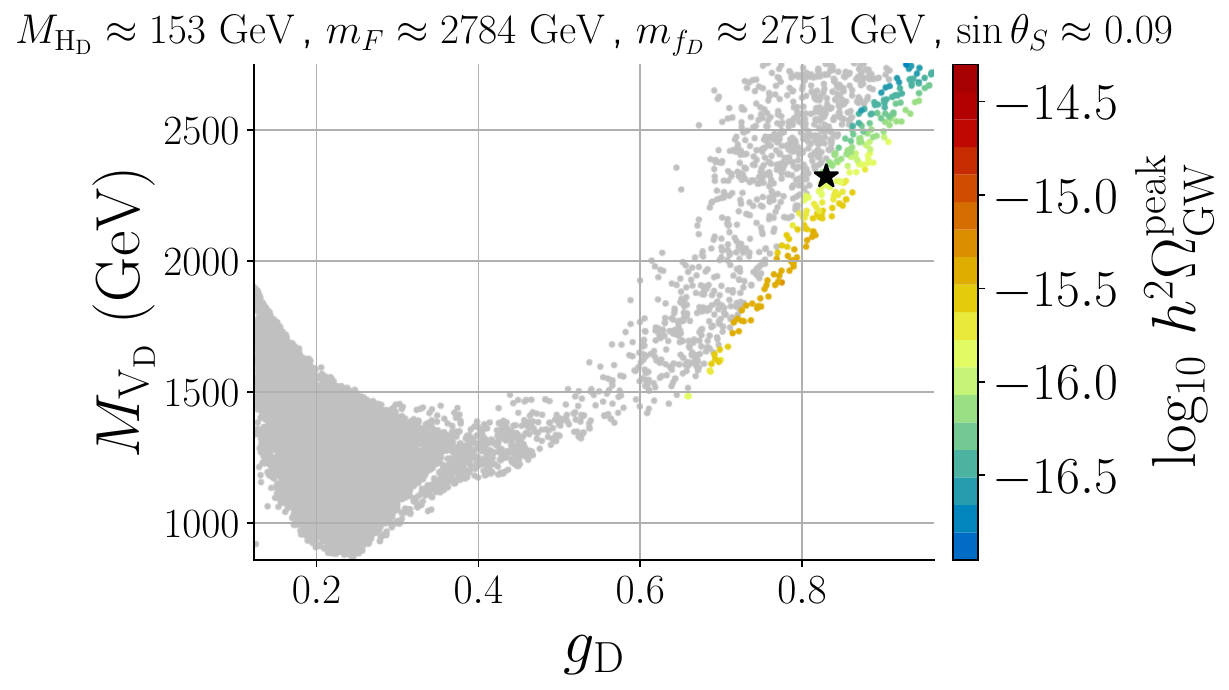}}
    \subfloat{\includegraphics[width=0.48\textwidth]{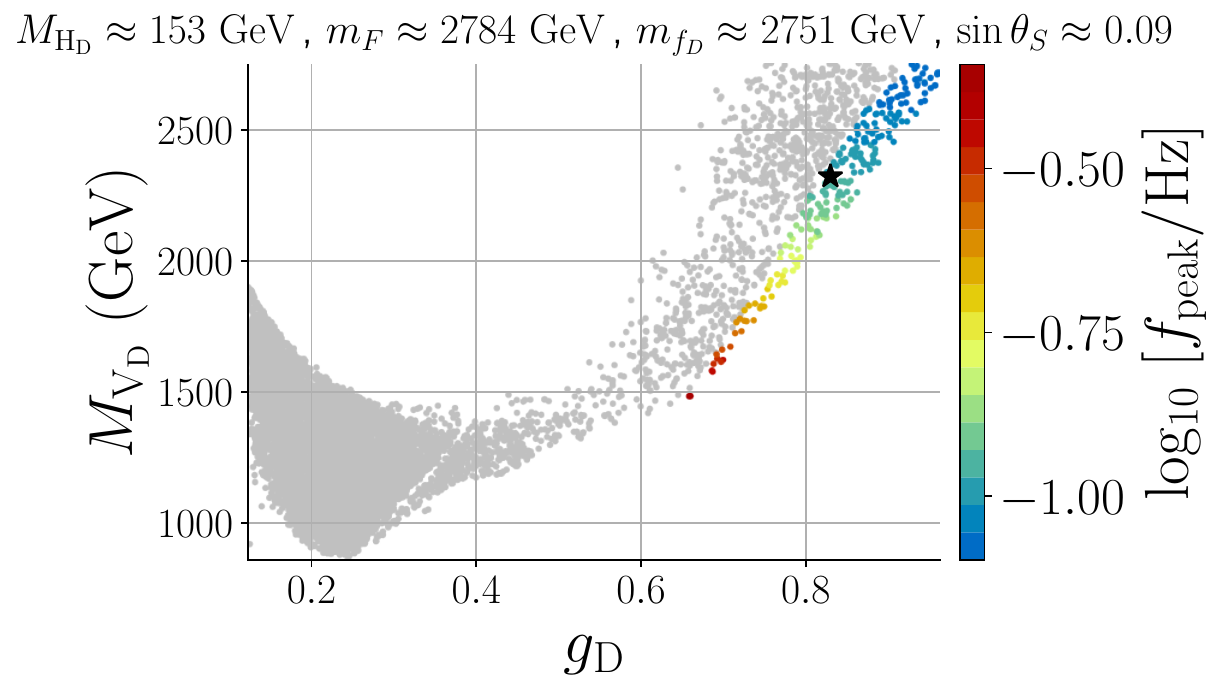}} \\
    \subfloat{\includegraphics[width=0.48\textwidth]{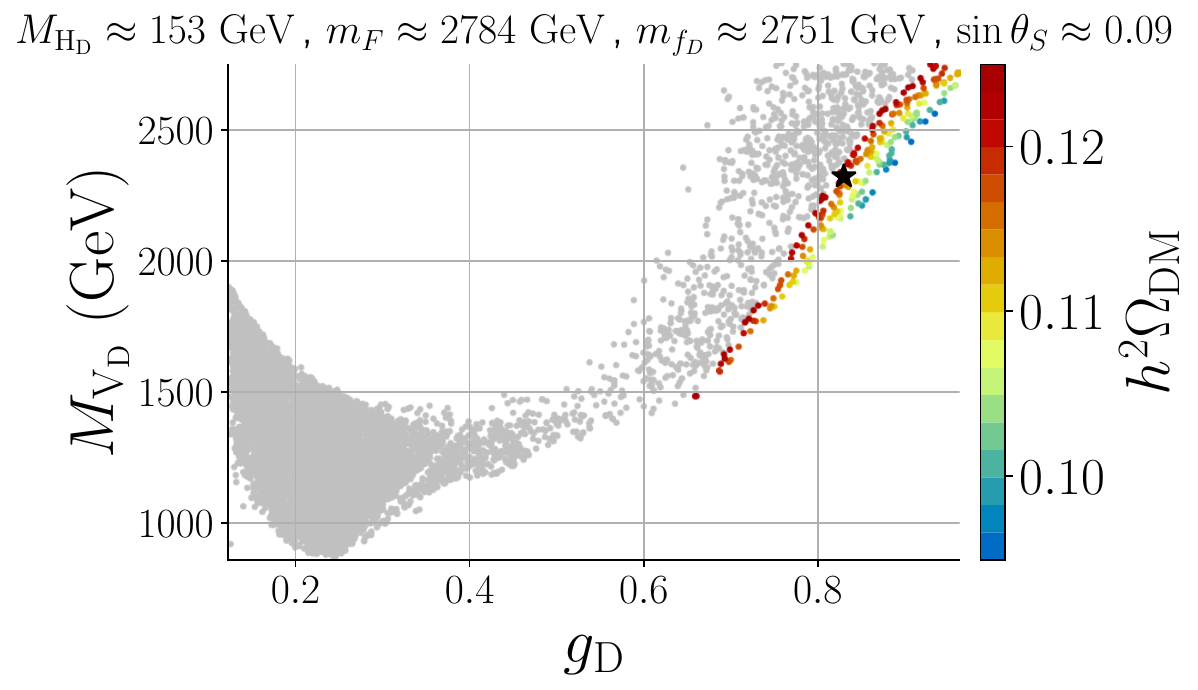}}
    \subfloat{\includegraphics[width=0.48\textwidth]{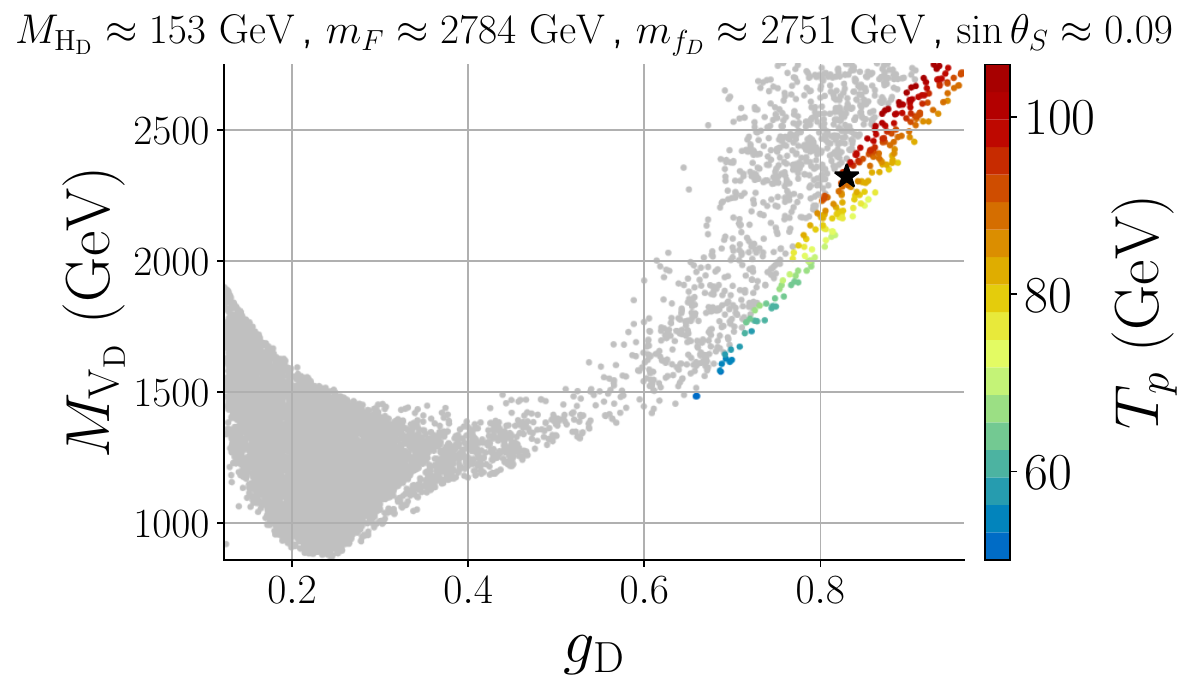}} \\
 	\subfloat{\includegraphics[width=0.48\textwidth]{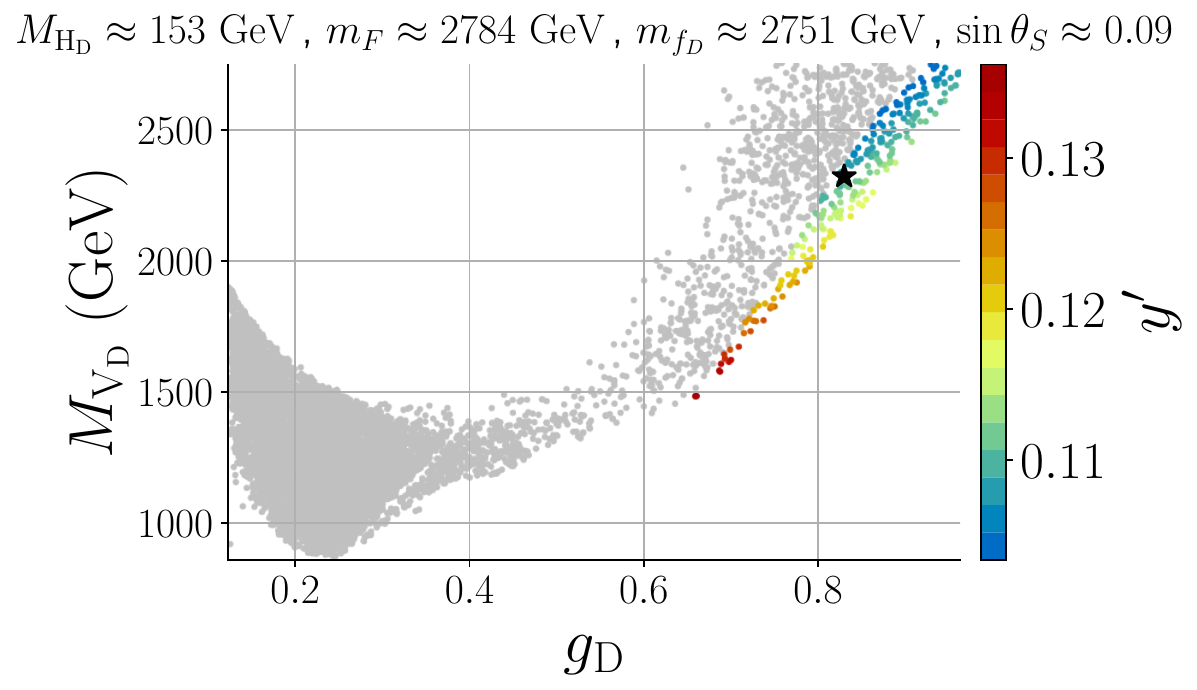}}
    \subfloat{\includegraphics[width=0.48\textwidth]{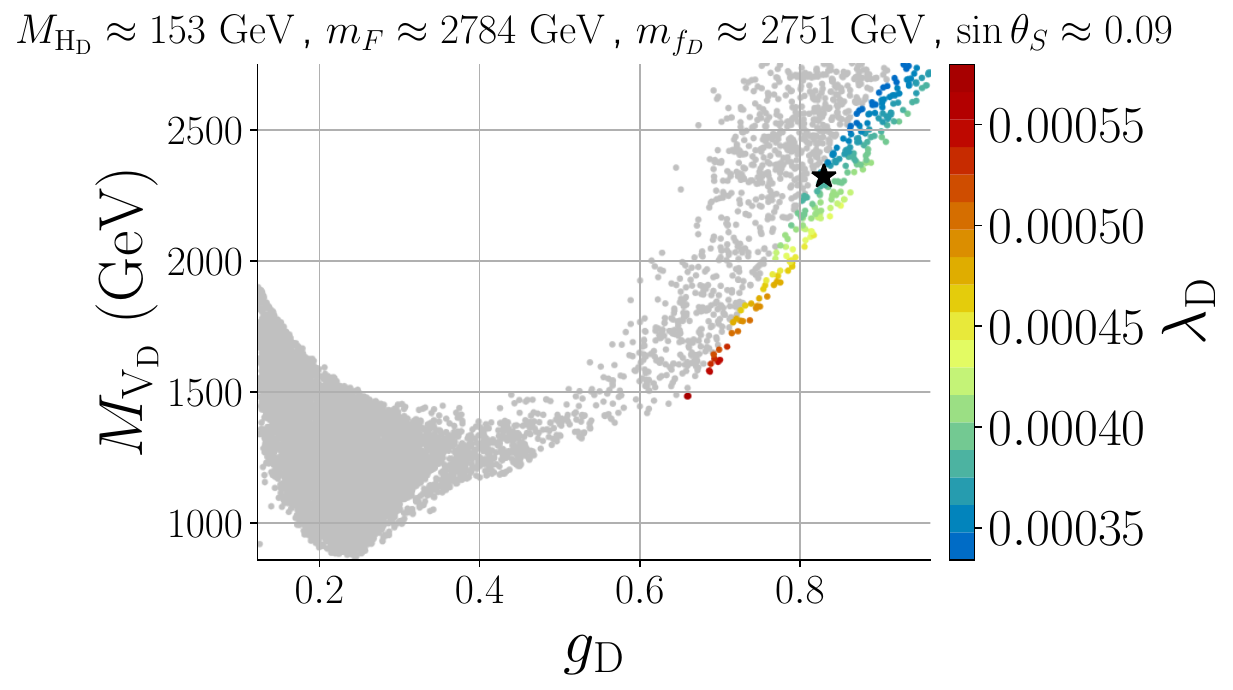}}
    \caption{\footnotesize 
    The colour map of various parameters for the 2D parameter scan for the benchmark point BP2 in \cref{tab:benchs} in Scenario III. Some of the representative fixed values are indicated in the title of each panel. All transitions follow the pattern $(v_h,0)\rightarrow (v_h, v_{\rm D})$.}
	\label{fig:2DParamScan_scenario3}
\end{figure}

The first example corresponds to the $(v,0) \to (v,v_\mathrm{D})$ transition pattern in \cref{tab:summary} and is shown in \cref{fig:2DParamScan_scenario3}. 
In this scenario, the dark Higgs mass is on the order of the EW scale, while the remaining BSM particles have masses between 2 TeV and 3 TeV, with a large scalar portal interaction of the order $\sin \theta_S \sim \mathcal{O}(0.1)$. This benchmark accounts for all of the DM abundance, falling below LISA's sensitivity range. Our goal here is to determine if, by moving within the 2D $(\g{D},M_\mathrm{V_D})$ plane, we can continuously connect this point, marked by a black star in \cref{fig:2DParamScan_scenario3}, to the early observability region, while simultaneously saturating the DM abundance. 

As we can see in the left panel of the third row in \cref{fig:2DParamScan_scenario3}, saturating the DM abundance can be achieved by decreasing $\g{D}$, independently of $M_\mathrm{V_D}$. However, note that the colour gradient in the two panels of the first row is identical, such that an increase in the strength of the phase transition $\alpha$ is accompanied by an increase in its inverse duration $\beta/H(T_p)$. This results in a balancing effect on the SGWB peak amplitude: while a growing $\alpha$ dictates an increase in $h^2 \Omega_\mathrm{GW}^\mathrm{peak}$, a growing $\beta/H(T_p)$ damps it while increasing the peak frequency. Consequently, the SGWB peak amplitude never enters the LISA sensitivity region and remains mostly between $10^{-16} \lesssim h^2 \Omega_\mathrm{GW}^\mathrm{peak} \lesssim 10^{-15}$ for frequencies between 0.1 Hz and 1 Hz. However, it is possible to saturate the DM relic abundance for $\g{D} \approx 0.75$ and a $\mathrm{V_D}$ mass between 1.7 TeV and 2 TeV, with SGWB predictions within reach of future planned experiments such as BBO. In this region, the percolation temperature is $T_p \approx 70~\mathrm{GeV}$, slightly below the EW scale, caused by moderate supercooling due to $\alpha \approx 1$. The fermion portal coupling is $y^\prime \approx 0.12$, with rather small $\lambda_\mathrm{D} \sim \mathcal{O}(10^{-4})$. 

 \begin{figure}[htbp]
    \centering
  	\subfloat{\includegraphics[width=0.48\textwidth]{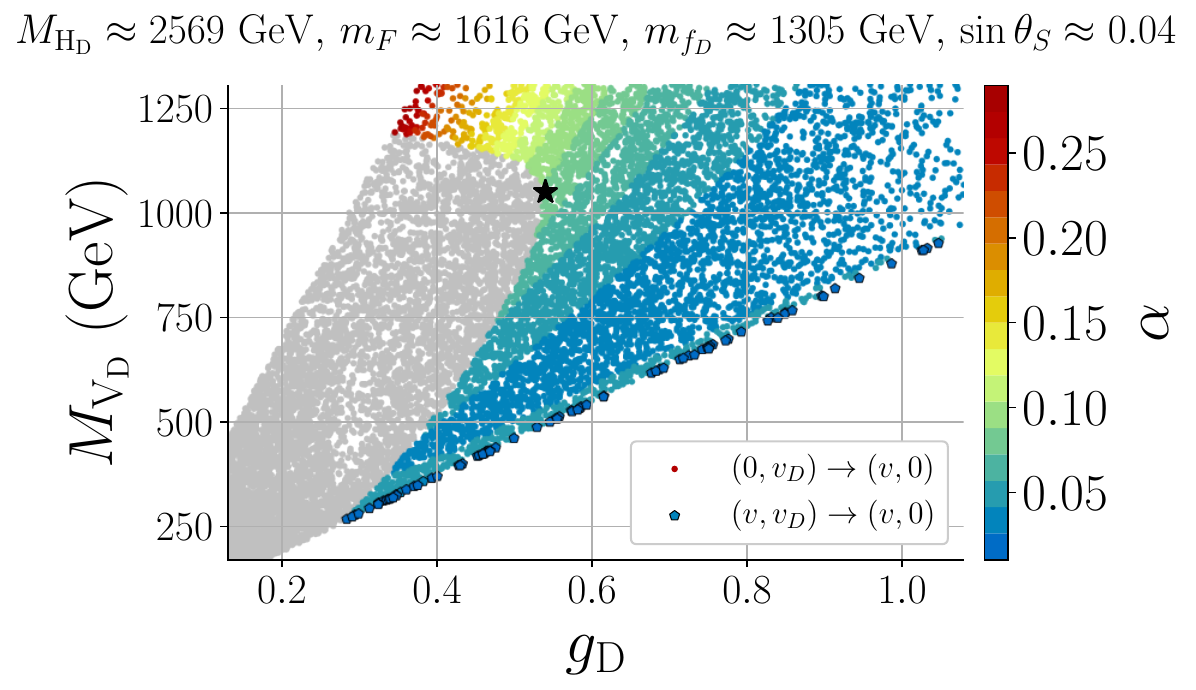}}
    \subfloat{\includegraphics[width=0.48\textwidth]{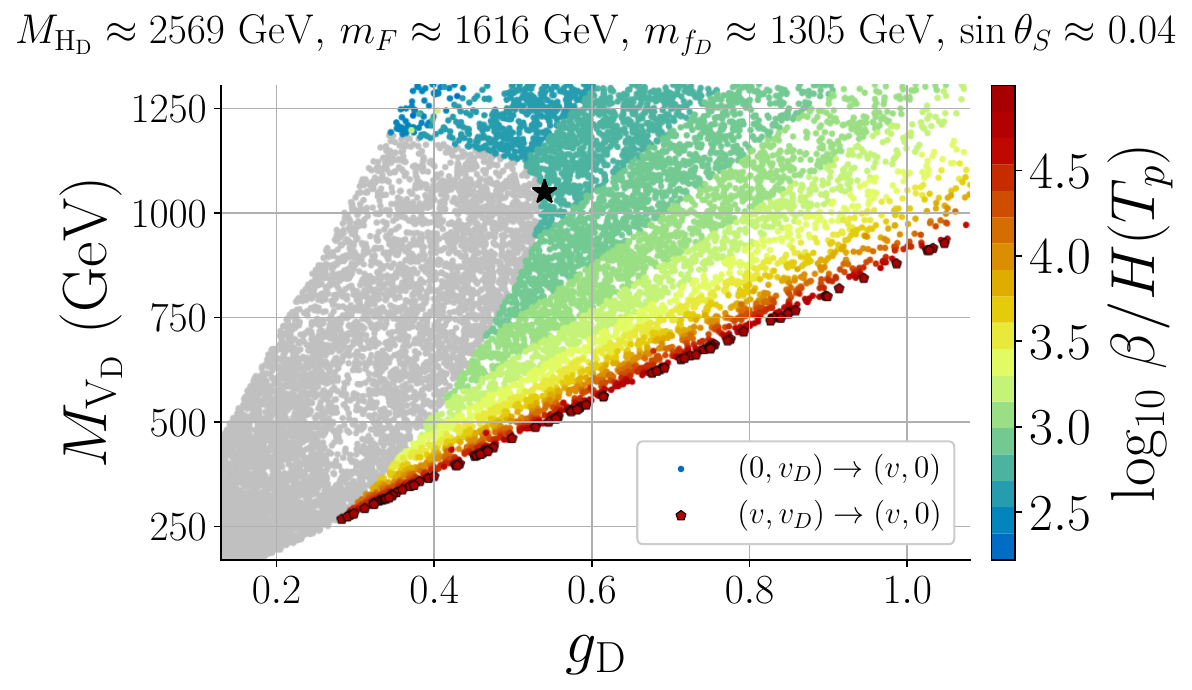}}
    \vspace*{-0.3cm}
    \\
    \subfloat{\includegraphics[width=0.48\textwidth]{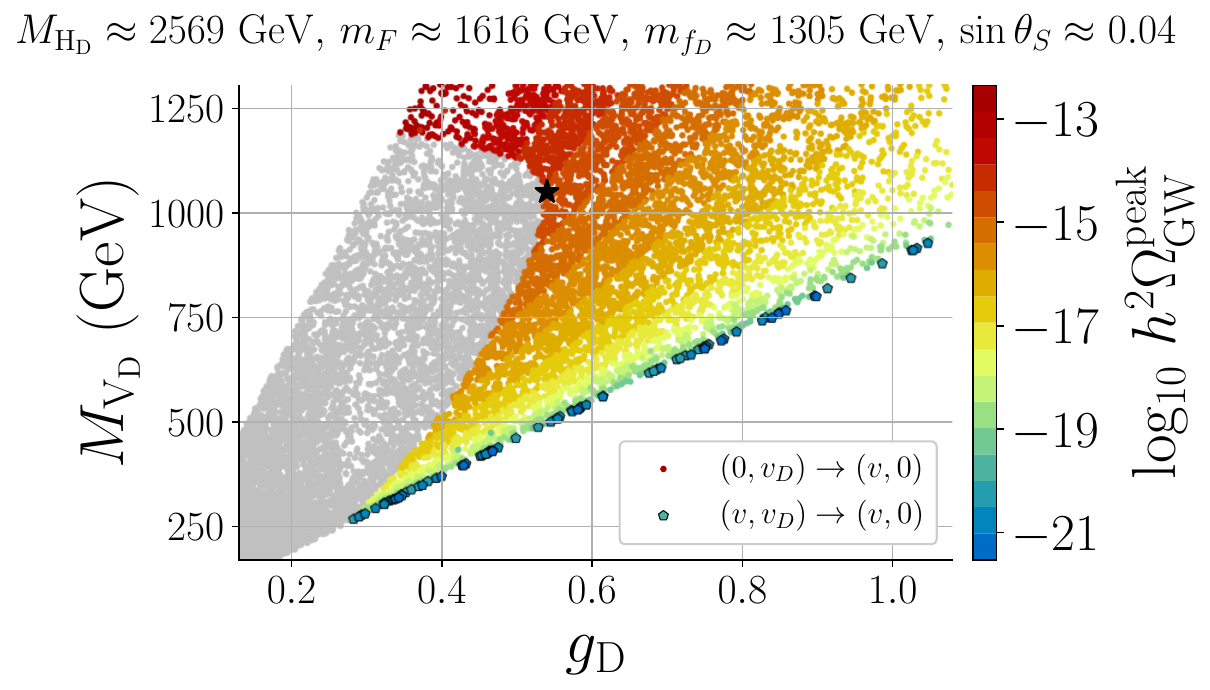}}
    \subfloat{\includegraphics[width=0.48\textwidth]{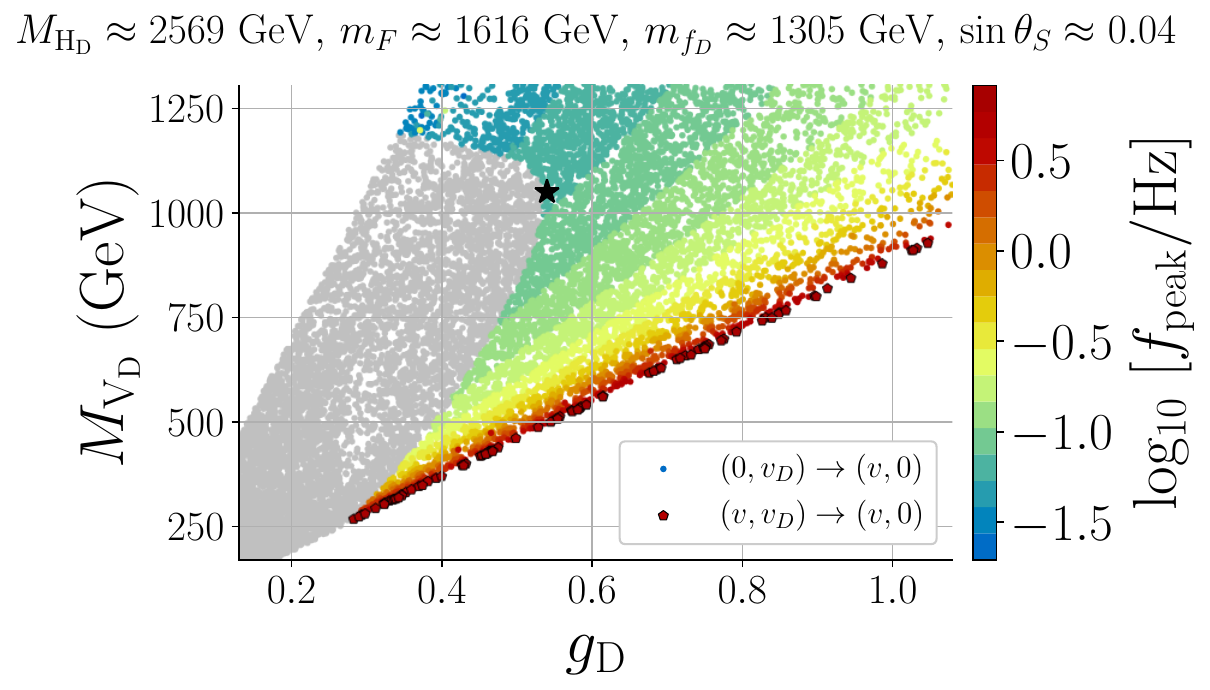}} 
    \vspace*{-0.3cm}
   \\
    \subfloat{\includegraphics[width=0.48\textwidth]{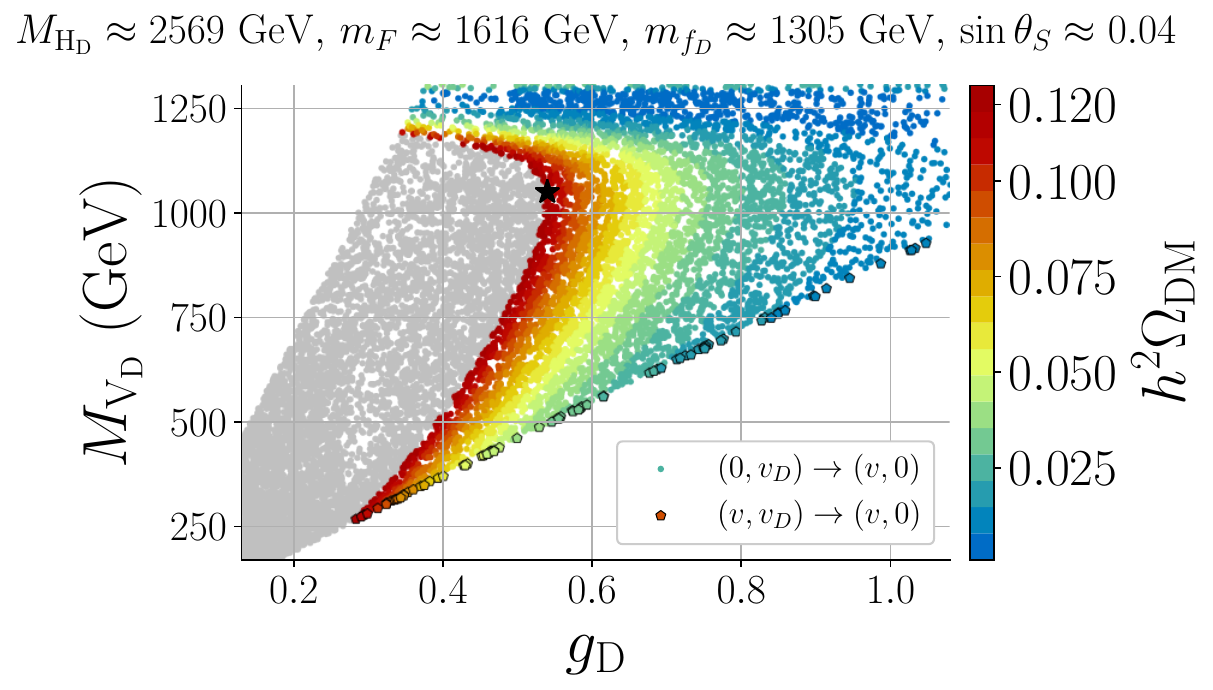}}
    \subfloat{\includegraphics[width=0.48\textwidth]{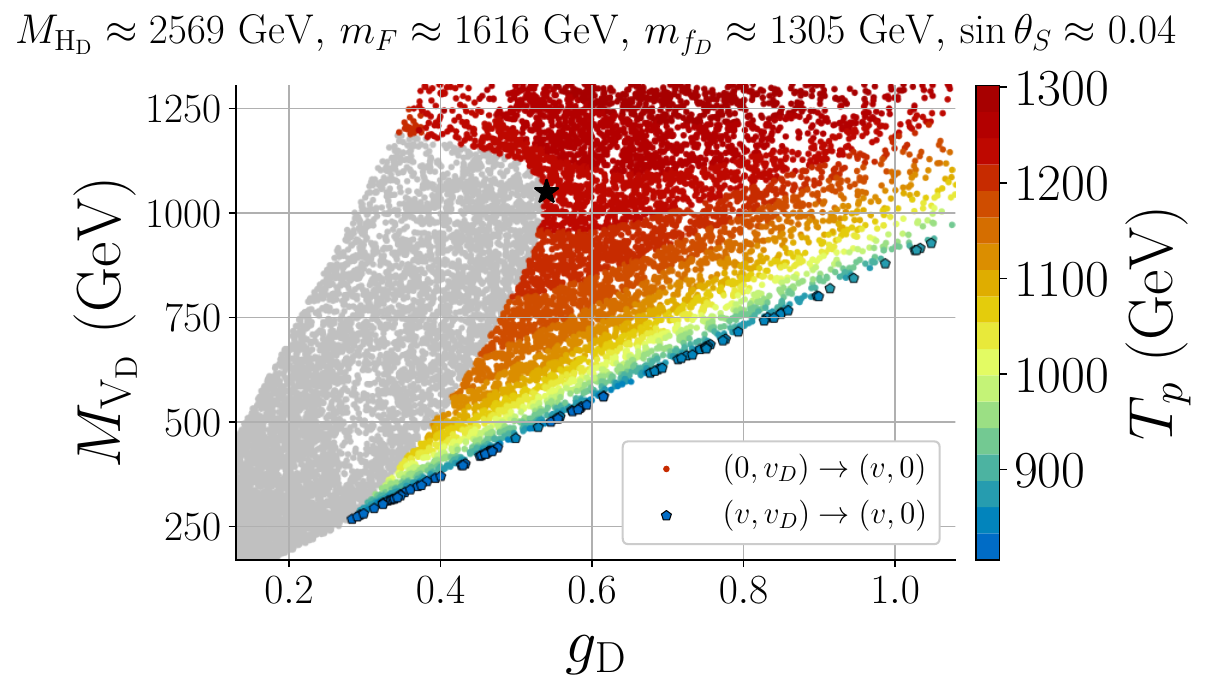}}     
    \vspace*{-0.3cm}
   \\
    \subfloat{\includegraphics[width=0.48\textwidth]{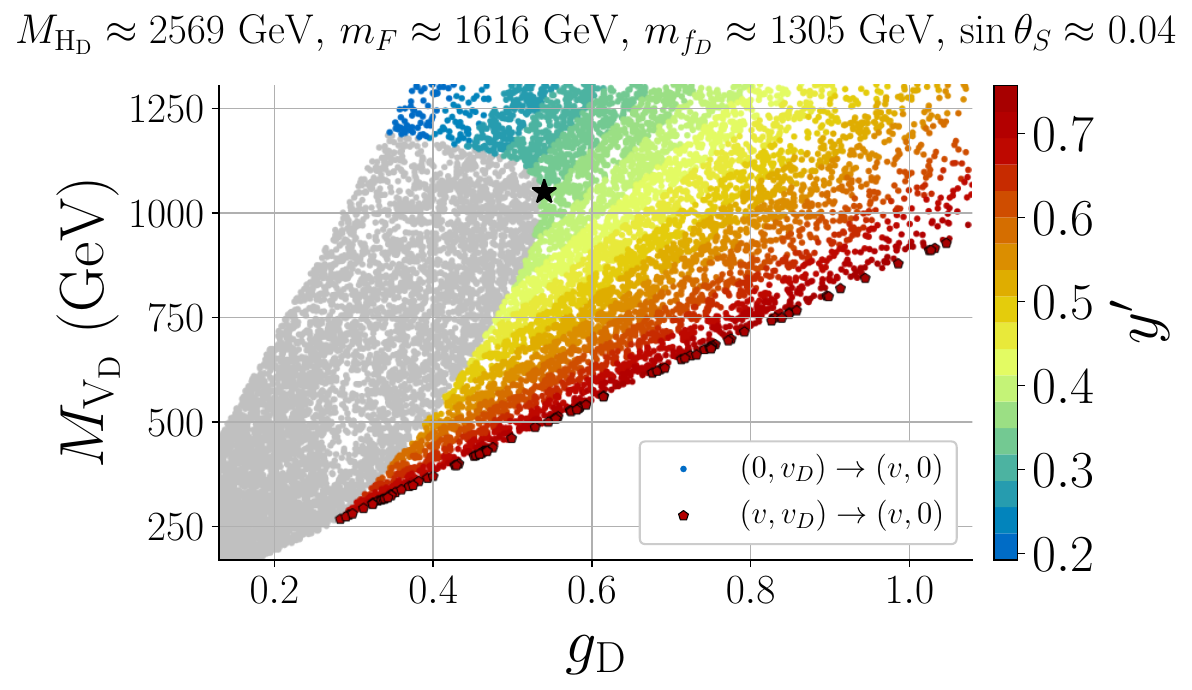}}
    \subfloat{\includegraphics[width=0.48\textwidth]{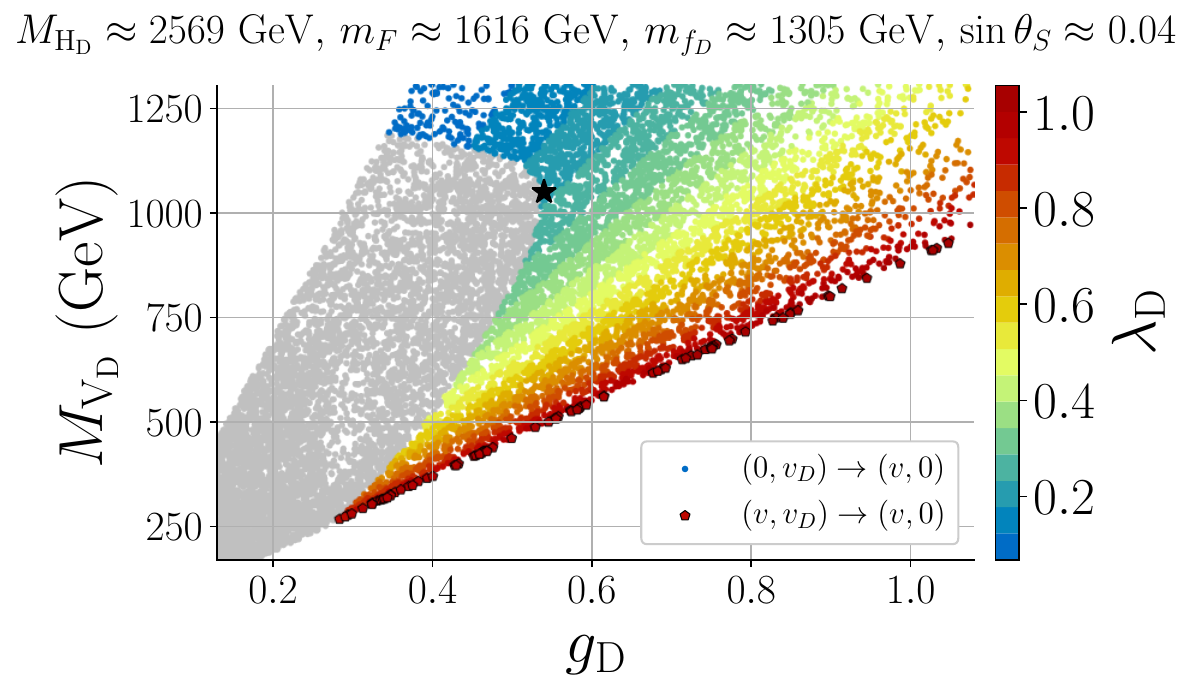}} 
    \caption{\footnotesize
    The colour map of various parameters for the 2D parameter scan for the benchmark point BP3 in \cref{tab:benchs} in Scenario III. Some of the representative fixed values are indicated in the title of each panel.}
	\label{fig:2DParamScan_scenario3_BigGW}
\end{figure}

The second benchmark we consider is that on the second column of \cref{tab:benchs} and involves the phase transition pattern $(0,v_\mathrm{D}) \to (v,0)$, as shown in the last column of \cref{tab:summary}. The key differences here include a TeV-scale dark Higgs, as well as an order-one portal Yukawa coupling $y^\prime$ and dark doublet self-interaction $\lambda_\mathrm{D}$. In \cref{fig:2DParamScan_scenario3_BigGW}, we present our results for the 2D scan performed around the black star in the figure panels.
In contrast to what was observed in the previous case, the panels in the first row show that as $\alpha$ increases, $\beta/H(T_p)$ decreases, both contributing to enhance the SGWB peak amplitude. Additionally, a smaller FOPT inverse duration results in a lower frequency. 

The point selected from the inclusive scan (black star) accounts for all of the DM relic abundance and has a SGWB peak amplitude below LISA's sensitivity. Decreasing the $\SU{2}{D}$ gauge coupling allows for saturation of $h^2 \Omega_\mathrm{DM}$ along the red stripe in the left plot of the third row. Conversely, we observe that only in the top-left corner of the viable parameter space, with $\g{D}$ just under $0.4$ and $M_\mathrm{V_D}$ around $1.25~\mathrm{TeV}$, does the SGWB peak amplitude enter the region covered by LISA. Therefore, by comparing the left panels of the two middle rows, we find a small overlap between the red stripe where $h^2 \Omega_\mathrm{DM} = 0.12$ and the dark red area featuring $h^2 \Omega_\mathrm{GW}^\mathrm{peak} \lesssim 10^{-13}$. For those points, we have $y^\prime \approx \lambda_\mathrm{D} \approx 0.2$.

Note that for this benchmark, we observe a smooth shift to a different FOPT pattern, specifically from $(0,v_\mathrm{D}) \to (v,0)$ to $(v,v_\mathrm{D}) \to (v,0)$, highlighted with black circles. These points populate a stripe along the bottom edge of the viable parameter space, where the portal Yukawa coupling and quartic self-interaction reach their maximum values of $y^\prime \approx 0.7$ and $\lambda_\mathrm{D} \approx 1$. However, the predicted SGWB peak amplitude reaches values on the order of $h^2 \Omega_\mathrm{GW}^\mathrm{peak} \sim \mathcal{O}(10^{-21})$ and a peak frequency of approximately $f_\mathrm{peak} \approx 10~\mathrm{Hz}$, which are too weak to be detected by GW interferometers.
\newpage
\subsection{Collider signatures and multi-messenger interplay in FPVDM}

This section is devoted to collider phenomenology and its interplay with GW and DM observables. In particular, we highlight the remarkable fact that benchmark points selected by GW observability are simultaneously testable at the LHC and HL-LHC through distinctive multi-top final states.

So far, we have discussed in detail the strongly FOPT and the resulting GW signal originating from the dark gauge sector. Scenario III, however, presents an even more intriguing case, as it features a remarkable interplay of qualitatively distinct signatures, including GW signals, DM direct detection prospects, and collider signatures with multi-top final states at the LHC, which we elaborate upon in this section.
By requiring a GW SNR ratio at either LISA/BBO exceeding 10, \textit{i.e.} $\mathrm{SNR_{LISA}} > 10$ or $\mathrm{SNR_{BBO}} > 10$, and ensuring compatibility with the observed DM relic density, $h^2 \Omega_\mathrm{DM} < 0.12$ and DM direct detection bounds, our comprehensive parameter scan yields only four viable benchmark points. These points are summarised in Table~\ref{tab:SNR10_benchmarks}.
A particularly striking outcome is that all benchmark points BM1--BM4 satisfying the GW observability requirement are simultaneously testable at the HL-LHC through their collider signatures. This non-trivial correlation between GW sensitivity and collider accessibility constitutes one of the key results of this work.

\begin{table}[htb!]
\centering
\begin{tabular}{lcccc}
\toprule
{} & BM1 & BM2 & BM3 & BM4 \\
\midrule
\textbf{$g_{\rm D}$}
& 1.074 & 1.004 & 0.977 & 0.954 \\
\textbf{$\lambda_{\rm D}$} 
& $4.812\times 10^{-5}$ & 0.00411 & $0.00227$ & 0.00215 \\
\textbf{$\lambda_{\rm HD}$} 
& $-2.488\times 10^{-10}$ & $-8.240\times 10^{-11}$ & $1.464\times 10^{-5}$ & $1.951\times 10^{-12}$ \\
\textbf{$y'$} 
& 0.6769 & $0.142$ & $0.290$ & $0.232$ \\
\textbf{$y_t$} 
& 0.9986 & $0.989$ & $1.050$ & $1.014$ \\
\textbf{$\sin\theta_S$} 
& $2.29\times 10^{-9}$ & $2.395\times 10^{-9}$ & $5.13\times 10^{-4}$ & $1.853\times 10^{-10}$ \\
\textbf{$M_{V_{\rm D}}\mathrm{[GeV]}$ } 
& 312.6 & 480.342 & 1359.868 & 1064.723 \\
\textbf{$m_{f_\mathrm{D}}\mathrm{[GeV]}$} 
& 1770 & 1747.016 & $1594.319$ & 1582.937 \\
\textbf{$m_{F}\mathrm{[GeV]}$} 
& 1792 & 1749.673 & $1694.404$ & 1625.273 \\
\textbf{$M_{H_{\rm D}}\mathrm{[GeV]}$} 
& 5.709 & 86.669& 187.617 & 146.360 \\
\textbf{$T_c\mathrm{[GeV]}$} & 15.268 & $35.872$ & $49.933$ & $47.518$ \\
\textbf{$T_n\mathrm{[GeV]}$} 
& 15.265 & $34.807$ & $49.172$ & $46.465$ \\
\textbf{$T_p\mathrm{[GeV]}$} 
& 15.264 & $34.807$ & $49.172$ & $46.464$ \\
\textbf{$\alpha$} 
& 82.31 & $1.144$ & $1.442$ & $1.38$ \\
\textbf{$\beta/H(T_p)$} 
& 539.09 & $3860.759$ & $6332.000$ & $5768.412$ \\
\textbf{$f_\mathrm{peak}/\mathrm{Hz}$} 
& 0.00152 & $0.00961$ & $0.0226$ & $0.0198$ \\
\textbf{$h^2\Omega^\mathrm{GW}_\mathrm{peak}$} & $1.50\times 10^{-11}$ & $6.289\times 10^{-14}$ & $3.052\times 10^{-14}$ & $3.513\times 10^{-14}$ \\
\textbf{$h^2\Omega_\mathrm{DM}$} 
& 0.000949 & 0.00286 & 0.0229 & 0.0159 \\
\texttt{DD\_pval} (LZ2024) 
& 0.64 & $0.93$ & 0.89 & 0.98 \\
\texttt{DD\_factor} (LZ2024) 
& 5.2 & 33.1 & 21.5 & 141 \\
\textbf{$\mathrm{SNR}_\mathrm{LISA}$} & 26.27 & $0.178$ & $0.0275$ & $0.0404$ \\
\textbf{$\mathrm{SNR}_\mathrm{BBO}$} & $55.72$ & $111.117$ & $256.308$ & $223.925$ \\
\bottomrule
\end{tabular}
\caption{\footnotesize Benchmark points from Scenario III satisfying the requirement of a GW signal-to-noise ratio at LISA $\mathrm{SNR_{LISA}} > 10$ or BBO $\mathrm{SNR_{BBO}} > 10$ and consistent with the DM relic density constraint $h^2 \Omega_\mathrm{DM} < 0.12$ together with DM direct detection bounds. These points represent a distinctive interplay of GW signatures, DM direct detection prospects, and collider signals with multi-top final states.}
\label{tab:SNR10_benchmarks}
\end{table}

These points, labelled BM1--BM4, provide clear and appealing examples of the multi-signature phenomenology that can be expected from this model. Notably, all these points are consistent with the LZ2024 DM direct detection constraints. However, BM1 lies relatively close to the current LZ2024 sensitivity, with \texttt{DD\_factor} value of 5.2.

Looking ahead, the DARWIN experiment~\cite{Baudis:2024darwin} is expected to improve upon the LZ2024 sensitivity by approximately two orders of magnitude within the next decade. This advancement implies that all BM1--BM4 benchmark points can be simultaneously probed by both GW and DM direct detection experiments. This highlights the strong complementarity between GW observatories and DM direct detection experiments in probing the parameter space selected by a strong FOPT.

These points are characterised by non-vanishing values of the $y'$ coupling of the order $\mathcal{O}(0.1-1)$; a very light scalar $\mathrm{H_D}$, with mass near and below the EW scale and satisfying $M_{\mathrm{H_D}} \ll M_{\mathrm{V_D}}$; a relatively large dark gauge coupling $g_{\rm D}$ of order one; VL fermions with masses $m_F$ in the $1.5$--$2$~TeV range; and scalar mixing angles $\sin\theta_S$ with absolute values in the $10^{-9}$--$10^{-4}$ range. This parameter space is highly characteristic of the interplay between GW signals, DM direct detection, and collider signatures within Scenario~III.

An extremely intriguing aspect of the identified parameter space is the presence of a unique six-top final state signature, arising from $F\bar{F}$ pair production at the LHC or future hadron colliders such as FCC-hh~\cite{FCC:2018evy}. Importantly, the VL fermion masses associated with observable GW signals at LISA and BBO are not very large. As illustrated in Fig.~\ref{fig:mF_correlation_GWs}, the GW-sensitive region corresponds to $m_F \lesssim 2.3~\mathrm{TeV}$. This explains why all benchmark points selected by GW observability are simultaneously testable at the HL-LHC.

Specifically, the $F\bar{F}$ final state naturally leads to a cascade decay chain 
$$
F\bar{F} \;\to\; (t\, V')\, (\bar{t}\, V') \;\to\; (t\, t\bar{t})\, (\bar{t}\, t\bar{t}) \;\to\; t\bar{t}t\bar{t}t\bar{t} \ ,
$$
provided the kinematic conditions $M_{\mathrm{V_D}} > 2 m_t$ and $M_F > M_{V'} + m_t$ are satisfied.
For  $M_{\mathrm{V_D}}> 2 m_t$ and $m_F > M_{V'}+m_t$
the probability of $F\bar{F}\to 6t$ is about 100\%,
which makes the $6t$  signature unique and generic for almost entire the
parameter space relevant to GW signal.

\begin{figure}[htb]
\centering
    \begin{tabular}{c c}       
        \includegraphics[trim={0cm 13cm 0 0cm},clip,width=0.76\textwidth]{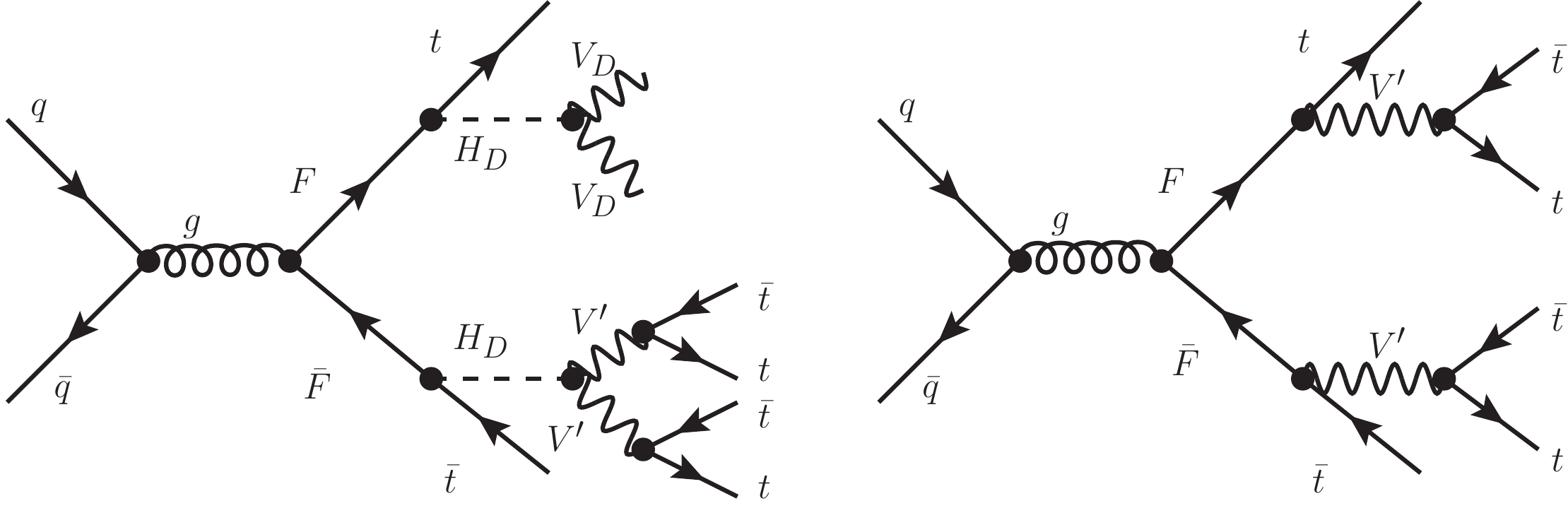} &\\
        (a) \hspace*{5cm} (b)&\\
                \includegraphics[trim={0cm 0cm 0 13cm},clip,width=0.76\textwidth]{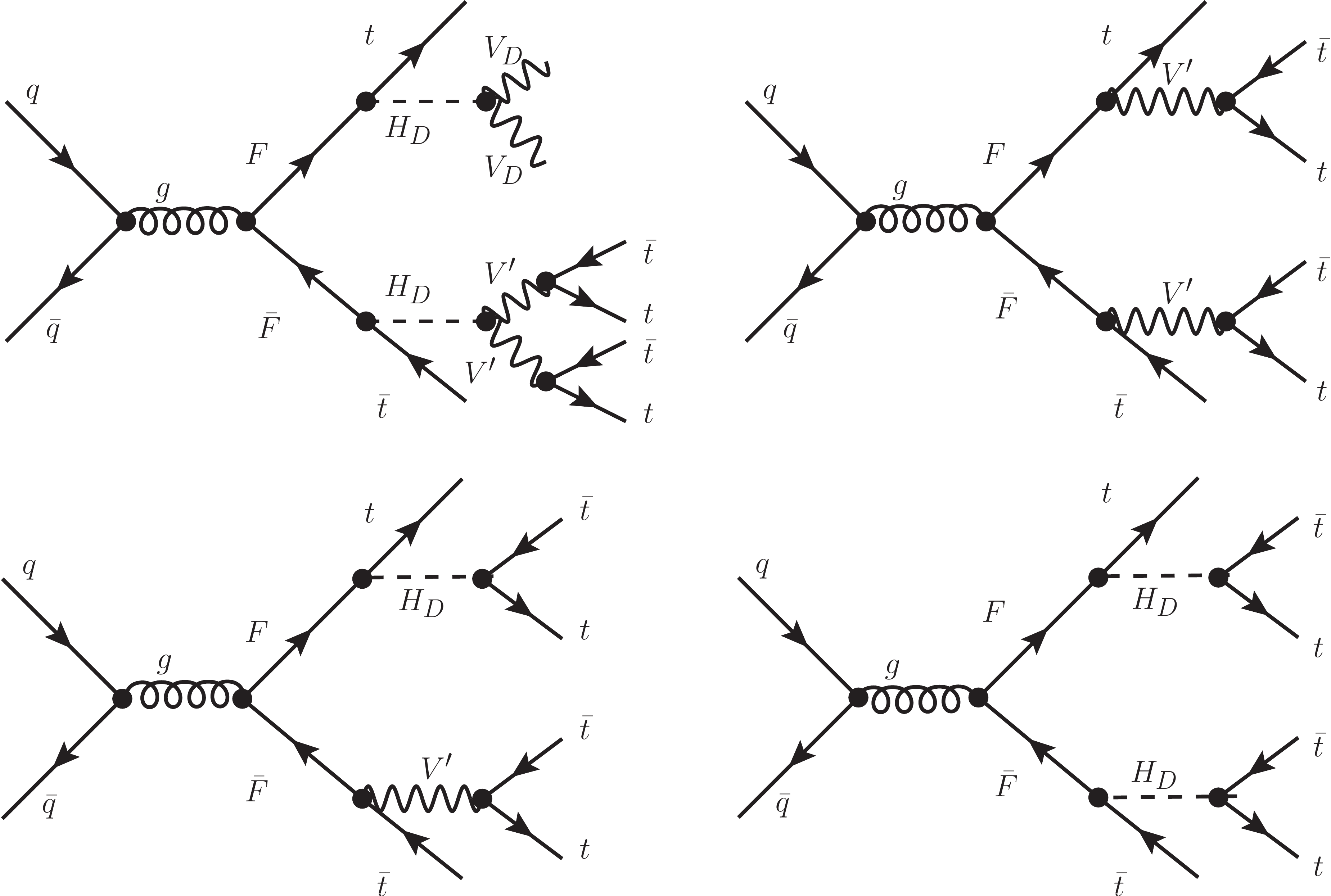} &\\
        (c) \hspace*{5cm} (d)&\\
        \end{tabular}
\caption{\footnotesize Representative Feynman diagrams contributing to the $pp \to F\bar{F} \to 6t$ final state. Diagram (b) shows the dominant contribution, where $F$ decays into a top quark and a heavy vector boson $V'$ which subsequently decays into a $t\bar{t}$ pair. Diagrams (a), (c), and (d) involve additional exchanges of the light scalar $\mathrm{H_D}$ and are suppressed due to the smallness of $M_{\mathrm{H_D}}$.\label{fig:6mu-diags} }
\end{figure}
\begin{figure}[htbp]
  \centering
  \includegraphics[width=0.43\textwidth]{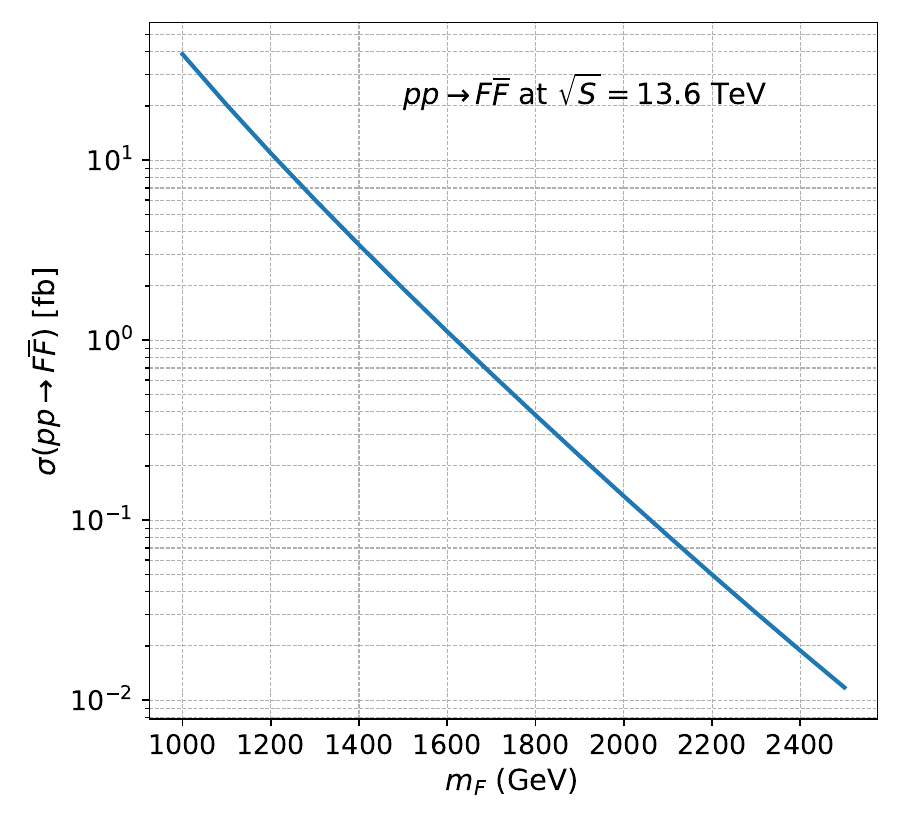}%
  \includegraphics[width=0.57\textwidth]{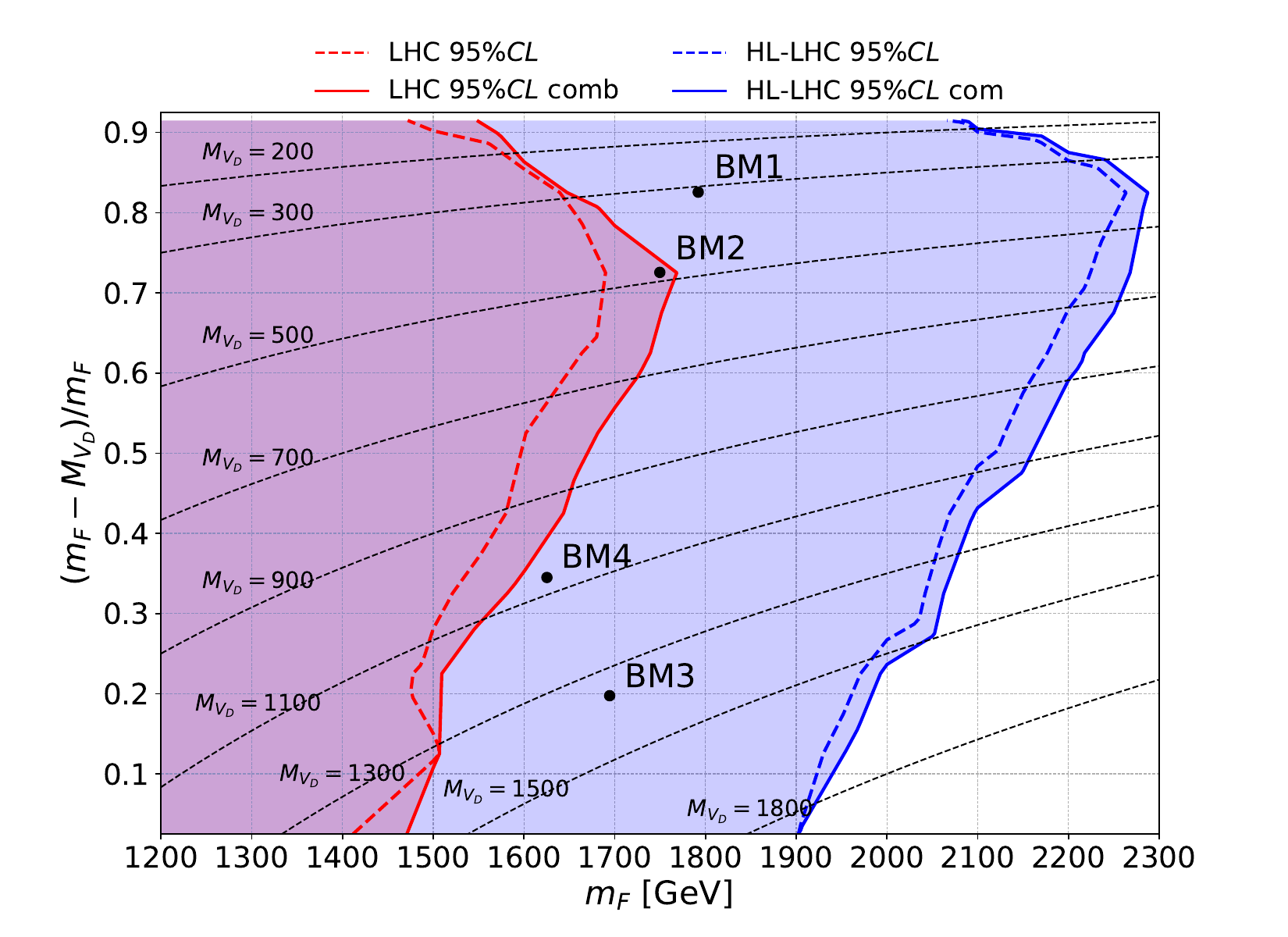}
  \caption{\footnotesize Left: Leading-order cross section for $pp \to F\bar{F}$ production as a function of $m_F$ at $\sqrt{s} = 13.6$~TeV. Right: Projected exclusion regions at 95\% CL for the six-top final state. See text for details.}
  \label{fig:6t-LHC}
\end{figure}

\begin{wrapfigure}{r}{0.6\textwidth}
    \vspace{-10pt}
    \includegraphics[width=\linewidth]{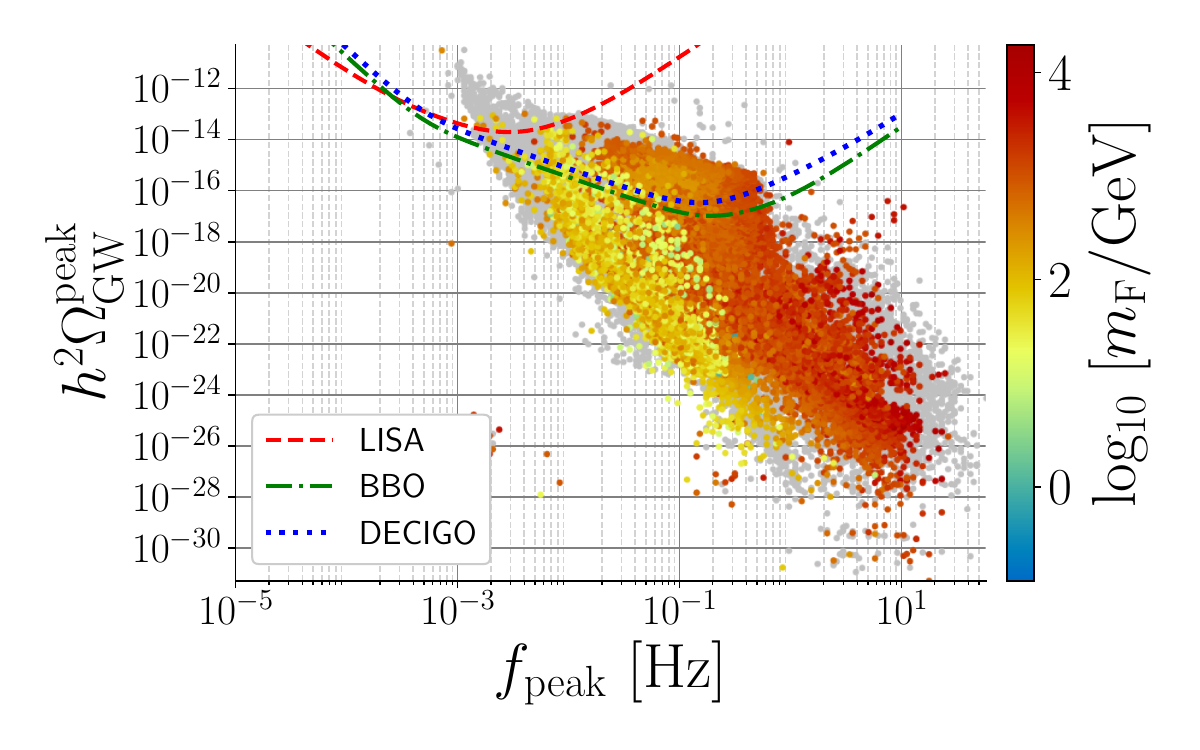}
    \caption{\footnotesize Peak SGWB amplitude $h^2\Omega_\mathrm{GW}^\mathrm{peak}$ as a function of the peak frequency $f_\mathrm{peak}$ with the mass of the VL fermion, $m_F$, in the colour axis.}
    \label{fig:mF_correlation_GWs}
\end{wrapfigure}

     
The cross section for $pp \to F\bar{F}$ production at the LHC at a centre of mass energy of $\sqrt{s} = 13.6$~TeV is shown in~\cref{fig:6t-LHC}(left) as a function of $m_F$. We evaluated the cross section and generated parton-level events at tree level for the $pp \to F\bar{F} \to 6t$ process using the \texttt{CalcHEP} package~\cite{Belyaev:2012qa}, version~\texttt{3.9.2}. We employed the \texttt{NNPDF40\_lo\_as\_01180} parton distribution function set via the \texttt{LHAPDF} library~\cite{Buckley:2014ana}, and chose the QCD factorisation and renormalisation scale $Q = m_F$. The resulting cross section ranges from 39~fb at $m_F = 1$~TeV down to 0.012~fb at $m_F = 2.5$~TeV.
We do not perform here a detailed analysis of the QCD scale, PDF, or NLO corrections for this process, as these effects have already been studied in depth for generic VL quark (VLQ) pair production, which directly applies to the case of the $F$ fermion. At NLO in QCD, pair production of VLQs with $M_\mathrm{VLQ} = 1\text{--}2~\mathrm{TeV}$ receives sizeable corrections, enhancing the leading-order cross section by approximately $40\% \text{--} 60\%$~\cite{Fuks:2016ftf}. The residual theoretical uncertainty, dominated by scale variation, typically ranges from $\pm 10\%$ to $\pm 15\%$, with additional PDF uncertainties of $\pm 3\% \text{--} 6\%$. Therefore, our leading-order cross section results, and the derived LHC sensitivity to the FPVDM parameter space with a six-top final state, can be regarded as conservative and robust.

As this specific six-top signature from VLQs has not been previously explored, we performed a detector-level analysis to estimate both current and future HL-LHC sensitivity. The simulation chain included \texttt{CalcHEP}--\texttt{PYTHIA}~\cite{Sjostrand:2014zea}~\texttt{8.3} and \texttt{Delphes}~\cite{deFavereau:2013fsa}~\texttt{3.5}, orchestrated through the \texttt{CheckMATE}~\cite{Dercks:2016npn}~\texttt{2.1} framework. As input, we used parton-level LHE files generated in \texttt{CalcHEP} on a grid in the $M_{\mathrm{V_D}}$--$m_F$ plane.

\texttt{CheckMATE} provides validated implementations of numerous ATLAS and CMS analyses targeting final states with multiple top quarks and $b$-jets, making it especially suitable for recasting our six-top VLQ signature. In particular, we employed the following analysis modules: the CMS inclusive $M_{T2}$-based SUSY search (cms\_sus\_19\_005)~\cite{CMS:2019zmd}; the ATLAS multi-jet + $E_T^\text{miss}$ search (atlas\_2010\_14293)~\cite{ATLAS:2021yyr}; the ATLAS all-hadronic stop search (atlas\_1908\_03122)~\cite{ATLAS:2019vcq}; the ATLAS gluino-mediated stop production search with same-sign or four-lepton final states (atlas\_2101\_01629)~\cite{ATLAS:2021twp}; the CMS search targeting gluino decays to top/bottom quarks with missing energy (cms\_1908\_04722)~\cite{CMS:2019xai}; the ATLAS search for stop production in multi-$b$-jet final states (atlas\_1709\_04183)~\cite{ATLAS:2017msx}; the ATLAS four-top-quark search via effective operators (atlas\_2106\_09609)~\cite{ATLAS:2021kty}; and the inclusive ATLAS SUSY search in high jet multiplicity final states (atlas\_2004\_14060)~\cite{ATLAS:2020syg}.

We emphasise that benchmark point BM2 already lies within the sensitivity of current LHC data when statistically orthogonal ATLAS and CMS signal regions are properly combined, as demonstrated in this section. This underlines the importance of combined analyses and indicates that existing data may already be probing the GW-motivated parameter space.

To improve and extend the \texttt{CheckMATE} recasting for our six-top signature, we implemented a dedicated procedure to evaluate both the current LHC sensitivity and a first forecast of the HL-LHC potential. This involved the following steps:

\begin{itemize}
    \item We developed a Python-based statistical module that uses the expected number of signal and background events in each signal region to reproduce the \texttt{CheckMATE} exclusion limits. The statistical inference follows the CL$_s$ likelihood method described in \cref{app:pyhf}.
    
    \item For each model point, we performed a statistical combination of the best-performing signal regions from both ATLAS and CMS analyses. Where applicable, we combined mutually orthogonal signal regions within the same analysis — for instance, between the 0-lepton and 1-lepton channels in CMS (cms\_sus\_19\_005 and cms\_1908\_04722) and ATLAS (atlas\_2101\_01629 and atlas\_2004\_14060) — to maximise sensitivity while avoiding double counting.
    
    \item To estimate the HL-LHC sensitivity, we performed a rescaling of both signal and background yields, assuming a tenfold increase in integrated luminosity. While this provides only a rough approximation, it offers a useful first look at the future potential.
\end{itemize}

All relevant ATLAS and CMS analyses employed were based on the full Run-2 data sets of 139~fb$^{-1}$ and 137~fb$^{-1}$ respectively.
Our results for current and projected LHC sensitivity are presented in~\cref{fig:6t-LHC}(right). The dashed red line indicates the 95\% CL exclusion contour based on the most sensitive single signal region. The solid red line and corresponding red-shaded area show the result of statistically combining orthogonal signal regions from ATLAS and CMS, demonstrating that such a combination can extend the exclusion reach by up to 100~GeV. For example, at $M_{\mathrm{V_D}} = 500$~GeV, the exclusion limit on $m_F$ improves from 1680~GeV to 1760~GeV with this combination.

The $y$-axis in the plot shows the dimensionless quantity $(m_F - M_{\mathrm{V_D}}) / m_F$ to clearly illustrate the mass gap dependence. Also shown are dashed black contours corresponding to fixed DM masses. The exclusion reach is maximised near $M_{\mathrm{V_D}} = 500$~GeV and weakens for both smaller and larger values of $M_{\mathrm{V_D}}$. For small DM masses, the top quarks from $V^\prime$ decays are softer, reducing signal efficiency under hard selection cuts. Conversely, for large DM masses, the top quarks from $F \to V^\prime t$ decays are softer, again leading to reduced sensitivity.
The HL-LHC projection is shown in blue, with shading to indicate the gain from combining signal regions. The reach is strongest near $M_{\mathrm{V_D}} = 400$~GeV, excluding up to $m_F = 2280$~GeV. The combination of signal regions provides an additional $\sim$50~GeV gain in sensitivity at HL-LHC.
The positions of our benchmark points BM1–BM3 are indicated on the plot. The HL-LHC will be able to exclude BM1, while BM2 and BM3 lie just beyond the projected reach. We anticipate that a dedicated experimental analysis could probe these points as well.

We stress that this novel $6t$ signature from VLQ pair production already reaches exclusion limits near 1.7~TeV — comparable to the best current bounds from ATLAS dedicated searches for $T \to bW$ decays~\cite{ATLAS:2024tvlq}. This highlights the significant potential of this signature for discovery and motivates the ATLAS and CMS collaborations to explore it in future analyses.

In combination with projections from GW and direct DM detection experiments, our study illustrates the broad experimental coverage of this scenario across multiple frontiers within the next decade. Taken together, the results presented in this section reveal a highly non-trivial and predictive interplay between GW signals, collider signatures, and DM direct detection. Remarkably, all benchmark points selected by GW observability lie within the reach or near-reach of the HL-LHC, highlighting the strong complementarity of these experimental probes.


\subsection{Comments on the impact of theoretical uncertainties}\label{subsection:varying_observables}

We conclude this work by addressing the theoretical uncertainty associated with the SGWB prediction. We focus on two benchmark cases from Scenario II: a) with strong supercooling ($\alpha \gtrsim 10$) and b) mild supercooling ($\alpha \sim 1$). Both of which are within the reach of LISA, with $\mathrm{SNR \approx 100}$, and are consistent with all DM constraints. Benchmark a) is characterised by the free parameters $M_{\mathrm{V_D}} = 1713.16~\mathrm{GeV}$, $M_{\mathrm{H_D}} = 9.78~\mathrm{GeV}$, $g_{\rm D} = 1.79$ and $\sin\theta_S = -4.52 \times 10^{-7}$, whereas benchmark b) is characterised by $M_{\mathrm{V_D}} = 455.71~\mathrm{GeV}$, $M_{\mathrm{H_D}} = 12.55~\mathrm{GeV}$ $g_{\rm D} = 1.86$ and $\sin\theta_S = -1.20 \times 10^{-5}$. A summary for the observables of interest for each of these benchmarks are shown in Tabs.~\ref{tab:var_params_largeAlpha1} (for benchmark a) and \ref{tab:var_params_smallAlpha} (for b).

We first analyze the renormalization scale. The primary advantage of the 3D EFT approach over the 4D effective potential (without RG running of the couplings) lies in a significant reduction of theoretical uncertainties in predicting the SGWB spectrum. It is known that the peak amplitude is highly sensitive to the renormalization scale in the standard 4D approach, where a small variation in the scale can lead to orders of magnitude shifts in the amplitude (see, $e.g.$, Fig.~1 of \cite{Croon:2020cgk}). To estimate the error associated with the 3D approach employed in our simulations, we varied the renormalization scale by a factor of two\footnote{Specifically, we varied the $\kappa$ parameter in the definition of the matching hard scale within the range $\kappa = [1, 2]$, resulting in $\mu_{\mathrm{4D}} = [\pi T, 2\pi T]$.}.

The main results are shown in the second row of Tabs.~\ref{tab:var_params_largeAlpha1} and \ref{tab:var_params_smallAlpha}. The use of a 3D EFT with one-loop effective potential, one-loop coupling matching and two-loop mass matching, has proven to be a significant improvement compared to the 4D approach. The error associated with this variation in the renormalization scale resulted in approximately a $60\%$ error on $h^2 \Omega^{\mathrm{peak}}_{\mathrm{GW}}$ for the case with strong supercooling, whereas in the benchmark with mild supercooling the error on amplitude can span roughly an order of magnitude, that while much larger, it still represents a substantial improvement over the non-RG improved 4D method.

These ranges can be understood based on how the thermodynamic parameters $\beta/H(T_p)$ and $\alpha$ vary. Concretely, they vary substantially by the renormalisation scale, which becomes more strongly noticeable for the scenario with large $\alpha$ (see Tab.~\ref{tab:var_params_largeAlpha1}). This is contrasted with the phase transitions temperatures, $T_c$ and $T_p$, whose variations are much milder. Do note that, in the limit of large supercooling, the dependency of $\alpha$ on the peak amplitude becomes unimportant as it scales as $\Omega_\mathrm{GW}^\mathrm{peak} \propto [\alpha/(1 + \alpha)] [\beta/H(T_p)]^{-1}$, making the uncertainties on $\beta/H(T_p)$ and $T_p$ more relevant. On the other hand, for $\alpha \sim 1$ its uncertainty is important and can quickly scale up and lead to large SGWB uncertainties. This is then reflected on a larger uncertainty for $\Omega_\mathrm{GW}^\mathrm{peak}$ for benchmark of Tab.~\ref{tab:var_params_smallAlpha} when compared to that of Tab.~\ref{tab:var_params_largeAlpha1}.

The examples provided in Tabs.~\ref{tab:var_params_largeAlpha1} and \ref{tab:var_params_smallAlpha} demonstrates that our determination of the SGWB parameters are relatively robust to changes in the renormalisation scale, especially when compared to the 4D method. The dependence on the renormalisation scale can be further minimised if the 3D EFT thermal potential is performed at two-loop order \cite{Croon:2020cgk}. 

\begin{table}[htb!]
	\centering
    \captionsetup{justification=raggedright}
    \resizebox{\textwidth}{!}{%
	\begin{tabular}{c|c|c|c|c|c|c|c}
		\toprule
		-& $T_p$ (GeV) & $T_c$ (GeV) & $\alpha$ & $\beta/H(T_p)$ & $h^2 \Omega^{\mathrm{peak}}_{\mathrm{GW}}$ & $f_{\mathrm{peak}}$ (Hz)  \\
		\midrule
        Ref. & $13.53$ & $22.93$ & $20.75$ & $436.58$ & $6.72\times 10^{-12}$ & $5.66\times 10^{-4}$ \\
        $\Delta \mu_{\mathrm{4D}}$ & $14.38 \pm 0.85$ & $27.77 \pm 4.84$ & $1558.85 \pm 1538.1$ & $345.14 \pm 91.44$ & $(1.46 \pm 0.88)\times 10^{-11}$ & $(4.68 \pm 0.98) \times 10^{-4}$ \\
        $\Delta v_w$ & $13.50 \pm 0.03$ & $-$ & $21.00 \pm 0.25$ & $428.51 \pm 8.07$ & $(4.83 \pm 1.89) \times 10^{-12}$ & $(7.13 \pm 1.47)\times 10^{-4}$ \\
        $\Delta T$ & $-$ & $-$ & $25.01 \pm 14.21$ & $466.77 \pm 118.72$ & $(7.82 \pm 4.25)\times 10^{-12}$ & $(5.90 \pm 2.04)\times 10^{-4}$ \\
		\bottomrule
	\end{tabular}
    }
	\caption{\footnotesize Effect of theoretical uncertainties on the defining parameters of the GW spectra, for a benchmark with large supercooling. Namely the renormalisation scale (second row), the bubble wall velocity (in the third row) and the temperature (in the fourth row). In the fist row, we show the values we obtain for the benchmark indicated in the text, without any variation of the parameters. For each row, only the first variable as indicated in the first column is varied, with the others remaining fixed.}
	\label{tab:var_params_largeAlpha1}
\end{table}

\begin{table}[htb!]
	\centering
    \captionsetup{justification=raggedright}
    \resizebox{\textwidth}{!}{%
	\begin{tabular}{c|c|c|c|c|c|c|c}
		\toprule
		-& $T_p$ (GeV) & $T_c$ (GeV) & $\alpha$ & $\beta/H(T_p)$ & $h^2 \Omega^{\mathrm{peak}}_{\mathrm{GW}}$ & $f_{\mathrm{peak}}$ (Hz)  \\
		\midrule
        Ref. & $15.81$ & $22.64$ & $1.02$ & $657.85$ & $5.95\times 10^{-13}$ & $9.97\times 10^{-4}$ \\
        $\Delta \mu_{\mathrm{4D}}$ & $(17.53 \pm 1.72)$ & $(31.00 \pm 8.36)$ & $(3.27 \pm 2.25)$ & $468.23 \pm 189.62$ & $(6.53 \pm 5.93) \times 10^{-12}$ & $(7.56 \pm 2.41)\times 10^{-4}$ \\
        $\Delta v_w$ & $15.79 \pm 0.02$ & $-$ & $1.025 \pm 0.005$ & $654.86 \pm 2.99$ & $(4.62 \pm 1.34)\times 10^{-13}$ & $(12.79 \pm 2.82) \times 10^{-4}$ \\
        $\Delta T$ & $-$ & $-$ & $1.19 \pm 0.58$ & $836.84 \pm 307.12$ & $(4.28 \pm 3.71)\times 10^{-13}$ & $-$ \\
		\bottomrule
	\end{tabular}
    }
	\caption{\footnotesize Effect of theoretical uncertainties on the defining parameters of the GW spectra, for a benchmark with mild supercooling. Namely the renormalisation scale (second row), the bubble wall velocity (in the third row) and the temperature (in the fourth row). In the fist row, we show the values we obtain for the benchmark indicated in the text, without any variation of the parameters. For each row, only the first variable as indicated in the first column is varied, with the others remaining fixed.}
	\label{tab:var_params_smallAlpha}
\end{table}

For completeness, we also investigate the impact of the bubble wall velocity $v_w$ (third row of Tabs.~\ref{tab:var_params_largeAlpha1} and \ref{tab:var_params_smallAlpha}). This uncertainty was assessed by treating $v_w$ as a free parameter and varying it between 0.6 and 1.0. As expected for strong phase transitions the impact is minimal. However, this may not hold true if the transition is weak. In such cases, the dependence on $v_w$ could be stronger \cite{Ai:2023see}.

As noted in \cite{Athron:2023rfq,Athron:2022mmm}, current state-of-the-art simulations for the SGWB spectral ansatz are typically conducted at a fixed temperature, making  the correct choice of temperature an unknown. It is then typical to assume the percolation (nucleation) temperature for supercooled (non-supercooled) cases \cite{Athron:2022mmm}. Here, we estimate the uncertainty associated with varying the FOPT-defining temperature at which GW observables are computed. The results are presented in the last rows of Tabs.~\ref{tab:var_params_largeAlpha1} and \ref{tab:var_params_smallAlpha}. Our approach involved varying $T_p$ by approximately $10\%$. This relatively narrow range is necessary because the nucleation of bubbles is suppressed as the temperature decreases. Conversely, as the temperature increases, the action begins to diverge as it approaches the critical temperature, where the percolation condition fails. The variation of the temperature can potentially have a big impact on the spectrum, given that all geometric parameters ($f_\mathrm{peak}$ and $h^2\Omega_{\mathrm{peak}}^\mathrm{GW}$) and thermodynamic parameters ($\alpha$ and $\beta/H(T_p)$) are related to it. In particular, the $\alpha$ parameter is proportional to the inverse fourth power of $T_p$, the inverse time duration depends both on the temperature and the derivative of the action at this temperature choice and the peak frequency scales linearly with the temperature. Consequently, the error can be sizeable, and even larger than those from varying renormalisation scale and bubble wall velocity, underscoring the importance of an appropriate temperature choice.

As a final note, we comment on additional uncertainties arising from the gauge dependence of the effective potential. While our calculations were performed in the Landau gauge rather than in generic $R_\xi$ gauges, a prior analysis in \cite{Croon:2020cgk} investigated this aspect and reported an uncertainty of approximately $\mathcal{O}(10^{-3})$ (see Tab.~3 of \cite{Croon:2020cgk}) which are expected to be less significant in comparison to those discussed above. We would also like to highlight the recent work in \cite{Lewicki:2024xan}, which found that the predictions of the 3D effective potential are robust, with the underlying model parameters reconstructed with an accuracy of $\mathcal{O}(0.1\%)$ when considering a two-loop 3D effective potential and matching. Although the errors in reconstruction are small, this level of precision would still be competitive with the experimental uncertainties for an expected signal with an SNR of $\mathcal{O}(10)$.

\section{Conclusions}\label{sec:conclusion}

While particle colliders are crucial tools in particle physics, the search for DM and the need to explore fundamental physics beyond the current reach of colliders necessitate the use of alternative and complementary approaches. GW cosmology provides a compelling and promising avenue for investigation, offering access to energy regimes that are currently inaccessible to experiments based on particle colliders. Driven by this, we present an analysis of the thermal history of a non-Abelian vector DM model, focusing on its interactions with the SM via Yukawa and Higgs portal couplings.


We study the impact of BSM couplings on SGWB predictions by considering three distinct scenarios. Scenario I represents a fully secluded dark sector, featuring a pure vector-scalar DM model that is completely decoupled from the SM sector. Scenario II incorporates the SM, with the dark sector coupled to the SM via the Higgs portal interaction.  Finally, scenario III includes both Higgs and fermion portal interactions, encompassing the full FPVDM framework. For the phase-transition analysis in each of these scenarios, we have constructed a dimensionally reduced thermal EFT up to 1-loop order in the potential and 2-loop/1-loop order in the matching of masses and couplings, respectively.  This approach, as has been shown in the literature, helps to mitigate the theoretical uncertainties associated with the calculation of thermodynamic quantities.

To investigate the potential for generating observable SGWB signals in each of the scenarios described above, we have carried out a comprehensive scan of the parameter space of the model. This scan has allowed us to identify the relevant combinations of model parameters that can induce FOPTs resulting in a SGWB with peak amplitude and peak frequency within the reach of future experiments such as LISA. We emphasise that the parameter-space regions shown in our scatter plots correspond to points satisfying the consistency requirements imposed in this work, including the B\"{o}deker--Moore criterion, $T_n > T_f$, and the assumption of standard thermal freeze-out for DM production. Points outside these regions should therefore not be interpreted as excluded. They may instead correspond to phenomenologically viable scenarios requiring a more sophisticated treatment, such as a detailed analysis of bubble-wall dynamics and/or alternative DM production mechanisms, including bubble filtering or freeze-in, which are beyond the scope of the present work. With this in mind, the main conclusions that can be drawn from this analysis are summarised in the following bullet points:
\begin{itemize}
    \item For scenario I, the strength of the phase transition is primarily determined by the dark gauge coupling, $\g{D}$. Lower values of $\g{D}$ result in larger GW amplitudes, as shown in the top-left panel of \cref{fig:GW_plots_spectra_modelI}.
    \item  The mass scale of the dark vector boson in scenario I is directly related to the GWs peak frequency. Larger vector boson masses correspond to higher peak frequencies, as shown in the top-right panel of \cref{fig:GW_plots_spectra_modelI}.
    \item The observation of a SGWB signal by LISA, with a peak frequency near 1 mHz, is also possible for scenario II, which incorporates a Higgs portal coupling between the dark sector and the SM sector. This hypothetical observation leads to constraints on several of the model's parameters. Specifically, the dark gauge coupling is predicted to be near $\g{D} \approx 1.7$,  the mass of the dark sector Higgs boson near $M_\mathrm{H_D} \approx 10~\mathrm{GeV}$, while the mass of the dark vector boson is constrained to the range $0.1 \lesssim M_\mathrm{V_D}/\mathrm{TeV} \lesssim 4$; see \cref{fig:GW_scenarioII,fig:GW_plots_spectra}.
    \item Remaining in scenario II, LISA will be sensitive to a specific type of phase transition, denoted as $(v,0) \to (v,v_\mathrm{D})$. Our analysis shows that the DM relic abundance can be accommodated whenever the dark gauge boson masses in the approximate range 3–4 TeV falling outside the observability region. These features of the model's parameter space are shown in the bottom panels of \cref{fig:GW_and_DM_scenarioII_A}, in combination with the results shown in \cref{fig:GW_scenarioII,fig:GW_plots_spectra}.
    \item Our analysis reveals that the predicted peak frequencies of the SGWB differ substantially between scenario II and scenario III, providing a potential means of distinguishing between these two models using future LISA data. Specifically, in scenario III, the peak frequency is predicted to fall in the range of 1 to 10 mHz, which is well within the sensitivity range of LISA. Moreover, if we consider the currently projected sensitivity reach of BBO and DECIGO, the peak frequency in scenario III can extend up to 1 Hz. In contrast, in scenario II, the predicted peak frequency falls within the mHz range and is only marginally detectable by the BBO and DECIGO. These differences in the predicted peak frequencies are shown in Fig.~\ref{fig:overlap_scII_scIII}.
\begin{figure}[htbp]
    \centering
    \includegraphics[width=0.6\textwidth]{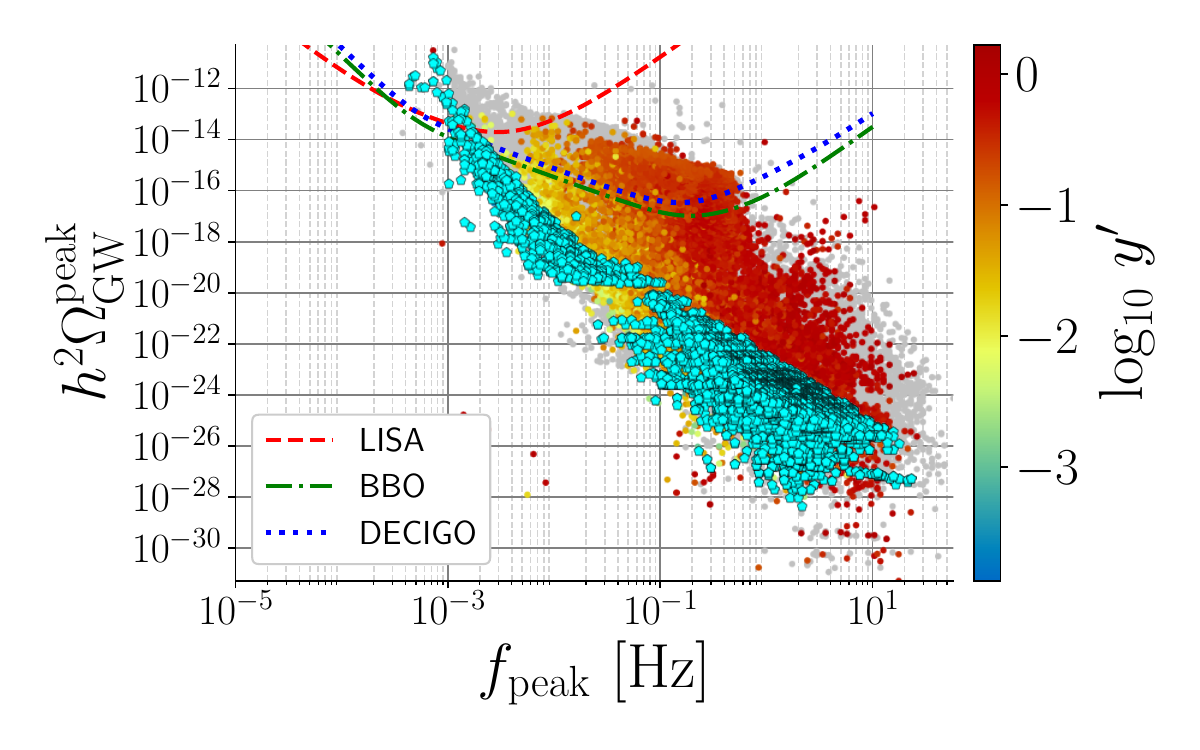}
    \caption{\footnotesize Scatter plot depicting the peak amplitude $h^2\Omega_{\mathrm{GW}}^\mathrm{peak}$ as a function of the peak frequency $f_\mathrm{peak}$ (in Hz), with the colour bar representing the Yukawa fermionic portal $y^\prime$. Scenario II points are marked by cyan pentagons.}
    \label{fig:overlap_scII_scIII}
\end{figure}

    \item In the context of scenario III, which includes both Higgs and fermion portal couplings between the dark and SM sectors, the detection of a SGWB signal by LISA would suggest that the phase transition follows the pattern $(v,0) \to (v,v_\mathrm{D})$. Moreover, such a detection would place constraints on several of the model parameters. Specifically, the dark gauge coupling would be constrained to the range $1 \lesssim \g{D} \lesssim 2$, the mass of the dark vector boson to the range $1 \lesssim M_\mathrm{V_D}/\mathrm{TeV} \lesssim 10$, the mass of the dark scalar boson to the range $10 \lesssim M_\mathrm{H_D}/\mathrm{GeV} \lesssim 100$, and the fermion portal coupling $y'$ would be required to have a value greater than 0.01. These constraints are illustrated in \cref{fig:GW_plots_phasepattern,fig:GW_plots_spectra_scenarioIII_params}.
    \item Our analysis shows that, within the sensitivity range of LISA, vector DM constitutes at least $40\%$ of the total relic abundance in most cases; see \cref{fig:GW_plots_spectraDM_scenarioIII}.
\end{itemize}

In scenario I, the dark vector boson mass and gauge coupling predominantly determine the SGWB peak frequency and amplitude, respectively. While scenarios II and III exhibit more complex behaviour due to the increased freedom, by performing dedicated scans in the $(\g{D},M_\mathrm{V_D})$ plane with the remaining parameters fixed it is possible to uncover such underlying structures. For two benchmark points in \cref{tab:benchs}, we identified optimal parameter combinations optimising the SGWB observability and accounting for all DM. For BP1, points that explain DM are within BBO and DECIGO sensitivity; see \cref{fig:2DParamScan_scenario3}. For BP2, while maximising both $h^2 \Omega_\mathrm{GW}^\mathrm{peak} \approx 10^{-13}$ and $h^2 \Omega_\mathrm{DM} \approx 0.12$, the peak frequency (0.02–0.03 Hz) lies in a region with $\mathrm{SNR} < 1$ at LISA but well within reach of future space-based interferometers; see \cref{fig:2DParamScan_scenario3_BigGW}.  

A key prediction of this work, within the context of the full FPVDM model, is the identification of a specific combination of model parameters that is well-suited for generating a SGWB signal detectable at LISA consistent with DM constraints. This preferred parameter space region is characterised by a dark vector boson mass scale of approximately $M_\mathrm{V_D} \sim \mathcal{O}(100-1000)~\mathrm{GeV}$, a dark gauge coupling constrained to $g_D \approx 1-1.8$, and a dark Higgs boson mass of order $M_\mathrm{H_D} \sim \mathcal{O}(10)~\mathrm{GeV}$. This is very similar to scenario II.

Beyond the cosmological observables, our study identifies a novel collider signature: the six-top final state arising from the pair production of VL fermions, that offers a complementary and experimentally accessible probe of the same parameter space. We have shown that this $6t$ signature, recast using multiple ATLAS and CMS analyses via the \texttt{CheckMATE} framework and validated with our custom statistical pipeline, is already constrained by current LHC data up to $m_F \sim 1.7$~TeV. Projecting to the HL-LHC, this sensitivity extends to approximately $2.3$~TeV. Notably, this collider reach covers the same region of parameter space that yields a detectable stochastic GW background at LISA and satisfies the observed DM abundance. This triple complementarity between collider, GW, and cosmological probes underscores the robustness of the FPVDM framework and highlights the importance of pursuing diverse experimental strategies to uncover the structure of hidden sectors.

\acknowledgments
We thank Andreas Ekstedt for email exchanges on various bug fixes regarding the correct implementation of VL fermions in the \texttt{DRAlgo} package.
We thank Alexander Pukhov for numerous detailed consultations on the \texttt{micrOMEGAs} package and for providing us with an unofficial new release of \texttt{micrOMEGAs}, which includes the recasting  the latest results from LZ collaboration.
A.P.M.~expresses gratitude to the CERN TH Department for supporting scientific visits, which have contributed to the development of the work presented in this article.  
J.G. and A.P.M. were supported by the Center for Research and Development in Mathematics and Applications (CIDMA) under the Portuguese Foundation for Science and Technology (FCT - Funda\c{c}\~{a}o para a Ci\^{e}ncia e a Tecnologia) Multi-Annual Financing Program for R\&D Units.
J.G.~and A.P.M.~are also supported by LIP and FCT, reference LA/P/0016/2020. 
J.G. is also directly funded by FCT through the doctoral program grant with the reference 2021.04527.BD (\url{https://doi.org/10.54499/2021.04527.BD}).
R.P.~and J.G.~are supported in part by the Swedish Research Council grant, contract number 2016-05996, as well as by the European Research Council (ERC) under the European Union's Horizon 2020 research and innovation programme (grant agreement No 668679).
A.B. acknowledges support from the STFC Consolidated Grant ST/L000296/1 and a partial support through the NExT Institute. A.B. also acknowledges support from  the Leverhulme Trust RPG-2022-057 grant.  N.T. is supported by NSRF via the Program
Management Unit for Human Resources \& Institutional Development, Research and Innovation [grant number B13F670063].
A.B., N.T. and J.G. acknowledge the use of the IRIDIS High-Performance Computing Facility and associated support services at the University of Southampton in completing this work.

\section*{Note added}

While this paper was being completed, a related study appeared in~\cite{Benincasa:2025tdr}. Similar to us, the authors investigated GW signatures from FOPTs in a dark non-Abelian $\mathrm{SU(2)_D}$ model, while simultaneously examining the viability of the associated vector DM candidate (considering both DM relic density and direct detection constraints). The model considered in~\cite{Benincasa:2025tdr} is identical to our Scenario~II, and our results for both the GW spectrum and the DM relic abundance are in qualitative agreement. However, our analysis differs in the treatment of the thermal effective potential when studying the phase transitions: we employ the dimensional reduction technique, whereas~\cite{Benincasa:2025tdr} uses a purely four-dimensional approach without RG-improvement of couplings or fields. Additionally, our Scenarios I and III were not studied in \cite{Benincasa:2025tdr}.

\cleardoublepage
\appendix

\section{Sound-wave efficiency factor}\label{app:efficiency_factor}

The calculation of the efficiency factors for the production of GWs from sound waves are based directly on the formalism introduced in \cite{Espinosa:2010hh}. We first begin by splitting into three distinct regions:
\begin{itemize}
    \item For subsonic deflagrations ($v_w < c_s$) we have
    \begin{equation}
        \kappa_\mathrm{SW} = \frac{c_s^{11/5} \kappa_A \kappa_B}{(c_s^{11/5} - v_w^{11/5})\kappa_B + v_w c_s^{6/5} \kappa_A}\,,
    \end{equation}
    where we have defined
    \begin{equation}
        \begin{aligned}
            & \kappa_A = \frac{6.9v_w^{6/5}\alpha}{1.36 - 0.037\sqrt{\alpha} + \alpha} \,, \\
            & \kappa_B = \frac{\alpha^{2/5}}{0.017 + (0.997 + \alpha)^{2/5}}\,.
        \end{aligned}
    \end{equation}
    \item For supersonic deflagrations ($c_s < v_w < v_J$) we have
    \begin{equation}
        \kappa_\mathrm{SW} = \kappa_B + (v_w + c_s)\delta\kappa + \frac{(v_w - c_s)^3}{(v_J - c_s)^3}(\kappa_C - \kappa_B - (v_J - c_s)\delta \kappa)\,,
    \end{equation}
    where we have defined
    \begin{equation}
    \begin{aligned}
        &\kappa_C = \frac{\sqrt{\alpha}}{0.135 + \sqrt{0.98 + \alpha}} \,, \\
        & \delta \kappa = -0.9 ~\mathrm{ln}\Big[\frac{\sqrt{\alpha}}{1 + \sqrt{\alpha}}\Big] \,, \\
        &v_J = \frac{\sqrt{(2/3)\alpha + \alpha^2} + \sqrt{1/3}}{1 + \alpha} \,.
    \end{aligned}
    \end{equation}
    \item For detonations ($v_w > v_J$)
    \begin{equation}
        \kappa_\mathrm{SW} = \frac{(v_J - 1)^3 v_J^{5/2} v_w^{-5/2} \kappa_C \kappa_D}{[(v_J - 1)^3 - (v_w - 1)^3] v_J^{5/2} \kappa_C + (v_w - 1)^3 \kappa_D}
    \end{equation}
    where
    \begin{equation}
        \kappa_D = \frac{\alpha}{0.73 + 0.083 \sqrt{\alpha} + \alpha}\,.
    \end{equation}
\end{itemize}
For extra details on the derivation of these expressions, see \cite{Espinosa:2010hh}.

\section{Effective potential from dimensional reduction}\label{sec:DR_eff}

In what follows, we provide a brief discussion on DR. It is important to note that the calculations presented in the remainder of this section have been verified to agree with the output from \texttt{DRAlgo}. For a more detailed discussion, we refer readers to \cite{Croon:2020cgk,Brauner:2016fla}. For simplicity, we will outline the procedure for Scenario~I only.

The 4D model action can be schematically expressed as
\begin{equation}
\label{eqn:4D_lagragian_action}
    \mathcal{S}_\mathrm{4D} = \mathcal{S}_{\mathrm{gauge}} + \mathcal{S}_{\mathrm{scalar}} + \mathcal{S}_{\mathrm{ghost}} + \mathcal{S}_{\mathrm{gauge-fix}} + {\mathrm{CT}}\,,
\end{equation}
where CT denotes the counter-terms associated with the theory parameters. It is important to note that we do not include fermions in Scenario~I. Each component of \cref{eqn:4D_lagragian_action} can be expanded as follows
\begin{equation}\label{eqn:each_piece_4D}
    \begin{aligned}
        & \mathcal{S}_{\mathrm{gauge}}     = \frac{1}{4}\int d^4x~V^i_{\mu\nu} V_i^{\mu\nu} \,, \\
        & \mathcal{S}_{\mathrm{scalar}}    = \int d^4x \Big[ D_\mu \Phi_{\rm{D}}^\dagger D_\mu \Phi_{\rm{D}} - \mu_{\rm D}^2 \Phi_{\rm{D}}^\dagger \Phi_{\rm{D}} + \lambda_{\rm D} (\Phi_{\rm{D}}^\dagger \Phi_{\rm{D}})^2 \Big] \,, \\
        & \mathcal{S}_{\mathrm{ghost}}     = \int d^4x~\partial_\mu \bar{\mathcal{V}}^c D_\mu \mathcal{V}^c  \,, \\
        & \mathcal{S}_{\mathrm{gauge-fix}} = \int d^4x~\frac{1}{2\xi_D} \qty(\partial_\mu V^c_{\mu} )^2 \,.
    \end{aligned}
\end{equation}
Here, the covariant derivative is defined as
\begin{equation}\label{eqn:Fmunu_Dmu}
    D_\mu = \partial_\mu - i\g{D}\frac{\sigma_i}{2}V^i_\mu\,.
\end{equation}
In these expressions, $\sigma_a$ represents the Pauli matrices, $\mathcal{V}$ is the ghost field and $\xi_D$ is the gauge parameter. Although we have explicitly written the gauge fixing term for a generic $R_\xi$ gauge, all computations are performed in the Landau gauge, where $\xi_D \rightarrow 0$. In the $\overline{\mathrm{MS}}$-scheme, the counter-terms take the following form
\begin{equation}\label{eqn:counter_terms4D}
    \begin{aligned}
        & \delta \g{D}^2 = - \frac{43 \g{D}^4}{96\pi^2 \epsilon}\,, \\
        & \delta \lambda_{\rm D} = \frac{1}{256 \pi ^2\epsilon} \qty[3\qty(3\g{D}^4 - 24\g{D}^2\lambda_{\rm D} + 64\lambda_{\rm D}^2) ]\,, \\
        & \delta \mu_{\rm D}^2 = -\frac{3\qty(\g{D}^2 - 8\lambda_{\rm D})\mu_{\rm D}^2}{64 \pi^2 \epsilon } \,,
    \end{aligned}
\end{equation}
with $\epsilon$ denoting the dimensional regularisation parameter. From these expressions, we first derive the $\beta$-functions for $\g{D}$, $\lambda_{\mathrm{D}}$, and $\mu^2_{\mathrm{D}}$, which are presented in \cref{app:rges_4d}. It is important to note that the counter-terms remain applicable at both zero and finite temperatures, as the ultraviolet behaviour of the theory is unaffected by finite temperature corrections.

The next step involves computing the one-loop two-, three-, and four-point functions and matching them to the 3D effective theory. This process is carried out in the high-temperature limit, where the dark $\SU{2}{D}$ symmetry remains unbroken. Consequently, the computation of the correlators is performed at this symmetry level. This approach significantly simplifies the calculations, as it eliminates the need to manage various mixing matrices. In this section, we will focus on detailing the calculation for the temporal couplings only. The procedure for the transverse couplings is analogous. With only the $\SU{2}{D}$ gauge group present, there is a single $(V_0)^2$ two-point function (equivalent to the thermal Debye mass) and two four-point functions: $(V_0)^4$ and $(V_0)^2 (\Phi_{\rm{D}}^\dagger \Phi_{\rm{D}})$. For detailed pedagogical calculations, we direct the reader to Refs.~\cite{Croon:2020cgk,Brauner:2016fla,Laine:2016hma,Kajantie:1995dw}. All correlators are also available in \cref{app:debye_masses} and \cref{app:soft_matching_couplings}. In this context, we only need two bosonic master loop integrals, as provided in \cite{Brauner:2016fla}\footnote{Additional master integrals may be necessary if fermions are considered. However, in this simplified model, fermionic content is neglected, and no extra formulas are required.}.
\begin{equation}\label{eqn:master_integrals}
	\begin{aligned}
		& I^{4b}_{100} =~{\mathclap{\displaystyle\int}\mathclap{\textstyle\sum}}_{P^\prime} \frac{1}{P^2} = \frac{T^2}{12} \qty(\frac{\mu_\mathrm{4D}}{4\pi T})^\epsilon \qty(1 + 2\qty[\ln 2\pi + \gamma_E - \frac{\zeta^\prime(2)}{\zeta(2)}]\epsilon + \mathcal{O}(\epsilon^2)  ),\\
		& I^{4b}_{200} =~{\mathclap{\displaystyle\int}\mathclap{\textstyle\sum}}_{P^\prime} \frac{1}{(P^2)^2} = \frac{1}{16\pi^2} \qty(\frac{\mu_\mathrm{4D}}{4\pi T})^{2\epsilon} \qty( \frac{1}{\epsilon} + 2\gamma_E + \mathcal{O}(\epsilon))\,.
	\end{aligned}
\end{equation}
where $\gamma_E = 0.577$ is the Euler-Mascheroni constant, $\zeta$ is the Riemann zeta function and $\mu_{\mathrm{4D}}$ is the renormalisation scale. At leading order in $T^2$, these integral functions scale as $I_{100}^{4b} \sim T^2/12 + \mathcal{O}(\epsilon)$ and $I_{200}^{4b} \sim 1 + \mathcal{O}(\epsilon)$. Additionally, to simplify the notation, we denote
\begin{equation}\label{eqn:integral_measure}  
{\mathclap{\displaystyle\int}\mathclap{\textstyle\sum}}_{P^\prime} \equiv T \sum_{\omega_n \neq 0} \qty(\frac{e^{\gamma_E}\mu^2}{4\pi})^\epsilon \int \frac{d^d\textbf{p}}{(2\pi)^d} \,,
\end{equation}
where, as usual, we utilise dimensional regularisation working in $d = 3 - 2\epsilon$ dimensions, and $P$ represents the Euclidean four-momentum defined as $P = (m_n, \textbf{p})$. In this context, the Matsubara frequency for bosons is given by $m_n = \omega_n = 2n\pi T$, where $n$ is an integer. With this definition, the two-point function $(V_0)^2$ corresponds to the one-loop corrections to the $V_\mu$ field. Considering the particle content and group charges, the following five diagrams are necessary:
\\
\begin{equation}\label{eqn:debye_mass}
	\centering
	\begin{aligned}
		&: (V_0)^2 : ~=
        \smash{\parbox{85pt}{\includegraphics{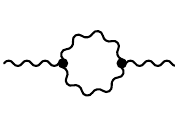}}}
        ~+~ 
        \smash{\parbox{85pt}{\includegraphics{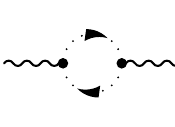}}}
        ~+~
        \smash{\parbox{85pt}{\includegraphics{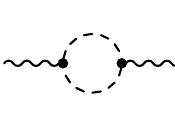}}}
        ~+~ \\[3.5em]
        &\hspace*{5.5em} \smash{\parbox{85pt}{\includegraphics{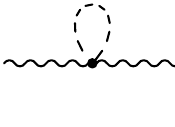}}} ~+~
        \smash{\parbox{85pt}{\includegraphics{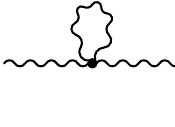}}}\,.
	\end{aligned}
\end{equation}
Detailed steps for computing these integrals can be found in \cite{Croon:2020cgk}. In this discussion, we will bypass these steps and directly utilise the results found in the appendices of \cite{Brauner:2016fla}. Only the longitudinal components acquire thermal masses; here, we consider\footnote{Generic results are available in \cite{Brauner:2016fla}. It is straightforward to verify that summing over all transverse components results in the thermal mass vanishing. This occurs because Lorentz invariance is broken only in the longitudinal direction and not in the transverse direction \cite{Croon:2020cgk}.} $\mu = \nu = 0$. Summing all contributions, we obtain:
\begin{equation}
    \begin{aligned}
    :(V_0)^2:~=& -\qty[ \frac{1}{6} \g{D}^2 \Big( {(d-3) d+16} I^{4b}_{200} P^2-6 (d-1) (2 d-1) I^{4b}_{100}\Big)] \\
    &= \frac{5}{6}\g{D}^2 T^2 \equiv \mu^2_{\mathrm{SU(2)_D}} \,.
    \end{aligned}
\end{equation}
This represents the result for the leading order (LO) Debye mass, denoted as $\mu^2_{\mathrm{SU(2)_D}}$ in this article. For the NLO, we would need to consider the two-loop contributions. While we do not present them here, the results can be found in \cref{app:debye_masses}. In analogy to the one-loop case, there are also master formulas available for two-loop calculations.

For the $:(V_0)^4:$ correlator, we follow the same procedure. The one-loop corrections read
\\[2em]
\begin{equation}\label{eqn:quartic_correlator}
    \centering
    \begin{aligned}
        &: (V_0)^4 :~ =
            \smash{\parbox{85pt}{\includegraphics{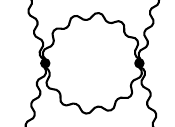}}}       
            ~+~
            \smash{\parbox{85pt}{\includegraphics{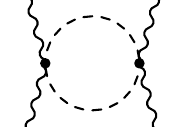}}} 
            ~+~
            \smash{\parbox{85pt}{\includegraphics{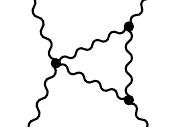}}} 
            ~+~ \\[5.0em]
            &\hspace*{5.5em}
            \smash{\parbox{85pt}{\includegraphics{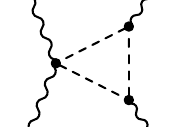}}}
            ~+~
            \smash{\parbox{85pt}{\includegraphics{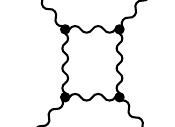}}}
            ~+~
            \smash{\parbox{85pt}{\includegraphics{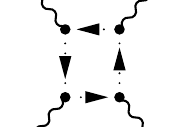}}}
            ~+~ \\[5.0em]
            &\hspace*{5.5em}
            \smash{\parbox{85pt}{\includegraphics{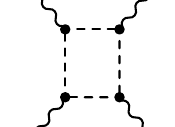}}}
            \\[2.0em]
            &\hspace*{5.0em}
            = \frac{1}{6} (d-3) (d-1) (8 d-7) \g{D}^4 I^{4b}_{200} = \frac{17 \g{D}^4}{24 \pi^2} \equiv \lambda_{V,1}\,,
    \end{aligned}
\end{equation}
which we refer to as $\lambda_{V,1}$ in the remainder of this article. The calculation of the 2- and 4-point correlators was simplified by the fact that these couplings do not exist at tree-level, meaning that wave-function renormalization does not contribute. In contrast, the $(V_0)^2 (\Phi_{\rm{D}}^\dagger \Phi_{\rm{D}})$ operator is gauge-invariant and therefore already present at the tree level, requiring additional details. Generally, the relationship between the 4D fields and the 3D fields can be described by the following equation
\begin{equation}\label{eqn:3D_to_4D_fields}
    \Phi^2_{\mathrm{3D}} = \frac{1}{T} \qty(1 + \frac{d\Pi}{dP}\Bigr|_{\substack{P=0}} - \delta Z_\Phi) \Phi^2_{\mathrm{4D}}\,,
\end{equation}
where $\Pi$ is the self-energy of the field, and $\delta Z_\Phi$ is the wave-function renormalization counter-term. In dimensional regularization, $\delta Z_\Phi$ is determined by the momentum-dependent $1/\epsilon$ poles of the self-energy contribution. The necessary contributions have already been calculated in \cite{Brauner:2016fla}. Specifically, we need to utilise their expressions (3.1) and (3.4) which lead to
\begin{equation}\label{eqn:Z_factors}
    \delta Z_V = \frac{25\g{D}^2}{96\pi^2} \frac{1}{\epsilon}\,, \quad \delta Z_{\Phi_{\rm{D}}} = \frac{9\g{D}^2}{64\pi^2} \frac{1}{\epsilon}\,.
\end{equation}
Based on the generic formula from \cref{eqn:3D_to_4D_fields}, along with the field renormalization factors in \cref{eqn:Z_factors} and the self-energy expressions found in \cite{Brauner:2016fla}, we find the following relationships for the fields
\begin{equation}\label{eqn:field_relations}
    \begin{aligned}
        & V^2_{\mathrm{3D}, 0} = \frac{V^2_{\mathrm{3D}, 0}}{T} \qty(1 - \frac{25 \g{D}^2L_b}{96\pi^2} + \frac{3\g{D}^2}{16\pi^2}) \,, \\
        & V^2_{\mathrm{3D}, r} = \frac{V^2_{\mathrm{3D}, r}}{T} \qty(1 - \frac{25 \g{D}^2 L_b}{96\pi^2} + \frac{2\g{D}^2}{48\pi^2}) \,, \\
        & |\Phi_{\rm{D}}|^2_{\mathrm{3D}} = \frac{[\Phi_{\rm{D}}]^2_{\mathrm{4D}}}{T} \qty(1 - \frac{9\g{D}^2 L_b}{64\pi^2}) \,, 
    \end{aligned}
\end{equation}
where we define $L_b = \ln{\mu^2_{\mathrm{4D}}/T^2} + 2 \gamma_E - 2\ln{4\pi}$.
With these relationships established, we can proceed to evaluate the correlator, starting with the contributions from the following one-loop Feynman diagrams
\\[2em]
\begin{equation}\label{eqn:mixed}
    \centering
    \begin{aligned}
        &\Gamma_{\Phi_{\rm{D}}^2 V^2} ~ =
            \smash{\parbox{85pt}{\includegraphics{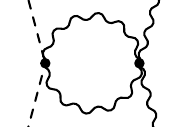}}}        
            ~+~
            \smash{\parbox{85pt}{\includegraphics{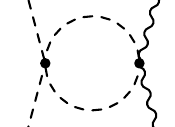}}}  
            ~+~
            \smash{\parbox{85pt}{\includegraphics{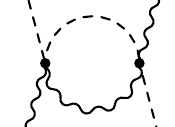}}}
            ~+~ \\[5.0em]
            &\hspace*{5.5em}
            \smash{\parbox{85pt}{\includegraphics{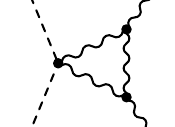}}}
            ~+~ 
            \smash{\parbox{85pt}{\includegraphics{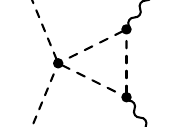}}}
            \\[3.0em]
            &\hspace*{5.0em}
            = \frac{1}{8}\g{D}^2 \qty(4 + (25 - 8d)d\g{D}^2 I^{4b}_{200} - 24(d-3)\lambda_{\rm D} I^{4b}_{200})  \,.
    \end{aligned}
\end{equation}
Meanwhile, the tree-level contribution is simply $\Gamma_{0} = -(1/2) \g{D}^2$. With these results in hand, we can match to the 3D theory by equating the correlators of the 3D and 4D theories. Let $\lambda_{S,1}$ be the coupling in the 3D theory, such that
\begin{equation}\label{eqn:matching_3D}
    \lambda_{S,1} |\Phi_{\rm{D}}|^2_{\mathrm{3D}} V^2_{\mathrm{3D}, 0} = |\Phi_{\rm{D}}|^2_{\mathrm{3D}} V^2_{\mathrm{3D}, 0}  \qty(\Gamma_0 + \frac{1}{2}\delta g^2 + \Gamma_{\Phi_{\rm{D}}^2 V^2} )\,.
\end{equation}
After applying the relations in \cref{eqn:field_relations} and expanding up to fourth order in the couplings, we derive that
\begin{equation}\label{eqn:matching_3D_couplings}
    \lambda_{S,1} = \frac{\g{D}^2 T (\g{D}^2 (51 + 43 L_b) + 96 \pi^2 + 72 \lambda_{\rm D})}{192 \pi^2}\,.
\end{equation}
This completes the calculation of the temporal couplings. The same procedure applies to the transverse couplings, so we simply present the final results:
\begin{equation}\label{eqn:remainder_softcouplings}
    \begin{aligned}
        &[g^{\mathcal{S}}_{\rm D} ]^2 = \g{D}^2 T + \frac{\g{D}^4(4 + 43L_b)T}{96\pi^2}  \,, \\
        &\lambda_{\rm D}^{\mathcal{S}} = \frac{T\qty[\g{D}^4(6-9L_b) + 72\g{D}^2 L_b \lambda_{\rm D} + 64\lambda_{\rm D}(4\pi^2 - 3L_b\lambda_{\rm D})]}{256\pi^2} \,, \\
        & [\mu^{\mathcal{S}}_{\rm D}]^2  = \frac{T^2}{16}\qty(3\g{D}^2 + 8\lambda_{\rm D}) + \mu_{\rm D}^2 \,.
    \end{aligned}
\end{equation}
The discussed 3D theory, derived from the original 4D model, is commonly referred to in the literature as the soft effective theory. At the soft scale, the effective 3D action of the model is expressed as
\begin{equation}\label{eqn:action_soft}
\begin{aligned}
\mathcal{S}_{\mathrm{soft}} = &\int d^3x~\Big[\frac{1}{4} |F^a_{rs}|^2  + \frac{1}{2} (D_r V^a_0)^2 + (D_r \Phi_{\rm{D}})^\dagger (D_r \Phi_{\rm{D}}) +  [\mu^{\mathcal{S}}_{\rm D}]^2 \Phi^\dagger_{\rm D} \Phi_{\rm{D}} + \\ &\hspace*{3.7em} \frac{1}{2} \mu^2_{\mathrm{SU(2)_D}} V^a_0 V^a_0 + \lambda_{V,1} (V^a_0 V^a_0)^2 + \lambda_{S,1} (V^a_0 V^a_0)(\Phi_{\rm{D}}^\dagger \Phi_{\rm{D}}) + \\ &\hspace*{3.7em}  \lambda_{\rm D}^{\mathcal{S}} (\Phi^\dagger_{\rm D} \Phi_{\rm{D}})^2 \Big]\,,
\end{aligned}
\end{equation}
where $F^a_{rs} = \partial_r V_s^a - \partial_s V_r^a + g^{\mathcal{S}}_{\rm D} \epsilon^{abc} V^b_r V^c_s$, and the covariant derivative is $D_r = \partial_r - i g^{\mathcal{S}}_{\rm D} \frac{\sigma_a}{2}V^a_r$. For simplicity, the names of the 3D fields have been retained to match those of the 4D fields. It is important to note that in the 3D EFT, the static modes have been integrated out, so the indices $r$ and $s$ refer only to spatial coordinates.

Phase transitions are typically driven by the lighter fields, so we can further simplify the action by integrating the heavy temporal field $V_0$ out, leaving only the scalars and the transverse vectors. This stage of approximation is commonly referred to as the ultrasoft regime. At this point, the action is given by 
\begin{equation}\label{eqn:action_ultrasoft}
    \begin{aligned}
        \mathcal{S}_{\mathrm{ultrasoft}} = \int d^3x~\Big[\frac{1}{4} |F^c_{rs}|^2 + (D_r \Phi_{\rm{D}})^\dagger (D_r \Phi_{\rm{D}}) +  [\mu^{\mathcal{US}}_{\rm D}]^2 \Phi^\dagger_{\rm D} \Phi_{\rm{D}} + \lambda_{\rm D}^{\mathcal{US}} (\Phi^\dagger_{\rm D} \Phi_{\rm{D}})^2 \Big]\,.
    \end{aligned}
\end{equation}
The matching relations between the soft and ultrasoft regimes can be determined based on previous literature as follows \cite{Brauner:2016fla,Kajantie:1995dw},
\begin{equation}\label{eqn:ultrasoft_matching}
    \begin{aligned}
        &[g^{\mathcal{US}}_{\rm D}]^2 = [g_{\rm D}^\mathcal{S}]^2-\frac{[g_{\rm D}^\mathcal{S}]^4}{24 \pi \sqrt{\mu^2_{\mathrm{SU(2)_D}}}} \,, \\
        &\lambda_{\rm D}^{\mathcal{US}} = \lambda_{\rm D}^{\mathcal{S}}-\frac{3\lambda^2_{S,1}}{32\pi\sqrt{\mu^2_{\mathrm{SU(2)_D}}}} \,, \\
        & [\mu^{\mathcal{US}}_{\rm D}]^2= [\mu_{\rm D}^\mathcal{S}]^2 - \frac{3 \sqrt{\mu^2_{\mathrm{SU(2)_D}}} \lambda_{S,1}}{8\pi} \,.
    \end{aligned}
\end{equation}

All calculations discussed here were performed at the one-loop level. However, we also incorporate two-loop contributions to the masses computed using \texttt{DRAlgo}. The corresponding formulas relevant for this section, along with the two-loop results, are provided in the appendix. The ultrasoft action in \cref{eqn:action_ultrasoft} is employed for studying the phase transitions in our numerical analysis. This is accomplished by noting that the 3D effective potential connects to the 4D potential through $V^{4D}_{\mathrm{eff}} = T V^{3D}_{\mathrm{eff}}$, a relation that we have implemented in \texttt{CosmoTransitions} to determine the bounce solution.

\section{3D effective potential and matching conditions}\label{sec:3d_eff}

To simplify the notation, we make use of the following quantities,
\begin{equation}\label{eq:lb_lf_factors}
    L_b = \ln{\frac{\mu^2_{\mathrm{4D}}}{T^2}} + 2 \gamma_E - 2\ln{4\pi}\,, \quad L_f = L_b + 4\ln{2}\,, \quad A = 1.282 \dots
\end{equation}
where $\mu^2_{\mathrm{4D}}$ is the 4D hard matching scale ($\mu^2_{\mathrm{4D}} \equiv \pi^2 \kappa^2 T^2$ with $\kappa = 1$), $T$ is the temperature and $A$ is the Glaisher–Kinkelin constant. In what follows, the 3D effective theory is defined at the $\mu_{\mathrm{3D}}$ matching scale. We also adopt standard nomenclature for the various scales involved in the DR approach, namely, the \textit{hard} scale is the scale at which the original 4D thermal theory is defined and all particle content exists. Varying this parameter allows to estimate the impact of the renormalisation scale on the GW observables and it was analysed in \cref{subsection:varying_observables}. The \textit{soft} scale lives at 3D, and it is where the fermionic as well as the non-zero bosonic Matsubara modes have been integrated out, leaving only the longitudinal and transverse components of the vectors, as well as the scalar particles. At the final stage, the \textit{ultrasoft} scale, the temporal modes of the vectors are integrated out, such that at this scale only the massless spatial vectors and scalar fields remain. In what follows, all has been calculated in \texttt{DRAlgo}. For scenario I we have made the computations ourselves and found agreement. To properly reproduce the coming results, one should use version 1.1 or above, as previous versions output erroneous results for models with extra VL fermions.

\subsection{3D effective potential}\label{sec:effective_potential}

For the purpose of this work, we treat the effective potential up to one-loop in the thermal expansion and two-loop in the matching. The effective potential at 3D can divided into two parts, $V^{3D}_\mathrm{eff} = V^{3D}_{\mathrm{LO}} + V^{3D}_{\mathrm{NLO}}$. The effective potential is  given as
\begin{equation}\label{eqn:potential}
    \begin{aligned}
        &\mathrm{Scenario~1:}~V^{3D}_{\mathrm{LO}}(T) = -\frac{1}{2}[\mu^{\mathcal{US}}_{\rm D}]^2 \varphi_{\rm D}^2  + \frac{1}{4}\lambda^{\mathcal{US}}_{\rm D} \varphi_{\rm D}^4 \,, \\
        &\mathrm{Scenario~2/3:}~V^{3D}_{\mathrm{LO}}(T) = -\frac{1}{2}[\mu^{\mathcal{US}}_{\rm D}]^2 \varphi_{\rm D}^2 - \frac{1}{2}[\mu^{\mathcal{US}}_{\rm H}]^2 \varphi_{\rm H}^2  + \frac{1}{4}\lambda^{\mathcal{US}}_{\rm D} \varphi_{\rm D}^4 + \frac{1}{4}\lambda^{\mathcal{US}}_{\rm H} \varphi_{\rm H}^4 + \frac{1}{4}\lambda^{\mathcal{US}}_{\rm HD} \varphi_{\rm D}^2\varphi_{\rm H}^2\,, \\
        &V^{3D}_{\mathrm{NLO}}(T) = -\frac{1}{12\pi} \sum_{i \subset \mathrm{scl.}} M_i^{3}(\varphi_{\rm H}, \varphi_{\rm D}, T) -\frac{2}{12\pi} \sum_{i \subset \mathrm{vec.}} M_i^{3}(\varphi_{\rm H}, \varphi_{\rm D}, T) \,, \\
    \end{aligned}
\end{equation}
where $i$ sums over the scalar fields (first term of $V^{3D}_{\mathrm{NLO}}$) and the vector fields (second term of $V^{3D}_{\mathrm{NLO}}$). The NLO potential is identical between the different scenarios, with the only difference being the scalars/vectors that appear in the sum. The couplings marked with the superscript $\mathcal{US}$ are evaluated at the ultrasoft scale and are temperature dependent. The exact analytical expressions that match to the original 4D theory are shown in the next sections. Note that the fields here live in 3D space, such that each carries mass units of $[M^{1/2}]$. They relate with the 4D fields through the simple relation $\Phi^2_i = T\varphi^2_i$. The 4D potential relates to the 3D one as $V^{\mathrm{4D}}(T) = T [V^{3D}_{\mathrm{LO}}(T) + V^{3D}_{\mathrm{NLO}}(T)]$. The bounce action is evaluated using $V^{\mathrm{4D}}(T)$.

\subsection{Renormalisation group equations at 4D}\label{app:rges_4d}

Here, we define $\beta_p \equiv \partial p / \partial \ln \mu$.
\subsubsection*{\underline{Scenario I:}}
\begingroup
\allowdisplaybreaks
\begin{align}\label{eqn:beta_functions_ScenarioI}
    &\beta_{\g{D}^2} = -\frac{43 \g{D}^4}{48\pi^2}\,, \\
    &\beta_{\lambda_{\rm D}} = \frac{3\qty(3\g{D}^4 - 24\g{D}^2\lambda_{\rm D} + 64\lambda_{\rm D}^2)}{128\pi^2}\,, \\
    & \beta_{\mu^2_{\rm D}} = -\frac{3(3\g{D}^2 - 8\lambda_{\rm D})\mu_{\rm D}^2}{32\pi^2} \,,
\end{align}
\endgroup

\subsubsection*{\underline{Scenario II:}}

Scenario II beta functions are identical to those of scenario III in the limit where $m_{f_{\rm D}}, y^\prime \rightarrow 0$. The beta function for $\g{D}$ in scenario II is the same as in scenario I.

\subsubsection*{\underline{Scenario III:}}

\begingroup
\allowdisplaybreaks
\begin{align}\label{eqn:beta_functions_ScenarioIII}
    &\beta_{\g{D}^2} = -\frac{31 \g{D}^4}{48 \pi ^2}\,, \\
    &\beta_{\g{W}^2} = -\frac{19 \g{W}^4}{48 \pi ^2}\,, \\
    &\beta_{\g{Y}^2} = \frac{187 \g{Y}^4-384 \g{Y}^2 y^{\prime 2}}{144 \pi ^2} \,, \\
    &\beta_{g_S^2} = -\frac{17 g_S^4}{24 \pi ^2} \,, \\
    &\beta_{\lambda_{\rm D}} = \frac{9 \g{D}^4-72 \g{D}^2 \lambda_{\rm D} + 16 \left(\lambda_{\rm HD}^2 - 3y^{\prime 4}\right)+96 \lambda_{\rm D} \left(2\lambda_{\rm D} + y^{\prime 2} \right)}{128 \pi ^2} \,,\\
    &\beta_{\lambda_{\rm H}} = \frac{9 \g{W}^4 +6 \g{W}^2 \left(\g{Y}^2 -12 \lambda_{\rm H} \right) +3 \big[\g{Y}^4 - 8 \g{Y}^2 \lambda_{\rm H} -16 y_t^4 + 32 \lambda_{\rm H} \left(2 \lambda_{\rm H} + y_t^2\right)\big]+16 \lambda_{\rm HD}^2}{128 \pi ^2} \,, \\
    &\beta_{\lambda_{\rm HD}} = \frac{-3 \lambda_{\rm HD} \left(3 \g{D}^2+3 \g{W}^2 + \g{Y}^2 - 4 \big[2 (\lambda_{\rm D} +\lambda_{\rm H}) + y^{\prime 2} + y_t^2\big]\right) + 8 \lambda_{\rm HD}^2 - 24 y^{\prime 2} y_t^2}{32 \pi ^2} \,, \\
    &\beta_{y^\prime} = \frac{y^\prime \left(-27 \g{D}^2+24g_S^2+32 \g{Y}^2+54 y^{\prime 2} + 12 y_t^2\right)}{192 \pi ^2} \,, \\
    &\beta_{y_t} = \frac{y_t \left(-96 g_S^2-27\g{W}^2-17 \g{Y}^2+12 y^{\prime 2} + 54y_t^2\right)}{192 \pi ^2} \,, \\
    &\beta_{\mu^2_{\rm D}} = \frac{-9 \g{D}^2 \mu^2_{\rm D} + 8 \lambda_{\rm HD} \mu^2_{\rm H} -24 m_{f_{\rm D}}^2y^{\prime 2} +12 \mu^2_{\rm D} \left(2\lambda_{\rm D} + y^{\prime 2} \right)}{32 \pi^2} \,, \\
    &\beta_{\mu^2_{\rm H}} = \frac{8 \lambda_{\rm HD} \mu^2_{\rm D} - 3 \mu^2_{\rm H} \left(3\g{W}^2+\g{Y}^2-8 \lambda_{\rm H} - 4 y_t^2\right)}{32 \pi ^2}\,, \\
    &\beta_{m_{f_{\rm D}}} = -\frac{m_{f_{\rm D}} \left(27 \g{D}^2 + 48 g_S^2 + 16 \g{Y}^2-3 y^{\prime 2}\right)}{96 \pi ^2}\,.
\end{align}
\endgroup

\subsection{Debye masses}\label{app:debye_masses}

\subsubsection*{\underline{Scenario I:}}
\begingroup
\allowdisplaybreaks
\begin{align}
    &\label{eqn:Debye_1field_LO} \qty[\mu^2_{\mathrm{SU(2)_D}}]^{\mathrm{(LO)}} = \frac{5}{6}\g{D}^2 T^2\,, \\
    &\label{eqn:Debye_1field_NLO} \qty[\mu^2_{\mathrm{SU(2)_D}}]^{\mathrm{(NLO)}} = \frac{\g{D}^4 \qty(207 + 430L_b)T^2 + 72\g{D}^2\qty(T^2 \lambda_{\rm D} + 2\mu_{\rm D}^2)}{1152\pi^2}\,.
\end{align}
\endgroup

\subsubsection*{\underline{Scenario II:}}
\begingroup
\allowdisplaybreaks
\begin{align}\label{eqn:debye_masses_ScenarioIII}
    &\qty[\mu^2_{\mathrm{SU(2)_L}}]^{\mathrm{(LO)}} =  \mathrm{Eq.}~\eqref{eqn:debye_SU2L_LO} \,, \\
    &\qty[\mu^2_{\mathrm{SU(2)_L}}]^{\mathrm{(NLO)}} = \mathrm{Eq.}~\eqref{eqn:debye_SU2L_LO_2} \,, \\
    &\qty[\mu^2_{\mathrm{SU(2)_D}}]^{\mathrm{(LO)}} = \mathrm{Eq.}~\eqref{eqn:Debye_1field_LO}\,, \\
    &\qty[\mu^2_{\mathrm{SU(2)_D}}]^{\mathrm{(NLO)}} = \frac{\g{D}^2 \left[T^2 \left(\g{D}^2 (430 L_b + 207)+24 (3 \lambda_{\rm D}+\lambda_{\rm HD})\right)+144 \mu^2_{\rm D}\right]}{1152 \pi ^2}\,, \\
    &\qty[\mu^2_{\mathrm{SU(3)_C}}]^{\mathrm{(LO)}} = 2g_S^2 T^2\,, \\
    &\qty[\mu^2_{\mathrm{SU(3)_C}}]^{\mathrm{(NLO)}} = \frac{g_S^2 T^2 \left(24 g_S^2 (11 L_b - 4 L_f + 5)-27 \g{W}^2-11 \g{Y}^2-12 y_t^2\right)}{192 \pi ^2} \\
    &\qty[\mu^2_{\mathrm{U(1)_Y}}]^{\mathrm{(LO)}} = \frac{11}{6}\g{Y}^2 T^2\,, \\\nonumber
    &\qty[\mu^2_{\mathrm{U(1)_Y}}]^{\mathrm{(NLO)}} = \frac{\g{Y}^2}{2304 \pi ^2}\Big[288 \mu^2_{\rm H}-2 T^2 (528 g_S^2+81 \g{W}^2+\g{Y}^2 (22 L_b + 880 L_f - 465) - \\
    &\hspace*{8em} 24 (3 \lambda_{\rm H} + \lambda_{\rm HD})+66 y_t^2)\Big]
\end{align}
\endgroup

\subsubsection*{\underline{Scenario III:}}
\begingroup
\allowdisplaybreaks
\begin{align}
    &\qty[\mu^2_{\mathrm{SU(2)_L}}]^{\mathrm{(LO)}} = \frac{11}{6}\g{W}^2 T^2\,, \label{eqn:debye_SU2L_LO}\\\nonumber
    &\qty[\mu^2_{\mathrm{SU(2)_L}}]^{\mathrm{(NLO)}} = \frac{\g{W}^2 T^2}{1152 \pi ^2} \Big[\bigl\{-432 g_S^2+11 \g{W}^2 (86 L_b-48L_f + 57) + \\
        &\hspace*{8em} 3 \left(-9 \g{Y}^2+24 \lambda_{\rm H} +8 \lambda_{\rm HD} - 6 y_t^2\right)\bigl\} + 144 (\mu^2_{\rm H}/T^2)\Big]\,, \label{eqn:debye_SU2L_LO_2}\\
    &\qty[\mu^2_{\mathrm{SU(2)_D}}]^{\mathrm{(LO)}} = \frac{4}{3}\g{D}^2 T^2\,, \\\nonumber
    &\qty[\mu^2_{\mathrm{SU(2)_D}}]^{\mathrm{(NLO)}} = \frac{\g{D}^2 T^2}{1152 \pi ^2} \Big[\bigl\{-6 \left(48 g_S^2 +16 \g{Y}^2-4 (3 \lambda_{\rm D} + \lambda_{\rm HD})+ 3y^{\prime 2} \right)\bigl\} + \\
        &\hspace*{8em} \g{D}^2 (688 L_b -192 L_f + 345) + 144 (\mu^2_{\rm D}/T^2)\Big]\,, \\
    &\qty[\mu^2_{\mathrm{SU(3)_C}}]^{\mathrm{(LO)}} = \frac{7}{3}g_S^2 T^2\,, \\\nonumber
    &\qty[\mu^2_{\mathrm{SU(3)_C}}]^{\mathrm{(NLO)}} = \frac{g_S^2 T^2}{576 \pi ^2} \Bigl\{-54 \g{D}^2+g_S^2 (528 L_b-448 L_f+792 \mathrm{ln}(A) +484) - \\
        &\hspace*{8em}792 g_S^2 ~\mathrm{ln} \left(\frac{4 \pi  T}{\mu_{\mathrm{4D}} }\right)-81 \g{W}^2-65 \g{Y}^2-36 \left(y^{\prime 2}+y_t^2\right)\Bigl\} \\
    &\qty[\mu^2_{\mathrm{U(1)_Y}}]^{\mathrm{(LO)}} = \frac{49}{18}\g{Y}^2 T^2\,, \\\nonumber
    &\qty[\mu^2_{\mathrm{U(1)_Y}}]^{\mathrm{(NLO)}} = \frac{\g{Y}^2T^2}{10368 \pi ^2} \Big[(1296/T^2) \mu^2_{\rm H} - \bigl\{2592 \g{D}^2 + 9360 g_S^2 + 729 \g{W}^2 + \\
        &\hspace*{8em}54\left(-4 (3 \lambda_{\rm H} + \lambda_{\rm HD}) +32 y^{\prime 2} + 11y_t^2 + \g{Y}^2 [294 L_b + 18032 L_f - 12569]\right)\bigl\}\Big]
\end{align}
\endgroup

\subsection{Coupling matching between 3D-soft and 4D theories}\label{app:soft_matching_couplings}

Here, soft couplings are marked with the superscript $\mathcal{S}$, whereas temporal couplings follow the same conventions as \texttt{DRAlgo}. For temporal couplings, the associated operator is written before the coupling expression.

\subsubsection*{\underline{Scenario I:}}
3D soft couplings:
\begingroup
\allowdisplaybreaks
\begin{align}
    &[g^{\mathcal{S}}_{\rm D} ]^2 = \g{D}^2 T + \frac{\g{D}^4(4 + 43L_b)T}{96\pi^2}  \label{eqn:gD_soft}\,, \\
    &\lambda_{\rm D}^{\mathcal{S}} = \frac{T\qty[\g{D}^4(6-9L_b) + 72\g{D}^2 L_b \lambda_{\rm D} + 64\lambda_{\rm D}(4\pi^2 - 3L_b\lambda_{\rm D})]}{256\pi^2} \,. 
\end{align}
\endgroup
Temporal couplings:
\begingroup
\allowdisplaybreaks
\begin{alignat}{2}\label{eqn:temporal_couplings_Scenario1}
    &:(V^a_0)^4:  &&\lambda_{V,1} = \frac{17 \g{D}^4 T}{24 \pi ^2} \,, \\
    &:(V^a_0)^2(\Phi_{\rm{D}}^\dagger \Phi_{\rm{D}}): &&\lambda_{S,1} = \frac{\g{D}^2 T\qty(\g{D}^2(51 + 43L_b) + 96\pi^2 + 72\lambda_{\rm D})}{192\pi^2}\,. 
\end{alignat}
\endgroup

\subsubsection*{\underline{Scenario II:}}
3D soft couplings:
\begingroup
\allowdisplaybreaks
\begin{align}
    &[g^{\mathcal{S}}_{\rm D} ]^2 = \mathrm{Eq.}~\eqref{eqn:gD_soft} \,, \\
    &[g^{\mathcal{S}}_W]^2 = \mathrm{Eq.}~\eqref{eqn:gW_soft} \,, \\
    &[g^{\mathcal{S}}_Y]^2 =  \g{Y}^2 T-\frac{\g{Y}^4 T (L_b + 40 L_f)}{96\pi ^2}\,, \\
    &[g^{\mathcal{S}}_S]^2 = \frac{g_S^4 T (11 L_b-4 L_f+1)}{16 \pi^2}+g_S^2 T \,, \\
    &\lambda_{\rm D}^{\mathcal{S}} = \mathrm{Eq.}~\eqref{eqn:lambD_soft}~\mathrm{when}~y^\prime\rightarrow 0 \,, \\
    &\lambda_{\rm H}^{\mathcal{S}} = \mathrm{Eq.}~\eqref{eqn:lambH_soft} \\
    & \lambda_{\rm HD}^{\mathcal{S}} = \mathrm{Eq.}~\eqref{eqn:lambHD_soft}~\mathrm{when}~y^\prime\rightarrow 0
\end{align}
\endgroup
Temporal couplings:
\begingroup
\allowdisplaybreaks
\begin{alignat}{2}\label{eqn:temporal_couplings_Scenario3}
    &:(V^a_0)^4:  &&\lambda_{V,1} = \frac{17 \g{D}^4 T}{24 \pi ^2} \,, \\
    &:(G^a_0)^4:  &&\lambda_{V,3} = \frac{g_S^4 T}{2\pi^2} \,, \\
    &:(G^a_0)^2(A^a_0)^2: && \lambda_{V,4} = \mathrm{Eq.}~\eqref{eqn:su3su2_temp} \,,\\
    &:(A^a_0)^4: && \lambda_{V,5} = \mathrm{Eq.}~\eqref{eqn:su2_temp} \,,\\
    &:(G^a_0)^2(B_0)^2: && \lambda_{V,8} = -\frac{11g_S^2\g{Y}^2 T}{12\pi^2} \,,\\
    &:(A^a_0)^2(B_0)^2: && \lambda_{V,9} = \mathrm{Eq.}~\eqref{eqn:su2u1_temp} \,,\\
    &:(B_0)^4: && \lambda_{V,10} = -\frac{371\g{Y}^4 T}{72\pi^2}\,,\\
    &:(G^a_0)^2(\Phi_{\rm{H}}^\dagger \Phi_{\rm{H}}): && \lambda_{S,4} = \mathrm{Eq.}~\eqref{eqn:su3lamH_temp} \,,\\
    &:(V^a_0)^2(\Phi_{\rm{D}}^\dagger \Phi_{\rm{D}}): && \lambda_{S,5} = \frac{\g{D}^2 T \left(\g{D}^2 (43 L_b+51)+72 \lambda_{\rm D}+96 \pi ^2\right)}{192 \pi ^2} \,,\\\nonumber
    &:(\vec{A_0}\cdot\vec{\tau}\Phi_{\rm{H}})(B_0 \Phi^\dagger_{\rm H}): && \lambda_{S,6} = -\frac{\g{W} \g{Y} T}{384 \pi ^2} \Big[\g{W}^2 (43 L_b - 24 L_f + 12) - \\\nonumber
    & \hphantom{.} && \hspace*{8em} \g{Y}^2 (-44 + L_b + 40L_f) - \\
    & \hphantom{.} && \hspace*{8em} 24 \left(-2 \lambda_{\rm H} + 3 L_f y_t^2 + y_t^2-8 \pi ^2\right)\Big]\,, \\\nonumber
    &:(B_0)^2(\Phi_{\rm{H}} \Phi^\dagger_{\rm H}): && \lambda_{S,7} = -\frac{\g{Y}^2 T}{192 \pi ^2} \Big[-9 \g{W}^2+\g{Y}^2 (L_b + 40 L_f - 41) - \nonumber\\
    & \hphantom{.} && \hspace*{7.5em} -96\pi^2 + 68y_t^2 - 72\lambda_{\rm H} \Big] \,, \\
    & :(A^a_0)^2(\Phi_{\rm{H}} \Phi^\dagger_{\rm H}): && \lambda_{S,8} = \mathrm{Eq.}~\eqref{eqn:su2temp} \\
    & :(V^a_0)^2(\Phi_{\rm{D}} \Phi^\dagger_{\rm D}): && \lambda_{S,9} = \mathrm{Eq.}~\eqref{eqn:vphiD_temp}\\
    & :(A^a_0)^2(\Phi_{\rm{H}} \Phi^\dagger_{\rm H}): && \lambda_{S,10} = \mathrm{Eq.}~\eqref{eqn:AphiD_temp}\\
    & :(B_0)^2(\Phi_{\rm{D}} \Phi^\dagger_{\rm D}): && \lambda_{S,11} = \mathrm{Eq.}~\eqref{eqn:BphiD_temp}~\mathrm{when}~y^\prime \rightarrow 0
\end{alignat}
\endgroup

\subsubsection*{\underline{Scenario III:}}
3D soft couplings:
\begingroup
\allowdisplaybreaks
\begin{align}
    &[g^{\mathcal{S}}_{\rm D} ]^2 = \frac{\g{D}^4 T (43 L_b-12 L_f+4)}{96 \pi ^2}+\g{D}^2 T  \,, \\
    &[g^{\mathcal{S}}_W]^2 = \frac{\g{W}^4 T (43 L_b - 24 L_f + 4)}{96 \pi ^2} + \g{W}^2 T \,, \label{eqn:gW_soft} \\
    &[g^{\mathcal{S}}_Y]^2 = \frac{\g{Y}^2 T \Big[96 \left(4 L_f y^{\prime 2} + 3 \pi ^2\right) - \g{Y}^2 + (3 L_b + 184 L_f)\Big]}{288 \pi ^2}\,, \\
    &[g^{\mathcal{S}}_S]^2 = \frac{g_S^4 T (33 L_b-16 L_f+3)}{48 \pi ^2} + g_S^2 T \,, \\\nonumber
    &\lambda_{\rm D}^{\mathcal{S}} = \frac{T}{256 \pi ^2} \Big[\g{D}^4 (6-9 L_b)+72 \g{D}^2 \lambda_{\rm D} L_b - 16\lambda_{\rm HD}^2 L_b \\
        &\hspace*{6em}-32 \lambda_{\rm D} \left(6 \lambda_{\rm D} L_b + 3L_f y^{\prime 2} - 8 \pi ^2\right) + 48 L_f y^{\prime 4} \Big]\,, \label{eqn:lambD_soft}\\\nonumber
    &\lambda_{\rm H}^{\mathcal{S}} = \frac{T}{256\pi ^2} \Big[24 \lambda_{\rm H} \left(3 \g{W}^2 L_b + \g{Y}^2 L_b-4 L_f y_t^2\right)+(2-3 L_b) \left(3 \g{W}^4 +2 \g{W}^2 \g{Y}^2 + \g{Y}^4\right)+ \\
        &\hspace*{6em} 256 \pi ^2 \lambda_{\rm H} - 16 L_b \left(12\lambda_{\rm H}^2 + \lambda_{\rm HD}^2\right)+48 L_f y_t^4\Big]\label{eqn:lambH_soft} \\\nonumber
    & \lambda_{\rm HD}^{\mathcal{S}} = \frac{T}{64 \pi ^2} \Bigg(\lambda_{\rm HD} \left(3 L_b \big[3 \g{D}^2 + 3 \g{W}^2+\g{Y}^2-8 (\lambda_{\rm D} + \lambda_{\rm H})\big]-8\lambda_{\rm HD} L_b + 64 \pi ^2\right) - \\ 
        &\hspace*{6em}12 \lambda_{\rm HD} L_f \left(y^{\prime 2} + y_t^2\right)+24 L_f y^{\prime 2}y_t^2\Bigg) \label{eqn:lambHD_soft}
\end{align}
\endgroup
Temporal couplings:
\begingroup
\allowdisplaybreaks
\begin{alignat}{2}
    &:(V^a_0)^4:  &&\lambda_{V,1} = \frac{11 \g{D}^4 T}{24 \pi ^2} \,, \\
    &:(V^a_0)^2(G^a_0)^2:  &&\lambda_{V,2} = -\frac{\g{D}^2 g_S^2 T}{2 \pi ^2} \,, \\
    &:(G^a_0)^4:  &&\lambda_{V,3} = \frac{g_S^4 T}{2\pi^2} \,, \\
    &:(G^a_0)^2(A^a_0)^2: && \lambda_{V,4} = -\frac{3g_S^2\g{W}^2 T}{4\pi^2} \,,\label{eqn:su3su2_temp} \\
    &:(A^a_0)^4: && \lambda_{V,5} = \frac{5\g{W}^4 T}{24\pi^2} \,, \label{eqn:su2_temp}\\
    &:(V^a_0)^2(B_0)^2: && \lambda_{V,7} = -\frac{4\g{D}^2\g{Y}^2 T}{3\pi^2} \,,\\
    &:(G^a_0)^2(B_0)^2: && \lambda_{V,8} = -\frac{-65g_S^2\g{Y}^2 T}{36\pi^2} \,,\\
    &:(A^a_0)^2(B_0)^2: && \lambda_{V,9} = -\frac{3\g{W}^2\g{Y}^2T}{8\pi^2}\,, \label{eqn:su2u1_temp}\\
    &:(B_0)^4: && \lambda_{V,10} = -\frac{1625\g{Y}^4 T}{216\pi^2}\,,\\
    &:(G^a_0)^2(\Phi_{\rm{D}}^\dagger \Phi_{\rm{D}}): &&\lambda_{S,1} + \lambda_{S,2} = \frac{-g_S^2 T y^{\prime 2}}{4\pi^2} + \frac{g_S^2(-1 + L_f)y^{\prime 2}T}{4\pi^2}  \,,\\
    &:(\vec{V_0}\cdot\vec{\tau}\Phi_{\rm{D}})(B_0 \Phi^\dagger_{\rm D}): &&\lambda_{S,3} = -\frac{\g{D} \g{Y}\qty(-1 + L_f)T y^{\prime 2}}{4\pi^2} \,,\\
    &:(G^a_0)^2(\Phi_{\rm{H}}^\dagger \Phi_{\rm{H}}): && \lambda_{S,4} = -\frac{g_S^2 T y_t^2}{4\pi^2} \,, \label{eqn:su3lamH_temp} \\\nonumber
    &:(V^a_0)^2(\Phi_{\rm{D}}^\dagger \Phi_{\rm{D}}): && \lambda_{S,5} = \frac{\g{D}^2T}{192 \pi ^2} \Big[\g{D}^2 (43 L_b-12 L_f+63) + \\
    & \hphantom{.} && \hspace*{6.5em} 12 \left(6\lambda_{\rm D} - 3 y^{\prime 2} +8 \pi ^2\right)\Big] \,,\\\nonumber
    &:(\vec{A_0}\cdot\vec{\tau}\Phi_{\rm{H}})(B_0 \Phi^\dagger_{\rm H}): && \lambda_{S,6} = -\frac{\g{W} \g{Y} T}{1152 \pi ^2} \Big[3 \g{W}^2 (43 L_b - 24 L_f + 12) + \\\nonumber
    & \hphantom{.} && \hspace*{8em} \g{Y}^2 (-3 L_b-184 L_f+196) - \\
    & \hphantom{.} && \hspace*{8em} 72 \left(-2 \lambda_{\rm H} + 3 L_f y_t^2 + y_t^2-8 \pi ^2\right)\Big]\,, \\\nonumber
    &:(B_0)^2(\Phi_{\rm{H}} \Phi^\dagger_{\rm H}): && \lambda_{S,7} = -\frac{\g{Y}^2 T}{576 \pi ^2} \Big[-27 \g{W}^2+\g{Y}^2 (3 L_b + 184 L_f-187) - \\
    & \hphantom{.} && \hspace*{7.5em} 12\left(18 \lambda_{\rm H} - 17 y_t^2+24 \pi ^2\right)\Big] \,, \\\nonumber
    & :(A^a_0)^2(\Phi_{\rm{H}} \Phi^\dagger_{\rm H}): && \lambda_{S,8} = \frac{\g{W}^2 T}{192 \pi ^2} \Big[\g{W}^2 (43 L_b-24 L_f+75) + \\
    & \hphantom{.} && \hspace*{6.5em} 3 \bigl\{\g{Y}^2+4\left(6 \lambda_{\rm H} -3 y_t^2+8 \pi ^2\right)\bigl\}\Big] \label{eqn:su2temp} \\
    & :(V^a_0)^2(\Phi_{\rm{H}} \Phi^\dagger_{\rm H}): && \lambda_{S,9} = \frac{\g{D}^2 \lambda_{\rm HD} T}{8 \pi ^2} \label{eqn:vphiD_temp}\\
    & :(A^a_0)^2(\Phi_{\rm{D}} \Phi^\dagger_{\rm D}): && \lambda_{S,10} = \frac{\g{W}^2 \lambda_{\rm HD} T}{8 \pi ^2} \label{eqn:AphiD_temp}\\
    & :(B_0)^2(\Phi_{\rm{D}} \Phi^\dagger_{\rm D}): && \lambda_{S,11} = \frac{\g{Y}^2 T \left(3 \lambda_{\rm HD} + 16 (L_f-1)y^{\prime 2} \right)}{24 \pi ^2} \label{eqn:BphiD_temp}
\end{alignat}
\endgroup

\subsection{Scalar mass matching between 3D-soft and 4D theories}\label{sec:soft_matching_masses}

\subsubsection*{\underline{Scenario I}}
\begingroup
\allowdisplaybreaks
\begin{align}
    & [\mu^{\mathcal{S}}_{\rm D}]^2_{\mathrm{LO}}  = \frac{T^2}{16}\qty(3\g{D}^2 + 8\lambda_{\rm D}) + \mu_{\rm D}^2 \label{eqn:scalar_masses_LO_1} \\\nonumber
    & [\mu^{\mathcal{S}}_{\rm D}]^2_{\mathrm{NLO}} = \frac{1}{1536 \pi^2} \Big[ \g{D}^4 T^2 (-2916 ~\mathrm{ln}A - 141 L_b + 243 \gamma_E +167) + \\\nonumber
    & \hspace*{5em} 72 \g{D}^2 \left(\lambda_{\rm D} T^2 (-72 ~\mathrm{ln} A - 3 L_b + 6 \gamma_E +1)+3 L_b \mu^2_{\rm D} \right) - \\ \nonumber
    & \hspace*{5em} 576 \lambda_{\rm D} \left(\lambda_{\rm D} T^2 (\mathrm{ln}(A) -12 ~\mathrm{ln} A )+ L_b \mu_{\rm D}^2 \right) - \\
    & \hspace*{5em} 18 \left(13 [g_{\rm D}^\mathcal{S}]^4+16 [g_{\rm D}^\mathcal{S}]^2 (3 \lambda_{\rm D}^\mathcal{S}+2\lambda_{S,1} ) - 8\left(8 [\lambda_{\rm D}^\mathcal{S}]^2+\lambda_{S,1}^2\right)\right) ~\mathrm{ln} \left(\frac{\mu_{3\mathrm{d}}}{\mu_{4\mathrm{d}} }\right) \Big]\,.\label{eqn:scalar_masses_LO_2}
\end{align}
\endgroup

\subsubsection*{\underline{Scenario II}}
\begingroup
\allowdisplaybreaks
\begin{align}
    & [\mu^{\mathcal{S}}_{\rm D}]^2_{\mathrm{LO}}  = \mathrm{Eq.}~\eqref{eqn:mass_muD}~\mathrm{when}~y^\prime\rightarrow 0 \,, \\ 
    & [\mu^{\mathcal{S}}_{\rm H}]^2_{\mathrm{LO}}  = \mathrm{Eq.}~\eqref{eqn:mass_muH} \,, \\\nonumber
    & [\mu^{\mathcal{S}}_{\rm D}]^2_{\mathrm{NLO}} = \frac{1}{1536 \pi^2} \Big[\g{D}^4 T^2 (-2916 \mathrm{ln}(A)-141 L_b+243 \gamma_E + 167) + \\\nonumber
        & \hspace*{9em} 36 \g{D}^2 (T^2 (2 \lambda_{\rm D} (-72 \mathrm{ln}(A)-3 L_b+6 \gamma_E +1)+\lambda_{\rm HD} L_b)+6 L_b \mu^2_{\rm D}) - \\\nonumber
        & \hspace*{9em} 4 \Big(3 \g{W}^2 \lambda_{\rm HD} T^2 (144 \mathrm{ln}(A)+9 L_b-2)+144 \g{Y}^2 \lambda_{\rm HD} T^2 \mathrm{ln}(A) - \\\nonumber
        & \hspace*{9em} 1728 \lambda_{\rm D}^2 T^2 \mathrm{ln}(A)-288 \lambda_{\rm HD}^2 T^2 \mathrm{ln}(A)+12 \gamma_E  T^2 (\lambda_{\rm HD} (-3 \g{W}^2 - \\\nonumber
        & \hspace*{9em} \g{Y}^2+2 \lambda_{\rm HD})+12 \lambda_{\rm D}^2)+9 \g{Y}^2 \lambda_{\rm HD} L_b T^2-2 \g{Y}^2 \lambda_{\rm HD} T^2 + \\\nonumber
        & \hspace*{9em} 144 \lambda_{\rm D} L_b \mu^2_{\rm D}+48 \lambda_{\rm HD} L_b \mu^2_{\rm H}+24 \lambda_{\rm D} \lambda_{\rm HD} L_b T^2+24 \lambda_{\rm H} \lambda_{\rm HD} L_b T^2 - \\
        & \hspace*{9em} 4 \lambda_{\rm HD}^2 L_b T^2+18 \lambda_{\rm HD} L_b T^2 y_t^2-6 \lambda_{\rm HD} L_f T^2 y_t^2\Big)-6 \mathrm{ln} \left(\frac{\mu_{\mathrm{3D}}}{\mu_{\mathrm{4D}}  }\right)\mathcal{K}_1 \Big]\,, \\\nonumber
    & [\mu^{\mathcal{S}}_{\rm H}]^2_{\mathrm{NLO}} = \frac{1}{4608 \pi ^2} \Big[1728 T^2 \mathrm{ln}(A) \left(\lambda_{\rm HD} \left(2 \lambda_{\rm HD}-3 \g{D}^2\right)+12 \lambda_{\rm H}^2\right)+3 \g{W}^4 T^2 (-2916 \mathrm{ln}(A)- \\\nonumber
        & \hspace*{9em} 249 L_b+36 L_f+243 \gamma_E +191)-27 \g{W}^2 (2 \g{Y}^2 T^2 (-60 \mathrm{ln}(A)-4 L_b + \\\nonumber
        & \hspace*{9em} 5 \gamma_E +1)+T^2 \big[-8 \lambda_{\rm H} (-72 \mathrm{ln}(A)-3 L_b+6 \gamma_E +1)-4 \lambda_{\rm HD} L_b + \\\nonumber
        & \hspace*{9em} y_t^2 (-7 L_b+L_f+2)\big]-24 L_b \mu^2_{\rm H})+\g{Y}^4 T^2 (756 \mathrm{ln}(A)-147 L_b+60 L_f - \\\nonumber
        & \hspace*{9em} 63 \gamma_E +41)+\g{Y}^2 (T^2 \Big[72 \lambda_{\rm H} (-72 \mathrm{ln}(A)-3 L_b+6 \gamma_E +1)+36 \lambda_{\rm HD} L_b + \\\nonumber
        & \hspace*{9em} y_t^2 (47 L_b+55 L_f-66)\Big]+216 L_b \mu^2_{\rm H})-12 \Big(L_b \Big\{T^2 \Big[\lambda_{\rm HD} \{27 \g{D}^2 + \\\nonumber
        & \hspace*{9em} 24 (\lambda_{\rm D}+\lambda_{\rm H})-4 \lambda_{\rm HD}\}-9 y_t^4+54 \lambda_{\rm H} y_t^2\Big]+48 (3 \lambda_{\rm H} \mu^2_{\rm H}+\lambda_{\rm HD} \mu^2_{\rm D})\Big\} + \\\nonumber
        & \hspace*{9em} 6 T^2 \left(-\left((1+6 \gamma_E ) \g{D}^2 \lambda_{\rm HD}\right)+24 \gamma_E  \lambda_{\rm H}^2+4 \gamma_E  \lambda_{\rm HD}^2+L_f y_t^2 (3 \lambda_{\rm H}+2 \lambda_{\rm HD})\right) + \\
        & \hspace*{9em} 16 g_S^2 T^2 y_t^2 (L_b-4 L_f+3)+72 L_f \mu^2_{\rm H} y_t^2\Big)-18 \mathrm{ln} \left(\frac{\mu_{\mathrm{3D}}}{\mu_{\mathrm{4D}}  }\right)\mathcal{K}_2 \Big] \,,
\end{align}
where $\mathcal{K}_1$ correspond to the terms that multiply on the right hand-side of logarithm in \cref{eqn:muDsoft} and $\mathcal{K}_2$ is the equivalent terms in \cref{eqn:muHsoft}.
\endgroup

\subsubsection*{\underline{Scenario III}}
\begingroup
\allowdisplaybreaks
\begin{align}
    & [\mu^{\mathcal{S}}_{\rm D}]^2_{\mathrm{LO}}  = \frac{3 \g{D}^2 T^2}{16}+\mu^2_{\rm D}+\frac{1}{12} T^2 \left(6 \lambda_{\rm D} + 2 \lambda_{\rm HD}+3 y^{\prime 2} \right) \,, \label{eqn:mass_muD}\\ 
    & [\mu^{\mathcal{S}}_{\rm H}]^2_{\mathrm{LO}}  = \frac{1}{48} \left(9 \g{W}^2 T^2+3 \g{Y}^2 T^2+48 \mu^2_{\rm H} + 24 \lambda_{\rm H} T^2+8 \lambda_{\rm HD} T^2+12 T^2 y_t^2\right)\,, \label{eqn:mass_muH}\\ \nonumber
    & [\mu^{\mathcal{S}}_{\rm D}]^2_{\mathrm{NLO}} = \frac{1}{4608 \pi ^2} \Bigg[3 \g{D}^4 T^2 (-2916 ~\mathrm{ln} (A)+243 \gamma_E-195 L_b+18 L_f+179) + \\\nonumber
          & \hspace*{9em} 27 \g{D}^2 (T^2 \Big[8\lambda_{\rm D} (-72 ~\mathrm{ln} (A)+6 \gamma_E-3 L_b+1)+4 \lambda_{\rm HD} L_b + \\\nonumber
          & \hspace*{9em} y^{\prime 2} (7L_b-L_f-2)\Big]+24 L_b \mu^2_{\rm D}) + \\\nonumber 
          & \hspace*{9em} 4 \Big(432 T^2 ~\mathrm{ln} (A) \left(-\lambda_{\rm HD} \left(3\g{W}^2+\g{Y}^2\right)+12 \lambda_{\rm D}^2+2 \lambda_{\rm HD}^2\right)+ \\\nonumber
          & \hspace*{9em} 2 L_f (T^2 \left(y^{\prime 2} \left(51 g_S^2+8 \g{Y}^2-27 \lambda_{\rm D}+18 y_t^2\right)+9 \lambda_{\rm HD} \left(y_t^2-2 y^{\prime 2}\right)\right)+ \\\nonumber
          & \hspace*{9em} 216 m_{f_{\rm D}}^2 y^{\prime 2}-108 \mu^2_{\rm D} y^{\prime 2})+6 g_S^2 (7 L_b-4) T^2 y^{\prime 2}-81 \g{W}^2 \lambda_{\rm HD} L_b T^2 + \\\nonumber
          & \hspace*{9em} 108 \gamma_E \g{W}^2 \lambda_{\rm HD} T^2+18 \g{W}^2 \lambda_{\rm HD} T^2-27 \g{Y}^2 \lambda_{\rm HD} L_b T^2+32 \g{Y}^2 L_b T^2 y^{\prime 2} + \\\nonumber
          & \hspace*{9em} 36 \gamma_E \g{Y}^2 \lambda_{\rm HD} T^2+6 \g{Y}^2 \lambda_{\rm HD} T^2+16 \g{Y}^2 T^2 y^{\prime 2} - 432 \lambda_{\rm D} L_b \mu^2_{\rm D} - \\\nonumber
          & \hspace*{9em} 144 \lambda_{\rm HD} L_b \mu^2_{\rm H}-72 \lambda_{\rm D} \lambda_{\rm HD} L_b T^2-72 \lambda_{\rm H} \lambda_{\rm HD} L_b T^2+12 \lambda_{\rm HD}^2 L_b T^2 + \\\nonumber
          & \hspace*{9em} 27 L_b T^2 y^{\prime 4}-162 \lambda_{\rm D} L_b T^2 y^{\prime 2}+18 L_b T^2 y^{\prime 2} y_t^2-54 \lambda_{\rm HD} L_b T^2 y_t^2 \\\nonumber 
          & \hspace*{9em} -432 \gamma_E \lambda_{\rm D}^2 T^2-72 \gamma_E \lambda_{\rm HD}^2 T^2\Big) - \\\nonumber
          & \hspace*{9em} 18 ~\mathrm{ln} \left(\frac{\mu_{\mathrm{3D}}}{\mu_{\mathrm{4D}}}\right) \Bigl\{(39 [g^{\mathcal{S}}_{\rm D} ]^4+48 [g^{\mathcal{S}}_{\rm D} ]^2 (3 \lambda_{\rm D}^{\mathcal{S}}+2 \lambda_{S,5}) - \\\nonumber
          & \hspace*{9em} 8 \Big(-6 [g^{\mathcal{S}}_S]^2 (3 \lambda_{S,1}+5 \lambda_{S,2})-6 [g^{\mathcal{S}}_W]^2 (\lambda_{\rm HD}^{\mathcal{S}}+2 \lambda_{S,10}) - \\\nonumber
          & \hspace*{9em} 2 \lambda_{\rm HD}^{\mathcal{S}} \left([g^{\mathcal{S}}_Y]^2-2 \lambda_{\rm HD}^{\mathcal{S}}\right)+24 [\lambda_{\rm D}^{\mathcal{S}}]^2+3 \lambda_{S,1}^2+5 \lambda_{S,2}^2+\lambda_{S,11}^2 + \\
          & \hspace*{9em} 3 [2 \lambda_{S,3}^2+\lambda_{S,5}^2+\lambda_{S,10}^2]\Big)\Bigl\} \Bigg] \,, \label{eqn:muDsoft}\\\nonumber
   & [\mu^{\mathcal{S}}_{\rm H}]^2_{\mathrm{NLO}} = \frac{1}{13824 \pi ^2} \Bigg[-36 \Big(6 T^2 (2 (\gamma_E-12 ~\mathrm{ln} (A)) \left(\lambda_{\rm HD} \left(2 \lambda_{\rm HD}-3 \g{D}^2\right)+12 \lambda_{\rm H}^2\right) - \\\nonumber
          & \hspace*{10em} \g{D}^2 \lambda_{\rm HD})+L_b T^2 \Big(\lambda_{\rm HD} \left(27 \g{D}^2+24 (\lambda_{\rm D}+\lambda_{\rm H})-4 \lambda_{\rm HD}\right) - \\\nonumber
          & \hspace*{10em} 6 y^{\prime 2} \left(y_t^2-3 \lambda_{\rm HD}\right)-9 y_t^4+54 \lambda_{\rm H} y_t^2\Big)+16 g_S^2 T^2 y_t^2 (L_b-4 L_f+3) + \\\nonumber
          & \hspace*{10em} 48 L_b (3 \lambda_{\rm H} \mu_{\rm H}^2+\lambda_{\rm HD} \mu_{\rm D}^2)+6 L_f T^2 \left(y_t^2 (3 \lambda_{\rm H}+2 \lambda_{\rm HD})-y^{\prime 2} \left(\lambda_{\rm HD}+2 y_t^2\right)\right)+ \\\nonumber
          & \hspace*{10em} 72 L_f \mu_{\rm H}^2 y_t^2\Big)+9 \g{W}^4 T^2 (-2916 ~\mathrm{ln} (A)+243 \gamma_E-249 L_b+36 L_f+191) - \\\nonumber
          & \hspace*{10em} 81 \g{W}^2 (2 \g{Y}^2 T^2 (-60 ~\mathrm{ln} (A)+5 \gamma_E-4 L_b+1) + \\\nonumber
          & \hspace*{10em} T^2 \left(-8 \lambda_{\rm H} (-72 ~\mathrm{ln} (A)+6 \gamma_E-3 L_b+1)-4 \lambda_{\rm HD} L_b+y_t^2 (-7 L_b+L_f+2)\right) - \\\nonumber
          & \hspace*{10em} 24 L_b \mu_{\rm H}^2)+\g{Y}^4 T^2 (2268 ~\mathrm{ln} (A)-189 \gamma_E-729 L_b+276 L_f+187)+3 \g{Y}^2 \\\nonumber
          & \hspace*{10em} (T^2\big[72 \lambda_{\rm H} (-72 ~\mathrm{ln} (A)+6 \gamma_E-3 L_b+1)+36 \lambda_{\rm HD} L_b + \\\nonumber
          & \hspace*{10em} y_t^2 (47L_b+55 L_f-66)\big]+216 L_b \mu_{\rm H}^2) - \\\nonumber
          & \hspace*{10em} 54 ~\mathrm{ln} \left(\frac{\mu_{\mathrm{3D}}}{\mu_{\mathrm{4D}}}\right) (-8 \Big[-6 [g^{\mathcal{S}}_{\rm D} ]^2 (\lambda_{\rm HD}^{\mathcal{S}}+2 \lambda_{S,9})+8 \lambda_{S4} \left(\lambda_{S4}-6 [g^{\mathcal{S}}_S]^2\right) + \\\nonumber
          & \hspace*{10em} 24 [\lambda_{\rm H}^{\mathcal{S}}]^2+4 [\lambda_{\rm HD}^{\mathcal{S}}]^2+6 \lambda_{S6}^2+\lambda_{S,7}^2+3 \left(\lambda_{S,8}^2+\lambda_{S,9}^2\right)\Big]+39 [g^{\mathcal{S}}_W]^4 + \\
          & \hspace*{10em} 6 [g^{\mathcal{S}}_W]^2 \left(-3 [g^{\mathcal{S}}_Y]^2+24 \lambda_{\rm H}^{\mathcal{S}}+16 \lambda_{S,8}\right)-5 [g^{\mathcal{S}}_Y]^4+48 [g^{\mathcal{Y}}_D]^2 \lambda_{\rm H}^{\mathcal{S}}) \Bigg] \label{eqn:muHsoft}
\end{align}
\endgroup

\subsection{Coupling matching between 3D-ultrasoft and 3D soft theories}\label{app:Ultrasoft_matching_couplings}
Here, the ultrasoft couplings are marked with the superscript $\mathcal{US}$.
\subsubsection*{\underline{Scenario I}}
\begingroup
\allowdisplaybreaks
\begin{align}
    &[g^{\mathcal{US}}_{\rm D}]^2 = [g_{\rm D}^\mathcal{S}]^2-\frac{[g_{\rm D}^\mathcal{S}]^4}{24 \pi \mu_{\mathrm{SU(2)_D}}} \,, \\
    &\lambda_{\rm D}^{\mathcal{US}} = \lambda_{\rm D}^{\mathcal{S}}-\frac{3\lambda^2_{S,1}}{32\pi\mu_{\mathrm{SU(2)_D}}}\,,
\end{align}
\endgroup

\subsubsection*{\underline{Scenario II}}
\begingroup
\allowdisplaybreaks
\begin{align}
    &[g^{\mathcal{US}}_{\rm D}]^2 = \mathrm{Eq.}~\eqref{eqn:usoft_sud} \,, \\
    &[g^{\mathcal{US}}_{\rm W}]^2 = \mathrm{Eq.}~\eqref{eqn:usoft_su2} \,, \\
    &[g^{\mathcal{US}}_{\rm Y}]^2 = \mathrm{Eq.}~\eqref{eqn:usoft_u1} \,, \\
    &[g^{\mathcal{US}}_S]^2 = \mathrm{Eq.}~\eqref{eqn:usoft_su3} \,, \\
    &\lambda_{\rm D}^{\mathcal{US}} = \lambda_{\rm D}^{\mathcal{S}} - \frac{1}{32 \pi}\Bigg(\frac{3 \lambda_{S,10}^2}{\mu_{\mathrm{SU(2)_L}}}+\frac{3 \lambda_{S,5}^2}{\mu_{\mathrm{SU(2)_D}}}+\frac{\lambda_{S,11}^2}{\mu_{\mathrm{U(1)_Y}}}\Bigg)  \,, \\
    &\lambda_{\rm H}^{\mathcal{US}} = \mathrm{Eq.}~\eqref{eqn:usoft_lH} \,, \\
    &\lambda_{\rm HD}^{\mathcal{US}} = \lambda_{\rm HD}^{\mathcal{S}}-\frac{1}{16 \pi }\Bigg(\frac{3 \lambda_{S,8}\lambda_{S,10}}{\mu_{\mathrm{SU(2)_L}}}+\frac{3 \lambda_{S,5}\lambda_{S,9}}{\mu_{\mathrm{SU(2)_D}}} + \frac{\lambda_{S,7} \lambda_{S,11}}{\mu_{\mathrm{U(1)_Y}}}\Bigg) \,. 
\end{align}
\endgroup

\subsubsection*{\underline{Scenario III}}
\begingroup
\allowdisplaybreaks
\begin{align}
    &[g^{\mathcal{US}}_{\rm D}]^2 = [g^{\mathcal{S}}_{\rm D} ]^2-\frac{[g^{\mathcal{S}}_{\rm D} ]^4}{24 \pi \mu_{\mathrm{SU(2)_D}}} \,, \label{eqn:usoft_sud} \\
    &[g^{\mathcal{US}}_{\rm W}]^2 = [g^{\mathcal{S}}_W]^2-\frac{[g^{\mathcal{S}}_W]^4}{24 \pi \mu_{\mathrm{SU(2)_L}}} \label{eqn:usoft_su2} \,, \\
    &[g^{\mathcal{US}}_{\rm Y}]^2 = [g^{\mathcal{S}}_Y]^2 \label{eqn:usoft_u1} \,, \\
    &[g^{\mathcal{US}}_S]^2 = [g^{\mathcal{S}}_S]^2-\frac{[g^{\mathcal{S}}_S]^4}{16 \pi \mu_{\mathrm{SU(3)_C}}} \label{eqn:usoft_su3} \,, \\
    &\lambda_{\rm D}^{\mathcal{US}} = \lambda_{\rm D}^{\mathcal{S}} - \frac{1}{32 \pi}\Bigg(\frac{3 \lambda_{S,10}^2}{\mu_{\mathrm{SU(2)_L}}}+\frac{4 \lambda_{S,3}^2}{\mu_{\mathrm{SU(2)_D}}+\mu_{\mathrm{U(1)_Y}}}+\frac{3 \lambda_{S,5}^2}{\mu_{\mathrm{SU(2)_D}}}+\frac{3 \lambda_{S,1}^2+5 \lambda_{S,2}^2}{\mu_{\mathrm{SU(3)_C}}}+\frac{\lambda_{S,11}^2}{\mu_{\mathrm{U(1)_Y}}}\Bigg) \,, \\
    &\lambda_{\rm H}^{\mathcal{US}} = \lambda_{\rm H}^{\mathcal{S}} - \frac{1}{32 \pi } \Bigg(\frac{4 \lambda_{S,6}^2}{\mu_{\mathrm{SU(2)_L}}+\mu_{\mathrm{U(1)_Y}}}+\frac{3 \lambda_{S,8}^2}{\mu_{\mathrm{SU(2)_L}}}+\frac{3 \lambda_{S,9}^2}{\mu_{\mathrm{SU(2)_D}}}+\frac{8 \lambda_{S,4}^2}{\mu_{\mathrm{SU(3)_C}}}+\frac{\lambda_{S,7}^2}{\mu_{\mathrm{U(1)_Y}}}\Bigg)\,, \label{eqn:usoft_lH} \\
    &\lambda_{\rm HD}^{\mathcal{US}} = \lambda_{\rm HD}^{\mathcal{S}}-\frac{1}{16 \pi }\Bigg(\frac{3 \lambda_{S,8}\lambda_{S,10}}{\mu_{\mathrm{SU(2)_L}}}+\frac{3 \lambda_{S,5}\lambda_{S,9}}{\mu_{\mathrm{SU(2)_D}}}+\frac{(3 \lambda_{S,1}+5 \lambda_{S,2}) \lambda_{S,4}}{\mu_{\mathrm{SU(3)_C}}}+\frac{\lambda_{S,7} \lambda_{S,11}}{\mu_{\mathrm{U(1)_Y}}}\Bigg)\,.
\end{align}
\endgroup

\subsection{Scalar mass matching between 3D-ultra-soft and 3D soft theories}\label{app:soft_matching_scalar}
\subsubsection*{\underline{Scenario I}}
\begingroup
\allowdisplaybreaks
\begin{align}
    & [\mu^{\mathcal{US}}_{\rm D}]^2_{\mathrm{LO}} = [\mu_{\rm D}^\mathcal{S}]^2 - \frac{3 \mu_{\mathrm{SU(2)_D}} \lambda_{S,1}}{8\pi} \,,\\
    & [\mu^{\mathcal{US}}_{\rm D}]^2_{\mathrm{NLO}} =  -\frac{3}{128 \pi ^2} \Bigg[ \lambda_{S,1} \left(-4 [g_{\rm D}^\mathcal{S}]^2 +2 \lambda_{S,1} - 5\lambda_{V,1}\right) + \\ \nonumber
    & \hspace*{5em} 2 \left([g_{\rm D}^\mathcal{S}]^4-8 [g_{\rm D}^\mathcal{S}]^2 \lambda_{S,1}+2 \lambda_{S,1}^2\right) ~\mathrm{ln} \left(\frac{\mu_{3\mathrm{d}}}{2 \mu_{\mathrm{SU(2)_D}}}\right)\Bigg]\,,
\end{align}
\endgroup

\subsubsection*{\underline{Scenario II}}

\begingroup
\allowdisplaybreaks
\begin{align}
    & [\mu^{\mathcal{US}}_{\rm D}]^2_{\mathrm{LO}} = [\mu^{\mathcal{S}}_{\rm D}]^2- \frac{3 \mu_{\mathrm{SU(2)_L}} \lambda_{S,10} + 3 \mu_{\mathrm{SU(2)_D}} \lambda_{S,5} + 5 \lambda_{S,2}) + \mu_{\mathrm{U(1)_Y}}\lambda_{S,11} }{8 \pi} \,,\\
    & [\mu^{\mathcal{US}}_{\rm H}]^2_{\mathrm{LO}} = \mathrm{Eq.}~\eqref{eqn:usoft_muH} \,, \\\nonumber
    & [\mu^{\mathcal{US}}_{\rm D}]^2_{\mathrm{NLO}} = \frac{1}{128 \pi ^2} \Big[3 \lambda_{S,5} \left(4 [g^{\mathcal{S}}_{\rm D} ]^2-2 \lambda_{S,5}+5\lambda_{V,1}\right)-6 \Big([g^{\mathcal{S}}_{\rm D} ]^4-8 [g^{\mathcal{S}}_{\rm D} ]^2 \lambda_{S,5} + \\\nonumber
        &\hspace*{5.5em} 2 \lambda_{S,5}^2\Big) \mathrm{ln} \left(\frac{\mu_{3\mathrm{D}}}{2 \mu_{\mathrm{SU(2)_D}}}\right)+3 \lambda_{S,10} \left(4 [g^{\mathcal{S}}_W]^2-2 \lambda_{S,10}+\frac{8 \lambda_{V,4} \mu_{\mathrm{SU(3)_C}}}{\mu_{\mathrm{SU(2)_L}}}+5 \lambda_{V,5}\right) + \\\nonumber
        &\hspace*{5.5em} 12 \lambda_{S,10} \left(4 [g^{\mathcal{S}}_W]^2-\lambda_{S,10}\right) \mathrm{ln} \left(\frac{\mu_{3\mathrm{D}}}{2 \mu_{\mathrm{SU(2)_L}}}\right)+\frac{3 \lambda_{V,9} (\lambda_{S,11} \mu_{\mathrm{SU(2)_L}}+\lambda_{S,10} \mu_{\mathrm{U(1)_Y}})}{\mu_{\mathrm{SU(2)_L}} \mu_{\mathrm{U(1)_Y}}} + \\
       &\hspace*{5.5em} \frac{8 \lambda_{S,11} \lambda_{V,8} \mu_{\mathrm{SU(3)_C}}}{\mu_{\mathrm{U(1)_Y}}}+\lambda_{S,11} \lambda_{V,10}-4 \lambda_{S,11}^2 \mathrm{ln} \left(\frac{\mu_{3\mathrm{D}}}{2 \mu_{\mathrm{U(1)_Y}}}\right)-2 \lambda_{S,11}^2\Big]\,,\\ \nonumber
   & [\mu^{\mathcal{US}}_{\rm H}]^2_{\mathrm{NLO}} = \frac{1}{128 \pi ^2} \Big[12 \lambda_{S,9} \left(4 [g^{\mathcal{S}}_{\rm D} ]^2-\lambda_{S,9}\right) \mathrm{ln} \left(\frac{\mu_{3\mathrm{D}}}{2 \mu_{\mathrm{SU(2)_D}}}\right)-2 \Big(3 \lambda_{S,9} \left(\lambda_{S,9}-2 [g^{\mathcal{S}}_{\rm D} ]^2\right) + \\\nonumber
        &\hspace*{5.5em} 6 \lambda_{S,6}^2+\lambda_{S,7}^2\Big)+8 \lambda_{S,4} \Big[\frac{3 \lambda_{V,4} \mu_{\mathrm{SU(2)_L}}}        {\mu_{\mathrm{SU(3)_C}}}-2 (2 \left(\lambda_{S,4}-6 [g^{\mathcal{S}}_S]^2\right) \mathrm{ln} \left(\frac{\mu_{3\mathrm{D}}}{2 \mu_{\mathrm{SU(3)_C}}}\right)+\lambda_{S,4} - \\\nonumber
        &\hspace*{5.5em} 5 \lambda_{V,3})\Big]+48 [g^{\mathcal{S}}_S]^2 \lambda_{S,4}+12 [g^{\mathcal{S}}_W]^2 \lambda_{S,8}-6 \Big([g^{\mathcal{S}}_W]^4-8 [g^{\mathcal{S}}_W]^2 \lambda_{S,8} + \\\nonumber
        &\hspace*{5.5em} 2 \lambda_{S,8}^2\Big) \mathrm{ln} \left(\frac{\mu_{3\mathrm{D}}}{2 \mu_{\mathrm{SU(2)_L}}}\right)+\frac{24 \lambda_{S,8} \lambda_{V,4} \mu_{\mathrm{SU(3)_C}}}{\mu_{\mathrm{SU(2)_L}}}+\frac{3 \lambda_{S,7} \lambda_{V,9} \mu_{\mathrm{SU(2)_L}}}{\mu_{\mathrm{U(1)_Y}}} + \\\nonumber
        &\hspace*{5.5em} \frac{3 \lambda_{S,8} \lambda_{V,9} \mu_{\mathrm{U(1)_Y}}}{\mu_{\mathrm{SU(2)_L}}}+\frac{8 \lambda_{S,4} \lambda_{V,8} \mu_{\mathrm{U(1)_Y}}}{\mu_{\mathrm{SU(3)_C}}}+\frac{8 \lambda_{S,7} \lambda_{V,8} \mu_{\mathrm{SU(3)_C}}}{\mu_{\mathrm{U(1)_Y}}}+15 \lambda_{S,9}\lambda_{V,1} + \\\nonumber
        &\hspace*{5.5em} 15 \lambda_{S,8} \lambda_{V,5}+\lambda_{S,7} \lambda_{V,10}-24 \lambda_{S,6}^2 \mathrm{ln} \left(\frac{\mu_{3\mathrm{D}}}{\mu_{\mathrm{SU(2)_L}}+\mu_{\mathrm{U(1)_Y}}}\right) - \\\nonumber
        &\hspace*{5.5em} 4 \lambda_{S,7}^2 \mathrm{ln} \left(\frac{\mu_{3\mathrm{D}}}{2 \mu_{\mathrm{U(1)_Y}}}\right)-6 \lambda_{S,8}^2\Big]\,.
\end{align}
\endgroup

\subsubsection*{\underline{Scenario III}}
\begingroup
\allowdisplaybreaks
\begin{align}
    & [\mu^{\mathcal{US}}_{\rm D}]^2_{\mathrm{LO}} = [\mu^{\mathcal{S}}_{\rm D}]^2- \frac{3 \mu_{\mathrm{SU(2)_L}} \lambda_{S,10} + 3 \mu_{\mathrm{SU(2)_D}} \lambda_{S,5} + \mu_{\mathrm{SU(3)_C}}(3 \lambda_{S,1}+5 \lambda_{S,2}) + \mu_{\mathrm{U(1)_Y}}\lambda_{S,11} }{8 \pi}\,,\\
    & [\mu^{\mathcal{US}}_{\rm H}]^2_{\mathrm{LO}} = [\mu^{\mathcal{S}}_{\rm H}]^2-\frac{3 \mu_{\mathrm{SU(2)_L}} \lambda_{S,8} + 3 \mu_{\mathrm{SU(2)_D}} \lambda_{S,9} + 8 \mu_{\mathrm{SU(3)_C}} \lambda_{S,4} + \mu_{\mathrm{U(1)_Y}}\lambda_{S,7}}{8 \pi}\,, \label{eqn:usoft_muH} \\\nonumber
    & [\mu^{\mathcal{US}}_{\rm D}]^2_{\mathrm{NLO}} =  \frac{1}{128 \pi ^2} \Bigg[-6 [g_{\rm D}^\mathcal{S}]^4 ~\mathrm{ln} \left(\frac{\mu_{\mathrm{3D}}}{2 \mu_{\mathrm{SU(2)_D}}}\right)+12 [g_{\rm D}^\mathcal{S}]^2 \lambda_{S,5} \left(4 ~\mathrm{ln}\left(\frac{\mu_{\mathrm{3D}}}{2 \mu_{\mathrm{SU(2)_D}}}\right)+1\right) + \\\nonumber
        & \hspace*{5.5em}18 [g_S^\mathcal{S}]^2 \lambda_{S,1} \left(4 ~\mathrm{ln} \left(\frac{\mu_{\mathrm{3D}}}{2 \mu_{\mathrm{SU(3)_C}}}\right)+1\right)+30 [g_S^\mathcal{S}]^2 \lambda_{S,2} \left(4 ~\mathrm{ln} \left(\frac{\mu_{\mathrm{3D}}}{2 \mu_{\mathrm{SU(3)_C}}}\right)+1\right) + \\\nonumber
        & \hspace*{5.5em} 12 [\g{W}^\mathcal{S}]^2 \lambda_{S,10} \left(4 ~\mathrm{ln} \left(\frac{\mu_{\mathrm{3D}}}{2 \mu_{\mathrm{SU(2)_L}}}\right)+1\right) + \\\nonumber
        & \hspace*{5.5em} \frac{3 \lambda_{S,1} \left(3 \lambda_{V,4} \mu_{\mathrm{SU(2)_L}}+3 \lambda_{V,2} \mu_{\mathrm{SU(2)_D}}+10 \lambda_{V,3} \mu_{\mathrm{SU(3)_C}}+\lambda_{V,8} \mu_{\mathrm{U(1)_Y}}\right)}{\mu_{\mathrm{SU(3)_C}}} + \\\nonumber
        & \hspace*{5.5em} \frac{5 \lambda_{S,2} \left(3 \lambda_{V,4} \mu_{\mathrm{SU(2)_L}}+3 \lambda_{V,2} \mu_{\mathrm{SU(2)_D}}+10 \lambda_{V,3} \mu_{\mathrm{SU(3)_C}}+\lambda_{V,8} \mu_{\mathrm{U(1)_Y}}\right)}{\mu_{\mathrm{SU(3)_C}}} + \\\nonumber
        & \hspace*{5.5em} \lambda_{S,11} \left(\frac{3 \lambda_{V,9} \mu_{\mathrm{SU(2)_L}}+3 \lambda_{V,7} \mu_{\mathrm{SU(2)_D}}+8 \lambda_{V,8} \mu_{\mathrm{SU(3)_C}}}{\mu_{\mathrm{U(1)_Y}}} + \lambda_{V,10}\right) + \\\nonumber
        & \hspace*{5.5em} \frac{3 \lambda_{S,10} \left(5 \lambda_{V,5} \mu_{\mathrm{SU(2)_L}}+8 \lambda_{V,4} \mu_{\mathrm{SU(3)_C}}+\lambda_{V,9} \mu_{\mathrm{U(1)_Y}}\right)}{\mu_{\mathrm{SU(2)_L}}} + \\\nonumber
        & \hspace*{5.5em} \frac{3 \lambda_{S,5} \left(5 \lambda_{V,1} \mu_{\mathrm{SU(2)_D}}+8 \lambda_{V,2} \mu_{\mathrm{SU(3)_C}}+\lambda_{V,7} \mu_{\mathrm{U(1)_Y}}\right)}{\mu_{\mathrm{SU(2)_D}}}- \\\nonumber
        & \hspace*{5.5em} 6 \lambda_{S,10}^2 \left(2 ~\mathrm{ln} \left(\frac{\mu_{\mathrm{3D}}}{2 \mu_{\mathrm{SU(2)_L}}}\right)+1\right)-12 \lambda_{S,3}^2 \left(2 ~\mathrm{ln} \left(\frac{\mu_{\mathrm{3D}}}{\mu_{\mathrm{SU(2)_D}}+\mu_{\mathrm{U(1)_Y}}}\right)+1\right) - \\\nonumber 
        & \hspace*{5.5em} 6 \lambda_{S,5}^2 \left(2 ~\mathrm{ln} \left(\frac{\mu_{\mathrm{3D}}}{2 \mu_{\mathrm{SU(2)_D}}}\right)+1\right)-6 \lambda_{S,1}^2 \left(2 ~\mathrm{ln} \left(\frac{\mu_{\mathrm{3D}}}{2 \mu_{\mathrm{SU(3)_C}}}\right)+1\right) - \\ 
        & \hspace*{5.5em} 10 \lambda_{S,2}^2 \left(2 ~\mathrm{ln} \left(\frac{\mu_{\mathrm{3D}}}{2 \mu_{\mathrm{SU(3)_C}}}\right)+1\right)-2 \lambda_{S,11}^2 \left(2 ~\mathrm{ln} \left(\frac{\mu_{\mathrm{3D}}}{2 \mu_{\mathrm{U(1)_Y}}}\right)+1\right) \Bigg] \,,\\\nonumber 
    & [\mu^{\mathcal{US}}_{\rm H}]^2_{\mathrm{NLO}} = \frac{1}{128 \pi ^2} \Bigg[ 12 [g_{\rm D}^\mathcal{S}]^2 \lambda_{S,9} \left(4 ~\mathrm{ln} \left(\frac{\mu_{\mathrm{3D}}}{2 \mu_{\mathrm{SU(2)_D}}}\right)+1\right)+48 [g_S^\mathcal{S}]^2 \lambda_{S,4} \left(4 ~\mathrm{ln} \left(\frac{\mu_{\mathrm{3D}}}{2 \mu_{\mathrm{SU(3)_C}}}\right)+1\right) - \\\nonumber
        & \hspace*{5.5em}6 [\g{W}^\mathcal{S}]^4 ~\mathrm{ln} \left(\frac{\mu_{\mathrm{3D}}}{2 \mu_{\mathrm{SU(2)_L}}}\right)+12 [\g{W}^\mathcal{S}]^2 \lambda_{S,8} \left(4 ~\mathrm{ln} \left(\frac{\mu_{\mathrm{3D}}}{2 \mu_{\mathrm{SU(2)_L}}}\right)+1\right) + \\\nonumber
        & \hspace*{5.5em} \frac{8 \lambda_{S,4} \left(3 \lambda_{V,4} \mu_{\mathrm{SU(2)_L}}+3 \lambda_{V,2} \mu_{\mathrm{SU(2)_D}}+10 \lambda_{V,3} \mu_{\mathrm{SU(3)_C}}+\lambda_{V,8} \mu_{\mathrm{U(1)_Y}}\right)}{\mu_{\mathrm{SU(3)_C}}} + \\\nonumber
        & \hspace*{5.5em} \lambda_{S,7} \left(\frac{3 \lambda_{V,9} \mu_{\mathrm{SU(2)_L}}+3 \lambda_{V,7} \mu_{\mathrm{SU(2)_D}}+8 \lambda_{V,8} \mu_{\mathrm{SU(3)_C}}}{\mu_{\mathrm{U(1)_Y}}}+\lambda_{V,10}\right) + \\\nonumber
        & \hspace*{5.5em} \frac{3 \lambda_{S,8} \left(5 \lambda_{V,5} \mu_{\mathrm{SU(2)_L}}+8 \lambda_{V,4} \mu_{\mathrm{SU(3)_C}}+\lambda_{V,9} \mu_{\mathrm{U(1)_Y}}\right)}{\mu_{\mathrm{SU(2)_L}}} + \\\nonumber 
        & \hspace*{5.5em} \frac{3 \lambda_{S,9} \left(5 \lambda_{V,1} \mu_{\mathrm{SU(2)_D}}+8 \lambda_{V,2} \mu_{\mathrm{SU(3)_C}}+\lambda_{V,7} \mu_{\mathrm{U(1)_Y}}\right)}{\mu_{\mathrm{SU(2)_D}}} - \\\nonumber 
        & \hspace*{5.5em} 12 \lambda_{S,6}^2 \left(2 ~\mathrm{ln} \left(\frac{\mu_{\mathrm{3D}}}{\mu_{\mathrm{SU(2)_L}}+\mu_{\mathrm{U(1)_Y}}}\right)+1\right)-6 \lambda_{S,8}^2 \left(2 ~\mathrm{ln} \left(\frac{\mu_{\mathrm{3D}}}{2 \mu_{\mathrm{SU(2)_L}}}\right)+1\right) - \\\nonumber 
        & \hspace*{5.5em} 6 \lambda_{S,9}^2 \left(2 ~\mathrm{ln} \left(\frac{\mu_{\mathrm{3D}}}{2 \mu_{\mathrm{SU(2)_D}}}\right)+1\right)-16 \lambda_{S,4}^2 \left(2 ~\mathrm{ln} \left(\frac{\mu_{\mathrm{3D}}}{2 \mu_{\mathrm{SU(3)_C}}}\right)+1\right) - \\ 
        & \hspace*{5.5em} 2 \lambda_{S,7}^2 \left(2 ~\mathrm{ln} \left(\frac{\mu_{\mathrm{3D}}}{2 \mu_{\mathrm{U(1)_Y}}}\right)+1\right) \Bigg]\,.
\end{align}
\endgroup

\subsection{Vector and scalar field-dependent masses at ultra soft scale}\label{app:mass_matrices}

Here we write down the 3D field dependent masses in the ultrasoft limit in terms of the 3D field $\varphi_{\rm D}$ and the ultrasoft parameters, which appear directly in the NLO part of the effective potential, see \cref{eqn:potential}. They are calculated from LO effective potential. We start with scenario I. Here, for the scalar masses we have
\begin{align}
    & \mathcal{M}^2_{\varphi_1} = [\mu_{\rm D}^\mathcal{US}]^2 + \lambda_{\rm D}^{\mathcal{US}} \varphi_{\rm D}^2 \,, \\
    & \mathcal{M}^2_{\varphi_2} = [\mu_{\rm D}^\mathcal{US}]^2 + \lambda_{\rm D}^{\mathcal{US}} \varphi_{\rm D}^2 \,, \\
    & \mathcal{M}^2_{\varphi_3} = [\mu_{\rm D}^\mathcal{US}]^2 + 3\lambda_{\rm D}^{\mathcal{US}} \varphi_{\rm D}^2 \,, \\
    & \mathcal{M}^2_{\varphi_4} = [\mu_{\rm D}^\mathcal{US}]^2 + \lambda_{\rm D}^{\mathcal{US}} \varphi_{\rm D}^2 \,,
\end{align}
while for the vector bosons we have 
\begin{align}
    & \mathcal{M}^2_{\mathcal{V}_1} = \frac{1}{4} [g_{\rm D}^\mathcal{US}]^2 \varphi_{\rm D}^2 \,, \label{eqn:mass_vec_1} \\
    & \mathcal{M}^2_{\mathcal{V}_2} = \frac{1}{4} [g_{\rm D}^\mathcal{US}]^2 \varphi_{\rm D}^2 \,, \label{eqn:mass_vec_2}\\
    & \mathcal{M}^2_{\mathcal{V}_3} = \frac{1}{4} [g_{\rm D}^\mathcal{US}]^2 \varphi_{\rm D}^2 \,. \label{eqn:mass_vec_3}
\end{align}
For scenario II/III, the scalar masses are given by
\begin{align}
    & \mathcal{M}^2_{\varphi_1} = [\mu_{\rm D}^\mathcal{US}]^2 + \frac{1}{2}\lambda_{\rm HD}^{\mathcal{US}} \varphi_{\rm H}^2 + \lambda_{\rm D}^{\mathcal{US}} \varphi_{\rm D}^2 \,, \\
    & \mathcal{M}^2_{\varphi_2} = [\mu_{\rm D}^\mathcal{US}]^2 + \frac{1}{2}\lambda_{\rm HD}^{\mathcal{US}} \varphi_{\rm H}^2 + \lambda_{\rm D}^{\mathcal{US}} \varphi_{\rm D}^2 \,, \\
    & \mathcal{M}^2_{\varphi_3} = [\mu_{\rm D}^\mathcal{US}]^2 + \frac{1}{2}\lambda_{\rm HD}^{\mathcal{US}} \varphi_{\rm H}^2 + \lambda_{\rm D}^{\mathcal{US}} \varphi_{\rm D}^2 \,, \\
    & \mathcal{M}^2_{\varphi_4} = [\mu_{\rm H}^\mathcal{US}]^2 + \lambda_{\rm H}^{\mathcal{US}} \varphi_{\rm H}^2 + \frac{1}{2}\lambda_{\rm HD}^{\mathcal{US}} \varphi_{\rm D}^2 \,, \\
    & \mathcal{M}^2_{\varphi_5} = [\mu_{\rm H}^\mathcal{US}]^2 + \lambda_{\rm H}^{\mathcal{US}} \varphi_{\rm H}^2 + \frac{1}{2}\lambda_{\rm HD}^{\mathcal{US}} \varphi_{\rm D}^2 \,, \\
    & \mathcal{M}^2_{\varphi_6} = [\mu_{\rm H}^\mathcal{US}]^2 + \lambda_{\rm H}^{\mathcal{US}} \varphi_{\rm H}^2 + \frac{1}{2}\lambda_{\rm HD}^{\mathcal{US}} \varphi_{\rm D}^2 \,, \\
    & \mathcal{M}^2_{\varphi_7} = \mathrm{eig}(M^2_{\varphi_{\rm H} \varphi_{\rm D}}[1]) \,, \\
    & \mathcal{M}^2_{\varphi_8} = \mathrm{eig}(M^2_{\varphi_{\rm H} \varphi_{\rm D}}[2]) \,,
\end{align}
where $\mathcal{M}^2_{\varphi_7}$ and $\mathcal{M}^2_{\varphi_8}$ are eigenvalues of the mass matrix
\begin{equation}
   M^2_{\varphi_{\rm H} \varphi_{\rm D}} = 
   \begin{bmatrix}
       [\mu_{\rm H}^\mathcal{US}]^2 + 3\lambda_{\rm H}^{\mathcal{US}}\varphi_{\rm H}^2 + \frac{1}{2}\lambda_{\rm HD}^{\mathcal{US}}\varphi_{\rm D}^2 & \lambda_{\rm HD}^{\mathcal{US}}\varphi_{\rm H}\varphi_{\rm D} \\[0.5em]
       \lambda_{\rm HD}^{\mathcal{US}}\varphi_{\rm H}\varphi_{\rm D} & [\mu_{\rm D}^\mathcal{US}]^2 + \frac{1}{2}\lambda_{\rm HD}^{\mathcal{US}}\varphi_{\rm H}^2 + 3\lambda_{\rm D}^{\mathcal{US}}\varphi_{\rm D}^2 
    \end{bmatrix}\,.
\end{equation}
The vector masses in turn are given by
\begin{align}
    & \mathcal{M}^2_{\mathcal{V}_1} = \frac{1}{4} [g^{\mathcal{US}}_{\rm D}]^2\varphi_{\rm D}^2 \,, \\
    & \mathcal{M}^2_{\mathcal{V}_2} = \frac{1}{4} [g^{\mathcal{US}}_{\rm D}]^2\varphi_{\rm D}^2 \,, \\
    & \mathcal{M}^2_{\mathcal{V}_3} = \frac{1}{4} [g^{\mathcal{US}}_{\rm D}]^2\varphi_{\rm D}^2 \,, \\
    & \mathcal{M}^2_{\mathcal{V}_4} = \frac{1}{4} [g^{\mathcal{US}}_{\rm W}]^2\varphi_{\rm H}^2 \,, \\
    & \mathcal{M}^2_{\mathcal{V}_5} = \frac{1}{4} [g^{\mathcal{US}}_{\rm W}]^2\varphi_{\rm H}^2 \,, \\
    & \mathcal{M}^2_{\mathcal{V}_6} = \mathrm{eig}(M^2_{\mathcal{V}_W \mathcal{V}_Y}[1]) \,, \\
    & \mathcal{M}^2_{\mathcal{V}_7} = \mathrm{eig}(M^2_{\mathcal{V}_W \mathcal{V}_Y}[2]) \,,
\end{align}
where $\mathcal{M}^2_{\mathcal{V}_6}$ and $\mathcal{M}^2_{\mathcal{V}_7}$ are eigenvalues of the mass matrix
\begin{equation}
   M^2_{\mathcal{V}_W \mathcal{V}_Y} = 
   \begin{bmatrix}
        \frac{1}{4} [g^{\mathcal{US}}_{\rm W}]^2 \varphi_{\rm H}^2 & -\frac{1}{4} g^{\mathcal{US}}_{\rm W} g^{\mathcal{US}}_{\rm Y} \varphi_{\rm H}^2 \\[0.5em]
        -\frac{1}{4} g^{\mathcal{US}}_{\rm W} g^{\mathcal{US}}_{\rm Y} \varphi_{\rm H}^2 & \frac{1}{4} [g^{\mathcal{US}}_{\rm Y}]^2 \varphi_{\rm H}^2
    \end{bmatrix}\,.
\end{equation}

\subsection{Pressure in the ultrasoft limit}\label{app:pressure_US}
Here, the 4D pressure is related to the 3D pressure by $P = T (\mathcal{P}^{\mathcal{US}}_{\mathrm{LO}} + \mathcal{P}^{\mathcal{US}}_{\mathrm{NLO}})  $.
\subsubsection*{\underline{Scenario I}}
\begingroup
\allowdisplaybreaks
\begin{align}    &\mathcal{P}^{\mathcal{US}}_{\mathrm{LO}} = \frac{\Big[\mu^2_{\mathrm{SU(2)_D}}\Big]^{3/2}}{4\pi} \,, \\
    &\mathcal{P}^{\mathcal{US}}_{\mathrm{NLO}} = \frac{3}{2048 \pi ^2} \Bigg[\mu^2_{\mathrm{SU(2)_D}} \left(-64 [g_{\rm D}^\mathcal{US}]^2 \left(4 ~\mathrm{ln} \left(\frac{\mu_{3\mathrm{d}}}{2 \mu_{\mathrm{SU(2)_D}}}\right)+3\right)-5 \pi \lambda_{V,1}\right) \Bigg]\,,
\end{align}
\endgroup

\subsubsection*{\underline{Scenario II}}
\begingroup
\allowdisplaybreaks
\begin{align}
& \mathcal{P}^{\mathcal{US}}_{\mathrm{LO}} = \mathrm{Eq.}~\eqref{eqn:PLO}\,, \\
& \mathcal{P}^{\mathcal{US}}_{\mathrm{NLO}} = \mathrm{Eq.}~\eqref{eqn:PNLO}~+ \frac {3\pi \mu_{\mathrm{SU(2)_D}}}{1024\pi^2} \left(8 \lambda_{V,2} \mu_{\mathrm{SU(3)_C}}+\lambda_{V,7} \mu_{\mathrm{U(1)_Y}}\right) \,, 
\end{align}
\endgroup

\subsubsection*{\underline{Scenario III}}
\begingroup
\allowdisplaybreaks
\begin{align}
    & \mathcal{P}^{\mathcal{US}}_{\mathrm{LO}}  = \frac{1}{12\pi} \Bigg( 3 \Big[\mu^2_{\mathrm{SU(2)_L}}\Big]^{3/2} + 3 \Big[\mu^2_{\mathrm{SU(2)_D}}\Big]^{3/2}+ 8\Big[\mu^2_{\mathrm{SU(3)_C}}\Big]^{3/2} + \Big[\mu^2_{\mathrm{U(1)_Y}}\Big]^{3/2} \Bigg) \label{eqn:PLO}\\\nonumber
    & \mathcal{P}^{\mathcal{US}}_{\mathrm{NLO}} = -\frac{1}{2048 \pi ^2} \Bigg[192 [g^{\mathcal{US}}_{\rm D}]^2 \mu^2_{\mathrm{SU(2)_D}} \left(4 ~\mathrm{ln}\left(\frac{\mu_{\mathrm{3D}}}{2 \mu_{\mathrm{SU(2)_D}}}\right)+3\right) + \\\nonumber
        &\hspace*{5.0em} 768 [g^{\mathcal{US}}_S]^2 \mu^2_{\mathrm{SU(3)_C}} \left(4 ~\mathrm{ln}\left(\frac{\mu_{\mathrm{3D}}}{2 \mu_{\mathrm{SU(3)_C}}}\right)+3\right) + \\\nonumber
        &\hspace*{5.0em} 192 [g^{\mathcal{US}}_{\rm W}]^2 \mu^2_{\mathrm{SU(2)_L}} \left(4 ~\mathrm{ln}\left(\frac{\mu_{\mathrm{3D}}}{2 \mu_{\mathrm{SU(2)_L}}}\right)+3\right) + \\\nonumber
        &\hspace*{5.0em} \pi  \Big(16 \mu_{\mathrm{SU(3)_C}} \left(3 \lambda_{V,4} \mu_{\mathrm{SU(2)_L}}+\lambda_{V,8} \mu_{\mathrm{U(1)_Y}}\right)+6 \lambda_{V,9} \mu_{\mathrm{SU(2)_L}} \mu_{\mathrm{U(1)_Y}} + \\\nonumber
        &\hspace*{5.0em} 15 \lambda_{V,5} \mu^2_{\mathrm{SU(2)_L}}+6 \mu_{\mathrm{SU(2)_D}} \left(8 \lambda_{V,2} \mu_{\mathrm{SU(3)_C}}+\lambda_{V,7} \mu_{\mathrm{U(1)_Y}}\right)+15 \lambda_{V,1} \mu^2_{\mathrm{SU(2)_D}} + \\
        &\hspace*{5.0em} 80 \lambda_{V,3} \mu^2_{\mathrm{SU(3)_C}}+\lambda_{V,10} \mu^2_{\mathrm{U(1)_Y}}\Big) \Bigg] \,. \label{eqn:PNLO}
\end{align}
\endgroup


\section{Statistical Treatment of Signal Region Recasting}
\label{app:pyhf}

To quantify the exclusion reach for each parameter point in our model, we compute the expected 95\% CL upper limit on the signal yield, denoted as $s_{95}^{\mathrm{exp}}$. This is done using the Asimov dataset formalism and the CL$_s$ hypothesis testing approach, based on the profile likelihood ratio as described in Cowan \textit{et al.}~\cite{Cowan:2010js}. The statistical evaluation is performed using the \texttt{pyhf} package~\cite{pyhf-soft}, which provides a backend-independent, \texttt{JSON}-serialised implementation of likelihood construction and inference.

In our implementation, each signal region extracted from the \texttt{CheckMATE} recasting framework is modelled by a simplified likelihood consisting of:
\begin{itemize}
  \item a signal template, normalised to unity;
  \item a background expectation $b$, with associated uncertainty $\delta_b$, incorporated via a \texttt{normsys} nuisance parameter representing log-normal scaling;
  \item an observation given by the Asimov dataset, i.e. the expected number of events under the background-only hypothesis.
\end{itemize}

The numerical values for $s$, $\delta_s$, $b$, and $\delta_b$ are taken directly from the \texttt{CheckMATE} output files for each relevant analysis and signal region, based on detector-level event simulation. These quantities form the inputs to the statistical model built in \texttt{pyhf} for every parameter point in our scan.

The test statistic is evaluated using the asymptotic approximation, yielding the expected upper limit on the signal strength, $\mu_{95}^{\mathrm{exp}}$, from which the corresponding upper limit on the signal yield is derived as:
\begin{equation}
s_{95}^{\mathrm{exp}} = \mu_{95}^{\mathrm{exp}} \times s_{\mathrm{template}}.
\end{equation}

To define a conservative yet statistically meaningful exclusion criterion, we compute the ratio:
\begin{equation}
r_{\mathrm{exp}}^{\mathrm{cons}} = \frac{s - 1.64\,\delta_s}{s_{95}^{\mathrm{exp}}},
\end{equation}
where $s$ is the predicted signal yield at a given parameter point, and $\delta_s$ is the associated uncertainty. The factor 1.64 corresponds to a one-sided 95\% CL downward fluctuation under Gaussian statistics. A parameter point is deemed excluded at 95\% CL if $r_{\mathrm{exp}}^{\mathrm{cons}} > 1$.

This procedure is applied systematically to all signal regions relevant to our analysis, as implemented in \texttt{CheckMATE} and discussed in detail in the main text. In particular, it is used to:

\begin{itemize}
  \item reproduce and validate the individual signal region exclusions obtained from \texttt{CheckMATE};
  \item perform statistical combinations of mutually orthogonal signal regions from the same analysis (e.g. 0-lepton and 1-lepton channels in CMS SUS-19-005 or ATLAS 2101.01629);
  \item and combine the most sensitive signal regions across ATLAS and CMS to derive the most stringent overall exclusion limit.
\end{itemize}

This approach ensures consistency between detector-level simulation, recasting, and statistical interpretation, enabling a robust mapping between BSM signal predictions and current experimental constraints.

\bibliographystyle{JHEP}
\bibliography{Refs}

@article{Sakharov:1967dj,
    author = "Sakharov, A. D.",
    title = "{Violation of CP Invariance, C asymmetry, and baryon asymmetry of the universe}",
    doi = "10.1070/PU1991v034n05ABEH002497",
    journal = "Pisma Zh. Eksp. Teor. Fiz.",
    volume = "5",
    pages = "32--35",
    year = "1967"
}

@article{Harry:2006fi,
    author = "Harry, G. M. and Fritschel, P. and Shaddock, D. A. and Folkner, W. and Phinney, E. S.",
    title = "{Laser interferometry for the big bang observer}",
    doi = "10.1088/0264-9381/23/15/008",
    journal = "Class. Quant. Grav.",
    volume = "23",
    pages = "4887--4894",
    year = "2006",
    note = "[Erratum: Class.Quant.Grav. 23, 7361 (2006)]"
}

@article{Kawamura:2006up,
    author = "Kawamura, S. and others",
    editor = "Mio, N.",
    title = "{The Japanese space gravitational wave antenna DECIGO}",
    doi = "10.1088/0264-9381/23/8/S17",
    journal = "Class. Quant. Grav.",
    volume = "23",
    pages = "S125--S132",
    year = "2006"
}

@article{LISA:2017pwj,
    author = "Amaro-Seoane, Pau and others",
    collaboration = "LISA",
    title = "{Laser Interferometer Space Antenna}",
    eprint = "1702.00786",
    archivePrefix = "arXiv",
    primaryClass = "astro-ph.IM",
    month = "2",
    year = "2017"
}

@article{ATLAS:2020zms,
    author = "Aad, Georges and others",
    collaboration = "ATLAS",
    title = "{Search for heavy Higgs bosons decaying into two tau leptons with the ATLAS detector using $pp$ collisions at $\sqrt{s}=13$ TeV}",
    eprint = "2002.12223",
    archivePrefix = "arXiv",
    primaryClass = "hep-ex",
    reportNumber = "CERN-EP-2020-014",
    doi = "10.1103/PhysRevLett.125.051801",
    journal = "Phys. Rev. Lett.",
    volume = "125",
    number = "5",
    pages = "051801",
    year = "2020"
}

@article{ATLAS:2020jgy,
    author = "Aad, Georges and others",
    collaboration = "ATLAS",
    title = "{Search for the $HH \rightarrow b \bar{b} b \bar{b}$ process via vector-boson fusion production using proton-proton collisions at $\sqrt{s} = 13$ TeV with the ATLAS detector}",
    eprint = "2001.05178",
    archivePrefix = "arXiv",
    primaryClass = "hep-ex",
    reportNumber = "CERN-EP-2019-267",
    doi = "10.1007/JHEP07(2020)108",
    journal = "JHEP",
    volume = "07",
    pages = "108",
    year = "2020",
    note = "[Erratum: JHEP 01, 145 (2021), Erratum: JHEP 05, 207 (2021)]"
}

@article{ATLAS:2020tlo,
    author = "Aad, Georges and others",
    collaboration = "ATLAS",
    title = "{Search for heavy resonances decaying into a pair of Z bosons in the $\ell ^+\ell ^-\ell '^+\ell '^-$ and $\ell ^+\ell ^-\nu {{\bar{\nu }}}$ final states using 139 $\mathrm {fb}^{-1}$ of proton\textendash{}proton collisions at $\sqrt{s} = 13\,$TeV with the ATLAS detector}",
    eprint = "2009.14791",
    archivePrefix = "arXiv",
    primaryClass = "hep-ex",
    reportNumber = "CERN-EP-2020-153",
    doi = "10.1140/epjc/s10052-021-09013-y",
    journal = "Eur. Phys. J. C",
    volume = "81",
    number = "4",
    pages = "332",
    year = "2021"
}

@article{ATLAS:2021uiz,
    author = "Aad, Georges and others",
    collaboration = "ATLAS",
    title = "{Search for resonances decaying into photon pairs in 139 fb$^{-1}$ of $pp$ collisions at $\sqrt {s}$=13 TeV with the ATLAS detector}",
    eprint = "2102.13405",
    archivePrefix = "arXiv",
    primaryClass = "hep-ex",
    reportNumber = "CERN-EP-2020-248",
    doi = "10.1016/j.physletb.2021.136651",
    journal = "Phys. Lett. B",
    volume = "822",
    pages = "136651",
    year = "2021"
}

@article{Shaposhnikov:1986jp,
    author = "Shaposhnikov, M. E.",
    title = "{Possible Appearance of the Baryon Asymmetry of the Universe in an Electroweak Theory}",
    journal = "JETP Lett.",
    volume = "44",
    pages = "465--468",
    year = "1986"
}

@article{Farrar:1993hn,
    author = "Farrar, Glennys R. and Shaposhnikov, M. E.",
    title = "{Baryon asymmetry of the universe in the standard electroweak theory}",
    eprint = "hep-ph/9305275",
    archivePrefix = "arXiv",
    reportNumber = "CERN-TH-6732-93, CERN-TH-6734-93, RU-93-11",
    doi = "10.1103/PhysRevD.50.774",
    journal = "Phys. Rev. D",
    volume = "50",
    pages = "774",
    year = "1994"
}

@inproceedings{Elor:2022hpa,
    author = "Elor, Gilly and others",
    title = "{New Ideas in Baryogenesis: A Snowmass White Paper}",
    booktitle = "{Snowmass 2021}",
    eprint = "2203.05010",
    archivePrefix = "arXiv",
    primaryClass = "hep-ph",
    month = "3",
    year = "2022"
}

@article{Canetti_2012,
	doi = {10.1088/1367-2630/14/9/095012},
	url = {https://doi.org/10.1088%2F1367-2630%2F14%2F9%2F095012},
	year = 2012,
	month = {sep},
	publisher = {{IOP} Publishing},
	volume = {14},
	number = {9},
	pages = {095012},
	author = {Laurent Canetti and Marco Drewes and Mikhail Shaposhnikov},
	title = {Matter and antimatter in the universe},
	journal = {New Journal of Physics}
}

@article{Belyaev:2022shr,
    author = "Belyaev, Alexander and Deandrea, Aldo and Moretti, Stefano and Panizzi, Luca and Ross, Douglas A. and Thongyoi, Nakorn",
    title = "{Fermionic portal to vector dark matter from a new gauge sector}",
    eprint = "2204.03510",
    archivePrefix = "arXiv",
    primaryClass = "hep-ph",
    doi = "10.1103/PhysRevD.108.095001",
    journal = "Phys. Rev. D",
    volume = "108",
    number = "9",
    pages = "095001",
    year = "2023"
}

@article{Belyaev:2022zjx,
    author = "Belyaev, Alexander and Deandrea, Aldo and Moretti, Stefano and Panizzi, Luca and Ross, Douglas A. and Thongyoi, Nakorn",
    title = "{A fermionic portal to a non-abelian dark sector}",
    eprint = "2203.04681",
    archivePrefix = "arXiv",
    primaryClass = "hep-ph",
    doi = "10.3389/fphy.2024.1339886",
    journal = "Front. in Phys.",
    volume = "12",
    pages = "1339886",
    year = "2024"
}

@article{Kosowsky:1992rz,
    author = "Kosowsky, Arthur and Turner, Michael S. and Watkins, Richard",
    title = "{Gravitational waves from first order cosmological phase transitions}",
    reportNumber = "FERMILAB-PUB-91-333-A-REV, FERMILAB-PUB-91-333-A",
    doi = "10.1103/PhysRevLett.69.2026",
    journal = "Phys. Rev. Lett.",
    volume = "69",
    pages = "2026--2029",
    year = "1992"
}

@article{Bochkarev:1990fx,
    author = "Bochkarev, A. I. and Kuzmin, S. V. and Shaposhnikov, M. E.",
    title = "{Electroweak baryogenesis and the Higgs boson mass problem}",
    doi = "10.1016/0370-2693(90)90069-I",
    journal = "Phys. Lett. B",
    volume = "244",
    pages = "275--278",
    year = "1990"
}

@article{Cohen:1990py,
    author = "Cohen, Andrew G. and Kaplan, David B. and Nelson, Ann E.",
    title = "{WEAK SCALE BARYOGENESIS}",
    reportNumber = "NSF-ITP-90-72, UCSC-PTH-90-06, BUHEP-90-13",
    doi = "10.1016/0370-2693(90)90690-8",
    journal = "Phys. Lett. B",
    volume = "245",
    pages = "561--564",
    year = "1990"
}

@article{Cohen:1990it,
    author = "Cohen, Andrew G. and Kaplan, David B. and Nelson, Ann E.",
    title = "{Baryogenesis at the weak phase transition}",
    reportNumber = "NSF-ITP-90-85, UCSD-PTH-90-09, BUHEP-90-15",
    doi = "10.1016/0550-3213(91)90395-E",
    journal = "Nucl. Phys. B",
    volume = "349",
    pages = "727--742",
    year = "1991"
}

@article{Kajantie:1996mn,
    author = "Kajantie, K. and Laine, M. and Rummukainen, K. and Shaposhnikov, Mikhail E.",
    title = "{Is there a~ hot electroweak phase transition at $m_H \gtrsim m_W$?}",
    eprint = "hep-ph/9605288",
    archivePrefix = "arXiv",
    reportNumber = "CERN-TH-96-126, HD-THEP-96-15, IUHET-333",
    doi = "10.1103/PhysRevLett.77.2887",
    journal = "Phys. Rev. Lett.",
    volume = "77",
    pages = "2887--2890",
    year = "1996"
}

@article{Karsch:1996yh,
    author = "Karsch, F. and Neuhaus, T. and Patkos, A. and Rank, J.",
    editor = "Bernard, C. and Golterman, M. and Ogilvie, M. and Potvin, J.",
    title = "{Critical Higgs mass and temperature dependence of gauge boson masses in the SU(2) gauge Higgs model}",
    eprint = "hep-lat/9608087",
    archivePrefix = "arXiv",
    reportNumber = "FSU-SCRI-96C-79",
    doi = "10.1016/S0920-5632(96)00736-0",
    journal = "Nucl. Phys. B Proc. Suppl.",
    volume = "53",
    pages = "623--625",
    year = "1997"
}

@article{Linde:1981zj,
    author = "Linde, Andrei D.",
    title = "{Decay of the False Vacuum at Finite Temperature}",
    reportNumber = "LEBEDEV-81-265",
    doi = "10.1016/0550-3213(83)90072-X",
    journal = "Nucl. Phys. B",
    volume = "216",
    pages = "421",
    year = "1983",
    note = "[Erratum: Nucl.Phys.B 223, 544 (1983)]"
}

@article{Coleman:1977py,
    author = "Coleman, Sidney R.",
    title = "{The Fate of the False Vacuum. 1. Semiclassical Theory}",
    reportNumber = "HUTP-77-A004",
    doi = "10.1103/PhysRevD.16.1248",
    journal = "Phys. Rev. D",
    volume = "15",
    pages = "2929--2936",
    year = "1977",
    note = "[Erratum: Phys.Rev.D 16, 1248 (1977)]"
}

@article{Wainwright:2011kj,
    author = "Wainwright, Carroll L.",
    title = "{CosmoTransitions: Computing Cosmological Phase Transition Temperatures and Bubble Profiles with Multiple Fields}",
    eprint = "1109.4189",
    archivePrefix = "arXiv",
    primaryClass = "hep-ph",
    doi = "10.1016/j.cpc.2012.04.004",
    journal = "Comput. Phys. Commun.",
    volume = "183",
    pages = "2006--2013",
    year = "2012"
}

@article{Huber:2007vva,
    author = "Huber, Stephan J. and Konstandin, Thomas",
    title = "{Production of gravitational waves in the nMSSM}",
    eprint = "0709.2091",
    archivePrefix = "arXiv",
    primaryClass = "hep-ph",
    doi = "10.1088/1475-7516/2008/05/017",
    journal = "JCAP",
    volume = "05",
    pages = "017",
    year = "2008"
}

@phdthesis{JoseRedondo,
  author  = "Jose Miguel No Redondo",
  title   = "Aspects of Phenomenology and Cosmology in Hidden Sector Extensions of the Standard Model ",
  school  = "Universidad Autónoma de Madrid",
  year    = "2009"
}

@article{Athron:2022mmm,
    author = "Athron, Peter and Bal\'azs, Csaba and Morris, Lachlan",
    title = "{Supercool subtleties of cosmological phase transitions}",
    eprint = "2212.07559",
    archivePrefix = "arXiv",
    primaryClass = "hep-ph",
    doi = "10.1088/1475-7516/2023/03/006",
    journal = "JCAP",
    volume = "03",
    pages = "006",
    year = "2023"
}

@article{Freitas:2021yng,
    author = "Freitas, Felipe F. and Louren\c{c}o, Gabriel and Morais, Ant\'onio P. and Nunes, Andr\'e and Ol\'\i{}via, Joao and Pasechnik, Roman and Santos, Rui and Viana, Joao",
    title = "{Impact of SM parameters and of the vacua of the Higgs potential in gravitational waves detection}",
    eprint = "2108.12810",
    archivePrefix = "arXiv",
    primaryClass = "hep-ph",
    doi = "10.1088/1475-7516/2022/03/046",
    journal = "JCAP",
    volume = "03",
    number = "03",
    pages = "046",
    year = "2022"
}

@article{Schmitz:2020syl,
    author = "Schmitz, Kai",
    title = "{New Sensitivity Curves for Gravitational-Wave Signals from Cosmological Phase Transitions}",
    eprint = "2002.04615",
    archivePrefix = "arXiv",
    primaryClass = "hep-ph",
    reportNumber = "CERN-TH-2020-018",
    doi = "10.1007/JHEP01(2021)097",
    journal = "JHEP",
    volume = "01",
    pages = "097",
    year = "2021"
}

@article{Athron:2023rfq,
    author = "Athron, Peter and Morris, Lachlan and Xu, Zhongxiu",
    title = "{How robust are gravitational wave predictions from cosmological phase transitions?}",
    eprint = "2309.05474",
    archivePrefix = "arXiv",
    primaryClass = "hep-ph",
    doi = "10.1088/1475-7516/2024/05/075",
    journal = "JCAP",
    volume = "05",
    pages = "075",
    year = "2024"
}

@article{Guth:1981uk,
    author = "Guth, Alan H. and Weinberg, Erick J.",
    title = "{Cosmological Consequences of a First Order Phase Transition in the SU(5) Grand Unified Model}",
    reportNumber = "CU-TP-183",
    doi = "10.1103/PhysRevD.23.876",
    journal = "Phys. Rev. D",
    volume = "23",
    pages = "876",
    year = "1981"
}

@book{Stauffer_Aharony_2014, 
    place={Hoboken}, 
    title={Introduction to percolation theory revised second edition},
    publisher={Taylor and Francis},
    author={Stauffer, Dietrich and Aharony, Ammon}, 
    year={2014}}

@article{Ai:2023see,
    author = "Ai, Wen-Yuan and Laurent, Benoit and van de Vis, Jorinde",
    title = "{Model-independent bubble wall velocities in local thermal equilibrium}",
    eprint = "2303.10171",
    archivePrefix = "arXiv",
    primaryClass = "astro-ph.CO",
    reportNumber = "KCL-PH-TH/2023-19",
    doi = "10.1088/1475-7516/2023/07/002",
    journal = "JCAP",
    volume = "07",
    pages = "002",
    year = "2023"
}

@article{Addazi:2023ftv,
    author = "Addazi, Andrea and Marcian\`o, Antonino and Morais, Ant\'onio P. and Pasechnik, Roman and Viana, Jo\~ao and Yang, Hao",
    title = "{Gravitational echoes of lepton number symmetry breaking with light and ultralight Majorons}",
    eprint = "2304.02399",
    archivePrefix = "arXiv",
    primaryClass = "hep-ph",
    reportNumber = "CERN-TH-2023-054",
    doi = "10.1088/1475-7516/2023/09/026",
    journal = "JCAP",
    volume = "09",
    pages = "026",
    year = "2023",
    note = "[Erratum: JCAP 03, E01 (2024)]"
}

@article{Kierkla:2022odc,
    author = "Kierkla, Maciej and Karam, Alexandros and Swiezewska, Bogumila",
    title = "{Conformal model for gravitational waves and dark matter: a status update}",
    eprint = "2210.07075",
    archivePrefix = "arXiv",
    primaryClass = "astro-ph.CO",
    doi = "10.1007/JHEP03(2023)007",
    journal = "JHEP",
    volume = "03",
    pages = "007",
    year = "2023"
}

@article{Croon:2020cgk,
    author = "Croon, Djuna and Gould, Oliver and Schicho, Philipp and Tenkanen, Tuomas V. I. and White, Graham",
    title = "{Theoretical uncertainties for cosmological first-order phase transitions}",
    eprint = "2009.10080",
    archivePrefix = "arXiv",
    primaryClass = "hep-ph",
    reportNumber = "HIP-2020-26/TH",
    doi = "10.1007/JHEP04(2021)055",
    journal = "JHEP",
    volume = "04",
    pages = "055",
    year = "2021"
}

@article{Athron:2022jyi,
    author = "Athron, Peter and Balazs, Csaba and Fowlie, Andrew and Morris, Lachlan and White, Graham and Zhang, Yang",
    title = "{How arbitrary are perturbative calculations of the electroweak phase transition?}",
    eprint = "2208.01319",
    archivePrefix = "arXiv",
    primaryClass = "hep-ph",
    doi = "10.1007/JHEP01(2023)050",
    journal = "JHEP",
    volume = "01",
    pages = "050",
    year = "2023"
}

@article{Martin:2001vx,
    author = "Martin, Stephen P.",
    title = "{Two Loop Effective Potential for a General Renormalizable Theory and Softly Broken Supersymmetry}",
    eprint = "hep-ph/0111209",
    archivePrefix = "arXiv",
    reportNumber = "FERMILAB-PUB-01-348-T",
    doi = "10.1103/PhysRevD.65.116003",
    journal = "Phys. Rev. D",
    volume = "65",
    pages = "116003",
    year = "2002"
}

@article{McKeon:2015zxa,
    author = "McKeon, D. G. C.",
    title = "{Renormalization Scheme Dependence with Renormalization Group Summation}",
    eprint = "1503.03823",
    archivePrefix = "arXiv",
    primaryClass = "hep-th",
    doi = "10.1103/PhysRevD.92.045031",
    journal = "Phys. Rev. D",
    volume = "92",
    number = "4",
    pages = "045031",
    year = "2015"
}

@article{Hindmarsh:2017gnf,
    author = "Hindmarsh, Mark and Huber, Stephan J. and Rummukainen, Kari and Weir, David J.",
    title = "{Shape of the acoustic gravitational wave power spectrum from a first order phase transition}",
    eprint = "1704.05871",
    archivePrefix = "arXiv",
    primaryClass = "astro-ph.CO",
    reportNumber = "HIP-2017-02-TH, HIP-2017-02/TH",
    doi = "10.1103/PhysRevD.96.103520",
    journal = "Phys. Rev. D",
    volume = "96",
    number = "10",
    pages = "103520",
    year = "2017",
    note = "[Erratum: Phys.Rev.D 101, 089902 (2020)]"
}

@article{Hindmarsh:2015qta,
    author = "Hindmarsh, Mark and Huber, Stephan J. and Rummukainen, Kari and Weir, David J.",
    title = "{Numerical simulations of acoustically generated gravitational waves at a first order phase transition}",
    eprint = "1504.03291",
    archivePrefix = "arXiv",
    primaryClass = "astro-ph.CO",
    reportNumber = "HIP-2015-13-TH",
    doi = "10.1103/PhysRevD.92.123009",
    journal = "Phys. Rev. D",
    volume = "92",
    number = "12",
    pages = "123009",
    year = "2015"
}

@article{Ekstedt:2022bff,
    author = "Ekstedt, Andreas and Schicho, Philipp and Tenkanen, Tuomas V. I.",
    title = "{DRalgo: A package for effective field theory approach for thermal phase transitions}",
    eprint = "2205.08815",
    archivePrefix = "arXiv",
    primaryClass = "hep-ph",
    reportNumber = "HIP-2022-11/TH, NORDITA 2022-030",
    doi = "10.1016/j.cpc.2023.108725",
    journal = "Comput. Phys. Commun.",
    volume = "288",
    pages = "108725",
    year = "2023"
}

@article{Brauner:2016fla,
    author = "Brauner, Tom\'a\v{s} and Tenkanen, Tuomas V. I. and Tranberg, Anders and Vuorinen, Aleksi and Weir, David J.",
    title = "{Dimensional reduction of the Standard Model coupled to a new singlet scalar field}",
    eprint = "1609.06230",
    archivePrefix = "arXiv",
    primaryClass = "hep-ph",
    reportNumber = "HIP-2016-27-TH",
    doi = "10.1007/JHEP03(2017)007",
    journal = "JHEP",
    volume = "03",
    pages = "007",
    year = "2017"
}

@book{Laine:2016hma,
    author = "Laine, Mikko and Vuorinen, Aleksi",
    title = "{Basics of Thermal Field Theory}",
    eprint = "1701.01554",
    archivePrefix = "arXiv",
    primaryClass = "hep-ph",
    doi = "10.1007/978-3-319-31933-9",
    publisher = "Springer",
    volume = "925",
    year = "2016"
}

@article{NANOGrav:2023gor,
    author = "Agazie, Gabriella and others",
    collaboration = "NANOGrav",
    title = "{The NANOGrav 15 yr Data Set: Evidence for a Gravitational-wave Background}",
    eprint = "2306.16213",
    archivePrefix = "arXiv",
    primaryClass = "astro-ph.HE",
    doi = "10.3847/2041-8213/acdac6",
    journal = "Astrophys. J. Lett.",
    volume = "951",
    number = "1",
    pages = "L8",
    year = "2023"
}

@article{NANOGrav:2023hvm,
    author = "Afzal, Adeela and others",
    collaboration = "NANOGrav",
    title = "{The NANOGrav 15 yr Data Set: Search for Signals from New Physics}",
    eprint = "2306.16219",
    archivePrefix = "arXiv",
    primaryClass = "astro-ph.HE",
    reportNumber = "FERMILAB-PUB-23-589-T",
    doi = "10.3847/2041-8213/acdc91",
    journal = "Astrophys. J. Lett.",
    volume = "951",
    number = "1",
    pages = "L11",
    year = "2023"
}

@article{Fujikura:2023lkn,
    author = "Fujikura, Kohei and Girmohanta, Sudhakantha and Nakai, Yuichiro and Suzuki, Motoo",
    title = "{NANOGrav signal from a dark conformal phase transition}",
    eprint = "2306.17086",
    archivePrefix = "arXiv",
    primaryClass = "hep-ph",
    reportNumber = "UT-Komaba/23-6",
    doi = "10.1016/j.physletb.2023.138203",
    journal = "Phys. Lett. B",
    volume = "846",
    pages = "138203",
    year = "2023"
}

@article{Bringmann:2023opz,
    author = "Bringmann, Torsten and Depta, Paul Frederik and Konstandin, Thomas and Schmidt-Hoberg, Kai and Tasillo, Carlo",
    title = "{Does NANOGrav observe a dark sector phase transition?}",
    eprint = "2306.09411",
    archivePrefix = "arXiv",
    primaryClass = "astro-ph.CO",
    reportNumber = "DESY-23-077",
    doi = "10.1088/1475-7516/2023/11/053",
    journal = "JCAP",
    volume = "11",
    pages = "053",
    year = "2023"
}

@article{Gould:2021oba,
    author = "Gould, Oliver and Tenkanen, Tuomas V. I.",
    title = "{On the perturbative expansion at high temperature and implications for cosmological phase transitions}",
    eprint = "2104.04399",
    archivePrefix = "arXiv",
    primaryClass = "hep-ph",
    reportNumber = "NORDITA 2021-010",
    doi = "10.1007/JHEP06(2021)069",
    journal = "JHEP",
    volume = "06",
    pages = "069",
    year = "2021"
}

@article{Carena:2019une,
    author = "Carena, Marcela and Liu, Zhen and Wang, Yikun",
    title = "{Electroweak phase transition with spontaneous Z$_{2}$-breaking}",
    eprint = "1911.10206",
    archivePrefix = "arXiv",
    primaryClass = "hep-ph",
    reportNumber = "FERMILAB-PUB-19-667-T, FERMILAB-PUB-19-602-T",
    doi = "10.1007/JHEP08(2020)107",
    journal = "JHEP",
    volume = "08",
    pages = "107",
    year = "2020"
}

@article{Kainulainen:2019kyp,
    author = "Kainulainen, Kimmo and Keus, Venus and Niemi, Lauri and Rummukainen, Kari and Tenkanen, Tuomas V. I. and Vaskonen, Ville",
    title = "{On the validity of perturbative studies of the electroweak phase transition in the Two Higgs Doublet model}",
    eprint = "1904.01329",
    archivePrefix = "arXiv",
    primaryClass = "hep-ph",
    doi = "10.1007/JHEP06(2019)075",
    journal = "JHEP",
    volume = "06",
    pages = "075",
    year = "2019"
}

@article{Guo:2021qcq,
    author = "Guo, Huai-Ke and Sinha, Kuver and Vagie, Daniel and White, Graham",
    title = "{The benefits of diligence: how precise are predicted gravitational wave spectra in models with phase transitions?}",
    eprint = "2103.06933",
    archivePrefix = "arXiv",
    primaryClass = "hep-ph",
    doi = "10.1007/JHEP06(2021)164",
    journal = "JHEP",
    volume = "06",
    pages = "164",
    year = "2021"
}

@article{Gould:2023ovu,
    author = "Gould, Oliver and Tenkanen, Tuomas V. I.",
    title = "{Perturbative effective field theory expansions for cosmological phase transitions}",
    eprint = "2309.01672",
    archivePrefix = "arXiv",
    primaryClass = "hep-ph",
    reportNumber = "NORDITA 2023-037",
    month = "9",
    year = "2023"
}

@article{Borah:2021ocu,
    author = "Borah, Debasish and Dasgupta, Arnab and Kang, Sin Kyu",
    title = "{Gravitational waves from a dark U(1)D phase transition in light of NANOGrav 12.5~yr data}",
    eprint = "2105.01007",
    archivePrefix = "arXiv",
    primaryClass = "hep-ph",
    doi = "10.1103/PhysRevD.104.063501",
    journal = "Phys. Rev. D",
    volume = "104",
    number = "6",
    pages = "063501",
    year = "2021"
}

@article{Borah:2021ftr,
    author = "Borah, Debasish and Dasgupta, Arnab and Kang, Sin Kyu",
    title = "{A first order dark SU(2)$_{D}$ phase transition with vector dark matter in the light of NANOGrav 12.5 yr data}",
    eprint = "2109.11558",
    archivePrefix = "arXiv",
    primaryClass = "hep-ph",
    doi = "10.1088/1475-7516/2021/12/039",
    journal = "JCAP",
    volume = "12",
    number = "12",
    pages = "039",
    year = "2021"
}

@article{Lewicki:2021xku,
    author = "Lewicki, Marek and Pujol\`as, Oriol and Vaskonen, Ville",
    title = "{Escape from supercooling with or without bubbles: gravitational wave signatures}",
    eprint = "2106.09706",
    archivePrefix = "arXiv",
    primaryClass = "astro-ph.CO",
    doi = "10.1140/epjc/s10052-021-09669-6",
    journal = "Eur. Phys. J. C",
    volume = "81",
    number = "9",
    pages = "857",
    year = "2021"
}

@article{Marzo:2018nov,
    author = "Marzo, Carlo and Marzola, Luca and Vaskonen, Ville",
    title = "{Phase transition and vacuum stability in the classically conformal B\textendash{}L model}",
    eprint = "1811.11169",
    archivePrefix = "arXiv",
    primaryClass = "hep-ph",
    reportNumber = "KCL-PH-TH/2018-68",
    doi = "10.1140/epjc/s10052-019-7076-x",
    journal = "Eur. Phys. J. C",
    volume = "79",
    number = "7",
    pages = "601",
    year = "2019"
}

@article{Ellis:2020nnr,
    author = "Ellis, John and Lewicki, Marek and Vaskonen, Ville",
    title = "{Updated predictions for gravitational waves produced in a strongly supercooled phase transition}",
    eprint = "2007.15586",
    archivePrefix = "arXiv",
    primaryClass = "astro-ph.CO",
    reportNumber = "KCL-PH-TH/2020-40, CERN-TH-2020-129",
    doi = "10.1088/1475-7516/2020/11/020",
    journal = "JCAP",
    volume = "11",
    pages = "020",
    year = "2020"
}

@article{Kajantie:1995dw,
    author = "Kajantie, K. and Laine, M. and Rummukainen, K. and Shaposhnikov, Mikhail E.",
    title = "{Generic rules for high temperature dimensional reduction and their application to the standard model}",
    eprint = "hep-ph/9508379",
    archivePrefix = "arXiv",
    reportNumber = "CERN-TH-95-226, HU-TFT-95-50, IUHET-312",
    doi = "10.1016/0550-3213(95)00549-8",
    journal = "Nucl. Phys. B",
    volume = "458",
    pages = "90--136",
    year = "1996"
}

@article{Maggiore:1999vm,
    author = "Maggiore, Michele",
    title = "{Gravitational wave experiments and early universe cosmology}",
    eprint = "gr-qc/9909001",
    archivePrefix = "arXiv",
    reportNumber = "IFUP-TH-20-99",
    doi = "10.1016/S0370-1573(99)00102-7",
    journal = "Phys. Rept.",
    volume = "331",
    pages = "283--367",
    year = "2000"
}

@article{Figueroa:2012kw,
    author = "Figueroa, Daniel G. and Hindmarsh, Mark and Urrestilla, Jon",
    title = "{Exact Scale-Invariant Background of Gravitational Waves from Cosmic Defects}",
    eprint = "1212.5458",
    archivePrefix = "arXiv",
    primaryClass = "astro-ph.CO",
    doi = "10.1103/PhysRevLett.110.101302",
    journal = "Phys. Rev. Lett.",
    volume = "110",
    number = "10",
    pages = "101302",
    year = "2013"
}

@article{Caprini:2007xq,
    author = "Caprini, Chiara and Durrer, Ruth and Servant, Geraldine",
    title = "{Gravitational wave generation from bubble collisions in first-order phase transitions: An analytic approach}",
    eprint = "0711.2593",
    archivePrefix = "arXiv",
    primaryClass = "astro-ph",
    reportNumber = "CERN-PH-TH-2007-206, SACLAY-T07-142",
    doi = "10.1103/PhysRevD.77.124015",
    journal = "Phys. Rev. D",
    volume = "77",
    pages = "124015",
    year = "2008"
}

@article{Athron:2023xlk,
    author = "Athron, Peter and Bal\'azs, Csaba and Fowlie, Andrew and Morris, Lachlan and Wu, Lei",
    title = "{Cosmological phase transitions: from perturbative particle physics to gravitational waves}",
    eprint = "2305.02357",
    archivePrefix = "arXiv",
    primaryClass = "hep-ph",
    month = "5",
    year = "2023"
}

@article{Hindmarsh:2013xza,
    author = "Hindmarsh, Mark and Huber, Stephan J. and Rummukainen, Kari and Weir, David J.",
    title = "{Gravitational waves from the sound of a first order phase transition}",
    eprint = "1304.2433",
    archivePrefix = "arXiv",
    primaryClass = "hep-ph",
    reportNumber = "HIP-2013-07-TH",
    doi = "10.1103/PhysRevLett.112.041301",
    journal = "Phys. Rev. Lett.",
    volume = "112",
    pages = "041301",
    year = "2014"
}

@article{Kosowsky:1991ua,
    author = "Kosowsky, Arthur and Turner, Michael S. and Watkins, Richard",
    title = "{Gravitational radiation from colliding vacuum bubbles}",
    reportNumber = "FERMILAB-PUB-91-323-A",
    doi = "10.1103/PhysRevD.45.4514",
    journal = "Phys. Rev. D",
    volume = "45",
    pages = "4514--4535",
    year = "1992"
}

@article{Kosowsky:1992vn,
    author = "Kosowsky, Arthur and Turner, Michael S.",
    title = "{Gravitational radiation from colliding vacuum bubbles: envelope approximation to many bubble collisions}",
    eprint = "astro-ph/9211004",
    archivePrefix = "arXiv",
    reportNumber = "FERMILAB-PUB-92-295-A",
    doi = "10.1103/PhysRevD.47.4372",
    journal = "Phys. Rev. D",
    volume = "47",
    pages = "4372--4391",
    year = "1993"
}

@article{Cutting:2018tjt,
    author = "Cutting, Daniel and Hindmarsh, Mark and Weir, David J.",
    title = "{Gravitational waves from vacuum first-order phase transitions: from the envelope to the lattice}",
    eprint = "1802.05712",
    archivePrefix = "arXiv",
    primaryClass = "astro-ph.CO",
    reportNumber = "HIP-2018-4-TH",
    doi = "10.1103/PhysRevD.97.123513",
    journal = "Phys. Rev. D",
    volume = "97",
    number = "12",
    pages = "123513",
    year = "2018"
}

@article{Kamionkowski:1993fg,
    author = "Kamionkowski, Marc and Kosowsky, Arthur and Turner, Michael S.",
    title = "{Gravitational radiation from first order phase transitions}",
    eprint = "astro-ph/9310044",
    archivePrefix = "arXiv",
    reportNumber = "IASSNS-HEP-93-44, FERMILAB-PUB-93-235-A",
    doi = "10.1103/PhysRevD.49.2837",
    journal = "Phys. Rev. D",
    volume = "49",
    pages = "2837--2851",
    year = "1994"
}

@article{Lewicki:2020azd,
    author = "Lewicki, Marek and Vaskonen, Ville",
    title = "{Gravitational waves from colliding vacuum bubbles in gauge theories}",
    eprint = "2012.07826",
    archivePrefix = "arXiv",
    primaryClass = "astro-ph.CO",
    doi = "10.1140/epjc/s10052-021-09232-3",
    journal = "Eur. Phys. J. C",
    volume = "81",
    number = "5",
    pages = "437",
    year = "2021",
    note = "[Erratum: Eur.Phys.J.C 81, 1077 (2021)]"
}

@article{Caprini:2009yp,
    author = "Caprini, Chiara and Durrer, Ruth and Servant, Geraldine",
    title = "{The stochastic gravitational wave background from turbulence and magnetic fields generated by a first-order phase transition}",
    eprint = "0909.0622",
    archivePrefix = "arXiv",
    primaryClass = "astro-ph.CO",
    doi = "10.1088/1475-7516/2009/12/024",
    journal = "JCAP",
    volume = "12",
    pages = "024",
    year = "2009"
}

@article{RoperPol:2019wvy,
    author = "Roper Pol, Alberto and Mandal, Sayan and Brandenburg, Axel and Kahniashvili, Tina and Kosowsky, Arthur",
    title = "{Numerical simulations of gravitational waves from early-universe turbulence}",
    eprint = "1903.08585",
    archivePrefix = "arXiv",
    primaryClass = "astro-ph.CO",
    reportNumber = "NORDITA-2019-024",
    doi = "10.1103/PhysRevD.102.083512",
    journal = "Phys. Rev. D",
    volume = "102",
    number = "8",
    pages = "083512",
    year = "2020"
}

@article{Kahniashvili:2020jgm,
    author = "Kahniashvili, Tina and Brandenburg, Axel and Gogoberidze, Grigol and Mandal, Sayan and Roper Pol, Alberto",
    title = "{Circular polarization of gravitational waves from early-Universe helical turbulence}",
    eprint = "2011.05556",
    archivePrefix = "arXiv",
    primaryClass = "astro-ph.CO",
    reportNumber = "NORDITA-2020-102",
    doi = "10.1103/PhysRevResearch.3.013193",
    journal = "Phys. Rev. Res.",
    volume = "3",
    number = "1",
    pages = "013193",
    year = "2021"
}

@article{RoperPol:2021xnd,
    author = "Roper Pol, Alberto and Mandal, Sayan and Brandenburg, Axel and Kahniashvili, Tina",
    title = "{Polarization of gravitational waves from helical MHD turbulent sources}",
    eprint = "2107.05356",
    archivePrefix = "arXiv",
    primaryClass = "gr-qc",
    reportNumber = "NORDITA-2021-062",
    doi = "10.1088/1475-7516/2022/04/019",
    journal = "JCAP",
    volume = "04",
    number = "04",
    pages = "019",
    year = "2022"
}

@article{Auclair:2022jod,
    author = "Auclair, Pierre and Caprini, Chiara and Cutting, Daniel and Hindmarsh, Mark and Rummukainen, Kari and Steer, Dani\`ele A. and Weir, David J.",
    title = "{Generation of gravitational waves from freely decaying turbulence}",
    eprint = "2205.02588",
    archivePrefix = "arXiv",
    primaryClass = "astro-ph.CO",
    reportNumber = "HIP-2021-35/TH",
    doi = "10.1088/1475-7516/2022/09/029",
    journal = "JCAP",
    volume = "09",
    pages = "029",
    year = "2022"
}

@article{Caprini:2015zlo,
    author = "Caprini, Chiara and others",
    title = "{Science with the space-based interferometer eLISA. II: Gravitational waves from cosmological phase transitions}",
    eprint = "1512.06239",
    archivePrefix = "arXiv",
    primaryClass = "astro-ph.CO",
    reportNumber = "DESY-15-246",
    doi = "10.1088/1475-7516/2016/04/001",
    journal = "JCAP",
    volume = "04",
    pages = "001",
    year = "2016"
}

@article{Azatov:2019png,
    author = "Azatov, Aleksandr and Barducci, Daniele and Sgarlata, Francesco",
    title = "{Gravitational traces of broken gauge symmetries}",
    eprint = "1910.01124",
    archivePrefix = "arXiv",
    primaryClass = "hep-ph",
    reportNumber = "SISSA 28/2019/FISI",
    doi = "10.1088/1475-7516/2020/07/027",
    journal = "JCAP",
    volume = "07",
    pages = "027",
    year = "2020"
}

@article{Alves:2018jsw,
    author = "Alves, Alexandre and Ghosh, Tathagata and Guo, Huai-Ke and Sinha, Kuver and Vagie, Daniel",
    title = "{Collider and Gravitational Wave Complementarity in Exploring the Singlet Extension of the Standard Model}",
    eprint = "1812.09333",
    archivePrefix = "arXiv",
    primaryClass = "hep-ph",
    doi = "10.1007/JHEP04(2019)052",
    journal = "JHEP",
    volume = "04",
    pages = "052",
    year = "2019"
}

@article{Espinosa:2010hh,
    author = "Espinosa, Jose R. and Konstandin, Thomas and No, Jose M. and Servant, Geraldine",
    title = "{Energy Budget of Cosmological First-order Phase Transitions}",
    eprint = "1004.4187",
    archivePrefix = "arXiv",
    primaryClass = "hep-ph",
    reportNumber = "CERN-PH-TH-2010-027",
    doi = "10.1088/1475-7516/2010/06/028",
    journal = "JCAP",
    volume = "06",
    pages = "028",
    year = "2010"
}

@article{Ekstedt:2023sqc,
    author = "Ekstedt, Andreas and Gould, Oliver and Hirvonen, Joonas",
    title = "{BubbleDet: a Python package to compute functional determinants for bubble nucleation}",
    eprint = "2308.15652",
    archivePrefix = "arXiv",
    primaryClass = "hep-ph",
    doi = "10.1007/JHEP12(2023)056",
    journal = "JHEP",
    volume = "12",
    pages = "056",
    year = "2023"
}

@article{Krajewski:2024gma,
    author = "Krajewski, Tomasz and Lewicki, Marek and Zych, Mateusz",
    title = "{Bubble-wall velocity in local thermal equilibrium: hydrodynamical simulations vs analytical treatment}",
    eprint = "2402.15408",
    archivePrefix = "arXiv",
    primaryClass = "astro-ph.CO",
    doi = "10.1007/JHEP05(2024)011",
    journal = "JHEP",
    volume = "05",
    pages = "011",
    year = "2024"
}

@article{Lewicki:2024xan,
    author = "Lewicki, Marek and Merchand, Marco and Sagunski, Laura and Schicho, Philipp and Schmitt, Daniel",
    title = "{Impact of theoretical uncertainties on model parameter reconstruction from GW signals sourced by cosmological phase transitions}",
    eprint = "2403.03769",
    archivePrefix = "arXiv",
    primaryClass = "hep-ph",
    doi = "10.1103/PhysRevD.110.023538",
    journal = "Phys. Rev. D",
    volume = "110",
    number = "2",
    pages = "023538",
    year = "2024"
}

@article{Planck:2018vyg,
    author = "Aghanim, N. and others",
    collaboration = "Planck",
    title = "{Planck 2018 results. VI. Cosmological parameters}",
    eprint = "1807.06209",
    archivePrefix = "arXiv",
    primaryClass = "astro-ph.CO",
    doi = "10.1051/0004-6361/201833910",
    journal = "Astron. Astrophys.",
    volume = "641",
    pages = "A6",
    year = "2020",
    note = "[Erratum: Astron.Astrophys. 652, C4 (2021)]"
}

@article{Ekstedt:2024etx,
    author = "Ekstedt, Andreas and Schicho, Philipp and Tenkanen, Tuomas V. I.",
    title = "{Cosmological phase transitions at three loops: the final verdict on perturbation theory}",
    eprint = "2405.18349",
    archivePrefix = "arXiv",
    primaryClass = "hep-ph",
    reportNumber = "HIP-2024-15/TH",
    month = "5",
    year = "2024"
}

@article{DOnofrio:2015gop,
    author = "D'Onofrio, Michela and Rummukainen, Kari",
    title = "{Standard model cross-over on the lattice}",
    eprint = "1508.07161",
    archivePrefix = "arXiv",
    primaryClass = "hep-ph",
    reportNumber = "HIP-2015-30-TH",
    doi = "10.1103/PhysRevD.93.025003",
    journal = "Phys. Rev. D",
    volume = "93",
    number = "2",
    pages = "025003",
    year = "2016"
}

@article{Fodor:2004nz,
    author = "Fodor, Z. and Katz, S. D.",
    title = "{Critical point of QCD at finite T and mu, lattice results for physical quark masses}",
    eprint = "hep-lat/0402006",
    archivePrefix = "arXiv",
    reportNumber = "ITP-BUDAPEST-609, WUB-04-04",
    doi = "10.1088/1126-6708/2004/04/050",
    journal = "JHEP",
    volume = "04",
    pages = "050",
    year = "2004"
}

@article{Caprini:2024hue,
    author = "Caprini, Chiara and Jinno, Ryusuke and Lewicki, Marek and Madge, Eric and Merchand, Marco and Nardini, Germano and Pieroni, Mauro and Roper Pol, Alberto and Vaskonen, Ville",
    collaboration = "LISA Cosmology Working Group",
    title = "{Gravitational waves from first-order phase transitions in LISA: reconstruction pipeline and physics interpretation}",
    eprint = "2403.03723",
    archivePrefix = "arXiv",
    primaryClass = "astro-ph.CO",
    reportNumber = "LISA-COSWG-24-01, CERN-TH-2024-029",
    month = "3",
    year = "2024"
}

@article{Jinno:2022mie,
    author = "Jinno, Ryusuke and Konstandin, Thomas and Rubira, Henrique and Stomberg, Isak",
    title = "{Higgsless simulations of cosmological phase transitions and gravitational waves}",
    eprint = "2209.04369",
    archivePrefix = "arXiv",
    primaryClass = "astro-ph.CO",
    reportNumber = "DESY 22-148, IFT-UAM/CSIC-22-100, TUM-HEP-1416/22",
    doi = "10.1088/1475-7516/2023/02/011",
    journal = "JCAP",
    volume = "02",
    pages = "011",
    year = "2023"
}

@article{ParticleDataGroup:2024cfk,
    author = "Navas, S. and others",
    collaboration = "Particle Data Group",
    title = "{Review of particle physics}",
    doi = "10.1103/PhysRevD.110.030001",
    journal = "Phys. Rev. D",
    volume = "110",
    number = "3",
    pages = "030001",
    year = "2024"
}

@article{ATLAS:2021vrm,
    collaboration = "ATLAS",
    title = "{Combined measurements of Higgs boson production and decay using up to $139$ fb$^{-1}$ of proton-proton collision data at $\sqrt{s}= 13$ TeV collected with the ATLAS experiment}",
    reportNumber = "ATLAS-CONF-2021-053",
    year = "2021"
}

@inproceedings{Papaefstathiou:2022oyi,
    author = "Papaefstathiou, Andreas and Robens, Tania and White, Graham",
    title = "{Signal strength and W-boson mass measurements as a probe of the electro-weak phase transition at colliders - Snowmass White Paper}",
    booktitle = "{Snowmass 2021}",
    eprint = "2205.14379",
    archivePrefix = "arXiv",
    primaryClass = "hep-ph",
    reportNumber = "RBI-ThPhys-2022-20",
    month = "5",
    year = "2022"
}

@article{Marfatia:2021twj,
    author = "Marfatia, Danny and Tseng, Po-Yan",
    title = "{Correlated gravitational wave and microlensing signals of macroscopic dark matter}",
    eprint = "2107.00859",
    archivePrefix = "arXiv",
    primaryClass = "hep-ph",
    doi = "10.1007/JHEP11(2021)068",
    journal = "JHEP",
    volume = "11",
    pages = "068",
    year = "2021"
}

@article{Hambye:2008bq,
    author = "Hambye, Thomas",
    title = "{Hidden vector dark matter}",
    eprint = "0811.0172",
    archivePrefix = "arXiv",
    primaryClass = "hep-ph",
    reportNumber = "ULB-TH-08-35",
    doi = "10.1088/1126-6708/2009/01/028",
    journal = "JHEP",
    volume = "01",
    pages = "028",
    year = "2009"
}

@article{Dratopi,
  author = "Bertenstam \textit{et al.}, Mårten",
  title = "{Dratopi}",
  journal = "To appear",
  url = {https://gitlab.com/mb-1380649/dratopi},
  year = ""
}

@article{EPTA:2023fyk,
    author = "Antoniadis, J. and others",
    collaboration = "EPTA, InPTA:",
    title = "{The second data release from the European Pulsar Timing Array - III. Search for gravitational wave signals}",
    eprint = "2306.16214",
    archivePrefix = "arXiv",
    primaryClass = "astro-ph.HE",
    doi = "10.1051/0004-6361/202346844",
    journal = "Astron. Astrophys.",
    volume = "678",
    pages = "A50",
    year = "2023"
}

@article{Reardon:2023gzh,
    author = "Reardon, Daniel J. and others",
    title = "{Search for an Isotropic Gravitational-wave Background with the Parkes Pulsar Timing Array}",
    eprint = "2306.16215",
    archivePrefix = "arXiv",
    primaryClass = "astro-ph.HE",
    doi = "10.3847/2041-8213/acdd02",
    journal = "Astrophys. J. Lett.",
    volume = "951",
    number = "1",
    pages = "L6",
    year = "2023"
}

@article{Xu:2023wog,
    author = "Xu, Heng and others",
    title = "{Searching for the Nano-Hertz Stochastic Gravitational Wave Background with the Chinese Pulsar Timing Array Data Release I}",
    eprint = "2306.16216",
    archivePrefix = "arXiv",
    primaryClass = "astro-ph.HE",
    doi = "10.1088/1674-4527/acdfa5",
    journal = "Res. Astron. Astrophys.",
    volume = "23",
    number = "7",
    pages = "075024",
    year = "2023"
}

@article{Cowan:2010js,
  author    = "Cowan, Glen and Cranmer, Kyle and Gross, Eilam and Vitells, Ofer",
  title     = "Asymptotic formulae for likelihood-based tests of new physics",
  journal   = "Eur. Phys. J. C",
  volume    = "71",
  pages     = "1554",
  year      = "2011",
  doi       = "10.1140/epjc/s10052-011-1554-0",
  eprint    = "1007.1727",
  archivePrefix = "arXiv",
  primaryClass = "physics.data-an"
}

@article{pyhf-soft,
  author = "Heinrich, Lukas and Cranmer, Kyle and Feickert, Matthew and Stark, Giordon and others",
  title        = "{pyhf: pure-Python implementation of HistFactory-style statistical models}",
  month        = jul,
  year         = 2024,
  eprint    = "Zenodo, https://doi.org/10.5281/zenodo.1169739",
  url          = {https://doi.org/10.5281/zenodo.1169739}
}

@article{Benincasa:2025tdr,
    author = "Benincasa, Nico and Delle Rose, Luigi and Panizzi, Luca and Razzaq, Maimoona and Urzetta, Savio",
    title = "{Phase transitions and gravitational waves in a non-abelian vector dark matter scenario}",
    eprint = "2506.22248",
    archivePrefix = "arXiv",
    primaryClass = "hep-ph",
    month = "6",
    year = "2025"
}

@article{Baudis:2024darwin,
  author       = {Laura Baudis and DARWIN/XLZD Collaboration},
  title        = {DARWIN/XLZD: a future xenon observatory for dark matter and other rare interactions},
  journal      = {arXiv preprint arXiv:2404.19524},
  year         = {2024},
  eprint       = {2404.19524},
  url          = {https://arxiv.org/abs/2404.19524},
  note         = {DARWIN/XLZD will probe spin-independent WIMP-nucleon cross sections down to the neutrino floor}
}

@article{FCC:2018evy,
  author = {Abada, A. and others},
  title = {FCC-hh: The Hadron Collider},
  journal = {Eur. Phys. J. ST},
  volume = {228},
  number = {4},
  pages = {755-1107},
  year = {2019},
  doi = {10.1140/epjst/e2019-900087-0},
  archivePrefix = {arXiv},
  eprint = {1902.09960}
}

@article{Belyaev:2012qa,
    author = "Belyaev, Alexander and Christensen, Neil D. and Pukhov, Alexander",
    title = "{CalcHEP 3.4 for collider physics within and beyond the Standard Model}",
    eprint = "1207.6082",
    archivePrefix = "arXiv",
    primaryClass = "hep-ph",
    reportNumber = "PITT-PACC-1209",
    doi = "10.1016/j.cpc.2013.01.014",
    journal = "Comput. Phys. Commun.",
    volume = "184",
    pages = "1729--1769",
    year = "2013"
}

@article{Buckley:2014ana,
    author = {Buckley, Andy and Ferrando, James and Lloyd, Stephen and Nordstr\"om, Karl and Page, Ben and R\"ufenacht, Martin and Sch\"onherr, Marek and Watt, Graeme},
    title = "{LHAPDF6: parton density access in the LHC precision era}",
    eprint = "1412.7420",
    archivePrefix = "arXiv",
    primaryClass = "hep-ph",
    reportNumber = "GLAS-PPE-2014-05, MCNET-14-29, IPPP-14-111, DCPT-14-222",
    doi = "10.1140/epjc/s10052-015-3318-8",
    journal = "Eur. Phys. J. C",
    volume = "75",
    pages = "132",
    year = "2015"
}

@article{Fuks:2016ftf,
  author = {Fuks, Benjamin and Shao, Hua-Sheng},
  title = {QCD next-to-leading-order predictions matched to parton showers for vector-like quark models},
  journal = {Eur. Phys. J. C},
  volume = {77},
  number = {2},
  pages = {135},
  year = {2017},
  doi = {10.1140/epjc/s10052-017-4893-9},
  eprint = {1610.04622},
  archivePrefix = {arXiv},
  primaryClass = {hep-ph}
}

@article{Sjostrand:2014zea,
  author = {Torbjörn Sjöstrand and Stefan Ask and Jesper R. Christiansen and Richard Corke and Nishita Desai and Philip Ilten and Stephen Mrenna and Stefan Prestel and Christine O. Rasmussen and Peter Z. Skands},
  title = {{An introduction to PYTHIA 8.2}},
  journal = {Comput. Phys. Commun.},
  volume = {191},
  pages = {159--177},
  year = {2015},
  doi = {10.1016/j.cpc.2015.01.024},
  eprint = {1410.3012},
  archivePrefix = {arXiv},
  primaryClass = {hep-ph}
}

@article{deFavereau:2013fsa,
  author = {de Favereau, J. and Delaere, C. and Demin, P. and Giammanco, A. and Lemaître, V. and Mertens, A. and Selvaggi, M.},
  title = {{DELPHES 3, A modular framework for fast simulation of a generic collider experiment}},
  journal = {JHEP},
  volume = {02},
  pages = {057},
  year = {2014},
  doi = {10.1007/JHEP02(2014)057},
  eprint = {1307.6346},
  archivePrefix = {arXiv},
  primaryClass = {hep-ex}
}

@article{Dercks:2016npn,
  author = {Dercks, Daniel and Dreiner, Herbi K. and Kulkarni, Suchita and Marquard, Philip and Tattersall, Jamie},
  title = {{CheckMATE 2: From the model to the limit}},
  journal = {Comput. Phys. Commun.},
  volume = {221},
  pages = {383--418},
  year = {2017},
  doi = {10.1016/j.cpc.2017.08.021},
  eprint = {1611.09856},
  archivePrefix = {arXiv},
  primaryClass = {hep-ph}
}

@article{CMS:2019zmd,
  title = {Searches for physics beyond the standard model with the $M_{T2}$ variable in hadronic final states with and without disappearing tracks in pp collisions at $\sqrt{s}=13$ TeV},
  author = {CMS Collaboration},
  journal = {Eur. Phys. J. C},
  volume = {80},
  pages = {3},
  year = {2020},
  doi = {10.1140/epjc/s10052-019-7493-x},
  eprint = {1909.03460},
  archivePrefix = {arXiv},
  primaryClass = {hep-ex}
}

@article{ATLAS:2021yyr,
  title = {Search for squarks and gluinos in final states with jets and missing transverse momentum using 139 fb$^{-1}$ of $pp$ collisions at $\sqrt{s}=13$ TeV with the ATLAS detector},
  author = {ATLAS Collaboration},
  journal = {JHEP},
  volume = {02},
  pages = {143},
  year = {2021},
  doi = {10.1007/JHEP02(2021)143},
  eprint = {2010.14293},
  archivePrefix = {arXiv},
  primaryClass = {hep-ex}
}

@article{ATLAS:2019vcq,
  title = {Search for direct top squark pair production in final states with missing transverse momentum and two $b$-jets in $pp$ collisions at $\sqrt{s}=13$ TeV with the ATLAS detector},
  author = {ATLAS Collaboration},
  journal = {JHEP},
  volume = {06},
  pages = {046},
  year = {2020},
  doi = {10.1007/JHEP06(2020)046},
  eprint = {1908.03122},
  archivePrefix = {arXiv},
  primaryClass = {hep-ex}
}

@article{ATLAS:2021twp,
  title = {Search for supersymmetry in events with four or more leptons or two same-sign leptons and jets using 139 fb$^{-1}$ of $pp$ collisions at $\sqrt{s}=13$ TeV with the ATLAS detector},
  author = {ATLAS Collaboration},
  journal = {JHEP},
  volume = {12},
  pages = {142},
  year = {2021},
  doi = {10.1007/JHEP12(2021)142},
  eprint = {2101.01629},
  archivePrefix = {arXiv},
  primaryClass = {hep-ex}
}

@article{CMS:2019xai,
  title = {Search for supersymmetry in final states with multiple top and bottom quarks and missing transverse momentum in proton-proton collisions at $\sqrt{s}=13$ TeV},
  author = {CMS Collaboration},
  journal = {Phys. Rev. D},
  volume = {100},
  number = {1},
  pages = {012001},
  year = {2019},
  doi = {10.1103/PhysRevD.100.012001},
  eprint = {1908.04722},
  archivePrefix = {arXiv},
  primaryClass = {hep-ex}
}

@article{ATLAS:2017msx,
  title = {Search for direct top squark pair production in the all-hadronic $t\bar{t}$ plus missing transverse momentum final state in $pp$ collisions at $\sqrt{s}=13$ TeV with the ATLAS detector},
  author = {ATLAS Collaboration},
  journal = {JHEP},
  volume = {09},
  pages = {088},
  year = {2017},
  doi = {10.1007/JHEP09(2017)088},
  eprint = {1709.04183},
  archivePrefix = {arXiv},
  primaryClass = {hep-ex}
}

@article{ATLAS:2021kty,
  title = {Search for four-top-quark production using the single-lepton and opposite-sign dilepton final states in proton–proton collisions at $\sqrt{s}=13$ TeV with the ATLAS detector},
  author = {ATLAS Collaboration},
  journal = {Phys. Rev. D},
  volume = {104},
  number = {11},
  pages = {112009},
  year = {2021},
  doi = {10.1103/PhysRevD.104.112009},
  eprint = {2106.09609},
  archivePrefix = {arXiv},
  primaryClass = {hep-ex}
}

@article{ATLAS:2020syg,
  title = {Search for new phenomena in final states with large jet multiplicities and missing transverse momentum using $\sqrt{s}=13$ TeV proton–proton collisions recorded by ATLAS in Run 2 of the LHC},
  author = {ATLAS Collaboration},
  journal = {JHEP},
  volume = {10},
  pages = {062},
  year = {2020},
  doi = {10.1007/JHEP10(2020)062},
  eprint = {2004.14060},
  archivePrefix = {arXiv},
  primaryClass = {hep-ex}
}

@article{ATLAS:2024tvlq,
  author = {ATLAS Collaboration},
  title = {Search for pair production of vector-like top partners in final states with one lepton, jets and missing transverse momentum using the full Run 2 dataset collected with the ATLAS detector},
  year = {2024},
  eprint = {2401.17165},
  archivePrefix = {arXiv},
  primaryClass = {hep-ex},
  note = {Submitted to JHEP}
}

@article{Breitbach:2018ddu,
    author = "Breitbach, Moritz and Kopp, Joachim and Madge, Eric and Opferkuch, Toby and Schwaller, Pedro",
    title = "{Dark, Cold, and Noisy: Constraining Secluded Hidden Sectors with Gravitational Waves}",
    eprint = "1811.11175",
    archivePrefix = "arXiv",
    primaryClass = "hep-ph",
    reportNumber = "CERN-TH-2018-255, MITP/18-115",
    doi = "10.1088/1475-7516/2019/07/007",
    journal = "JCAP",
    volume = "07",
    pages = "007",
    year = "2019"
}

@article{Baker:2019ndr,
    author = "Baker, Michael J. and Kopp, Joachim and Long, Andrew J.",
    title = "{Filtered Dark Matter at a First Order Phase Transition}",
    eprint = "1912.02830",
    archivePrefix = "arXiv",
    primaryClass = "hep-ph",
    doi = "10.1103/PhysRevLett.125.151102",
    journal = "Phys. Rev. Lett.",
    volume = "125",
    number = "15",
    pages = "151102",
    year = "2020"
}

@article{Bodeker:2009qy,
    author = "Bodeker, Dietrich and Moore, Guy D.",
    title = "{Can electroweak bubble walls run away?}",
    eprint = "0903.4099",
    archivePrefix = "arXiv",
    primaryClass = "hep-ph",
    doi = "10.1088/1475-7516/2009/05/009",
    journal = "JCAP",
    volume = "05",
    pages = "009",
    year = "2009"
}

\end{document}